\documentclass[eqsecnum, aps, amsmath, amssymb, singlecolumn, superscriptaddress, 11pt ]{revtex4-1}

\usepackage{graphicx}% Include figure files
\usepackage{dcolumn}% Align table columns on decimal point
\usepackage[colorlinks=true,
            linkcolor=blue,
            urlcolor=blue,
            citecolor=blue]{hyperref}
\usepackage{bm}% bold math
\usepackage{enumerate}
\usepackage{lipsum}
\usepackage{threeparttable}
\usepackage{epstopdf}
\usepackage{float}
\usepackage{xcolor,colortbl}
\usepackage[caption = false]{subfig}
\usepackage{tabularx}
\usepackage{chngpage}

\begin{document}

%\preprint{APS/123-QED}

\title{Longitudinal and transversal resonant tunneling  of interacting bosons in a  two-dimensional Josephson junction: Mean-field and many-body dynamics}
\author{Anal Bhowmik}
\email{abhowmik@campus.haifa.ac.il}
\affiliation{Department of Mathematics, University of Haifa, Haifa 3498838, Israel}
\affiliation{Haifa Research Center for Theoretical Physics and Astrophysics, University of Haifa, Haifa 3498838, Israel}
\author{Ofir E. Alon}
\affiliation{Department of Mathematics, University of Haifa, Haifa 3498838, Israel}
 \affiliation{Haifa Research Center for Theoretical Physics and Astrophysics, University of Haifa, Haifa 3498838, Israel}

\date{\today}

%\collaboration{MUSO Collaboration}%\noaffiliation

%\author{Charlie Author}
 %\homepage{http://www.Second.institution.edu/~Charlie.Author}
%\affiliation{
 %Second institution and/or address\\
 %This line break forced% with \\
%}%
%\affiliation{
 %Third institution, the second for Charlie Author
%}%
%\author{Delta Author}
%\affiliation{%
 %Authors' institution and/or address\\
 %This line break forced with \textbackslash\textbackslash
%}%

%\collaboration{CLEO Collaboration}%\noaffiliation

%\date{\today}% It is always \today, today,
             %  but any date may be explicitly specified

\begin{abstract}
We unravel the out-of-equilibrium quantum dynamics of a few interacting bosonic clouds in a two-dimensional asymmetric double-well potential at the resonant tunneling scenario. At the single-particle level of resonant tunneling, particles tunnel under the barrier from, typically, the ground-state in the left well to an excited state in the right well, i.e., states of different shapes and properties are coupled when their one-particle energies coincide. In two spatial dimensions, two types of resonant tunneling processes are possible, to which we refer to as longitudinal and transversal resonant tunneling. Longitudinal resonant tunneling implies that the state in the right well is longitudinally-excited with respect to the state in the left well, whereas transversal resonant tunneling implies that the former is transversely-excited with respect to the latter. We show that interaction between bosons makes resonant tunneling phenomena in two spatial dimensions profoundly rich, and analyze these phenomena in terms of the loss of coherence of the junction and development of fragmentation, and coupling between transverse and longitudinal degrees-of-freedom and excitations. To this end, a detailed analysis of the tunneling dynamics is performed by exploring the time evolution of a few physical quantities, namely, the survival probability, occupation numbers of the reduced one-particle density matrix, and the many-particle position, momentum, and angular-momentum variances. To accurately calculate these physical quantities from the time-dependent many-boson wavefunction, we apply a well-established many-body method, the multiconfigurational time-dependent Hartree for bosons (MCTDHB), which incorporates quantum correlations exhaustively. By comparing the survival probabilities and variances at the mean-field and many-body levels of theory and investigating the development of fragmentation, we identify the detailed mechanisms of many-body longitudinal and transversal resonant tunneling in two dimensional asymmetric double-wells. In particular, we find that the position and momentum variances along the transversal direction are almost negligible at the longitudinal resonant tunneling, whereas they are substantial at the transversal resonant tunneling which is caused by the combination of the density and breathing mode oscillations. We show that the width of the interparticle interaction potential does not affect the qualitative physics of resonant tunneling dynamics, both at the mean-field and many-body levels. In general, we characterize the impact of the transversal and longitudinal degrees-of-freedom in the many-boson tunneling dynamics at the resonant tunneling scenarios.
\end{abstract}

%\pacs{Valid PACS appear here}% PACS, the Physics and Astronomy
                             % Classification Scheme.
%\keywords{Suggested keywords}%Use showkeys class option if keyword
                              %display desired
\maketitle
\section{INTRODUCTION}
 Tunneling is a purely quantum mechanical phenomenon which takes place in  classically-forbidden region,   originally  intended to account for $\alpha$-decay, fusion, and fission in nuclear physics \cite{Razavy2003, Gamow1928}. Apart from nuclear physics, tunneling occurs naturally in photoassociation and photodissociation processes, and  in  solid-state structures \cite{Vatasescu2000, Wagner1993, Glutsch2004,Josephson1962, Davis2002}.  Ultracold  quantum  gasses  have been the subject of research  to  simulate  solid-state  systems \cite{Smerzi1997, Albiezet2005, Gati2007, Morsch2006, Lewenstein2007, Lewenstein2012}. In the context of the study of  tunneling dynamics of ultracold atoms, a double-well potential with static barrier is a standard example. Most of the tunneling phenomena were investigated  with ultracold atoms  in one dimension, either in a double-well potential or in periodic optical lattices, and  focused on the atomic motion in the lowest  band, i.e.,  particularly with the ground state \cite{Grynberg2001, Salgueiro2007,  Liu2015,  Burchinati2017, Sakmann2009, Haldar2018}. Moreover, it was shown in a double-well potential that the inclusion of higher single-particle levels is fundamentally important for correlations  \cite{Dobrzyniecki2016, Dobrzyniecki2018}.  Josephson-like dynamics  of a Bose-Einstein condensate of rubidium atoms was investigated in the second Bloch band of an optical square lattice \cite{Vargas2021}. Unconventional orbital superfluidity in the $P$- and $F$-bands of a bipartite optical square lattice was achieved in \cite{Wirth2011, Olschlager2011}. Recently, we explored the tunneling dynamics of ultracold  bosons in a  symmetric double-well potential where the atomic motion lies in the ground as well as excited  bands \cite{Bhowmik2020}.

The interband quantum tunneling of ultracold atoms  between the ground and excited bands can  occur  by breaking the symmetry of the two wells of a symmetric double-well potential, say, by generating an asymmetric double-well potential.  This asymmetric double-well potential  can be  understood as a unit cell of a tilted optical lattice \cite{Kolovsky2018, Kolovsky2016, Kolovsky2010, Zenesini2009}.  Also, an external electromagnetic field can induce interband transitions as demonstrated in the spectroscopy of Wannier-Stark levels \cite{Wilkinson1996}. Here, we are interested in the interband quantum tunneling in an asymmetric double-well potential, typically describes in the tilted optical lattice.  Per definition, the interband tunneling in an asymmetric double-well potential occurs   when the energy of the ground state on one side of the double-well coincides with the energy of  an excited state on the other side. This  leads to  tunneling between those states which is resonantly enhanced by the energy matching, usually referred to as resonantly enhanced tunneling \cite{ Zenesini2008, Sias2007}. 
 
Resonantly enhanced tunneling phenomenon has been observed in various fields of research, such as in a nonlinear effect in a Mott insulator by creating  a particle-hole excitation \cite{Greiner2002},   in the presence of many-body coherences \cite{Lee2008}, in waveguide arrays \cite{Rosam2003}, in an optical lattice with an external magnetic field producing a Zeeman splitting of the energy levels \cite{Teo2002}, in the two-terminal current-voltage characteristics of a finite superlattice \cite{Tsu1973},  in  solid states systems such as superlattice \cite{Leo2003, Glutsch2004}, and in   accelerated  optical  lattice  potentials \cite{Kolovsky2016, Kolovsky2018}.

Hitherto, resonantly enhanced tunneling is studied when the resonant condition of energy matching of the two wells is along the direction of the barrier \cite{ Zenesini2008, Sias2007, Haldar2019}.   We name it longitudinal resonant tunneling. In  Ref. \cite{ Sias2007}, the condensates were prepared in a one-dimensional  optical lattice with  an additional Stark force and determined the tunneling rate by the Landau-Zener formula \cite{Landau1932a, Landau1932b, Zener1932, Morsch2001, Bharucha1997}. Zenesini \textit{et al. } experimentally investigated  the impact of atom-atom interactions on the resonantly enhanced tunneling process and presented a complementary  theoretical description of Landau-Zener tunneling for ultracold atoms in periodic potentials \cite{Zenesini2008}.  The out-of-equilibrium quantum mean-field and many-body dynamics of  interacting bosons  were recently explored in a one-dimensional  asymmetric double-well potential \cite{Haldar2019} using the multi-configurational time-dependent Hartree  for bosons (MCTDHB) method \cite{Streltsov2007, Alon2008, Lode2020}.  Precisely, the interacting bosons  were prepared in the ground state of the left well (harmonic potential) and the mechanism of the tunneling process in the resulting double-well was studied by analyzing the time-evolution  of the survival probability, depletion and fragmentation, and the many-particle position and momentum variances \cite{Haldar2019}.

Resonant tunneling in a two-dimensional (2D) double-well setup brings interesting new questions, especially on the role of transverse excitations. Also, in 2D there can be a resonant tunneling phenomenon which does not exist in one-dimension, namely,  transversal resonant tunneling when the energy-matching condition of the two wells of an asymmetric double-well is satisfied along the transverse direction of the barrier.   In this work, we focus on the longitudinal and transversal resonant tunneling for the initial bosonic structures of the ground and excited states in a 2D  asymmetric double-well potential. Here the bosons are loaded in the left well of the double-well potential and the barrier is formed along the $x$-direction. In order to investigate the tunneling dynamics in the longitudinal resonant scenario, we consider two different  structures of bosonic clouds, i.e., the ground and transversely excited ($y$-excited) states. Although, the ground state has a one-dimensional analog,  to create a transversely excited state, one requires at least a 2D geometry \cite{Bhowmik2020}. Therefore, to investigate the effect of the longitudinal resonant tunneling on the very basic state which includes transverse excitations is of fundamental interest. Moreover,  we investigate the impact of the transverse direction on the  tunneling dynamics of the ground state at the longitudinal resonant tunneling condition.    In the transversal resonant scenario, the bosonic structures are assumed to be the ground and longitudinally excited ($x$-excited) states. These two states have one-dimensional analogs but  the transversal resonant tunneling does exist only in  a 2D geometry. Therefore, the role of longitudinal excitations in transversal resonant tunneling can be examined. In order to accurately explore the out-of-equilibrium tunneling dynamics for both  resonant tunneling scenarios, we solve the underlying time-dependent many-boson Schr\"odinger equation using the  MCTDHB method \cite{ Alon2008, Lode2020}. We solve the quantum dynamics of all the bosonic clouds at the mean-field and many-body levels. 

Tunneling dynamics for all the bosonic structures are analyzed by  the time evolution of various physical quantities, namely, the survival probability, loss of coherence, depletion and fragmentation, and the many-particle position, momentum, and angular-momentum  variances.  Even when the bosons are fully condensed, the variance is a sensitive probe of  correlations \cite{Klaiman2015}. Therefore, to display the effects of the quantum correlations on the variances of different operators, we  compare the mean-field and many-body variances  in addition to the corresponding  comparison of the survival probabilities.  We notice that the   rate of growth of the quantum correlations depends on the shape of the bosonic clouds, presence of transverse excitations in the system, and the resonant tunneling condition. The interconnection between the density oscillations and the variances are discussed both at the longitudinal and transversal resonant tunneling scenarios.  We show that correlations have different consequences on the various quantities discussed in this work. In general, we ask how the transverse degrees-of-freedom, perpendicular to the junction, can influence the  time evolution of various physical quantities in a 2D asymmetric double-well potential at the resonant tunneling scenario. We find that the  time evolutions of the variances    behave completely differently in the longitudinal and transversal resonant tunneling conditions.

\section{Theoretical framework}
The dynamics of $N$ interacting  bosons in a two-dimensional trap can be described by the time-dependent many-body Shr\"{o}dinger equation, 
\begin{equation}\label{1}
\hat{H}\Psi = i\dfrac{\partial \Psi}{\partial t}, \hspace{1cm} \hat{H}(\textbf{r}_1. \textbf{r}_2, . . . , \textbf{r}_N)=\sum_{j=1}^{N} [\hat{T}(\textbf{r}_j)+\hat{V}(\textbf{r}_j)]+\sum_{j<k} \hat{W}(\textbf{r}_j-\textbf{r}_k),
\end{equation}
 where $\hat{T}(\textbf{r})$ and $\hat{V}(\textbf{r})$ represent the kinetic  and  potential energy terms, respectively. The interaction between the bosons is repulsive and considered as  a Gaussian function  \cite{Doganov2013} $W(\textbf{r}_1-\textbf{r}_1)=\lambda_0\dfrac{e^{-(\textbf{r}_1-\textbf{r}_2)^2/2\sigma^2}}{2\pi\sigma^2}$ with $\sigma=0.25\sqrt{\pi}$. In Appendix B, we investigate   the role of the width of the interparticle interaction potential and demonstrate explicitly that the mean-field and many-body physics found in this work do qualitatively not depend on this width. $\lambda_0$ is  the  interaction strength and $\Lambda$ the interaction parameter.  $\Lambda=\lambda_0(N-1)$ where $N$ is the number of bosons. Note that the model inter-bosons  interaction does not have  any qualitative impact on the physics described here.

In order to solve Eq.~\ref{1} in-principle numerically exactly, we use  the multi-configurational time-dependent Hartree  for bosons (MCTDHB) method \cite{Streltsov2007, Alon2008,  Sakmann2009, Grond2009, Grond2011, Streltsova2014, Klaiman2014, Fischer2015, Tsatsos2015, Weiner2017, Beinke2015, Schurer2015, Lode2016, Lode2017, Weiner2017, Lode2018, Klaiman2018, Haldar2018, Alon2018, Chatterjee2019, Haldar2019, Alon2019a, Alon2019b,Sakmann2010,Sakmann2014,Klaiman2015}. MCTDHB incorporates a variational optimal ansatz which is generated by distributing the $N$ bosons over $M$ time-dependent orbitals. The  MCTDHB wavefunction is described as \cite{Alon2008}
\begin{equation}\label{2}
|\Psi(t)\rangle =\sum_{\{\textbf{n}\}}C_\textbf{n}(t)|\textbf{n};t\rangle,
\end{equation}
where $C_n(t)$ is the expansion coefficients and $|\textbf{n};t\rangle= |n_1, n_2, . . ., n_M;t\rangle$. The number of time-dependent permanents $|\textbf{n};t\rangle$ is $\big(\begin{smallmatrix} N+M-1\\ N \end{smallmatrix}\big)$.  $M= 1$ in the set of permanents,  $|\textbf{n};t\rangle$, of Eq.~\ref{2} represents the Gross-Pitaevskii ansatz and solves the time-dependent Gross-Pitaevskii equation \cite{Dalfovo1999}. The accuracy of the wavefunction increases with $M$ and achieves   the convergence of different physical quantities of  interest discussed in this work, such as the survival probability, depletion and fragmentation, and the many-particle position, momentum, and angular-momentum variances. The numerical implementation of the real-time dynamics  employed in this work  can be found in \cite{Streltsov2015, Streltsov2019}.

\begin{figure*}[!h]
\centering
{\includegraphics[trim = 2.5cm 0.5cm 2.1cm 0.2cm, scale=.60]{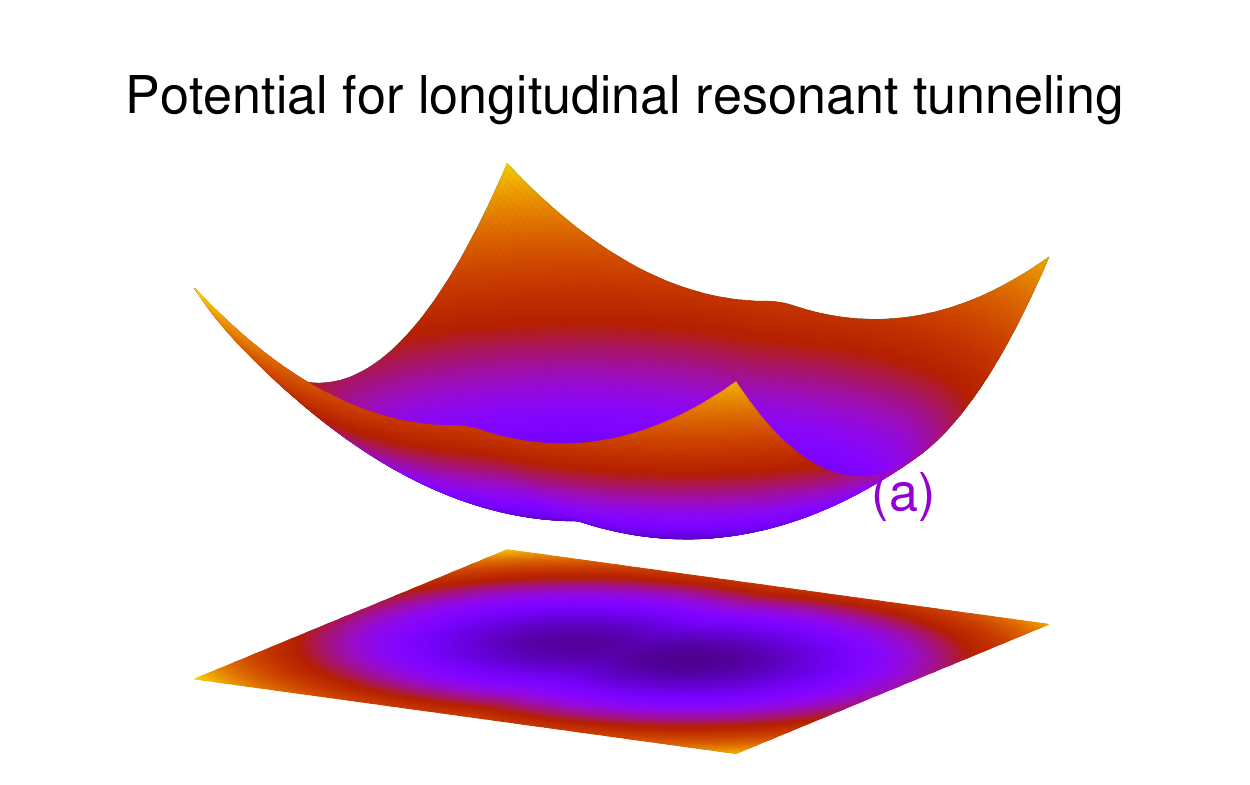}}
\hspace{1.5cm}
{\includegraphics[trim =  2.5cm 0.5cm 2.1cm 0.2cm, scale=.60]{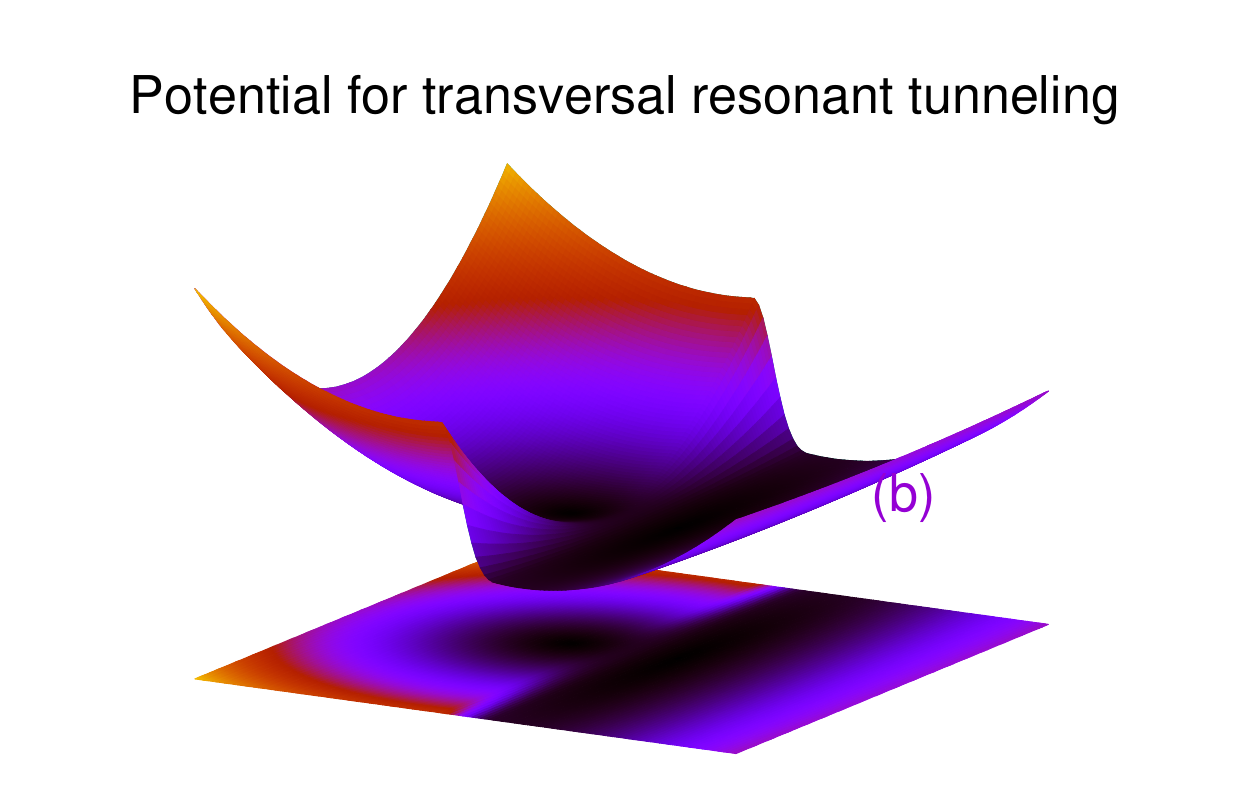}}\\
\vglue 0.25 truecm 
{\includegraphics[trim = 2.5cm 0.5cm 2.1cm 0.2cm, scale=.60]{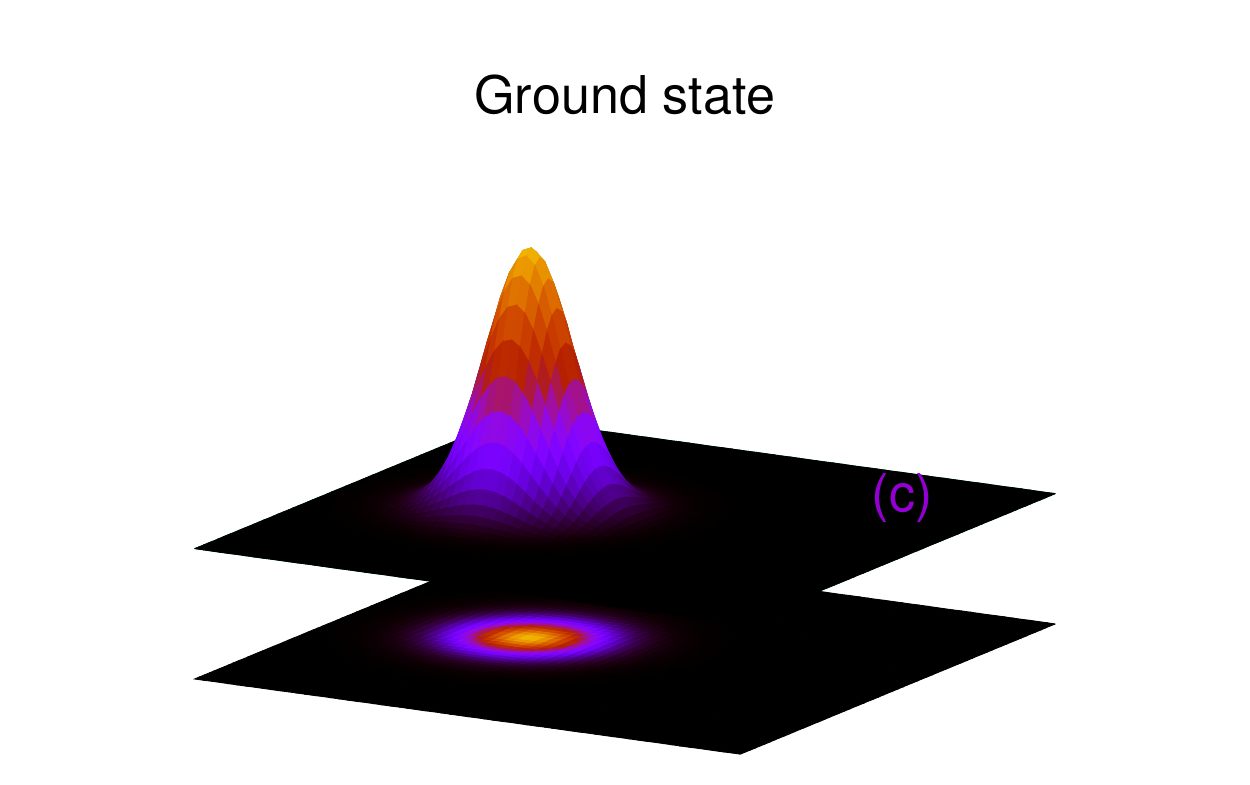}}
{\includegraphics[trim =  2.5cm 0.5cm 2.1cm 0.2cm, scale=.60]{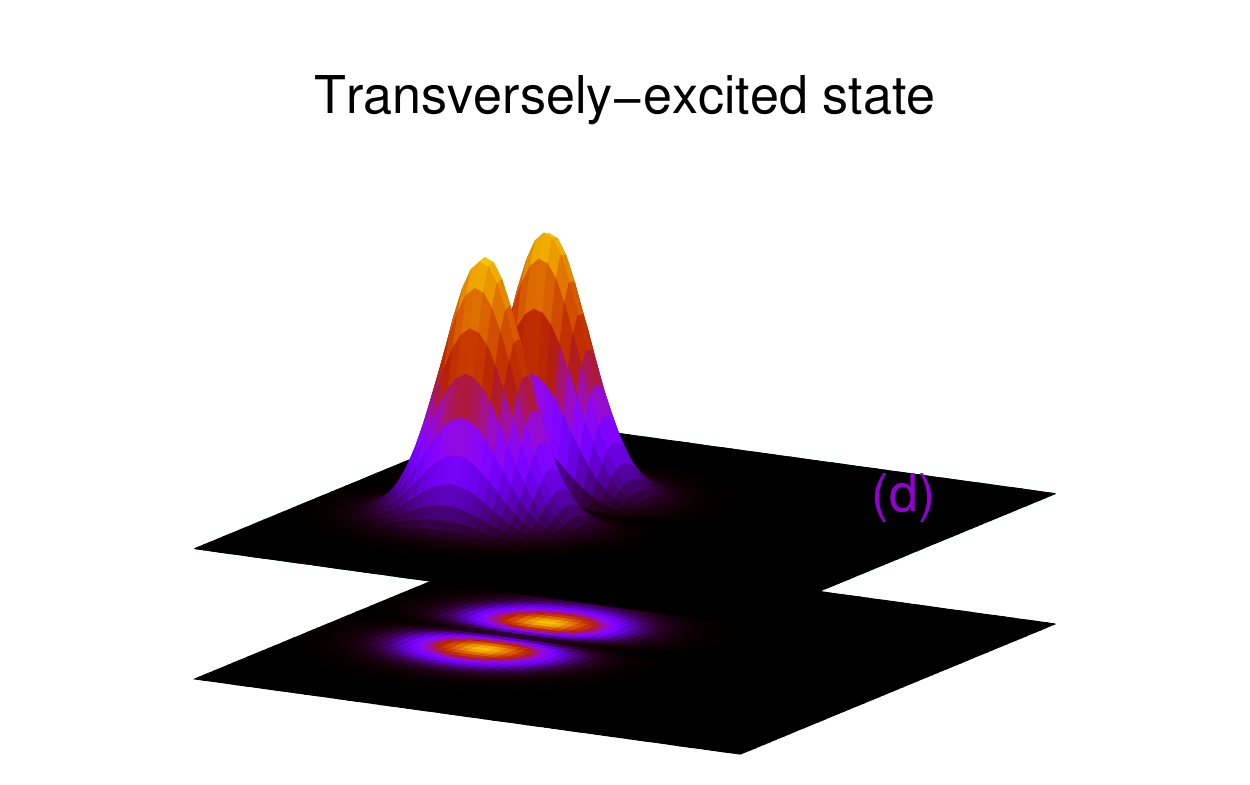}}
{\includegraphics[trim =  2.5cm 0.5cm 2.1cm 0.2cm, scale=.60]{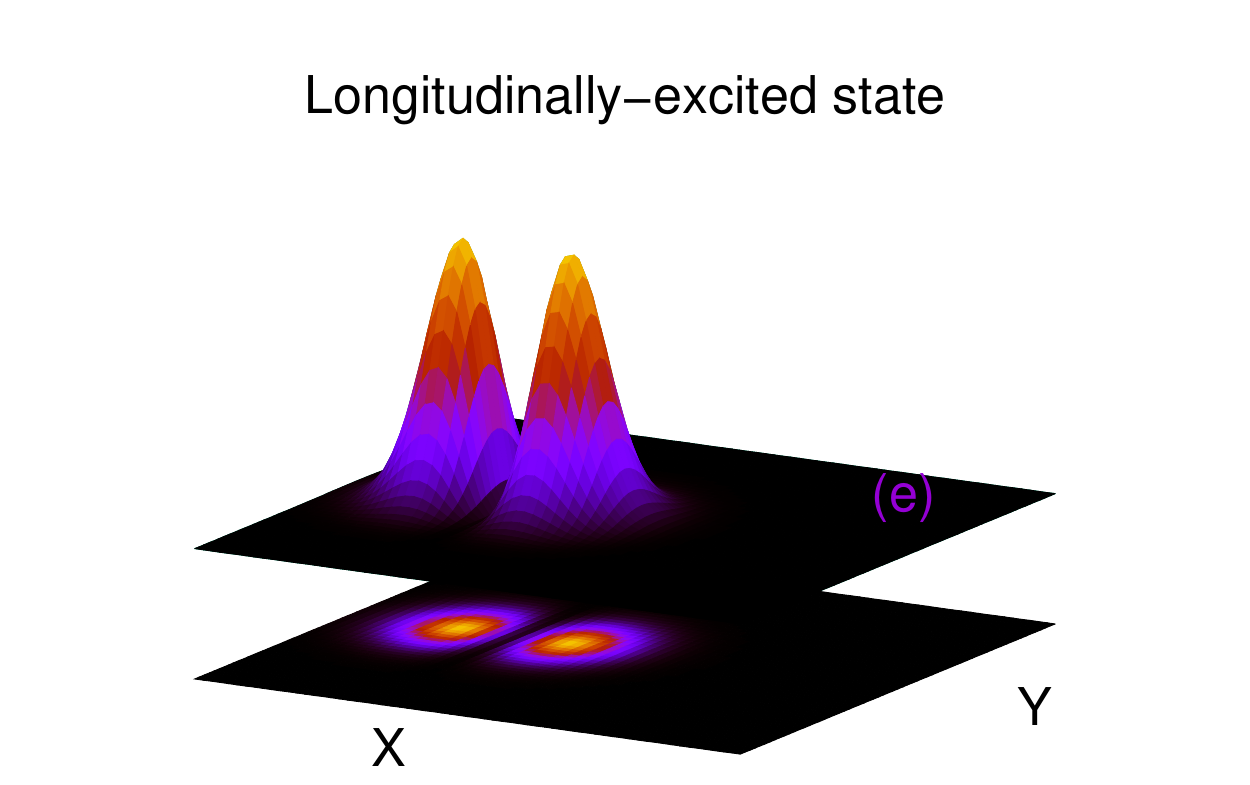}}\\
\caption{Schematic diagrams for  the asymmetric 2D  double-well potential for (a) longitudinal resonant tunneling, described in Eq.~\ref{3}  and (b) transversal resonant tunneling, described in Eq.~\ref{5}. Panels (c), (d), and (e) show the  initial density distributions for the ground, transversely-excited, and longitudinally-excited states, respectively. Panel (e) shows the $x$- and $y$-axes.}
\label{Fig1}
\end{figure*}

Here, we investigate the dynamics of a few bosonic clouds prepared as either the ground, transversely-excited or longitudinally-excited state for longitudinal and transversal resonant tunneling phenomena in two different double-well potentials. In case of longitudinal resonant tunneling, we consider that the bosons are initially prepared   in the left well,
\begin{equation}\label{2.1}
 V_L(x,y)=\dfrac{1}{2}(x+2)^2+\dfrac{1}{2}y^2-cx,
\end{equation}
 of a double-well potential, where $c$ is the asymmetry parameter. The bosons are taken in the state of  either as the non-interacting ground, $\Psi_G=\dfrac{1}{\sqrt{\pi}}F(x,y)$, or as the  transversely-excited, $\Psi_Y=\sqrt{\dfrac{2}{\pi}}y F(x,y)$, states, where $F(x,y)=exp[-\{(x+2)^2+y^2\}/2]$. Fig.~\ref{Fig1} shows the  initial density distributions of $\Psi_G$ and $\Psi_Y$. To explore the dynamics of the bosonic clouds, we suddenly quench the inter-particle interaction at $t=0$ from $\Lambda=0$ to $\Lambda=0.01\pi$ and simultaneously change  the trapping potential  from  $V_L(x,y)$ to the longitudinally-asymmetric 2D  double-well potential, $V_T(x,y)$.  Here the asymmetry implies the right well is lower than the left well. This is achieved by adding a linear term in the longitudinal direction. The mathematical form  of $V_T(x,y)$ is (see Fig.~\ref{Fig1}(a)):

\begin{equation}\label{3}
V_T(x,y)=
\begin{cases}
\dfrac{1}{2}(x+2)^2+\dfrac{1}{2}y^2-cx, \hspace{0.5cm}  x<-\dfrac{1}{2},  -\infty< y<\infty,  \\
     \dfrac{3}{2}(1-x^2)+\dfrac{1}{2}y^2-cx, \hspace{0.6cm} |x|\leq \dfrac{1}{2}, -\infty< y<\infty,  \\
      \dfrac{1}{2}(x-2)^2+\dfrac{1}{2}y^2-cx, \hspace{0.7cm}  x>+\dfrac{1}{2}  -\infty< y<\infty.
 \end{cases}      
\end{equation}

In order  to investigate the transversal resonant tunneling phenomenon, we assume the bosons are prepared either as the non-interacting $\Psi_G$  or longitudinally-excited state, $\Psi_X=\sqrt{\dfrac{2}{\pi}}(x+2)F(x,y)$, in the left well, $V_L^\prime(x,y)$,  of  a transversely-asymmetric 2D double-well potential, $V_T^\prime(x,y)$.  Here the asymmetry means the right well is wider than the left well. This is generated by making the transverse frequency of the trap spatially-dependent. $V_L^\prime(x,y)$ is functioned as

  \begin{equation}\label{4}
V_L^\prime(x,y)=
\begin{cases}
\dfrac{1}{2}(x+2)^2+\dfrac{1}{2}y^2, \hspace{3.6cm}  x<-1,  -\infty< y<\infty,  \\
     \dfrac{1}{2}(x+2)^2+\dfrac{1}{2}[S(x)]^2y^2, \hspace{1.5cm}  -1< x<+1,  -\infty< y<\infty,  \\        
      \dfrac{1}{2}(x+2)^2+\dfrac{1}{2}\omega_n^2y^2, \hspace{3.3cm}  x>+1,  -\infty< y<\infty.
  \end{cases}
\end{equation}
 Here, $S(x)=\left[1+(\omega_n-1)sin^2\dfrac{(x+1)\pi}{4}\right]$ is a switching function for the transversal frequency,  where $\omega_n$ is the minimal value of the transversal frequency. The initial density distribution of $\Psi_X$ is presented in  Fig.~\ref{Fig1}.  For investigating the time evolution, as described in the longitudinal resonant tunneling, we quench the  inter-particle interaction at $t=0$ from $\Lambda=0$ to $\Lambda=0.01\pi$. At the same time, we convert $V_L^\prime(x,y)$ to the transversely-asymmetric 2D double-well potential $V_T^\prime(x,y)$ (see Fig.~\ref{Fig1}(b)), where 

\begin{equation}\label{5}
V_T^\prime(x,y)=
\begin{cases}
\dfrac{1}{2}(x+2)^2+\dfrac{1}{2}y^2, \hspace{2.7cm}  x<-1,  -\infty< y<\infty, \\
     \dfrac{1}{2}(x+2)^2+\dfrac{1}{2}[S(x)]^2y^2, \hspace{0.5cm}  -1< x<-\dfrac{1}{2},  -\infty< y<\infty,  \\
     \dfrac{3}{2}(1-x^2)+\dfrac{1}{2}[S(x)]^2y^2, \hspace{1.6cm} |x|\leq \dfrac{1}{2}, -\infty< y<\infty, \\
     \dfrac{1}{2}(x-2)^2+\dfrac{1}{2}[S(x)]^2y^2, \hspace{1.1cm}  \dfrac{1}{2}< x<1,  -\infty< y<\infty,  \\      
      \dfrac{1}{2}(x-2)^2+\dfrac{1}{2}\omega_n^2y^2, \hspace{2.35cm}  x>+1,  -\infty< y<\infty
 \end{cases}     
\end{equation}

To accurately tackle the many-body physics involved in the two-dimensional bosonic Josephson junctions, we have performed  the MCTDHB computations  with different time-dependent orbitals for different initial bosonic structures. For $\Psi_G$ and $\Psi_X$, the many-body computations are performed with $M=6$ time-dependent orbitals, while for $\Psi_Y$ with $M=10$ time-dependent orbitals. The convergence with the orbital numbers of the quantities discussed in this work are shown in the supplemental materials. For the numerical solution of the longitudinal resonant tunnelling, the considered box size is  $[-10, 10)\times [-10, 10)$ with periodic boundary conditions having  a grid of $64\times 64$ points and for the transversal resonant tunneling, we increase the grid density to $128\times 128$.  Convergence of the results with respect to the number of grid points has been checked and presented in the supplemental materials. Adequacy of the 2D box size for both longitudinal and transversal scenarios have been checked. For the many-body dynamics of both the resonant tunnelings, the number of bosons and interaction parameter are chosen as $N=10$ and $\Lambda=0.01\pi$. The mean-field dynamics are also computed with  same interaction parameter $\Lambda$ to relate the results with the respective many-body ones. It is noted that as the considered interparticle interaction  in the dynamics is weak, the  preparation of the initial state in the relaxation process either with noninteracting  bosons or or with $\Lambda=0.01\pi$ does not affect the dynamical behavior of the properties discussed in this work.      The  consistency  of  preparation  of  the  initial state  is  documented  in  the  supplemental materials of \cite{Bhowmik2020}. The natural units $\hbar=m=1$ are employed throughout this work.

\section{Results and discussions}
Here, we divide the section into two parts, namely, longitudinal and transversal resonant tunneling. In each part, we investigate  the time evolution of various physical quantities, such as the survival probability, loss of coherence, and the variances of the position, momentum, and angular-momentum many-particle operators when the symmetry of the double-well potential is  broken and it reaches to the resonant tunneling condition. The quantities discussed here incorporate a detailed information of the time-dependent many-boson wave function, explicitly,  the density, reduced one-particle density matrix, and reduced two-particle density matrix.

\subsection{Longitudinal resonant tunneling}
At first we investigate how the longitudinal resonant tunneling phenomena of  $\Psi_G$ and $\Psi_Y$ are achieved by gradually increasing the asymmetry  parameter, $c$,  in the longitudinally-asymmetric 2D double-well potential described in Eq.~\ref{3}. Further, we discuss the time-evolution of the different physical quantities, mentioned above, for $\Psi_G$ and $\Psi_Y$  at the resonant tunneling condition and compare them with the corresponding results of symmetric 2D double-well potential, which can be obtained at $c=0$ from Eq.~\ref{3}. The investigation is performed and compared at the mean-field and many-body levels of theory.

In this work, $\Psi_G$ and $\Psi_Y$ are prepared in the left well of a  2D double-well potential.  When we gradually increase the value of $c$ starting from zero, the symmetry of the double-well potential breaks and the left well becomes the upper well and, consequently, the right well becomes the lower well. For some special values of $c$, the one-body  energy of the left well coincides with one of the higher one-body  energy levels of the right well, resulting  in an enhanced tunneling of bosons from the left well to the right well. This enhancement of tunneling is  usually referred to  as the resonant  tunneling in a double-well potential, see in this context  \cite{Zenesini2008, Sias2007}. From  Eq.~\ref{3}, we can find out that at any asymmetry parameter $c$,  the energy difference of one-body spectrum  between the two wells becomes 4$c$. Analytically, for harmonic left and right wells, one can realize that the resonant tunneling occurs when $4c$ will be equal to an integer.

\begin{figure*}[!h]
{\includegraphics[trim = 0.1cm 0.5cm 0.1cm 0.2cm, scale=.70]{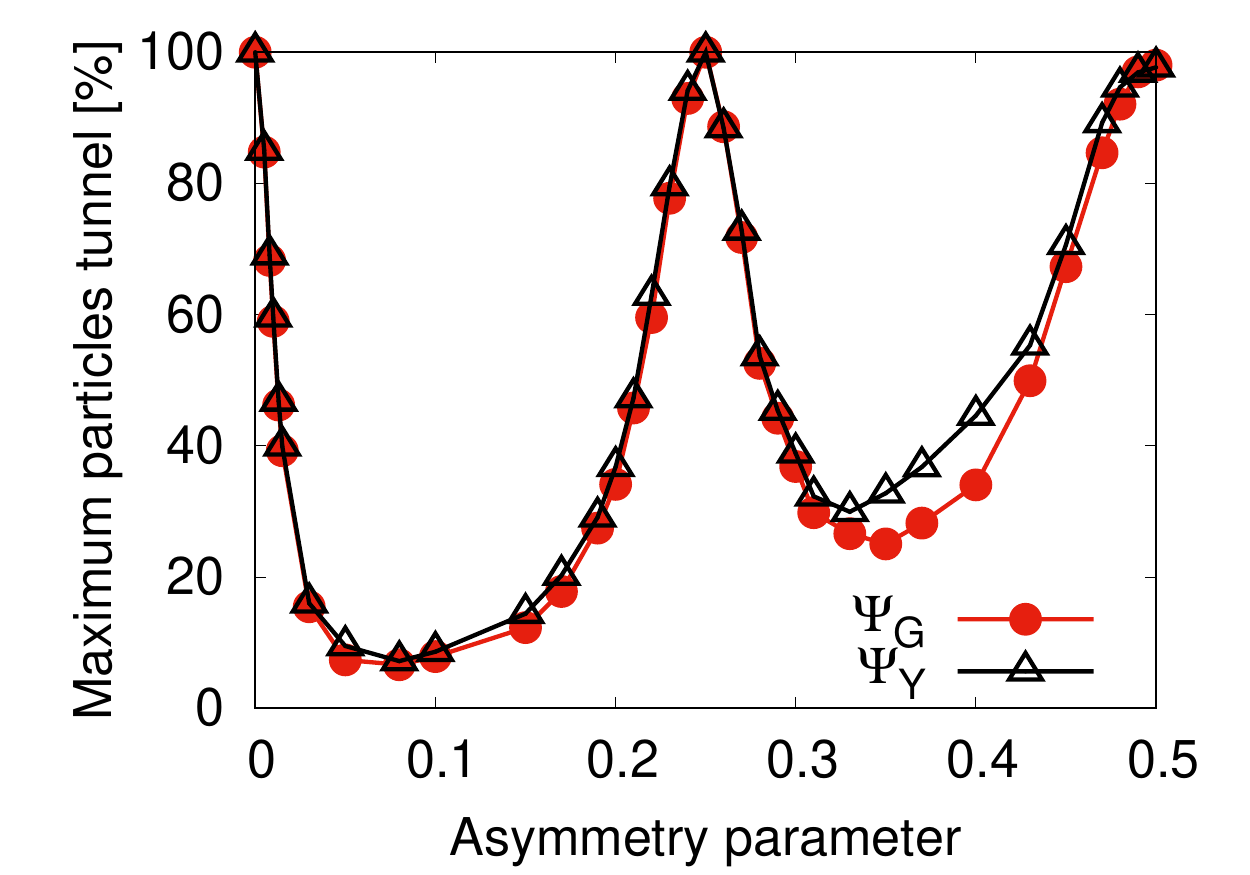}}\\
\caption{Variation of the maximal number of particles tunneling [in \%] from the left to the right well with the asymmetry parameter $c$ for the states $\Psi_G$ and $\Psi_Y$ at the mean-field level. The data is generated from the time-dependent solution of the Gross-Pitaeveskii equation. We show here dimensionless quantities.}
\label{Fig2}
\end{figure*}

Now, we show how the gradual increase of asymmetry between the two wells of a double-well potential influences  the tunneling of particles. In Fig.~\ref{Fig2}, we present, at the mean-field level, the variation of the maximal number of particles tunneling from the left to the right well   for the states $\Psi_G$ and $\Psi_Y$ with the asymmetry parameter $c$. This maximal number of particles in the right well is determined from  $\int\limits_{x=0}^{+\infty}\int\limits_{y=-\infty}^{+\infty}dx dy \dfrac{\rho(x,y;t)}{N}$, where  $\rho(x,y;t)$ is the density of the bosonic cloud.  Here, we start from the symmetric double-well potential. At $c=0$, when the bosons are allowed to evolve in time, we observe that 100\% of the bosons  can tunnel from the left to right well, signifying the delocalization of the one-particle eigen-functions in a symmetric double-well. Now, a small increase of asymmetry between the wells makes the one-particle state to be only partially delocalized over both wells,  leading to a suppression of tunneling of particles. Further increase of the asymmetry, beyond $c=0.1$, the one-particle state slowly begins to be delocalized in one of the wells, and the tunneling of particles increases. At $c=0.25$, we find well-delocalization  of the one-particle state in one of the well which leads to a complete tunneling of particles or specifically, refer to as resonant tunneling condition, when one-body  energy of the left well matches with the energy of one of the one-body excited states of the right well. Similarly, the second resonant tunneling appears at $c=0.5$. As the system is weakly interacting,  we find that the maximum tunneling of particles is $100\%$ both for $\Psi_G$ and $\Psi_Y$ at $c=0$, $c=0.25$, and $c=0.5$. Also, we observe a small difference of maximum tunneling of particles between the  results of $\Psi_G$ and $\Psi_Y$  at off-resonant tunneling. Here, it is noted that if the bosons are prepared in the right well (lower well), the gradual increase of asymmetry between the wells reduces the maximal number of bosons tunneling   to the left well. Further increase of asymmetry between the wells, at $c\gtrsim 0.1$, the bosons become trapped   in the right well, and produces only a mild breathing motion,  mainly due to the quenching the interaction. 

Next, we focus on   both the resonant tunneling values of $c$ and study how the many-body correlations affects  the dynamical behaviors of  the survival probability,  loss of coherence, and the many-particle position, momentum, and angular-momentum variances. Moreover, we will display the corresponding analysis of the results obtained for the symmetric double-well potential to  serve as a reference. In addition, we will present a comparison study of the mean-field and many-body results  both for $\Psi_G$ and $\Psi_Y$.

\begin{figure*}[!h]
{\includegraphics[trim = 0.1cm 0.5cm 0.1cm 0.2cm, scale=.60]{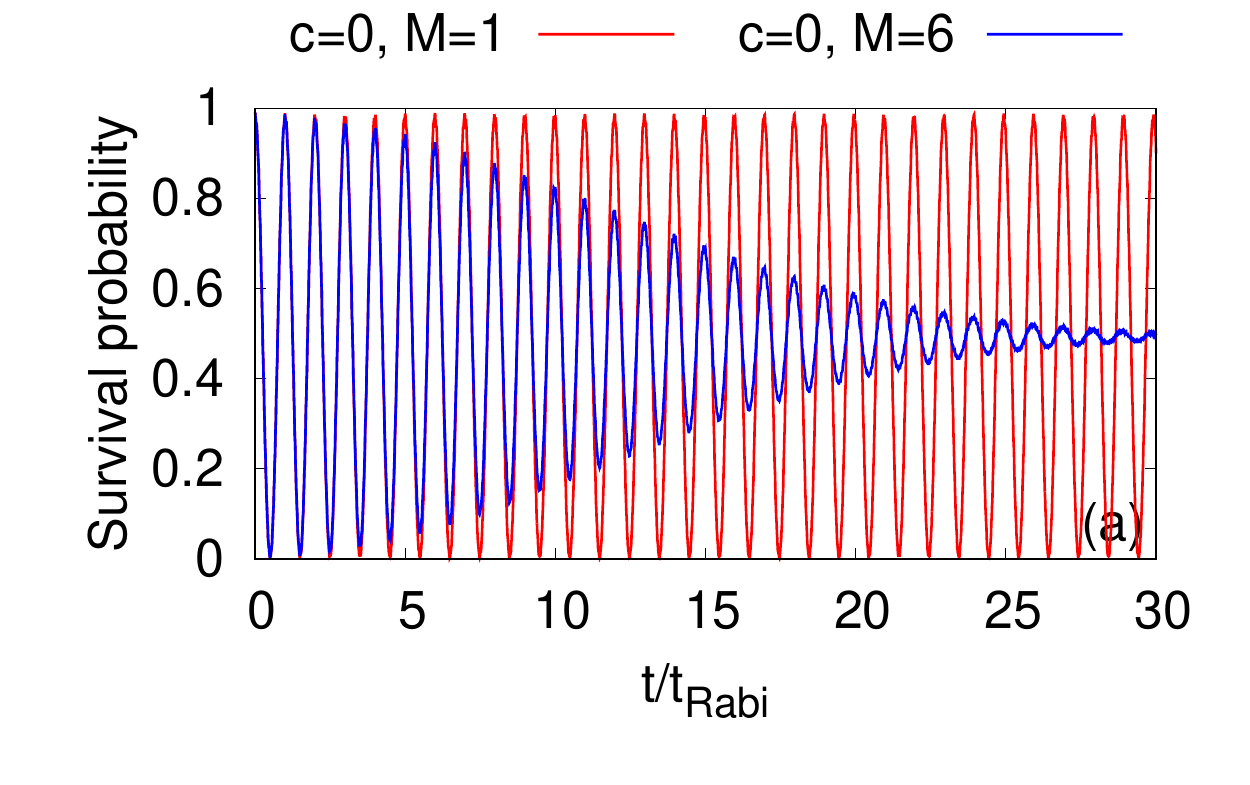}}
{\includegraphics[trim = 0.1cm 0.5cm 0.1cm 0.2cm, scale=.60]{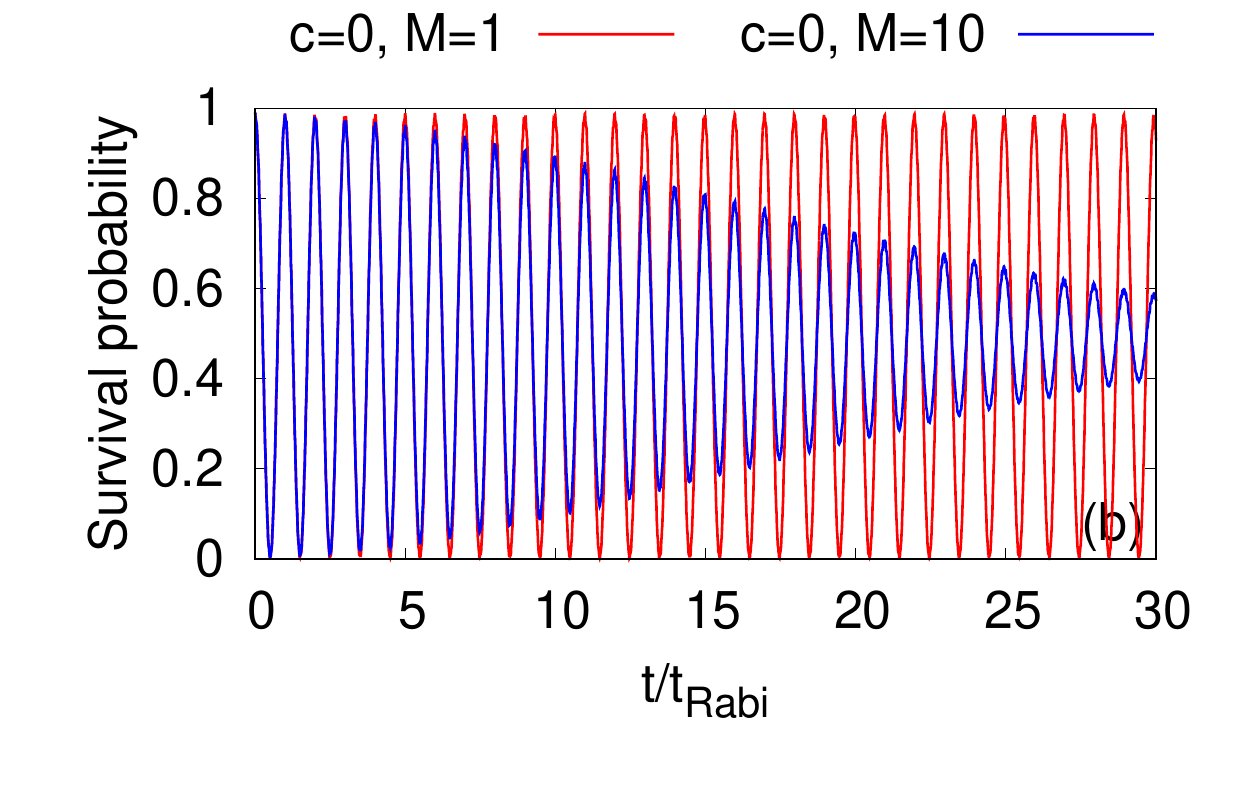}}\\
{\includegraphics[trim = 0.1cm 0.5cm 0.1cm 0.2cm, scale=.60]{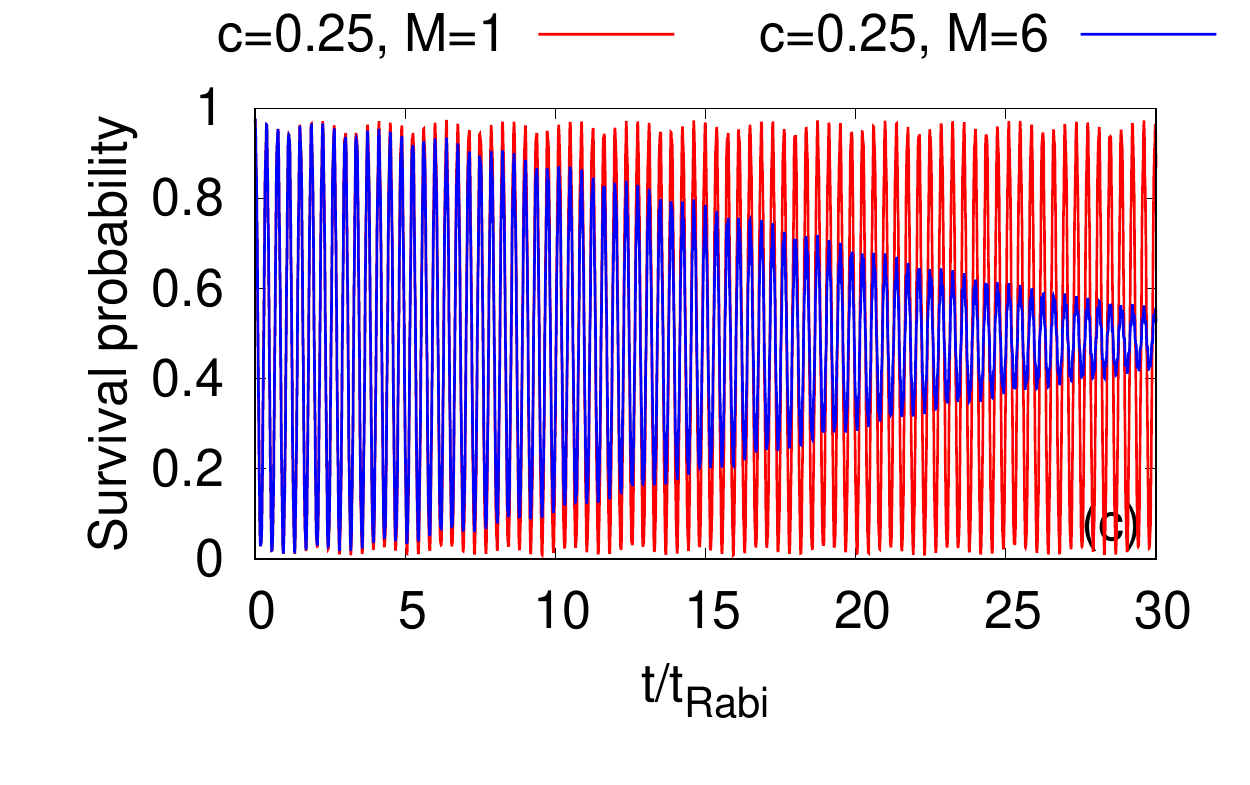}}
{\includegraphics[trim = 0.1cm 0.5cm 0.1cm 0.2cm, scale=.60]{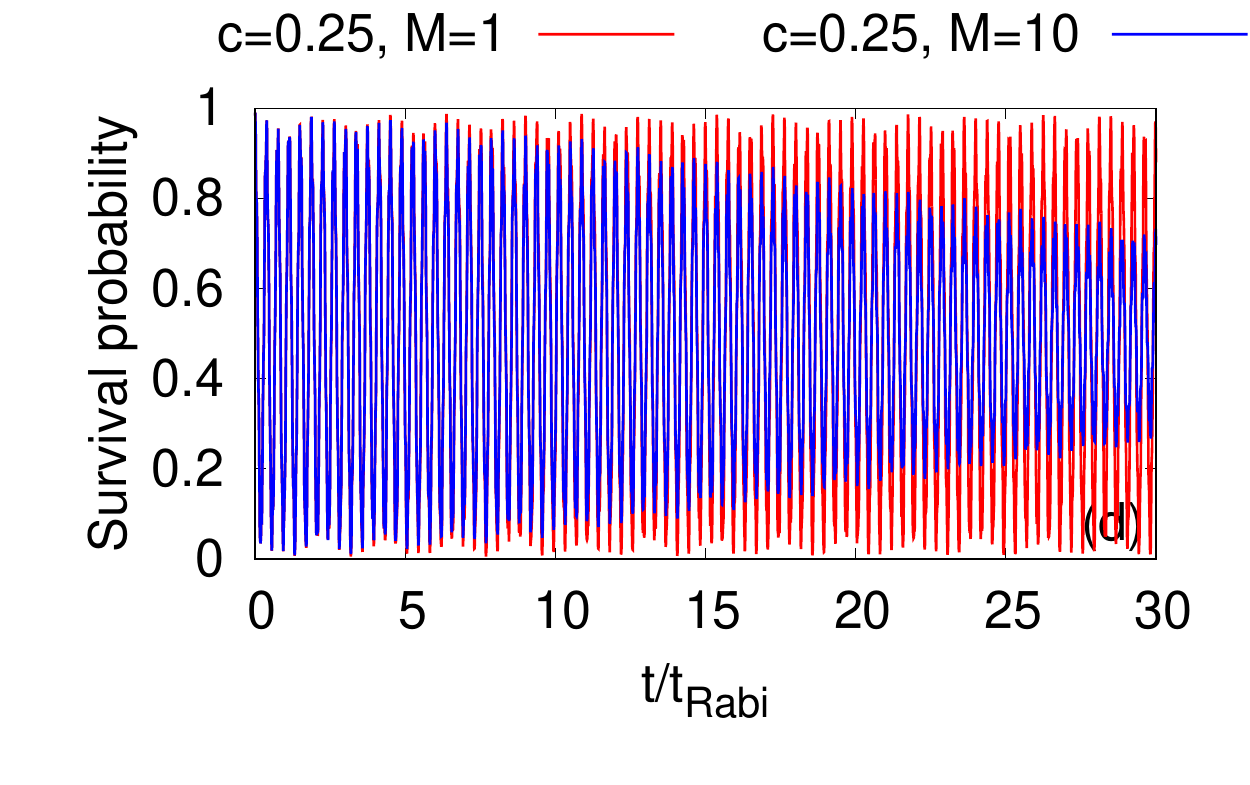}}\\
{\includegraphics[trim = 0.1cm 0.5cm 0.1cm 0.2cm, scale=.60]{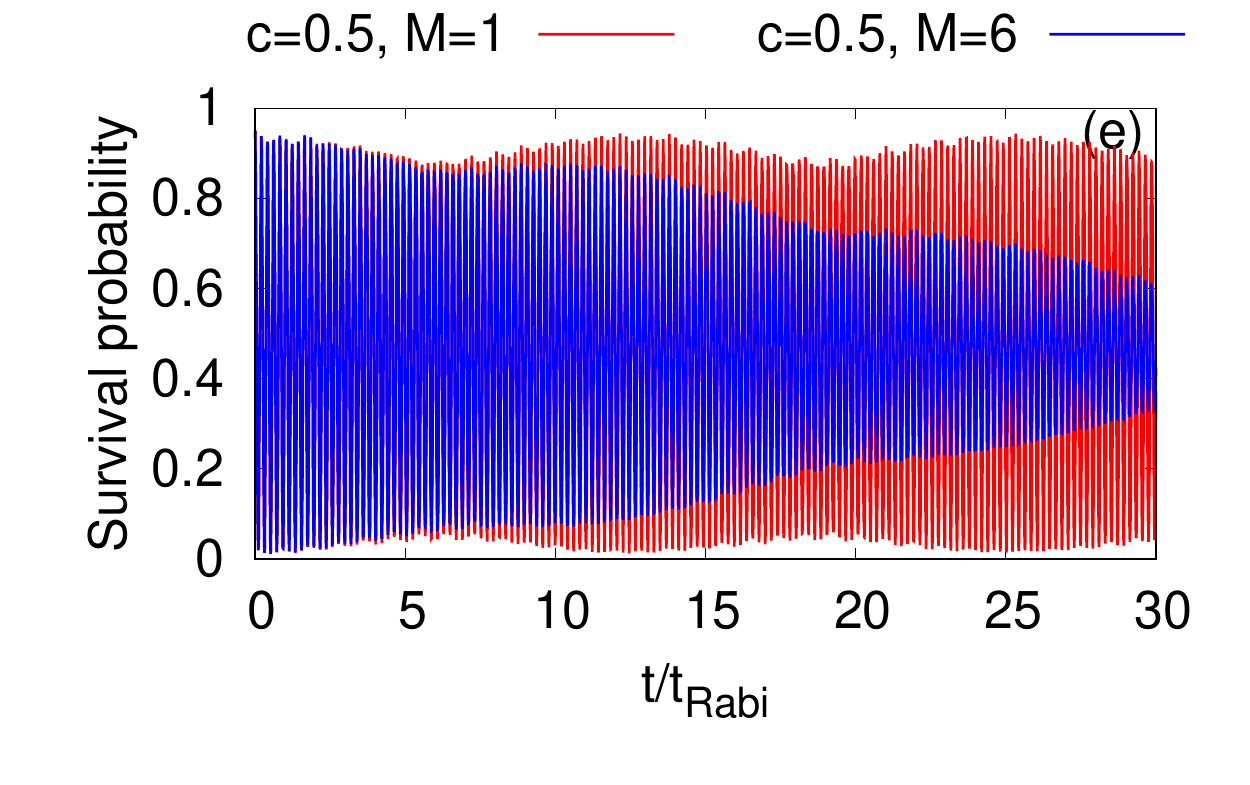}}
{\includegraphics[trim = 0.1cm 0.5cm 0.1cm 0.2cm, scale=.60]{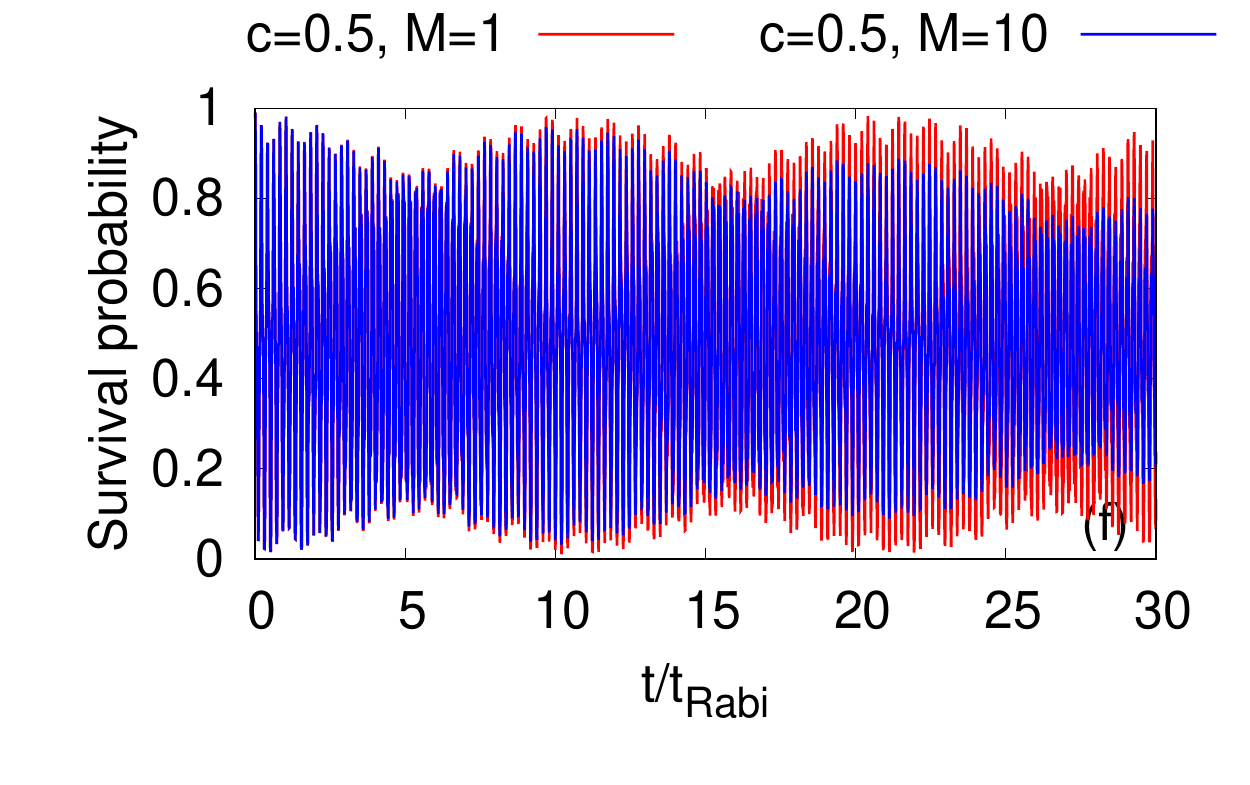}}\\
\caption{Dynamics of the survival probability of $\Psi_G$ (left column) and $\Psi_Y$ (right column) in the left well of longitudinally-asymmetric 2D double-well potential of $N=10$  bosons with the interaction parameter $\Lambda=0.01\pi$. The first row presents the results for the symmetric double-well potential.  Second and third rows  show the  survival probabilities at  the resonant tunneling with  asymmetry parameter, $c=0.25$ and 0.5, respectively.  $M=1$ signifies the mean-field results.   The many-body dynamics are computed with $M= 6$ time-dependent orbitals for $\Psi_G$   and $M= 10$ time-dependent orbitals for $\Psi_Y$.    We show here dimensionless quantities. Color codes are explained in each panel. }
\label{Fig3}
\end{figure*}

In order to adequately capture the time evolutions  of  $\Psi_G$ and $\Psi_Y$ between the two wells of a double-well potential, we examine   the survival probability in the left well, \break\hfill $P_L(t)=\int\limits_{x=-\infty}^{0}\int\limits_{y=-\infty}^{+\infty}dx dy \dfrac{\rho(x,y;t)}{N}$, where  $\rho(x,y;t)$ is the density of the bosonic cloud.  We compare between the results of the mean-field and many-body dynamics of $P_L(t)$ in Fig.~\ref{Fig3} for the asymmetry parameters, $c=0$, 0.25, and 0.5.  Here, we want to mention that, to have a proper comparison among all the results  presented in this paper, the time-scale for the dynamics  is set to be equal to the Rabi oscillations $(t_{Rabi})$ of the symmetric double-well trap (when $c=0$). We find  that $t_{Rabi} = 132.498$ for the symmetric 2D double-well potential \cite{Bhowmik2020}. 

In Fig.~\ref{Fig3}, we observe,  for both $\Psi_G$ and $\Psi_Y$,   back and forth of the density between  the left and right wells with the  essentially same frequency of oscillations at a particular value of $c$.  In the non-interacting system, at $t=0$,  $\Psi_G$ and $\Psi_Y$  lie approximately on the lowest band along the direction of the barrier and when there is an interaction quench, more bands are coupled. Both the frequency and amplitude of the tunneling oscillations of $\Psi_G$ are essentially similar to those of $\Psi_Y$ at the mean-field level. But, at the many-body level,  only the frequency of the tunneling oscillations remains very similar for $\Psi_G$ and $\Psi_Y$ at a particular value of $c$,  while the amplitudes show a substantial difference as time progresses. The difference in the many-body tunneling amplitude  between $\Psi_G$ and $\Psi_Y$  occurs as more states  start to couple  with time to $\Psi_Y$ in comparison to $\Psi_G$.      At $c=0.25$ and 0.5, the ground state energy of the left well matches with energies of the second (which is two-fold degenerate, $x\Psi_G$ and $y\Psi_G$) and third (which is three-fold degenerate, $xy\Psi_G$, $(x^2-1)\Psi_G$, and $(y^2-1)\Psi_G$) excited states of the right well, respectively. But, in the process of tunneling,  $\Psi_G$ couples to  $x\Psi_G$ and $(x^2-1)\Psi_G$ of the right well at $c=0.25$ and 0.5, respectively. Similarly,  $\Psi_Y$ couples respectively to $x\Psi_Y$ and $(x^2-1)\Psi_Y$ of the right well. As  $\Psi_G$ and $\Psi_Y$ mainly couple to those excited states which have excitations only along the $x$-direction, we call this tunneling  the longitudinal resonant tunneling.

For both $\Psi_G$ and $\Psi_Y$,   the short-time mean-field and many-body dynamics of $P_L(t)$ overlap  for the interaction strength taken here. This condition imitates the so-called infinite-particle limit of the time-dependent many-boson Schr\"odinger equation \cite{Erdos2007, Klaiman2016}. Contrary to the mean-field dynamics, one can find a signature of the quantum correlations in the many-body dynamics of $P_L(t)$ in terms of incomplete tunneling of the densities and,  consequently,   the amplitude of the oscillations of $P_L(t)$ gradually decays.  The decay in  the amplitude of the many-body $P_L(t)$ signifies a  collapse  in  the density oscillations,  which is more pronounced for $\Psi_G$ than for $\Psi_Y$. However, as $c$ increases, the decay rates of $\Psi_G$ and $\Psi_Y$ become smaller. For a certain value of $c$, the decay rate of $\Psi_G$ is larger compared to the corresponding rate of $\Psi_Y$, suggesting that in the tunneling process, the many-body effects develops in a different rate for different initial structures of the bosonic clouds. Explicitly, we observe that having transverse excitation delays the many-body process of the density oscillations collapse. Moreover, at $c=0.5$,  the time evolution of $P_L(t)$ exhibits a partial revival for both the states.

\begin{figure*}[!h]
{\includegraphics[trim = 0.1cm 0.5cm 0.1cm 0.2cm, scale=.60]{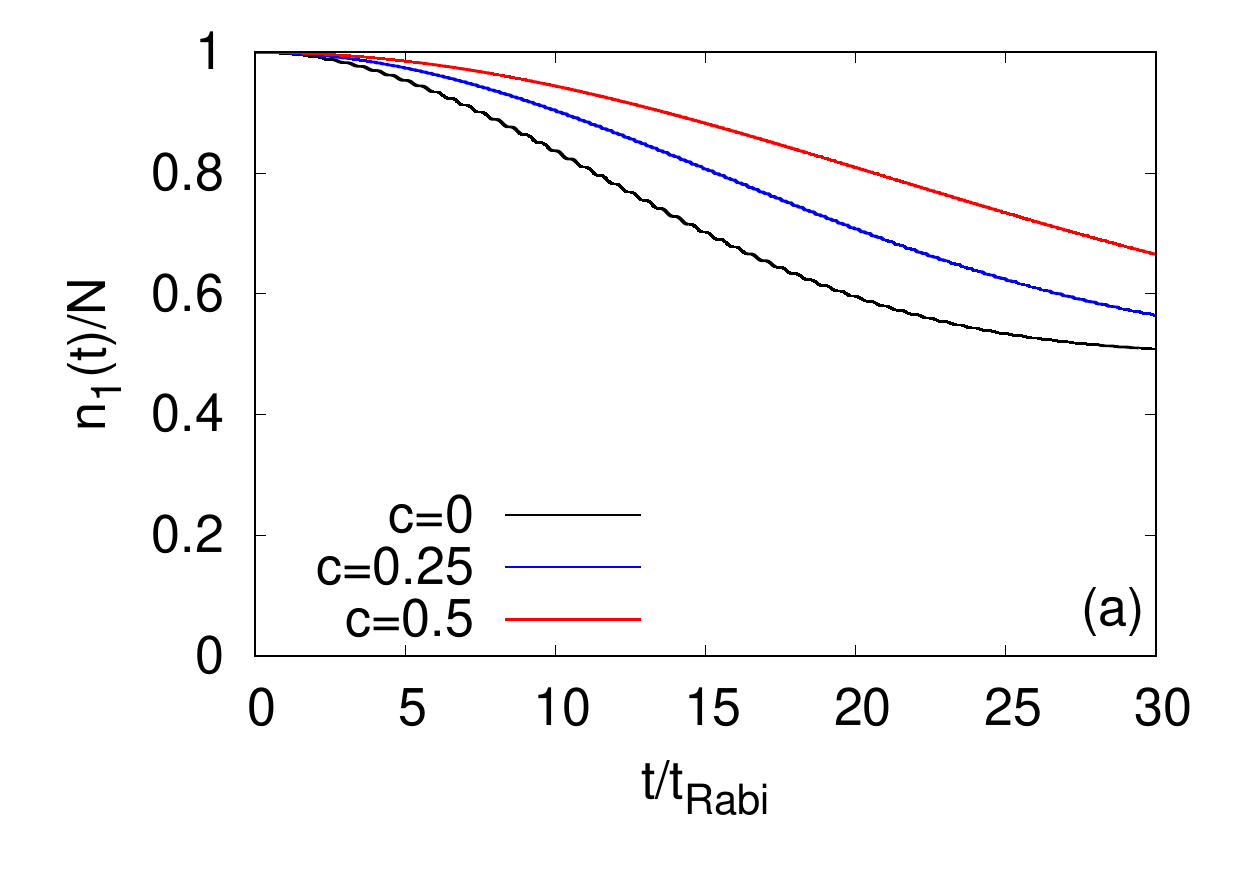}}
{\includegraphics[trim = 0.1cm 0.5cm 0.1cm 0.2cm, scale=.60]{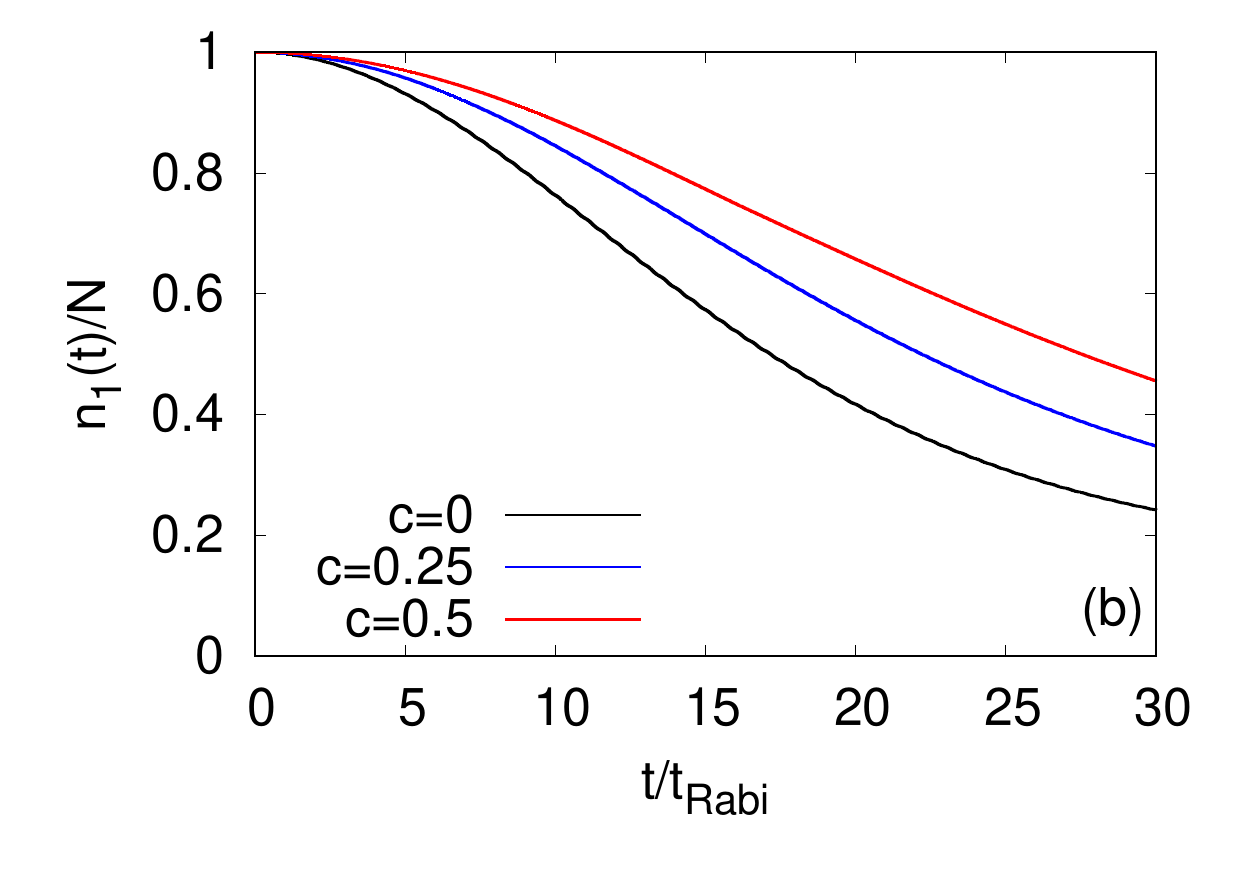}}\\
\caption{Time-dependent occupation per particle of the first natural orbital, $\dfrac{n_1(t)}{N}$,  in a longitudinally-asymmetric 2D double-well potential for $N=10$ bosons and  interaction parameter $\Lambda=0.01\pi$. Panels (a) and (b) are for $\Psi_G$ and $\Psi_Y$, respectively. The asymmetry parameters are  $c=0$, 0.25,  and 0.5.  The data have been obtained with $M= 6$ time-dependent orbitals for $\Psi_G$   and $M= 10$ time-dependent orbitals for $\Psi_Y$.    We show here dimensionless quantities. Color codes are explained in each panel.}
\label{Fig4}
\end{figure*}

Signature of a growing degree of quantum correlations is already found in terms of decay in the amplitude of  the time evolution of many-body $P_L(t)$. Now, we will discuss how this gradual increase of the many-body  correlations can affect the coherence of the condensates, i.e., bosonic clouds of $\Psi_G$ and $\Psi_Y$, when the resonant  tunneling occurs. To compare the results, we also show the loss of coherence for the tunneling of the above considered bosonic clouds  in the symmetric double-well potential as a reference.  In Fig.~\ref{Fig4}, we present the time evolution of the condensate fraction, $\dfrac{n_1(t)}{N}$, obtained by diagonalizing the reduced one-particle density matrix  of the time-dependent many-boson wave-function (Eq.~\ref{2}) \cite{Coleman2000, Sakmann2008}. The general feature of the occupation of the first natural orbital is found to be decreasing with time having a weak oscillatory background for all values of asymmetry parameters. It is observed that $\Psi_Y$ loses coherence faster than $\Psi_G$ for a particular value of $c$, suggesting that inclusion of the transverse excitation enhances fragmentation in $\Psi_Y$ \cite{Bhowmik2020}.  We find that the  rate of loss of coherence becomes slower when one move from  $c=0$ to the first resonant tunneling value at $c=0.25$ and further slower at the second resonant tunneling. 

In the context of occupations of the first natural orbital, it is worthwhile to mention  the occupancy of higher natural orbitals. We find that, as the time passes by and fragmentation of the condensate grows, the occupancy of all  higher natural orbitals gradually increases. The occupancy of  higher natural orbitals have direct impact on the different many-body quantities discussed later, i.e, on the mechanism of many-body resonant tunneling in 2D double-wells.  The detailed process of fragmentation with their convergences are discussed in the supplemental materials.

\begin{figure*}[!h]
{\includegraphics[trim = 0.1cm 0.5cm 0.1cm 0.2cm, scale=.60]{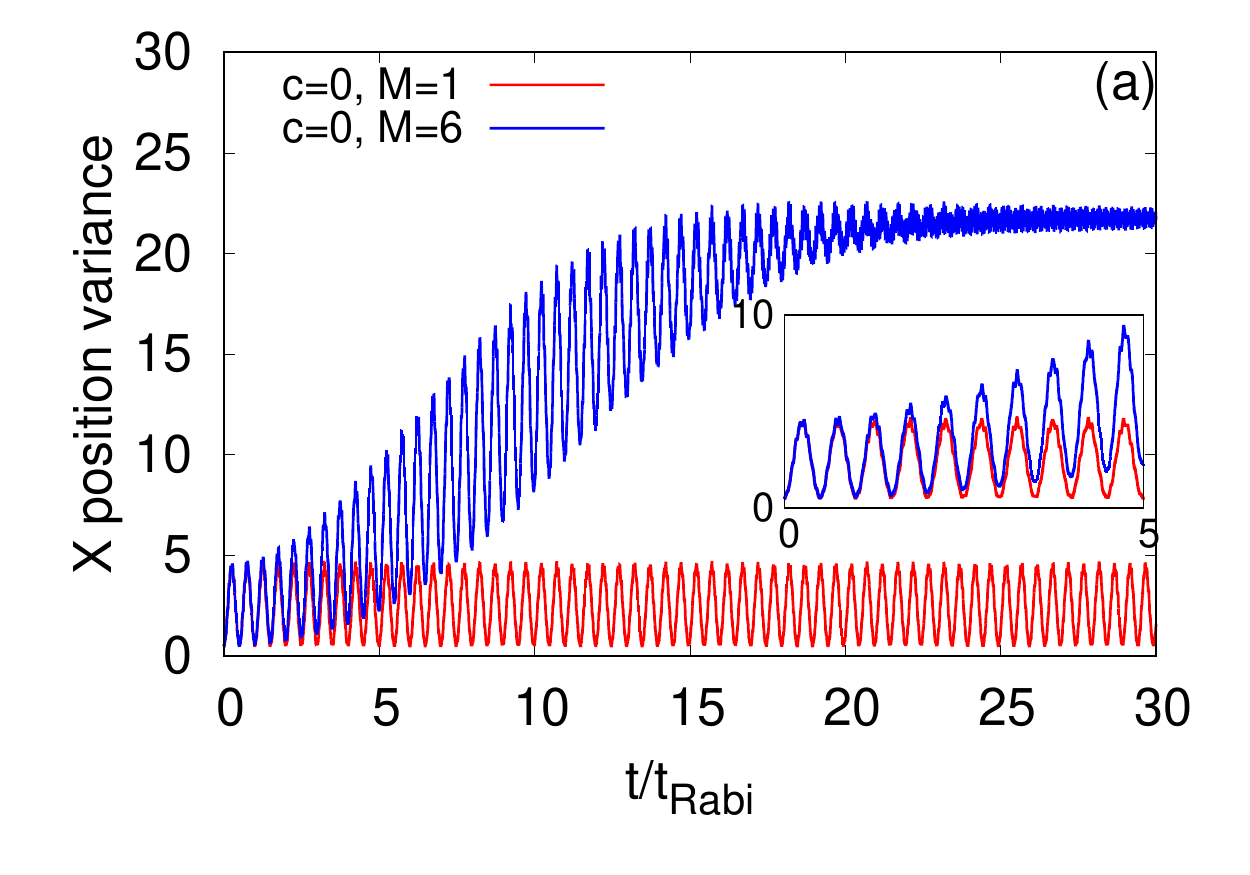}}
{\includegraphics[trim = 0.1cm 0.5cm 0.1cm 0.2cm, scale=.60]{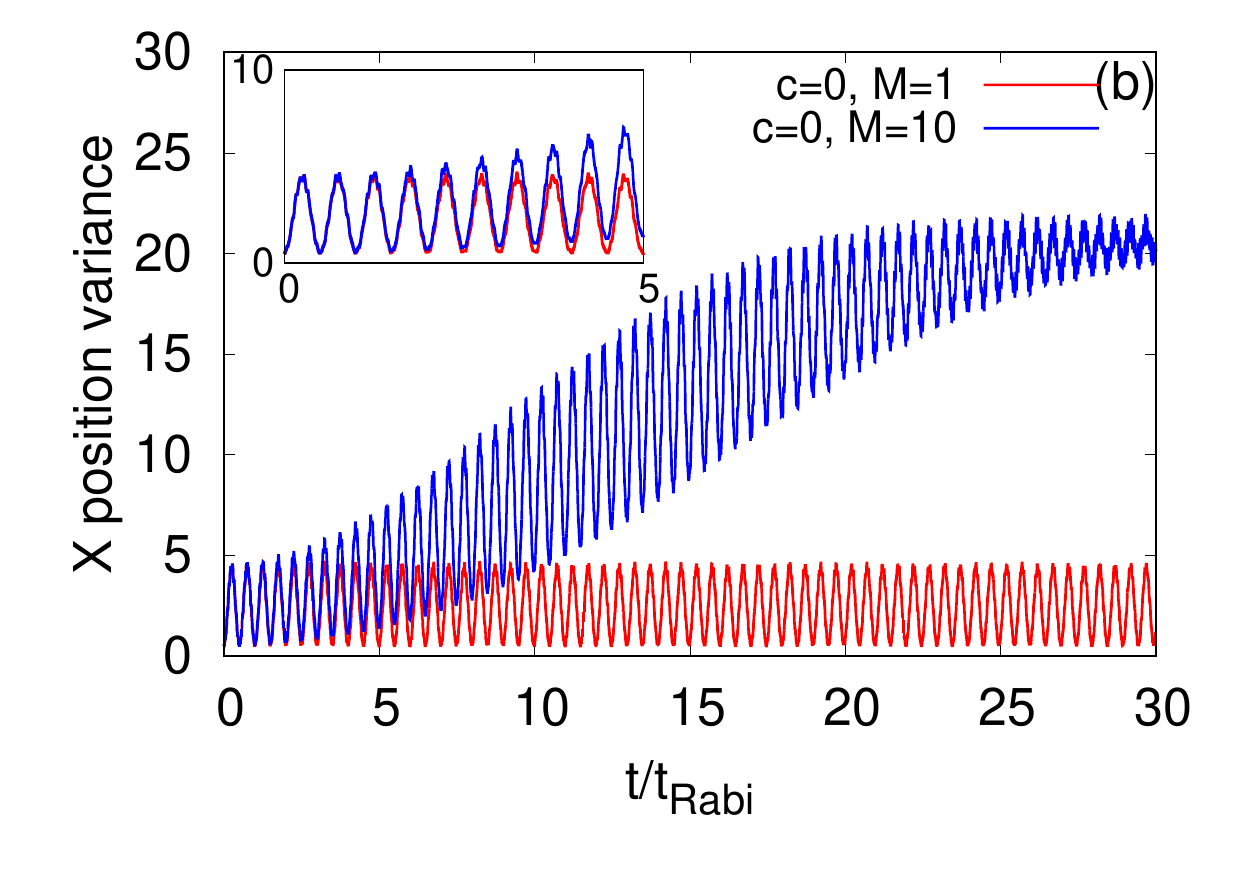}}\\
{\includegraphics[trim = 0.1cm 0.5cm 0.1cm 0.2cm, scale=.60]{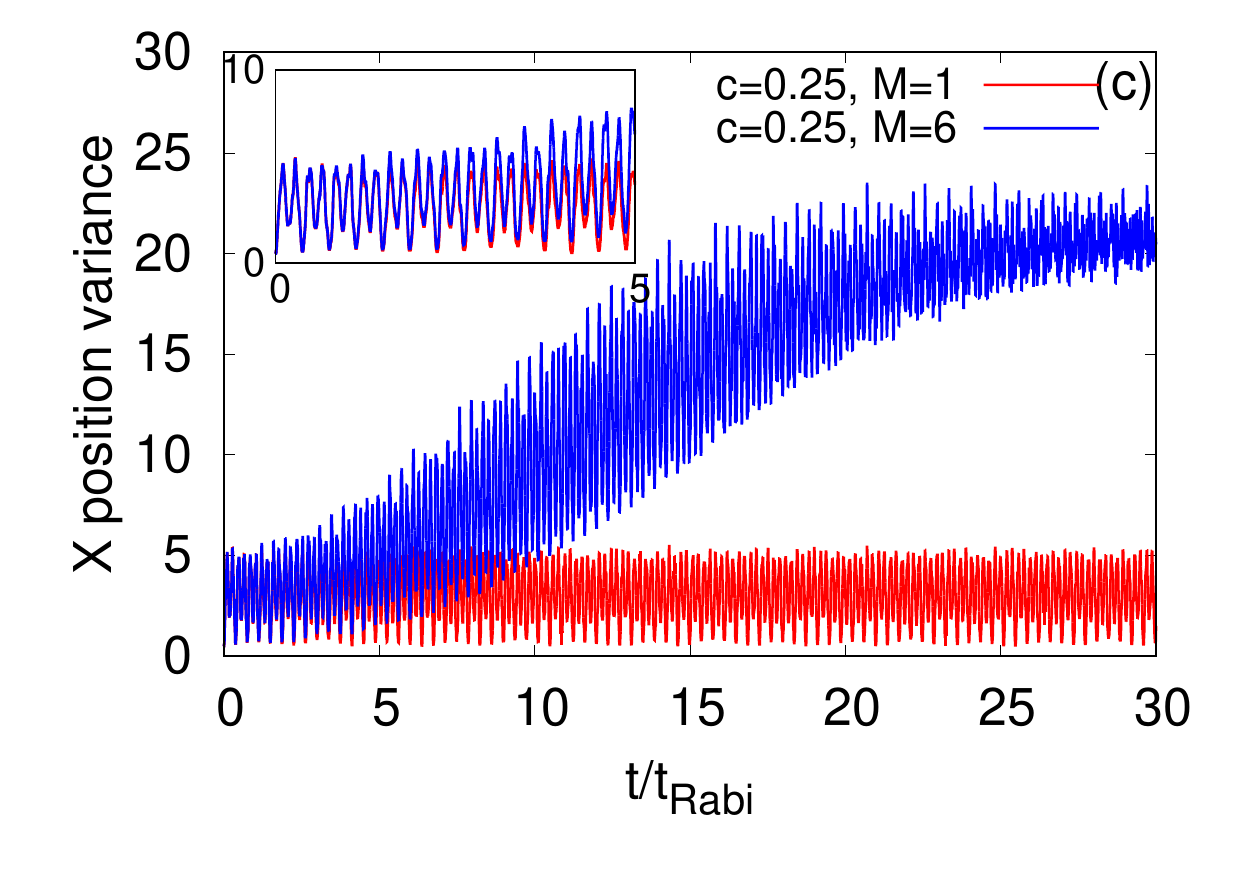}}
{\includegraphics[trim = 0.1cm 0.5cm 0.1cm 0.2cm, scale=.60]{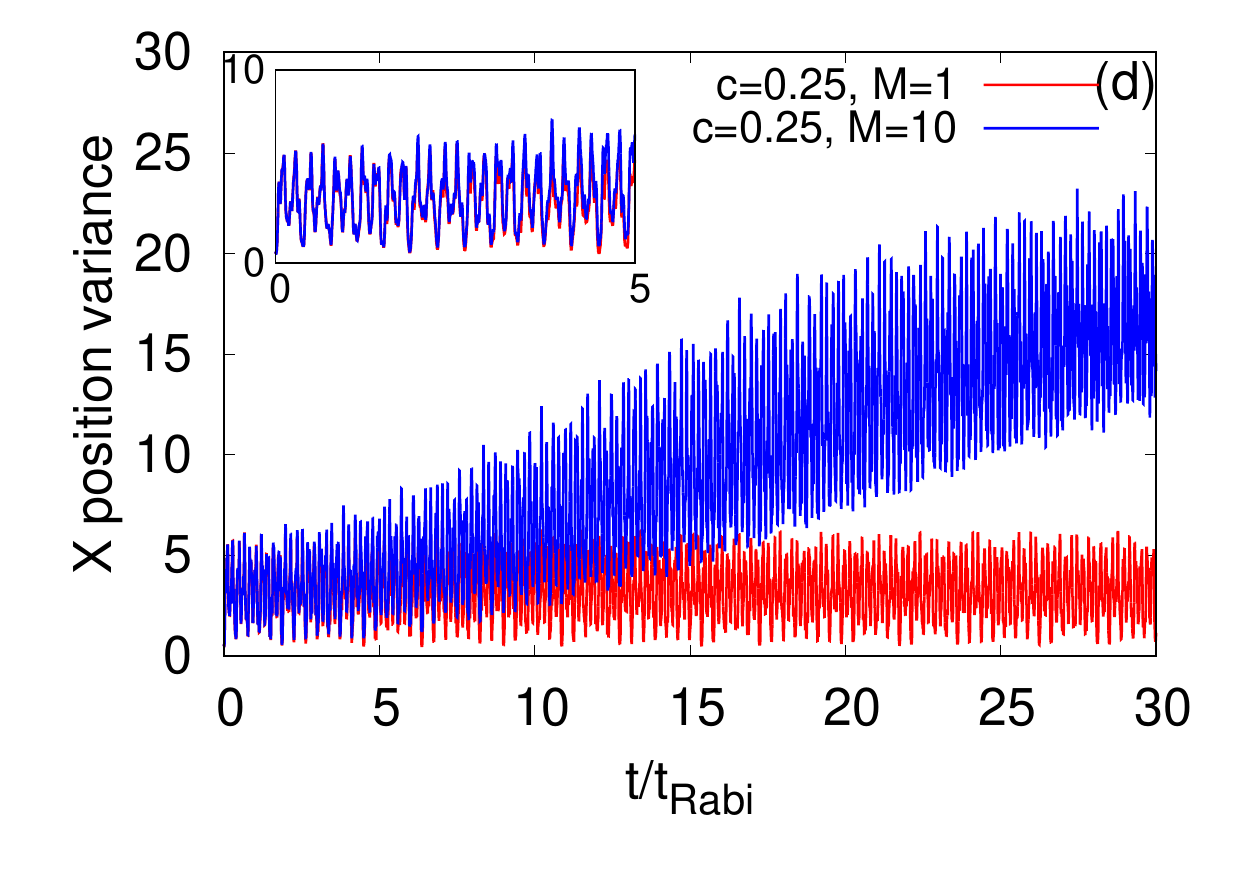}}\\
{\includegraphics[trim = 0.1cm 0.5cm 0.1cm 0.2cm, scale=.60]{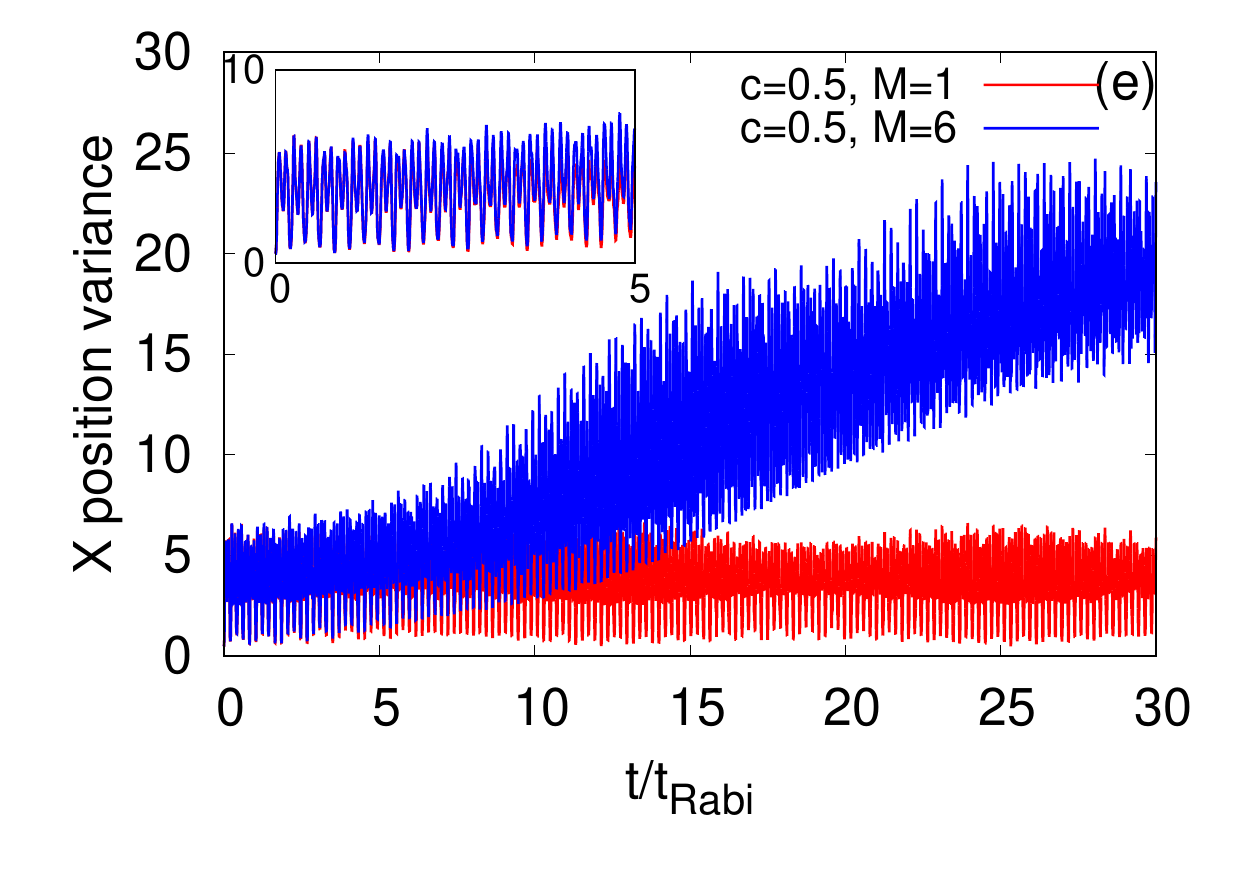}}
{\includegraphics[trim = 0.1cm 0.5cm 0.1cm 0.2cm, scale=.60]{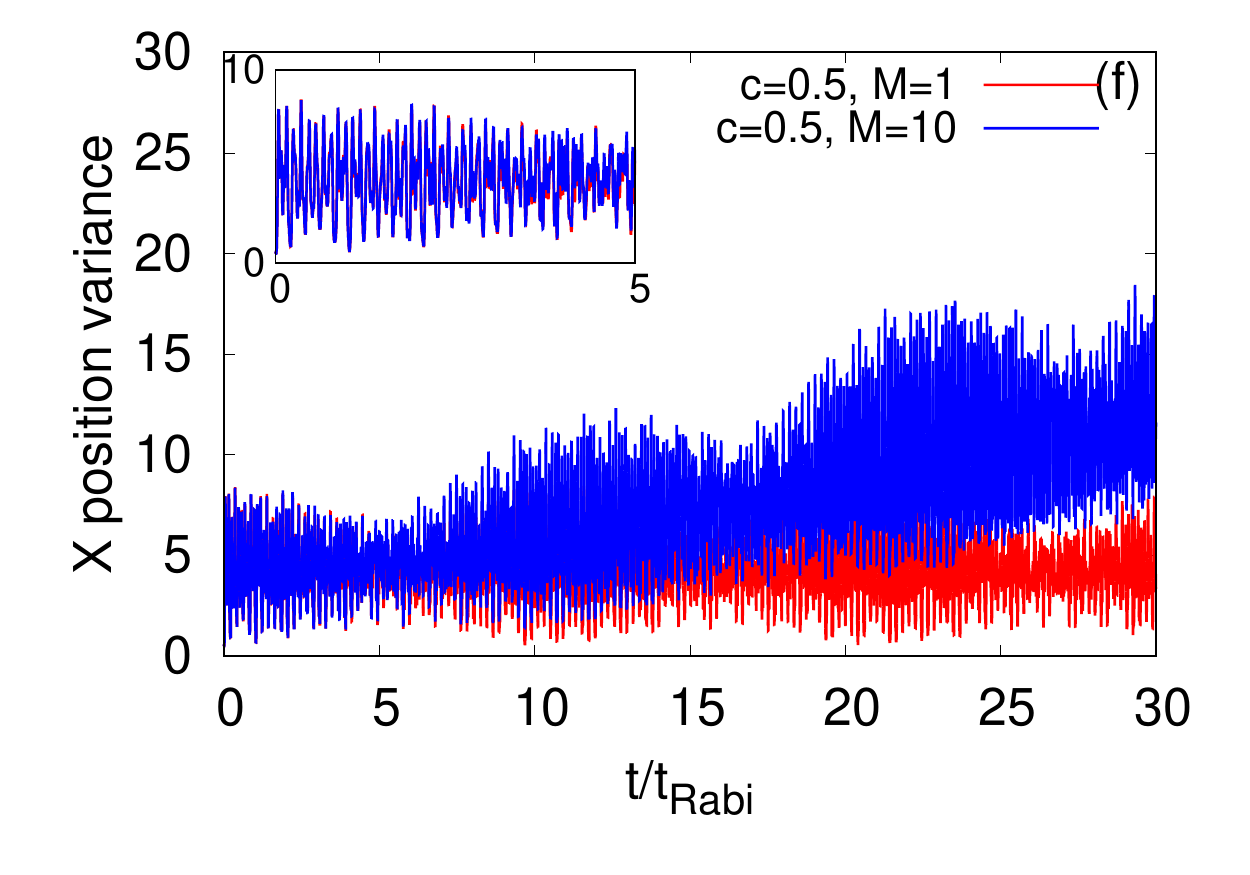}}\\
\caption{Time-dependent position variance  per particle along the $x$-direction, $\dfrac{1}{N}\Delta_{{\hat{X}}}^2(t)$, in a longitudinally-asymmetric 2D double-well potential for $\Psi_G$ (left column) and $\Psi_Y$ (right column) of $N=10$ bosons with the interaction parameter $\Lambda=0.01\pi$. The first row presents the results for the symmetric double-well potential.  The second and third rows  show $\dfrac{1}{N}\Delta_{{\hat{X}}}^2(t)$ at  the resonant tunneling with  asymmetry parameter, $c=0.25$ and 0.5, respectively.  $M=1$ signifies the mean-field results.  The many-body dynamics are computed with $M= 6$ time-dependent orbitals for $\Psi_G$   and $M= 10$ time-dependent orbitals for $\Psi_Y$.    From $t=0$ to 5$t_{Rabi}$  timescale of  $\dfrac{1}{N}\Delta_{{\hat{X}}}^2(t)$ is highlighted in the inset of each panel.  We show here dimensionless quantities. Color codes are explained in each panel.  }
\label{Fig5}
\end{figure*}

\begin{figure*}[!h]
{\includegraphics[trim = 0.1cm 0.5cm 0.1cm 0.2cm, scale=.60]{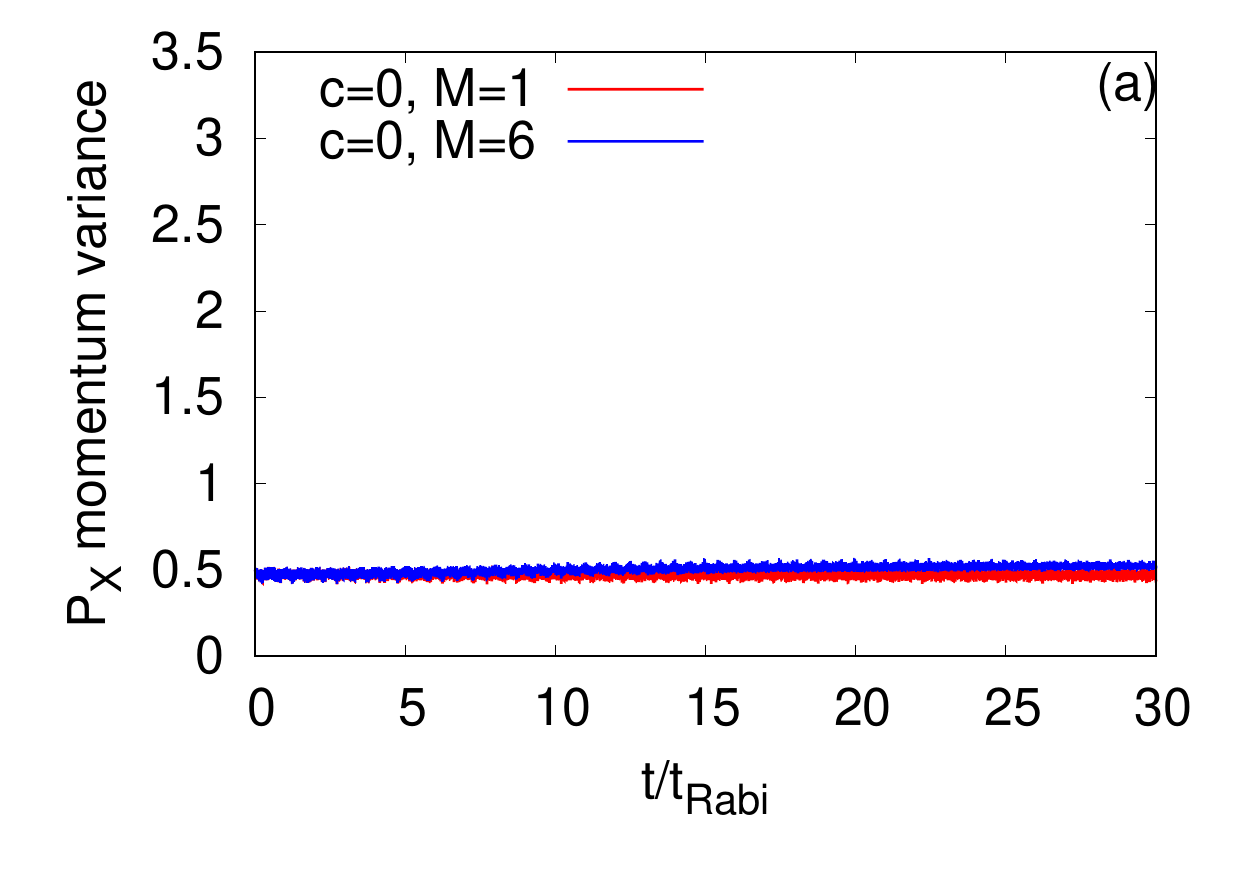}}
{\includegraphics[trim = 0.1cm 0.5cm 0.1cm 0.2cm, scale=.60]{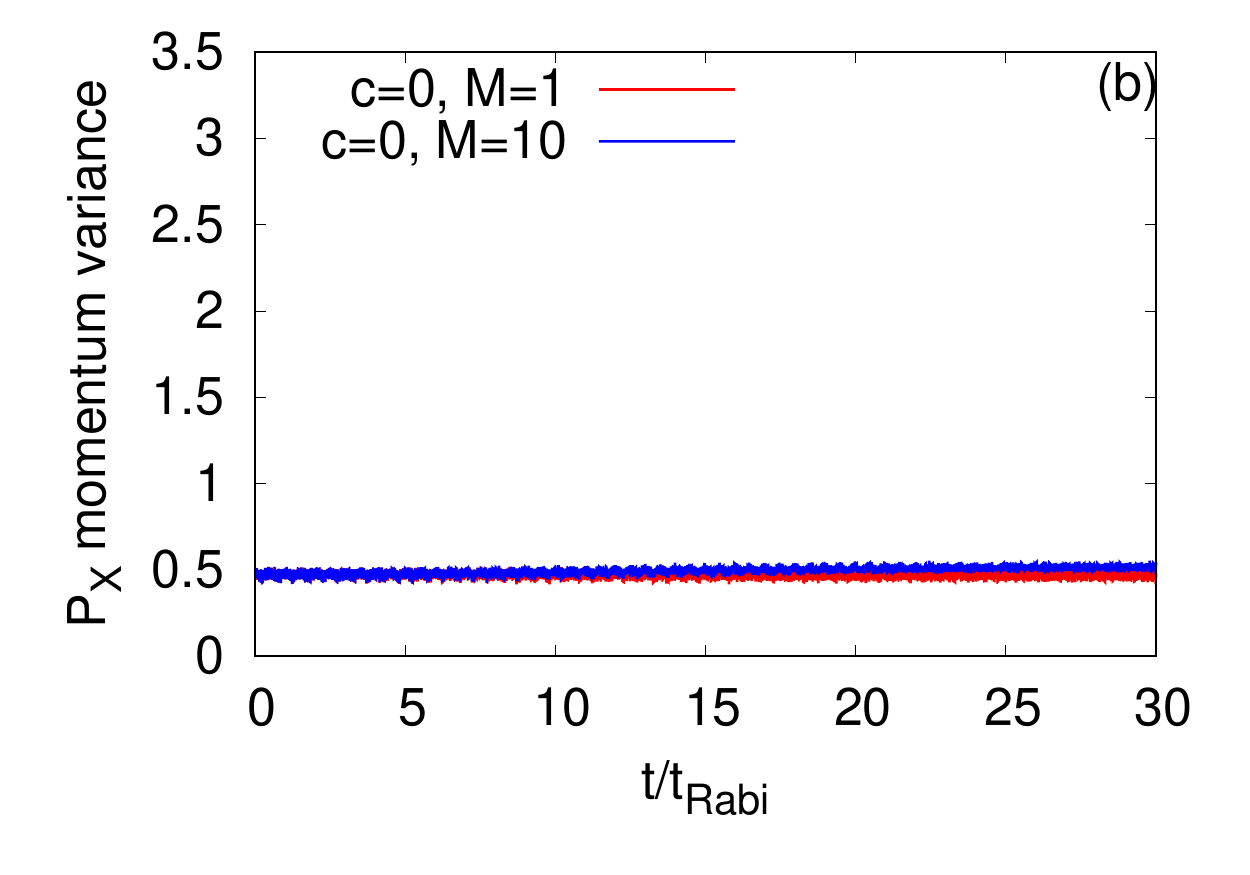}}\\
{\includegraphics[trim = 0.1cm 0.5cm 0.1cm 0.2cm, scale=.60]{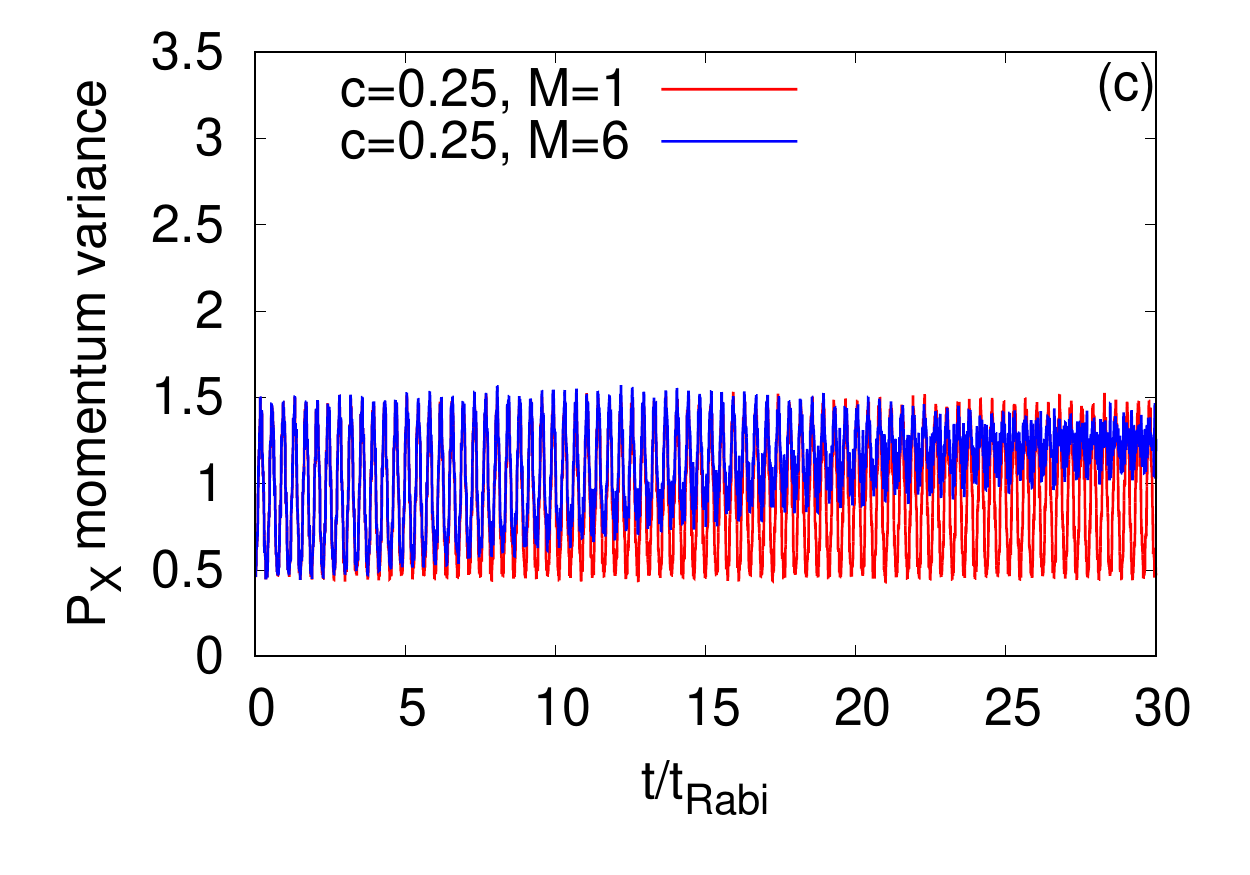}}
{\includegraphics[trim = 0.1cm 0.5cm 0.1cm 0.2cm, scale=.60]{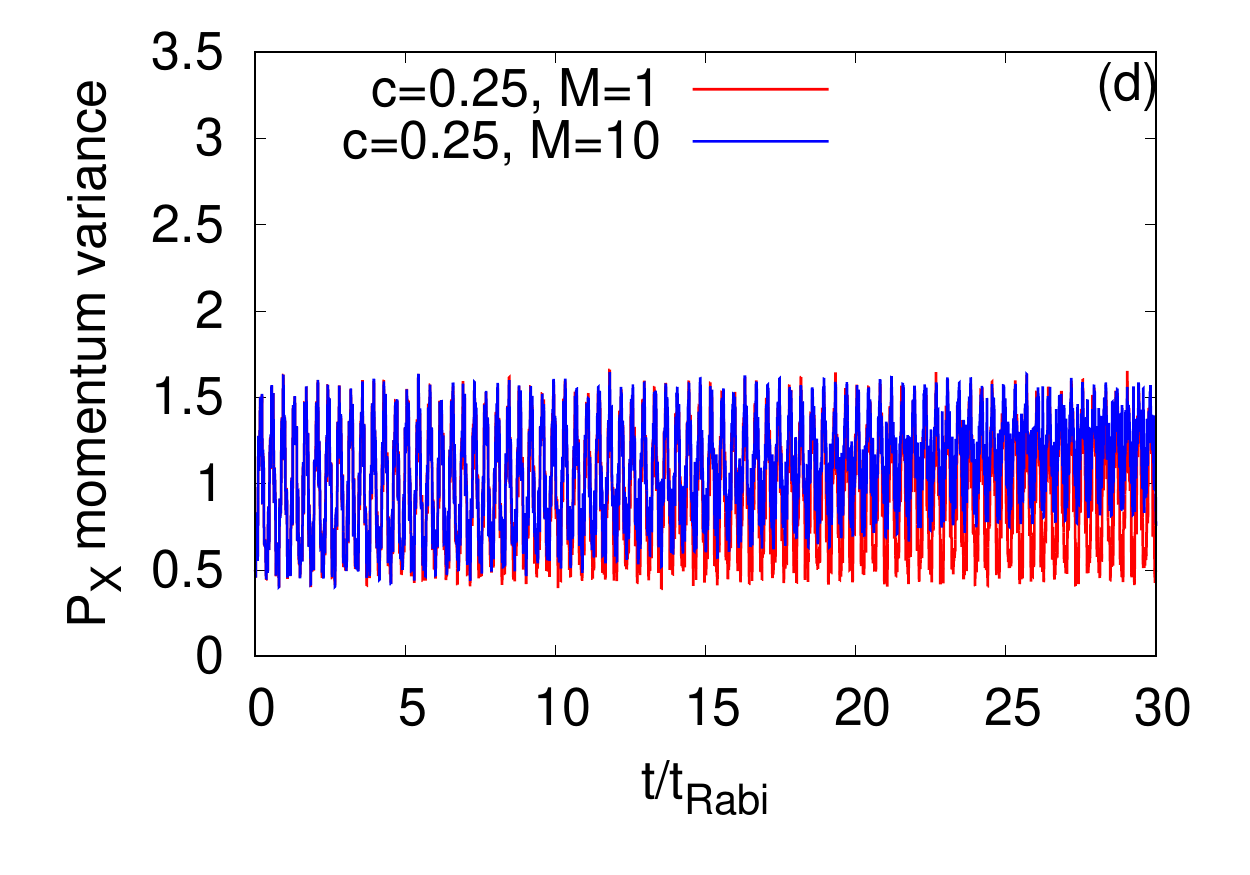}}\\
{\includegraphics[trim = 0.1cm 0.5cm 0.1cm 0.2cm, scale=.60]{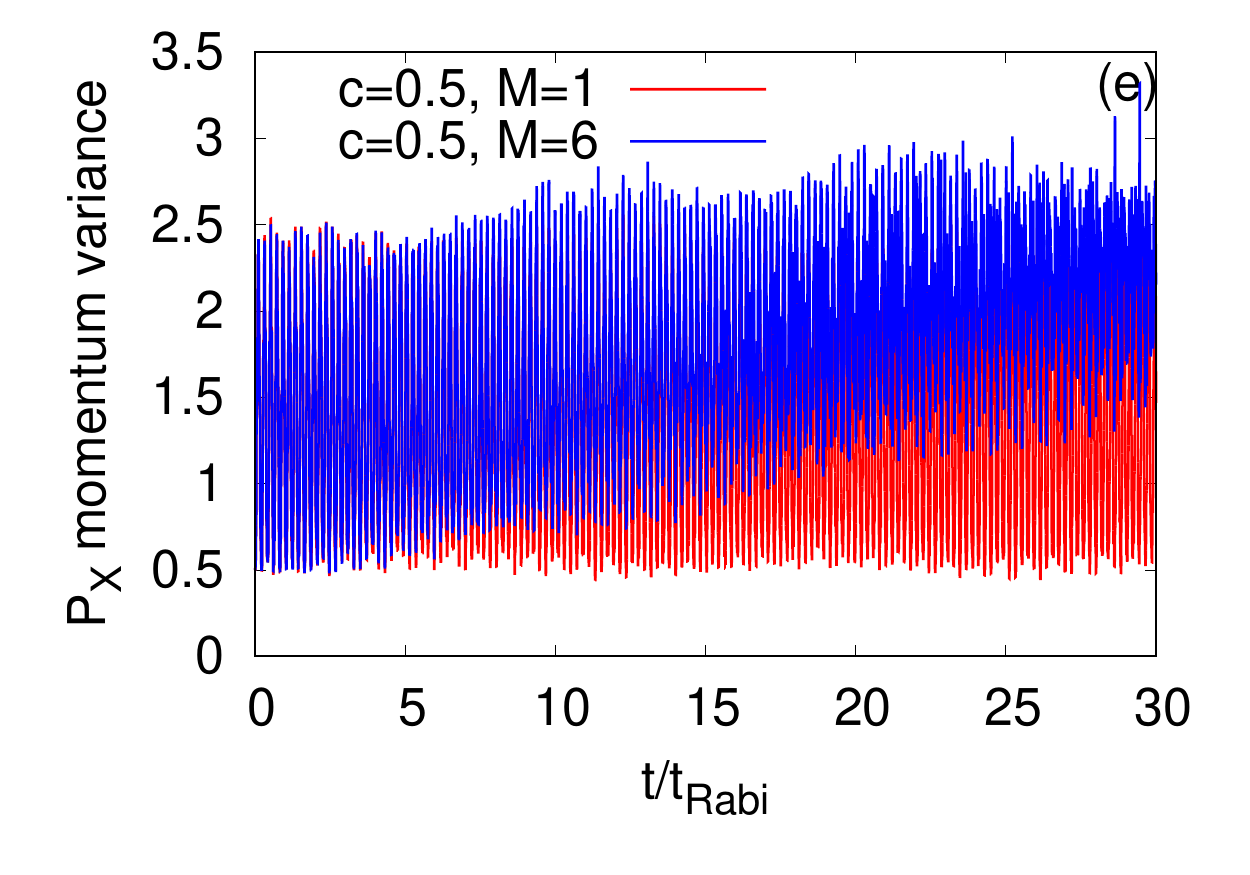}}
{\includegraphics[trim = 0.1cm 0.5cm 0.1cm 0.2cm, scale=.60]{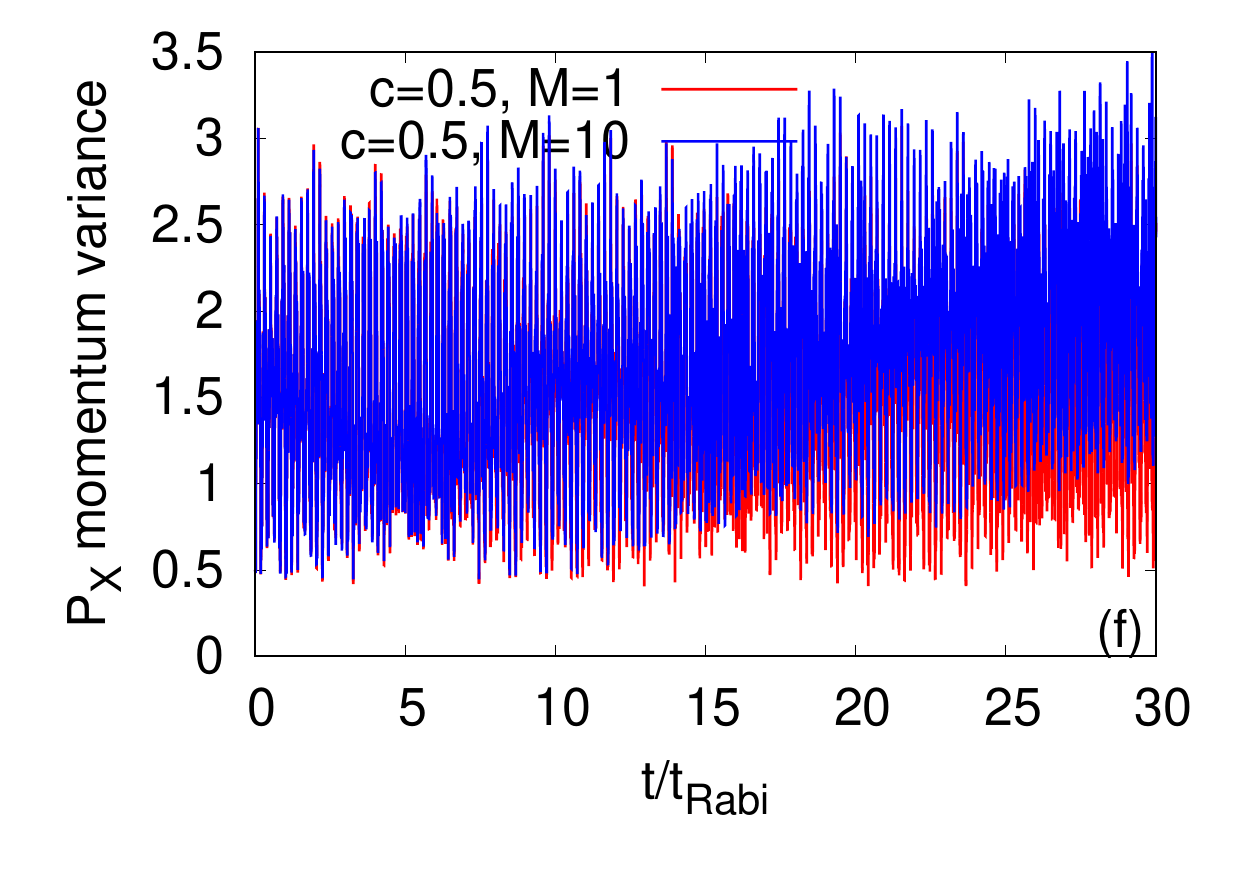}}\\
\caption{Time-dependent momentum variance  per particle along the $x$-direction, $\dfrac{1}{N}\Delta_{{\hat{P}_X}}^2(t)$, in a longitudinally-asymmetric 2D double-well potential for $\Psi_G$ (left column) and $\Psi_Y$ (right column) of $N=10$ bosons with the interaction parameter $\Lambda=0.01\pi$. The first row presents the results for the symmetric double-well potential. The second and third rows  show $\dfrac{1}{N}\Delta_{{\hat{P}_X}}^2(t)$ at  the resonant tunneling with  asymmetry parameter, $c=0.25$, and 0.5, respectively.  $M=1$ signifies the mean-field results.    The many-body dynamics are computed with $M= 6$ time-dependent orbitals for $\Psi_G$   and $M= 10$ time-dependent orbitals for $\Psi_Y$.      We show here dimensionless quantities. Color codes are explained in each panel.}
\label{Fig6}
\end{figure*}
\begin{figure*}[!h]
{\includegraphics[trim = 0.1cm 0.5cm 0.1cm 0.2cm, scale=.60]{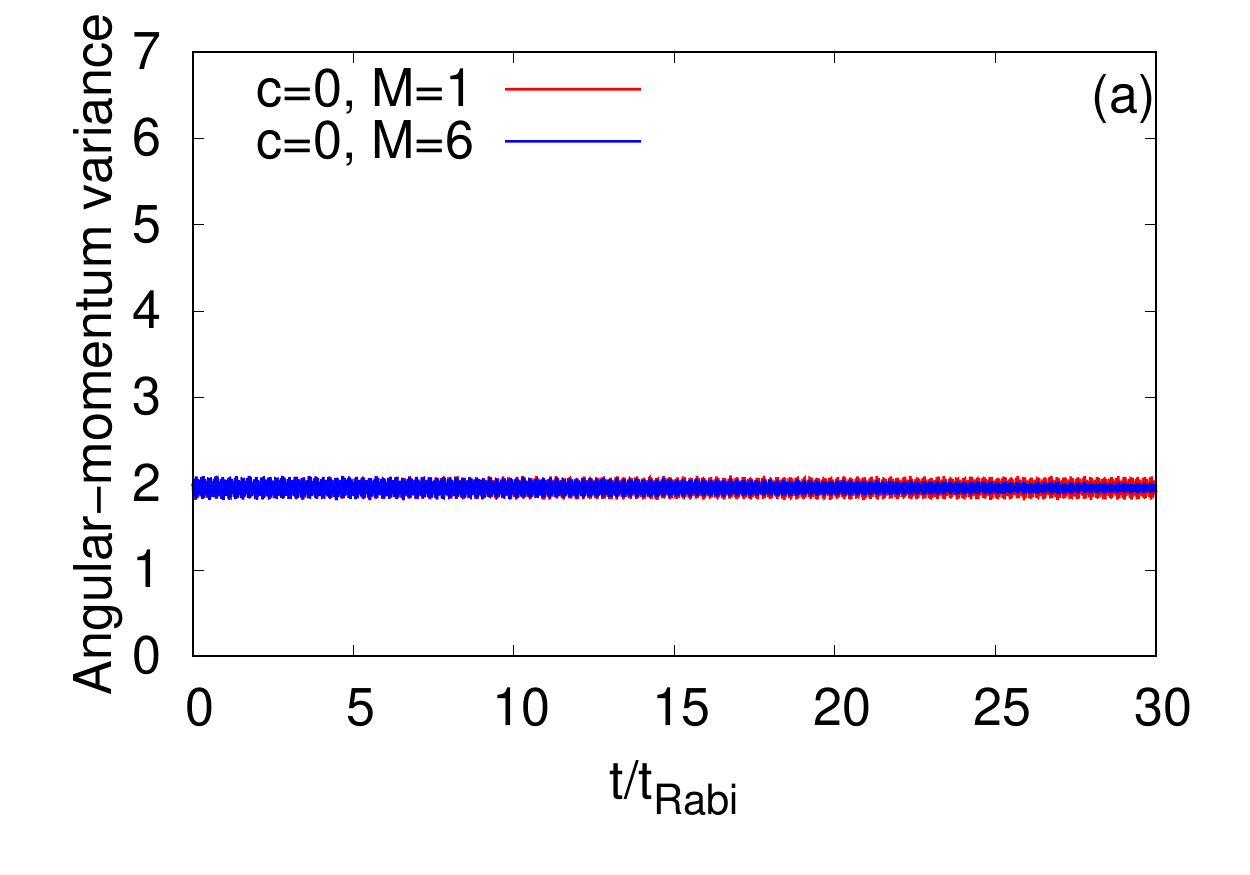}}
{\includegraphics[trim = 0.1cm 0.5cm 0.1cm 0.2cm, scale=.60]{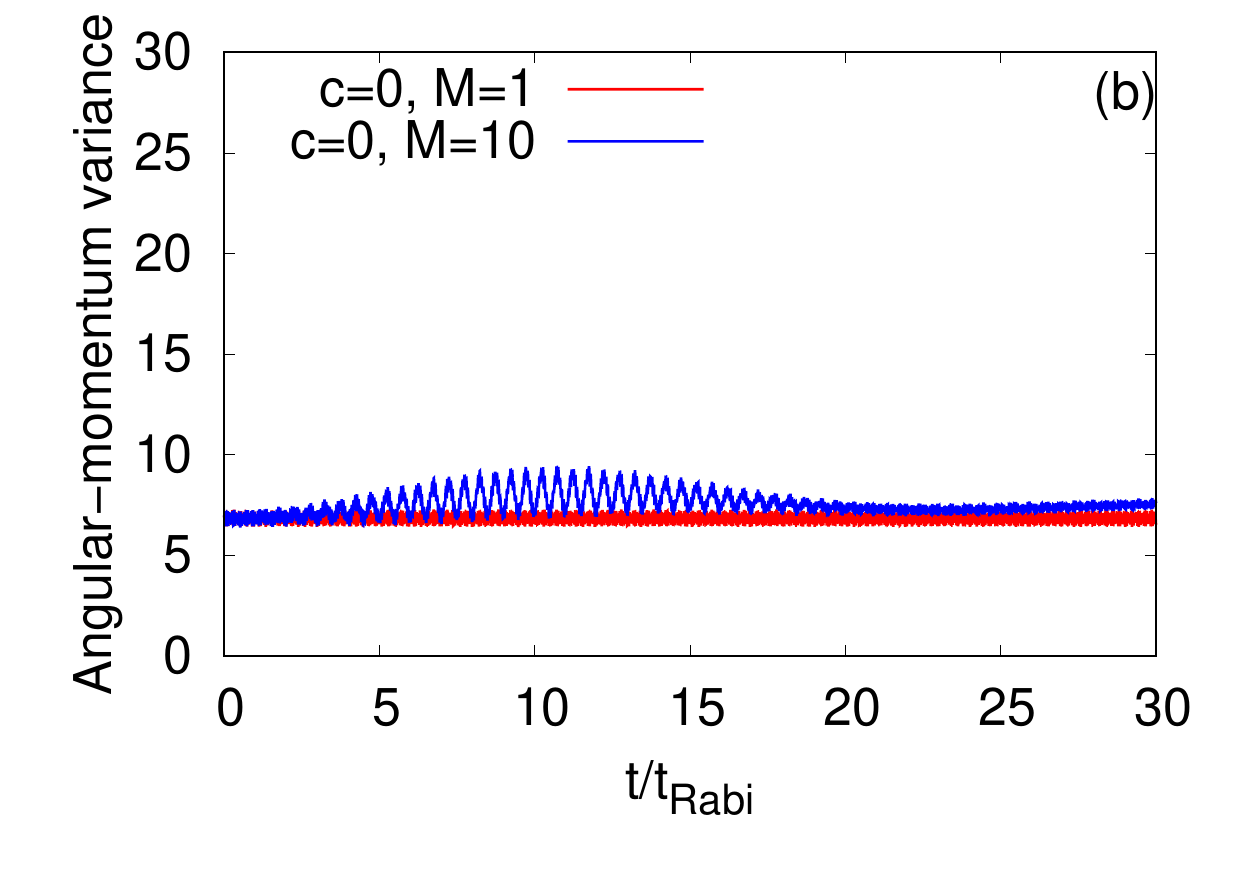}}\\
{\includegraphics[trim = 0.1cm 0.5cm 0.1cm 0.2cm, scale=.60]{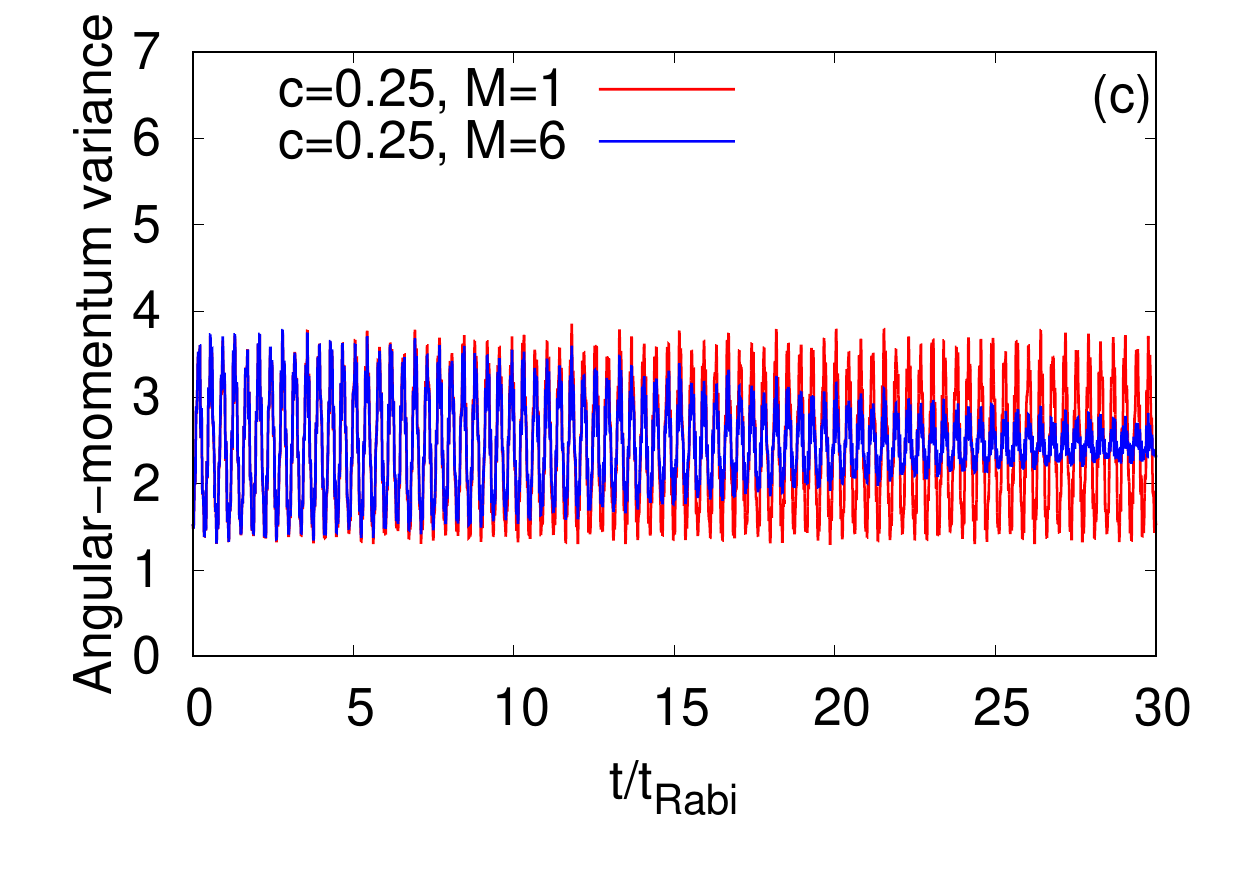}}
{\includegraphics[trim = 0.1cm 0.5cm 0.1cm 0.2cm, scale=.60]{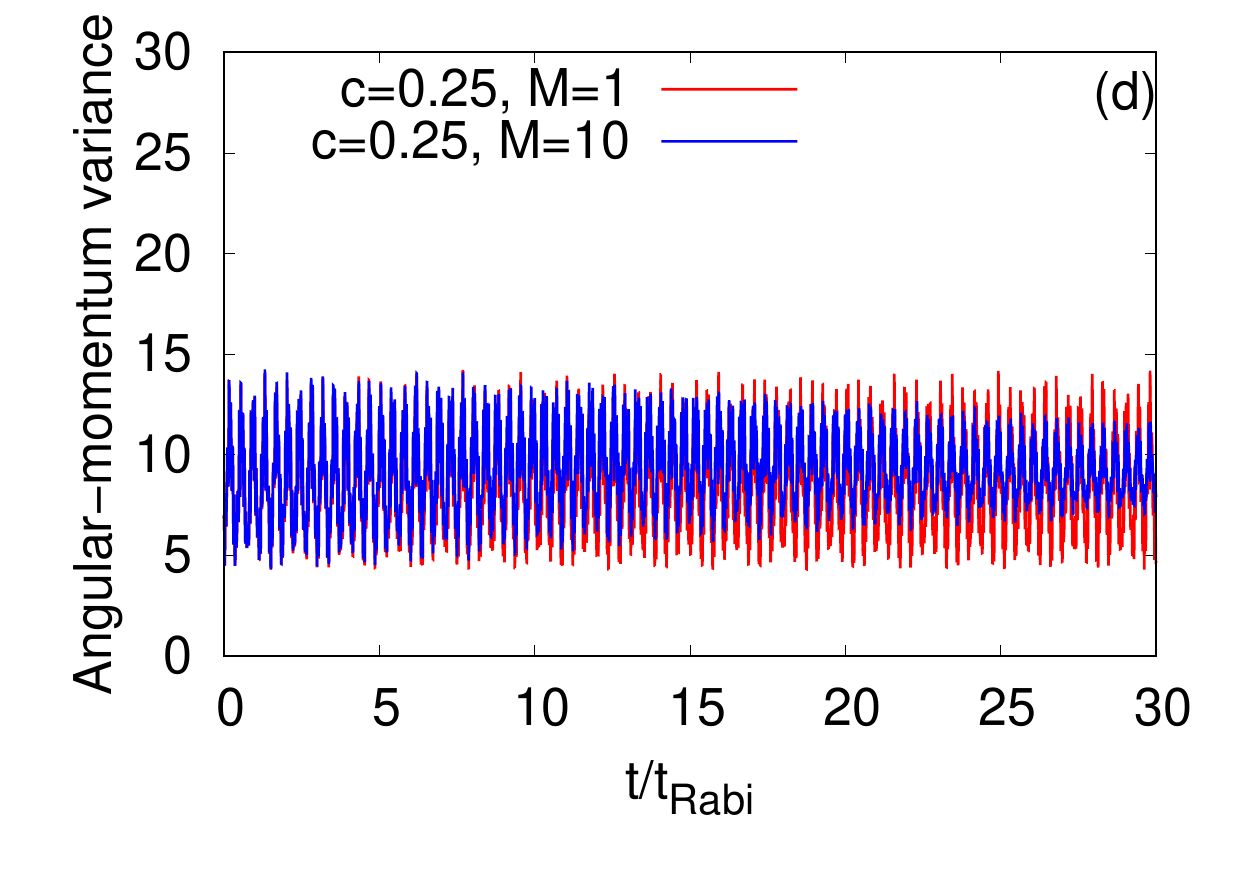}}\\
{\includegraphics[trim = 0.1cm 0.5cm 0.1cm 0.2cm, scale=.60]{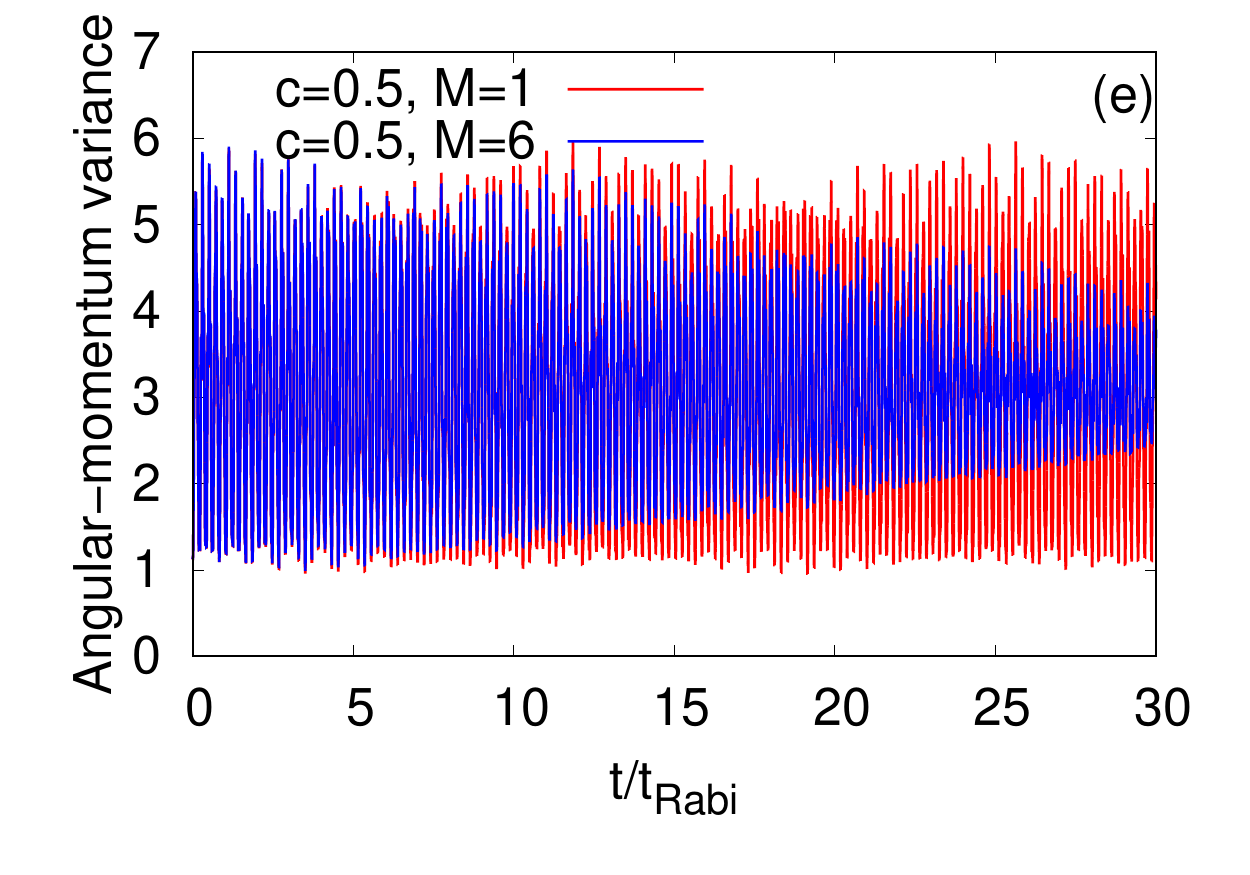}}
{\includegraphics[trim = 0.1cm 0.5cm 0.1cm 0.2cm, scale=.60]{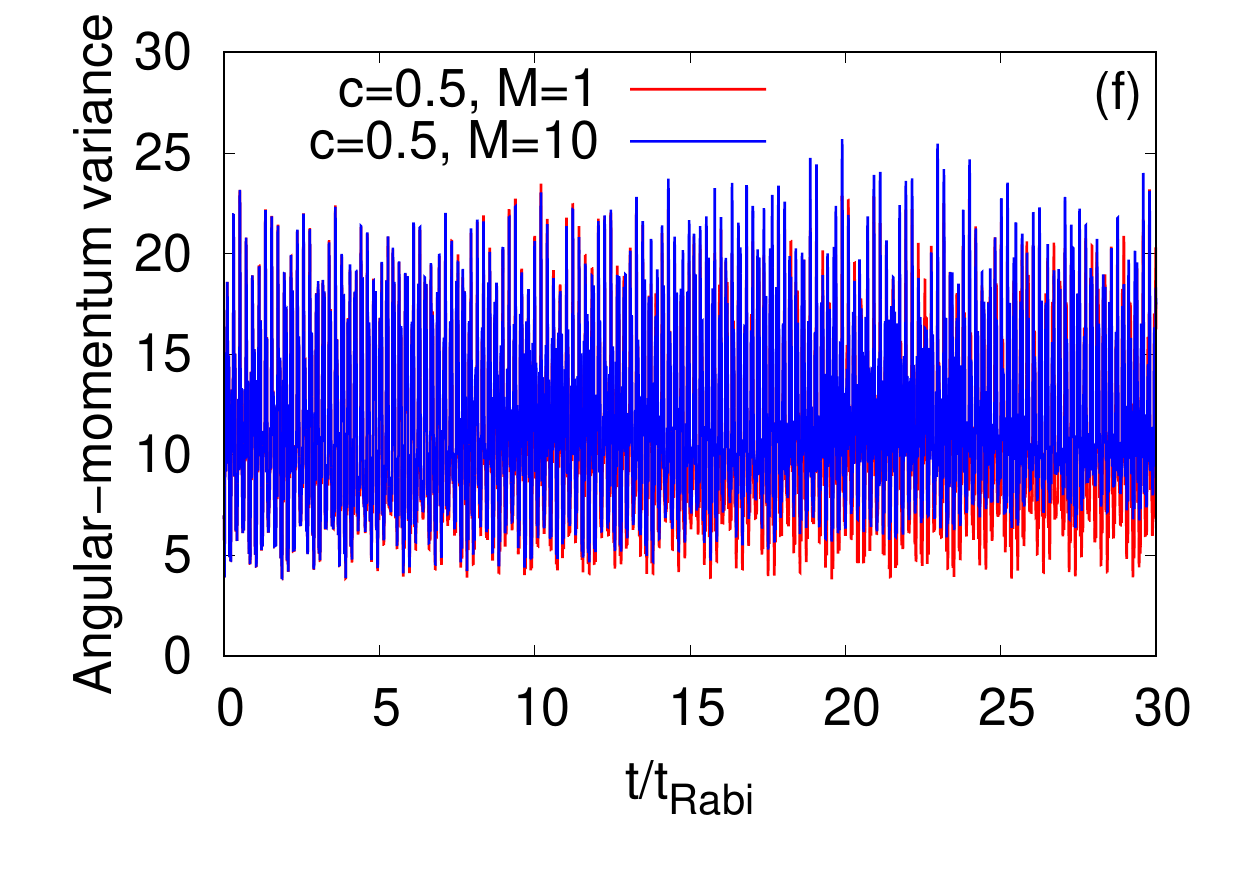}}\\
\caption{Time-dependent variance per particle of the $z$- component of the angular-momentum operator, $\dfrac{1}{N}\Delta_{\hat{L}_Z}^2(t)$, in a longitudinally-asymmetric 2D double-well potential for $\Psi_G$ (left column) and $\Psi_Y$ (right column) of $N=10$ bosons with the interaction parameter $\Lambda=0.01\pi$.  The first row presents the results for the symmetric double-well potential.  The second and third rows  show $\dfrac{1}{N}\Delta_{{\hat{L}_Z}}^2(t)$ at  the resonant tunneling with  asymmetry parameter, $c=0.25$ and 0.5, respectively.  $M=1$ signifies the mean-field results.   The many-body dynamics are computed with $M= 6$ time-dependent orbitals for $\Psi_G$   and $M= 10$ time-dependent orbitals for $\Psi_Y$.      We show here dimensionless quantities. Color codes are explained in each panel.}
\label{Fig7}
\end{figure*}
Having explicated the  development of many-body correlations and their effects on the time evolution of  the survival probability and coherence of the bosonic clouds, we find  further information of the time-dependent many-particle wavefunction and especially,  the signature of  the depletion and fragmentation on the condensate.  As the variance is a sensitive probe of correlation \cite{Klaiman2015},  we graphically analyze the dynamical behavior of the variances of a few fundamental quantum mechanical observables which are mostly influenced by the longitudinal resonant tunneling.   These are   the position and  momentum operators along the $x$-direction,  and the angular-momentum operator.

Fig.~\ref{Fig5} displays the time evolution of the many-particle position variance per particle along the $x$-direction, $\dfrac{1}{N}\Delta_{\hat{X}}^2(t)$,  in  the longitudinally-asymmetric 2D double-well potential with  asymmetry parameters, $c=0$, 0.25, and 0.5 for the bosonic clouds of $\Psi_G$ and $\Psi_Y$ (see  Appendix A for the mathematical form of the many-particle variance and its basic properties).  As a general feature, we find that the many-body and mean-field values of $\dfrac{1}{N}\Delta_{\hat{X}}^2(t)$  are oscillatory in nature. Due to the correlations, the average value of the  many-body $\dfrac{1}{N}\Delta_{\hat{X}}^2(t)$ deviates -- it rather significantly increases --  from the corresponding  mean-field results for all asymmetry parameters and bosonic clouds. Although, we found in Fig.~\ref{Fig4} that $\Psi_Y$ is always more fragmented than $\Psi_G$, the deviations in the values  of $\dfrac{1}{N}\Delta_{\hat{X}}^2(t)$ are always smaller for $\Psi_Y$ compared to $\Psi_G$ at a particular value of $c$,  apart for short-times at $c=0.5$.  Therefore, one can say that the many-body dynamics is complicated and  there is no one-to-one   correlation between the different many-body properties in the junction, such as fragmentation and  $\dfrac{1}{N}\Delta_{\hat{X}}^2(t)$, but it suggests how the transverse excitations participating in the dynamics can impact the  fluctuations in measuring an observable. From the insets of  Fig.~\ref{Fig5}(a) and (b), we find that the peaks of the oscillations of $\dfrac{1}{N}\Delta_{\hat{X}}^2(t)$ breaks into two sub-peaks which practically turn into  two  broad peaks at the resonant values of $c$.   Unlike the mean-field dynamics, the many-body $\dfrac{1}{N}\Delta_{\hat{X}}^2(t)$ oscillates with a  growing amplitude until it reaches an equilibrium value which is more evident at $c=0$ for both the states and at $c=0.25$ for $\Psi_G$. This equilibrium value of $\dfrac{1}{N}\Delta_{\hat{X}}^2(t)$ is reached when the density oscillations collapse as  observed in the time evolutions of $P_L(t)$ in  Fig.~\ref{Fig3}. At $c=0.5$, we notice that the amplitude of the oscillations of $\dfrac{1}{N}\Delta_{\hat{X}}^2(t)$ are also oscillatory, more pronouncedly for $\Psi_Y$, which is consistent with the density oscillations. In the dynamics of $\dfrac{1}{N}\Delta_{\hat{X}}^2(t)$, we observe two kinds of oscillations, i.e., small frequency with large
amplitude and high frequency with small amplitude oscillations. The former one comes from the density oscillations and the latter one  occurs due to the breathing-mode oscillations of the system. At the resonant values of $c$, the high frequency with small amplitude oscillations are more clearly visible.

Next, we consider the momentum variance per particle along the $x$-direction, $\dfrac{1}{N}\Delta_{{\hat{P}_X}}^2(t)$, and present its time evolution at the mean-field and many-body levels for $\Psi_G$ and $\Psi_Y$ in Fig.~\ref{Fig6}. We explicitly show  the possible quantum correlations effects on   the many-body dynamics of $\dfrac{1}{N}\Delta_{{\hat{P}_X}}^2(t)$ at the resonant tunneling condition and compare  with the corresponding  results obtained for the symmetric double-well potential. Recall that $\dfrac{1}{N}\Delta_{{\hat{P}_X}}^2(t)$ is relatively a more complex quantity than $\dfrac{1}{N}\Delta_{\hat{X}}^2(t)$ in the junction as the former one is controlled by and more sensitive to  the shape of the orbitals. In Fig.~\ref{Fig6}, we observe, also for $\dfrac{1}{N}\Delta_{{\hat{P}_X}}^2(t)$,  two  types of oscillations,  smaller amplitude with  higher frequency and larger amplitude with lower frequency. But there is a difference with respect to $\dfrac{1}{N}\Delta_{\hat{X}}^2(t)$. For the symmetric double-well potential, $\dfrac{1}{N}\Delta_{{\hat{P}_X}}^2(t)$ is dominated by the  breathing mode oscillations which produce small amplitude with high frequency oscillations. While at the first resonant tunneling, $c=0.25$, the density oscillations are more prominent producing the larger amplitude but lower frequency oscillations. But, at the second resonant tunneling, $c=0.5$, $\dfrac{1}{N}\Delta_{{\hat{P}_X}}^2(t)$ shows a mixture of the density and breathing mode oscillations. At $c=0.25$, the mean-field $\dfrac{1}{N}\Delta_{{\hat{P}_X}}^2(t)$ oscillates between 0.5 and 1.5, while at  $c=0.5$, between 0.5 and 2.5.  This reflects the ground  (left well) and excited (right well) orbitals participating in the dynamics.  One of the interesting features at the resonant tunneling condition is that  the minima values of the many-body  $\dfrac{1}{N}\Delta_{{\hat{P}_X}}^2(t)$ increases with the growing degree of fragmentation. Also, the amplitude of the many-body oscillations of $\dfrac{1}{N}\Delta_{{\hat{P}_X}}^2(t)$ eventually decays, which is clearly visible at $c=0.25$, when the density collapses. It suggests that in the long-time dynamics of $\dfrac{1}{N}\Delta_{{\hat{P}_X}}^2(t)$, the excited state in the right well slowly wins over the initial state prepared in the left well.

In this context, it is worthwhile to briefly discuss   the position and momentum variances along the $y$-direction, $\dfrac{1}{N}\Delta_{{\hat{Y}}}^2(t)$ and $\dfrac{1}{N}\Delta_{{\hat{P}_Y}}^2(t)$, respectively. We found in our work that the mean-field and many-body values of  $\dfrac{1}{N}\Delta_{{\hat{Y}}}^2(t)$ and $\dfrac{1}{N}\Delta_{{\hat{P}_Y}}^2(t)$  have very
small fluctuations, of the order of $10^{-3}$, atop the value of the non-interacting bosons,
i.e., 0.5 and 1.5 for both variances, for $\Psi_G$ and $\Psi_Y$, respectively (see supplemental materials). Also, with the change in asymmetry parameter, we do not observe any substantial changes in the values of position and momentum variances along the transverse direction. However, in addition to the motion along the $x$-direction,  the effect of the seemingly constant transverse degrees-of-freedom   lead to the existence of a purely two-dimensional quantity, such as the angular-momentum.

Now, we examine  the angular-momentum variance per particle, $\dfrac{1}{N}\Delta_{{\hat{L}_Z}}^2(t)$, and demonstrate the implications of the resonant tunneling on its dynamical behavior  in Fig.~\ref{Fig7}, see  Appendix A for the mathematical form of $\dfrac{1}{N}\Delta_{{\hat{L}_Z}}^2(t)$.  The results are compared at the mean-field and many-body levels for $\Psi_G$ and $\Psi_Y$.  At $t=0$, in case of the symmetric  double-well trap, $\dfrac{1}{N}\Delta_{{\hat{L}_Z}}^2(t)$  can be calculated analytically and is found to be $2$ and $7$ for $\Psi_G$ and $\Psi_Y$, respectively. At the first and second resonant tunneling, the initial values of $\dfrac{1}{N}\Delta_{{\hat{L}_Z}}^2(t)$ for $\Psi_G$  are reduced to 1.5 and and 1, respectively. But for  $\Psi_Y$,  $\dfrac{1}{N}\Delta_{{\hat{L}_Z}}^2(t)$ remains $7$ at $t=0$ for the resonant values of $c$.  For the symmetric double-well potential, we find that $\dfrac{1}{N}\Delta_{{\hat{L}_Z}}^2(t)$ of $\Psi_G$  oscillates  with  amplitude of fluctuations in the order of $10^{-1}$ at the mean-field level. A rather small difference, marking the collapse of the density oscillations, is present between the time evolutions of  $\dfrac{1}{N}\Delta_{{\hat{L}_Z}}^2(t)$ at the mean-field and many-body levels. Unlike for $\Psi_G$, interesting many-body features are found for  $\Psi_Y$ at $c=0$. It is noticed that the amplitude of the many-body $\dfrac{1}{N}\Delta_{{\hat{L}_Z}}^2(t)$  of $\Psi_Y$ initially grows and, as time passes by, it decays (at around $t=12t_{Rabi}$) for the symmetric double-well potential. The  maximal fluctuations on top of the baseline of $\dfrac{1}{N}\Delta_{{\hat{L}_Z}}^2(t)$ at the many-body level is 34\%  where as at the mean-field level it is 7\%. Contrary to the symmetric double-well potential, at  resonant tunneling, the angular-momentum variance shows a sudden increase in the amplitude both for $\Psi_G$ and  $\Psi_Y$. The mean-field and many-body dynamics of $\dfrac{1}{N}\Delta_{{\hat{L}_Z}}^2(t)$ are oscillatory in nature at the resonant tunneling values of $c$. At $c=0.25$, the mean-field $\dfrac{1}{N}\Delta_{{\hat{L}_Z}}^2(t)$ oscillates between 1.5 and about 3.5 for $\Psi_G$ and, between 5 and about 13  for $\Psi_Y$. At the second resonant tunneling, the amplitude of the mean-field of $\dfrac{1}{N}\Delta_{{\hat{L}_Z}}^2(t)$ further increases and oscillates between 1 and approximately 6 for $\Psi_G$, and between  5 and approximately 22 for $\Psi_Y$. We notice that $\Psi_Y$ has always larger fluctuations  of amplitude of $\dfrac{1}{N}\Delta_{{\hat{L}_Z}}^2(t)$ compared to $\Psi_G$ at a fixed value of $c$. The many-body $\dfrac{1}{N}\Delta_{{\hat{L}_Z}}^2(t)$ overlaps with the corresponding mean-field results at short time scales and shows a decay in amplitude as fragmentation grows, which is more evident at $c=0.25$. The rate of decay of $\dfrac{1}{N}\Delta_{{\hat{L}_Z}}^2(t)$ values is slower for $\Psi_Y$, which is consistent with the survival probability and  $\dfrac{1}{N}\Delta_{{\hat{P}_X}}^2(t)$.

Until now, we determined the effect of the transverse excitations on the  dynamics of $\Psi_G$ and $\Psi_Y$ at the resonant tunneling condition within the mean-field and many-body levels of theory. The effects of the transverse degrees-of-freedom are analyzed in terms of various physical quantities.  The results prove that the transverse direction  have substantial influences on the dynamics of the ground as well as excited states at the longitudinal resonant tunneling scenario in two spatial dimensions.

\subsection{Transversal resonant tunneling}

Proceeding to the transversal resonant tunneling, we want to emphasize  that this scenario does  only exist  in 2D, thereby bringing to the front  new degrees-of-freedom with the transverse direction, which we would like to explore. We have found that transverse excitations play a substantial role in longitudinal resonant tunneling. Here, we ask how and in what  capacity longitudinal excitations play a role in transversal resonant tunneling.

\begin{figure*}[!h]
{\includegraphics[trim = 0.1cm 0.5cm 0.1cm 0.2cm, scale=.70]{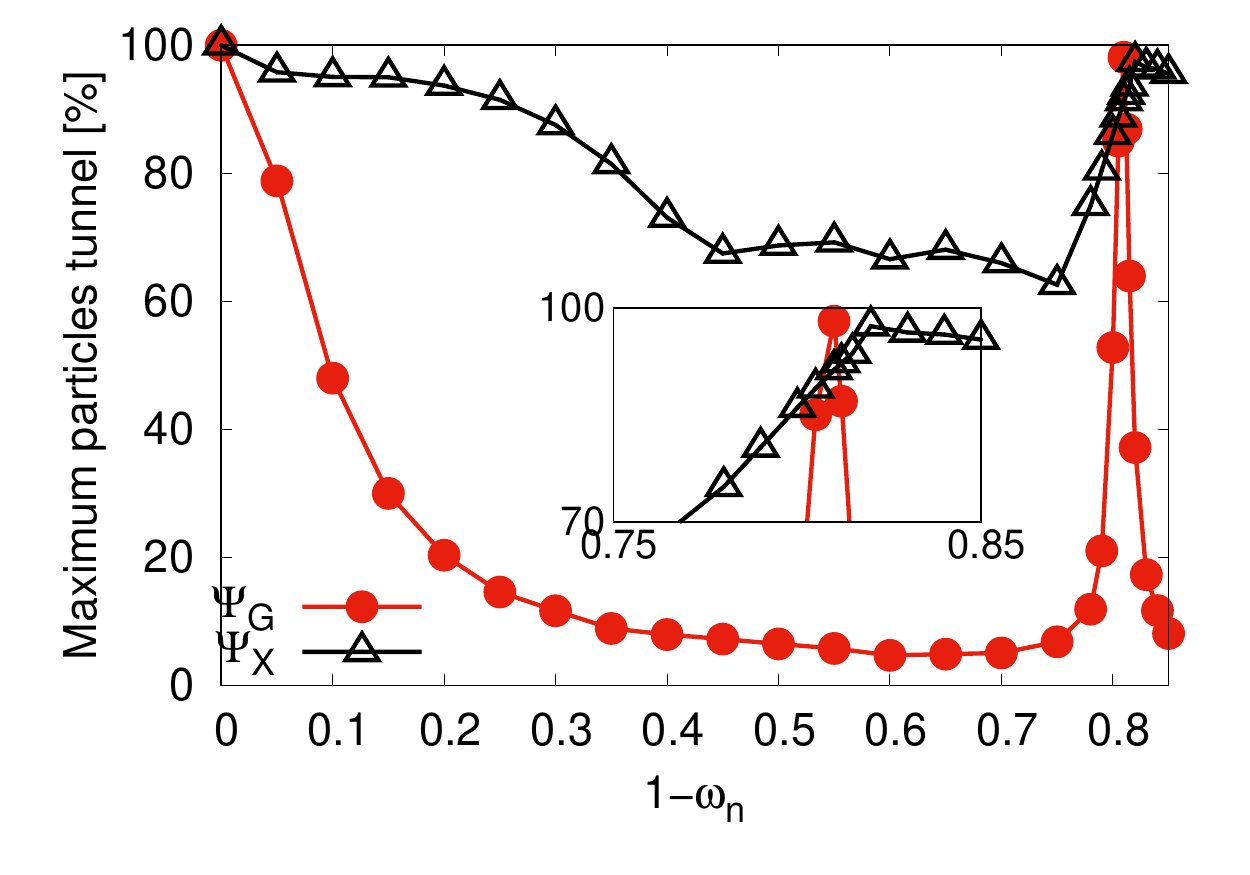}}\\
\caption{Variation of the maximal number of particles tunneling [in \%] from the left to the right well with the trapping $\omega_n$ of the right well  for $\Psi_G$ and $\Psi_X$ (see Fig.~\ref{Fig1}(b)). The  interaction parameter is  $\Lambda=0.01\pi$. The same plot is highlighted in the inset by enlarging the resonance region.   The data is generated from the time-dependent solution of the Gross-Pitaeveskii equation. We show here  dimensionless quantities.}
\label{Fig8}
\end{figure*}

At first,  we examine how the transversal resonant tunneling can be achieved for  two initial bosonic structures,  $\Psi_G$ and $\Psi_X$, in a transversely-asymmetric 2D double-well potential presented in Eq.~\ref{5}. Here, we gradually decrease  the transverse frequency in the right well, $\omega_n$, from an initial value ($\omega_n=1$) which represents a symmetric double-well potential. Similarly to the longitudinal resonant tunneling, here also, the initial states are prepared in the left well of the double-well potential. Fig.~\ref{Fig8} presents the maximal number of particles tunneling to the right well for $\Psi_G$ and $\Psi_X$. The maximal number is extracted from the time-dependent solution of the Gross-Pitaeveskii equation using $\int\limits_{x=0}^{+\infty}\int\limits_{y=-\infty}^{+\infty}dx dy \dfrac{\rho(x,y;t)}{N}$, where  $\rho(x,y;t)$ is the density of the bosonic cloud. One can see from Fig.~\ref{Fig8}, at $\omega_n=1$, that the maximal number of particles tunneling to the right well is 100\%.    Decreasing the value of $\omega_n$ reduces the number of maximum particles which can tunnel to the right well for both  initial bosonic clouds. The rate of decay of the maximal number of particles tunneling to the right well with $\omega_n$ is  smaller for $\Psi_X$ as it   essentially lies in the first excited band along the direction of the barrier and thus feels a smaller barrier when it tunnels. Further decrease of $\omega_n$, the maximal  number of  particles tunneling to the right well grows. When the one-body  energy of the left well coincides with one of the  one-body higher energy levels of the right well, again the maximal number of particles tunneling to the right well reaches to almost 100\%. We find only one resonant tunneling for the span of $\omega_n$ considered here.  For $\Psi_G$ and $\Psi_X$, the transversal resonant tunneling occurs at $\omega_n=0.19$ and $\omega_n=0.18$, respectively. Also, it can be observed from the figure that for slightly off-resonant values of $\omega_n$, the maximal number of particles tunneling to the right well significantly decreases in comparison to the respective value of $\omega_n$ at resonant tunneling for $\Psi_G$, but  it varies much weaker with  $\omega_n$  for $\psi_X$.

\begin{figure*}[!h]
{\includegraphics[trim = 3.5cm 0.5cm 1.5cm 0.2cm, scale=.95]{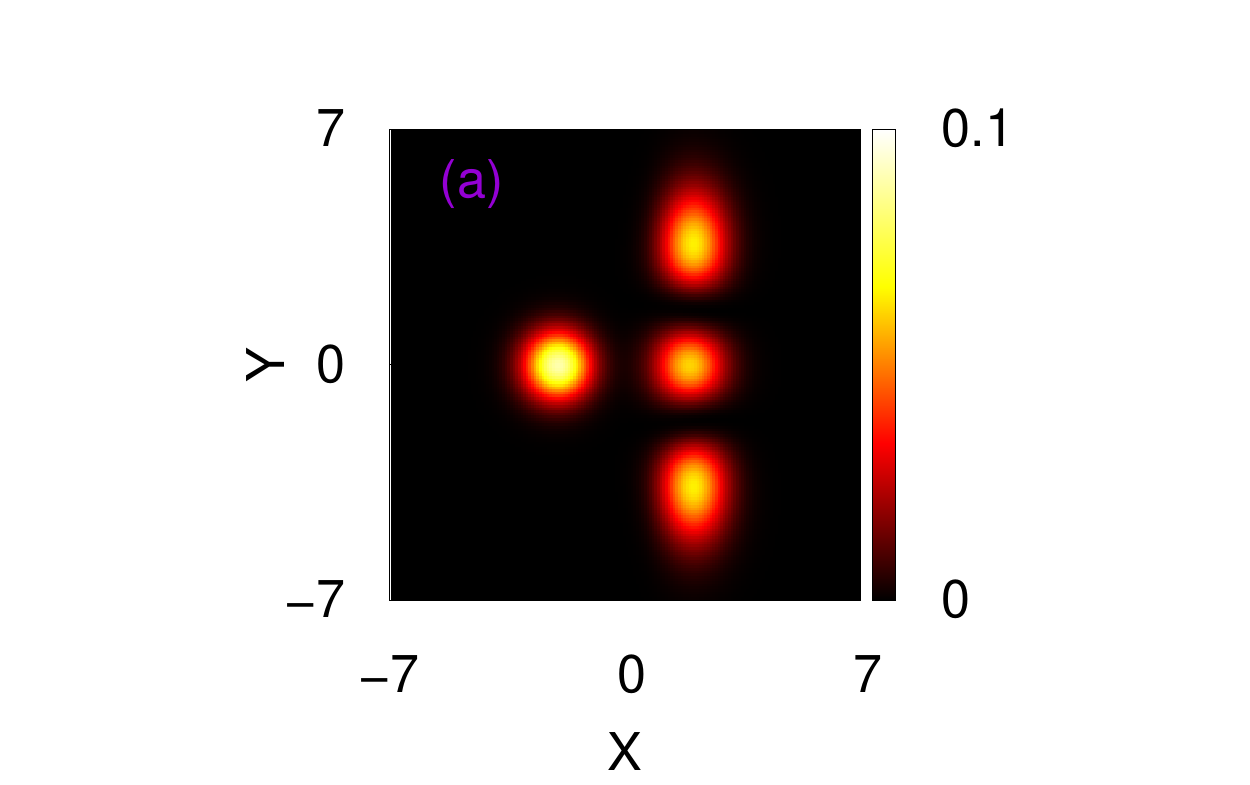}}
{\includegraphics[trim = 3.5cm 0.5cm 1.5cm 0.2cm, scale=.95]{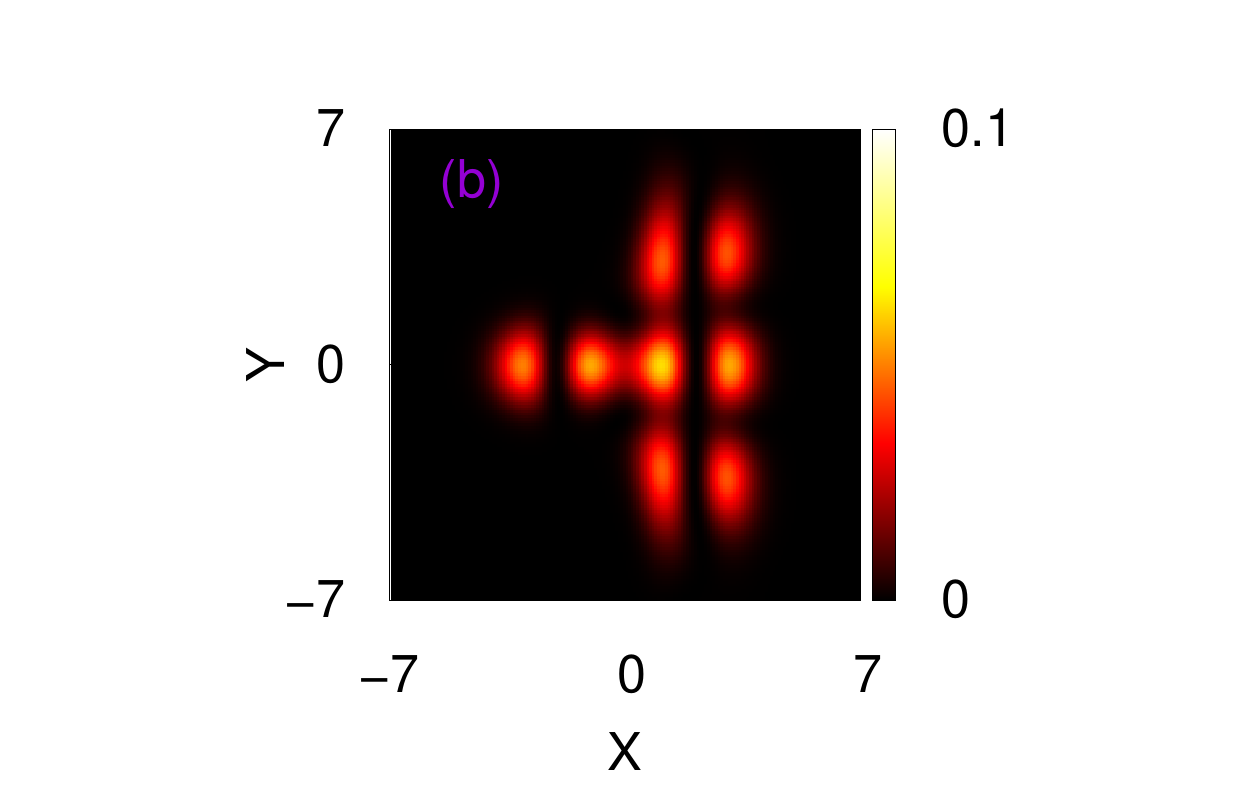}}\\
\caption{The densities per particle for the initial  states (a) $\Psi_G$ and (b) $\Psi_X$. The interaction parameter is $\Lambda=0.01\pi$. The plots are generated at  $t\approx t_{Rabi}$ from the time-dependent solution of the Gross-Pitaeveskii equation.  We show here  dimensionless quantities.}
\label{Fig9}
\end{figure*}

The transversal resonant tunneling for $\Psi_G$ found in Fig.~\ref{Fig8} is when $\Psi_G$ of the left well couples to $(y^2-1)\Psi_G$ of the right well.  Similarly,  $\Psi_X$ of the left well couples to $(y^2-1)\Psi_X$ of the right well to facilitate the transversal resonant tunneling of $\Psi_X$, see Fig.~\ref{Fig9}. The resonant tunneling channels $\Psi_G \rightarrow y\Psi_G$ and $\Psi_X \rightarrow y\Psi_X$ are symmetry forbidden, at least at the mean-field level, as the   time-dependent orbital becomes odd with respect to $y\rightarrow -y$. Here we select three  values of $\omega_n$ which are either at the resonant tunneling value or close to it, say, $\omega_n=0.18$, 0.19, and 0.20, and investigate the overall dynamical response through  the surviaval probability, loss of coherence, and the variances of position, momentum, and angular-momentum operators. To show the effects of the many-body correlations on different physical quantities, we compare the results of many-body dynamics with corresponding results at the mean-field level.

\begin{figure*}[!h]
{\includegraphics[trim = 0.1cm 0.5cm 0.1cm 0.2cm, scale=.60]{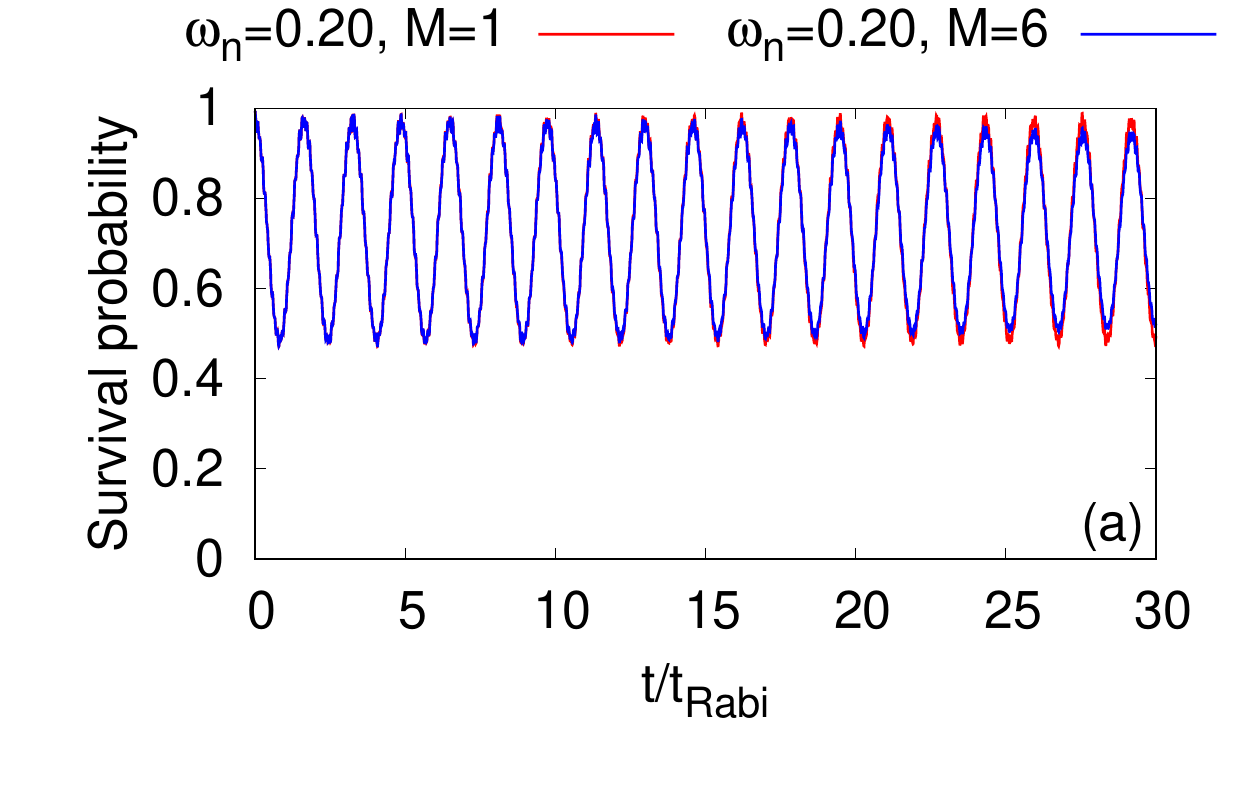}}
{\includegraphics[trim = 0.1cm 0.5cm 0.1cm 0.2cm, scale=.60]{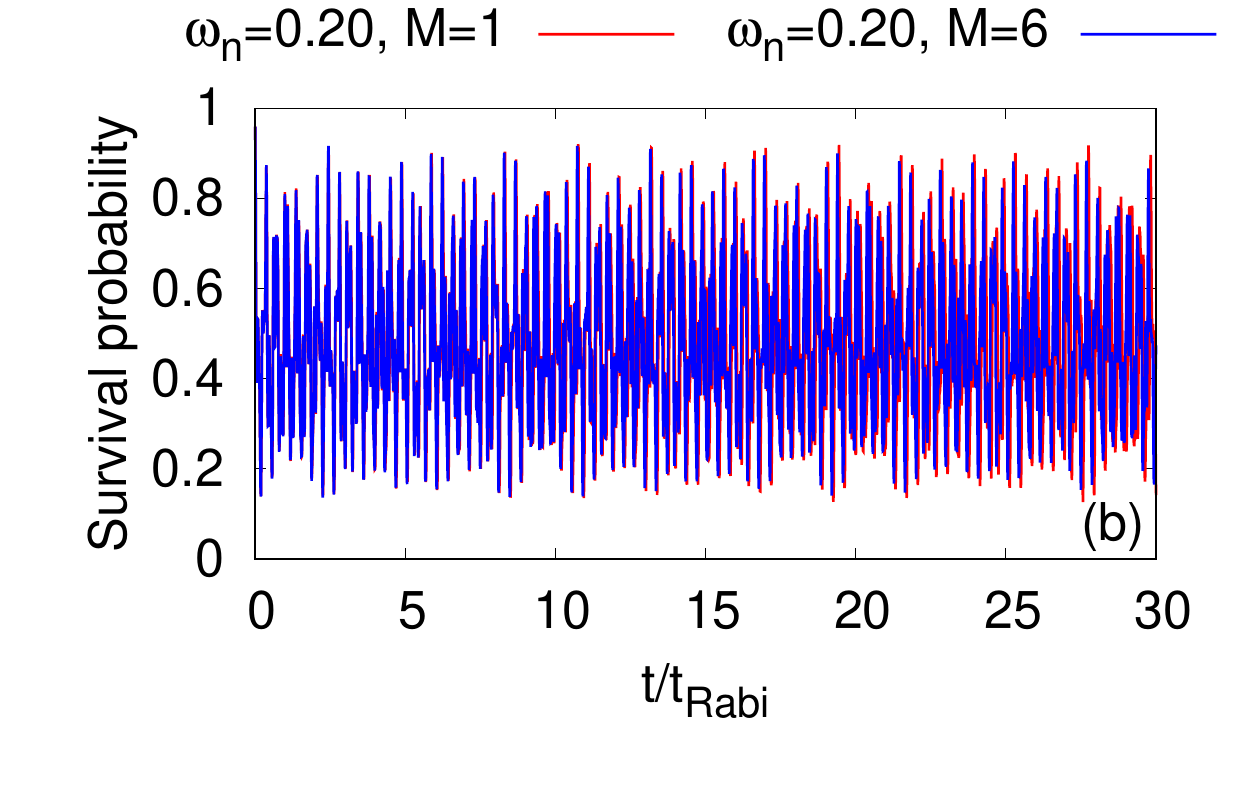}}\\
{\includegraphics[trim = 0.1cm 0.5cm 0.1cm 0.2cm, scale=.60]{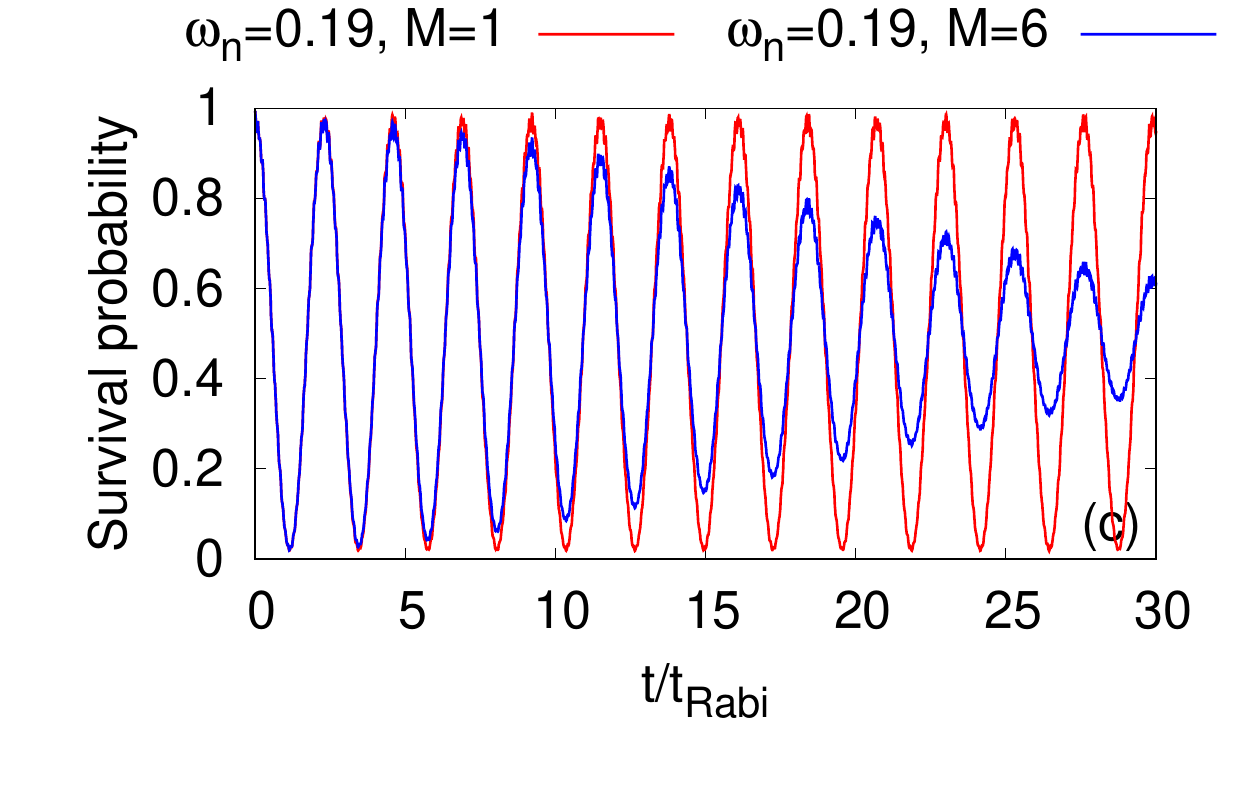}}
{\includegraphics[trim = 0.1cm 0.5cm 0.1cm 0.2cm, scale=.60]{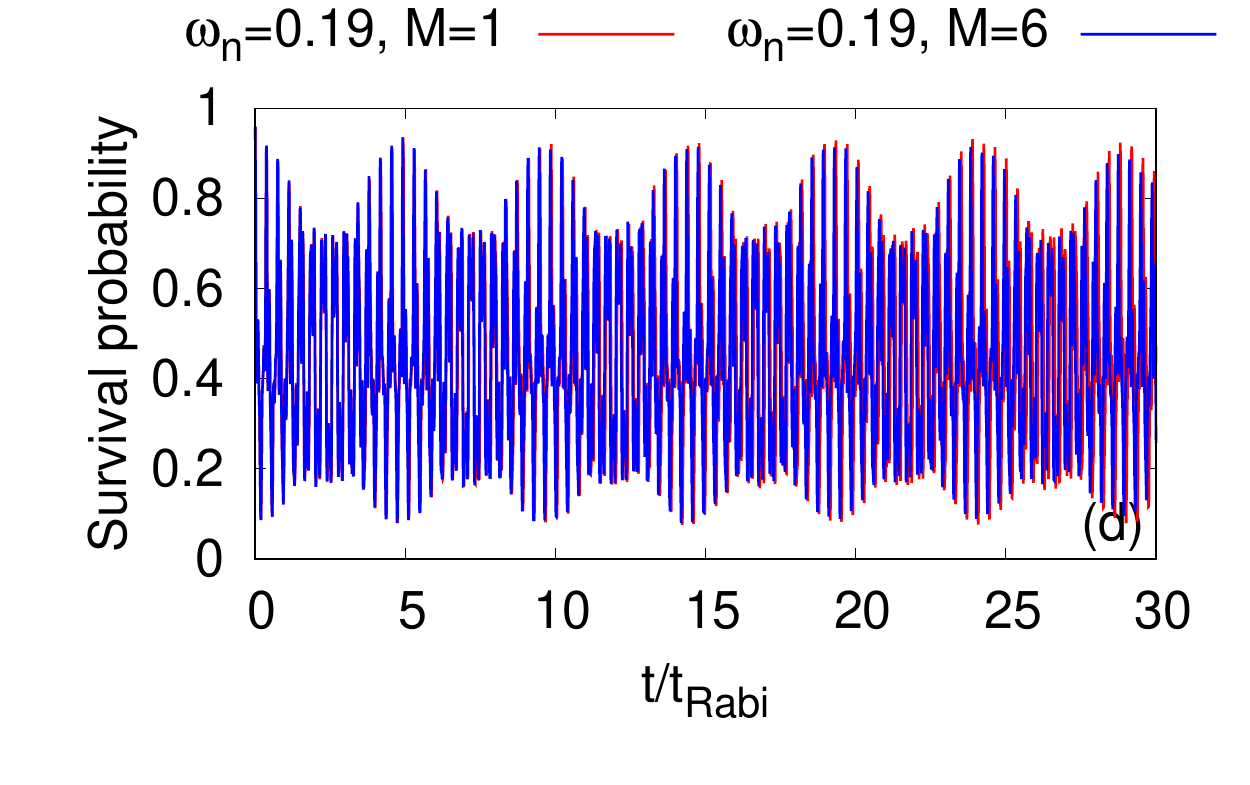}}\\
{\includegraphics[trim = 0.1cm 0.5cm 0.1cm 0.2cm, scale=.60]{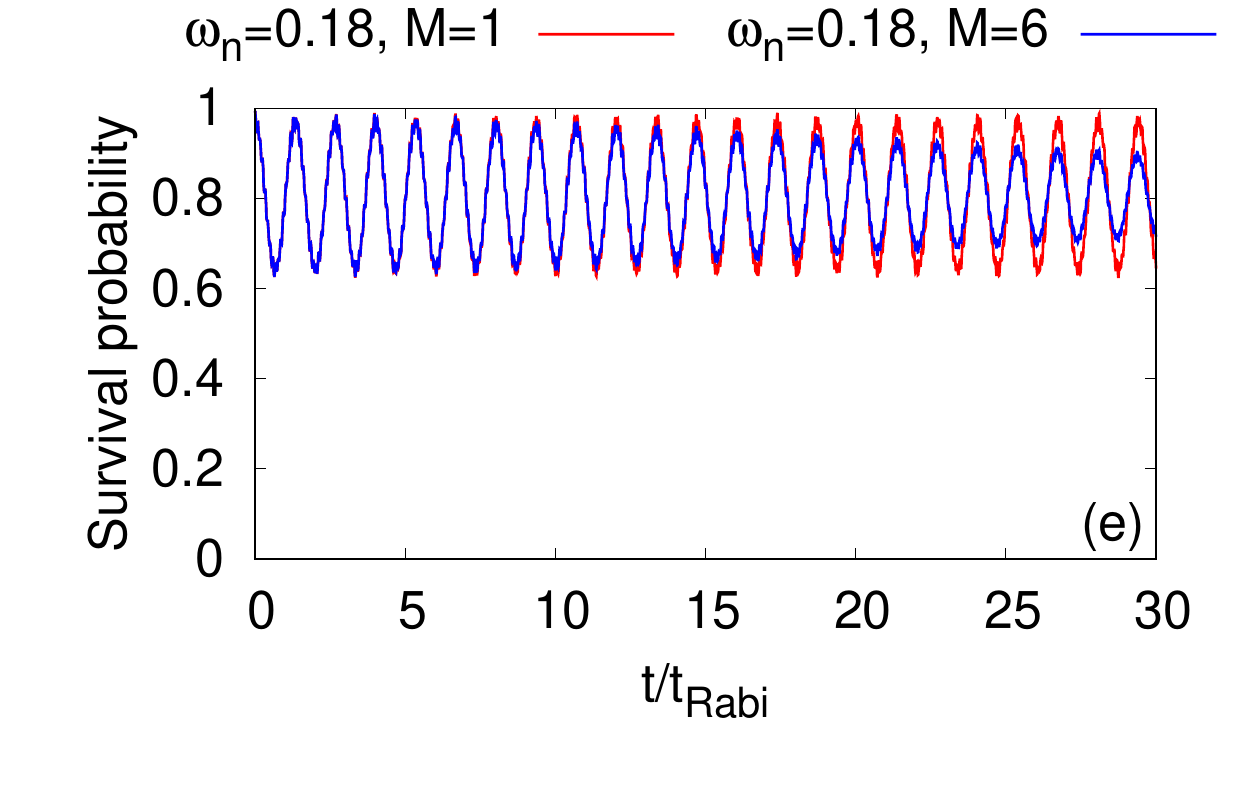}}
{\includegraphics[trim = 0.1cm 0.5cm 0.1cm 0.2cm, scale=.60]{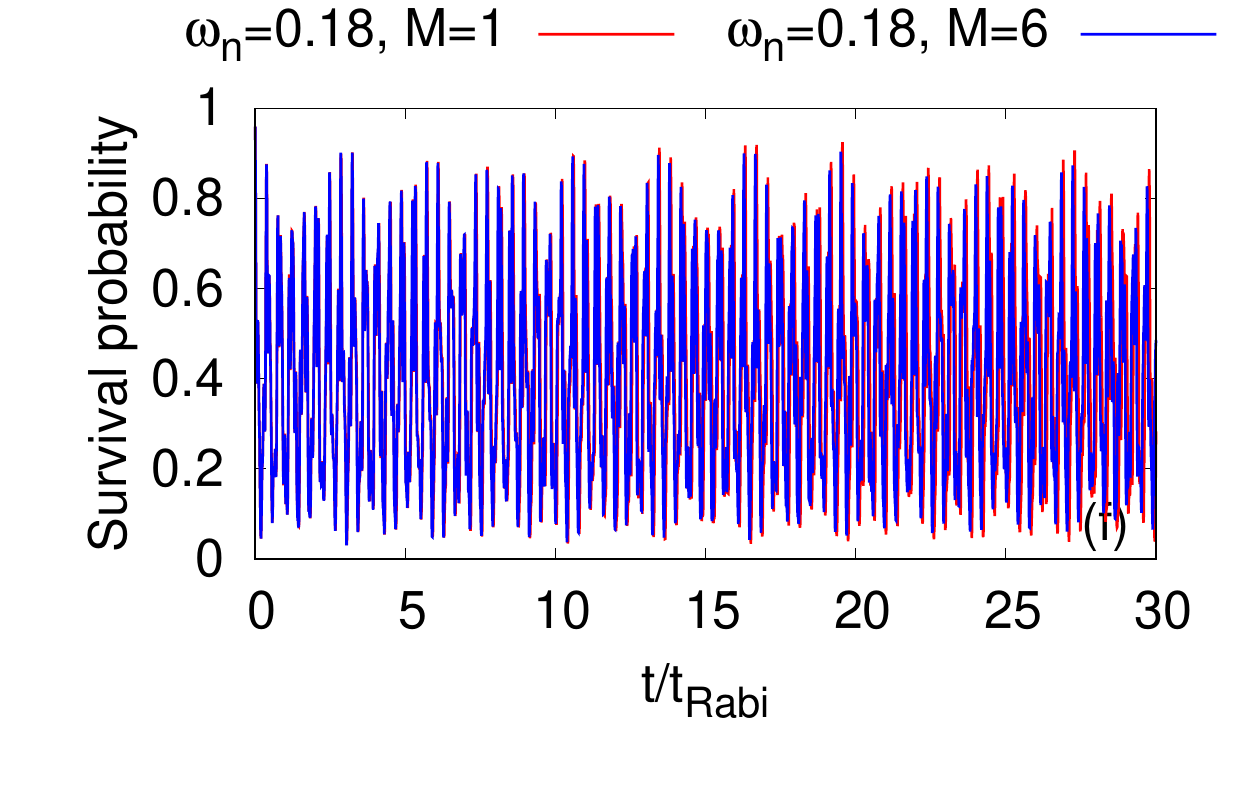}}\\
\caption{Dynamics of the survival probability of $\Psi_G$ (left column) and $\Psi_X$ (right column) in the left well of a transversely-asymmetric 2D double-well potential of $N=10$ bosons  with the interaction parameter $\Lambda=0.01\pi$. First, second, and third rows  show the  results at  frequency, $\omega_n=0.20$, 0.19, and 0.18, respectively.  $M=1$ signifies the mean-field results.   The many-body dynamics are computed   with $M= 6$ time-dependent orbitals.   We show here  dimensionless quantities. Color codes are explained in each panel. }
\label{Fig10}
\end{figure*}

Now, we start with the discussion of the  density oscillations in terms of the survival probability in the left well for both the states $\Psi_G$ and $\Psi_X$. Fig.~\ref{Fig10}(c) and (f) show the dynamical behaviors of the survival probabilities at the resonant tunneling conditions for  $\Psi_G$ and $\Psi_X$, respectively. We notice that the frequencies of tunneling of the density  between the left and right wells are different for $\Psi_G$ but are essentially same for $\Psi_X$ for different $\omega_n$ considered here. At $\omega_n=0.18$, 0.19, and 0.20, the tunneling frequencies for $\Psi_G$ are 1.41$t_{Rabi}$, 2.30$t_{Rabi}$, and 1.55$t_{Rabi}$, respectively.  While for $\Psi_X$, the tunneling frequency is around 0.38$t_{Rabi}$ for all  three $\omega_n$ values.

At a fixed value of  $\omega_n$, both the frequency and amplitude of the survival probability at the mean-field and many-body levels are practically the same for $\Psi_X$. Whereas for $\Psi_G$, although the frequency of the survival probability  are very similar at the mean-field and many-body levels, the amplitudes of the survival probability show a significant deviation which is a maximal at $\omega_n=0.19$. The gradual decay at the  many-body  level of dynamics of the survival probability  signifies the growing degree of quantum correlations which will be quantitatively discussed later  in terms of the loss of coherence in the system.     All in all,   Fig.~\ref{Fig10} reflects that correlations develops  faster for $\Psi_G$ compared to $\Psi_X$.

\begin{figure*}[!h]
{\includegraphics[trim = 0.1cm 0.5cm 0.1cm 0.2cm, scale=.60]{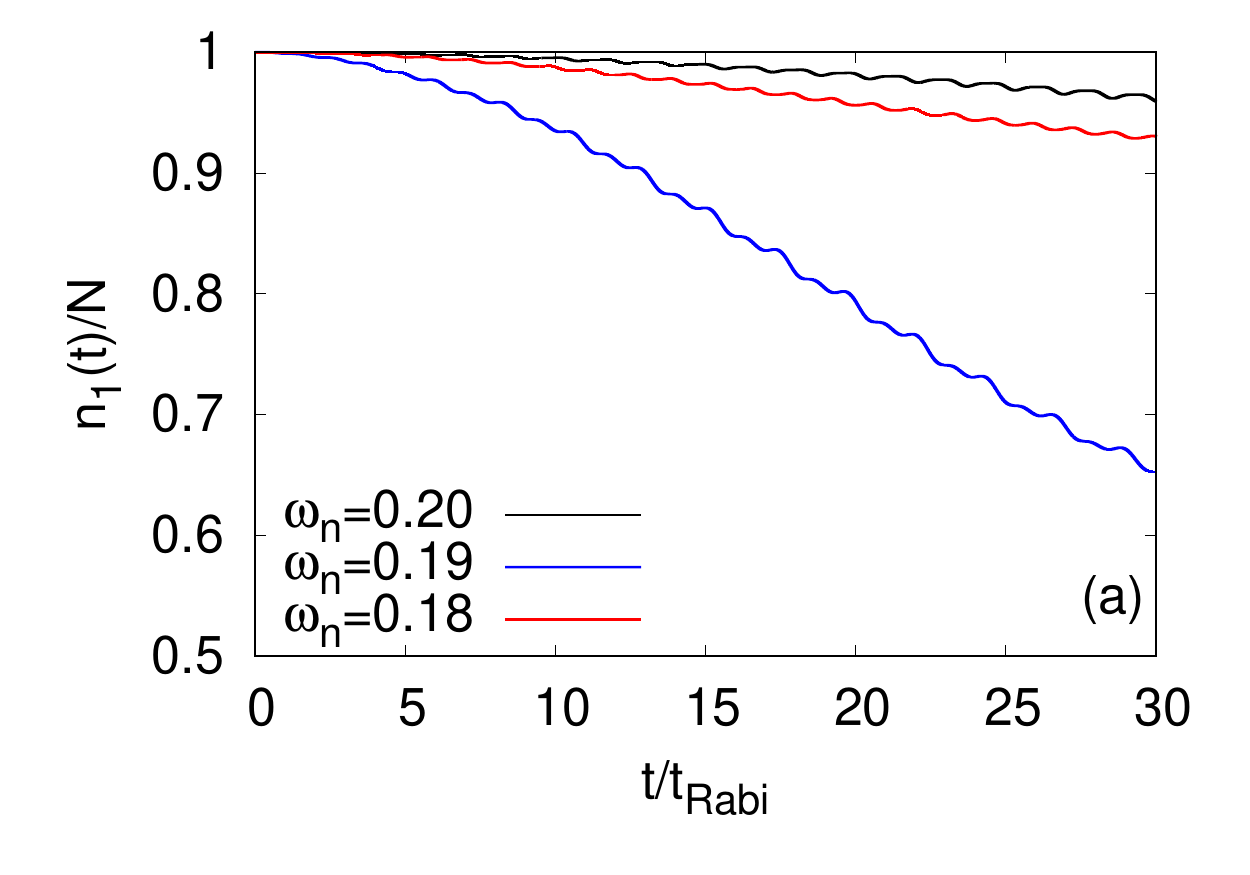}}
{\includegraphics[trim = 0.1cm 0.5cm 0.1cm 0.2cm, scale=.60]{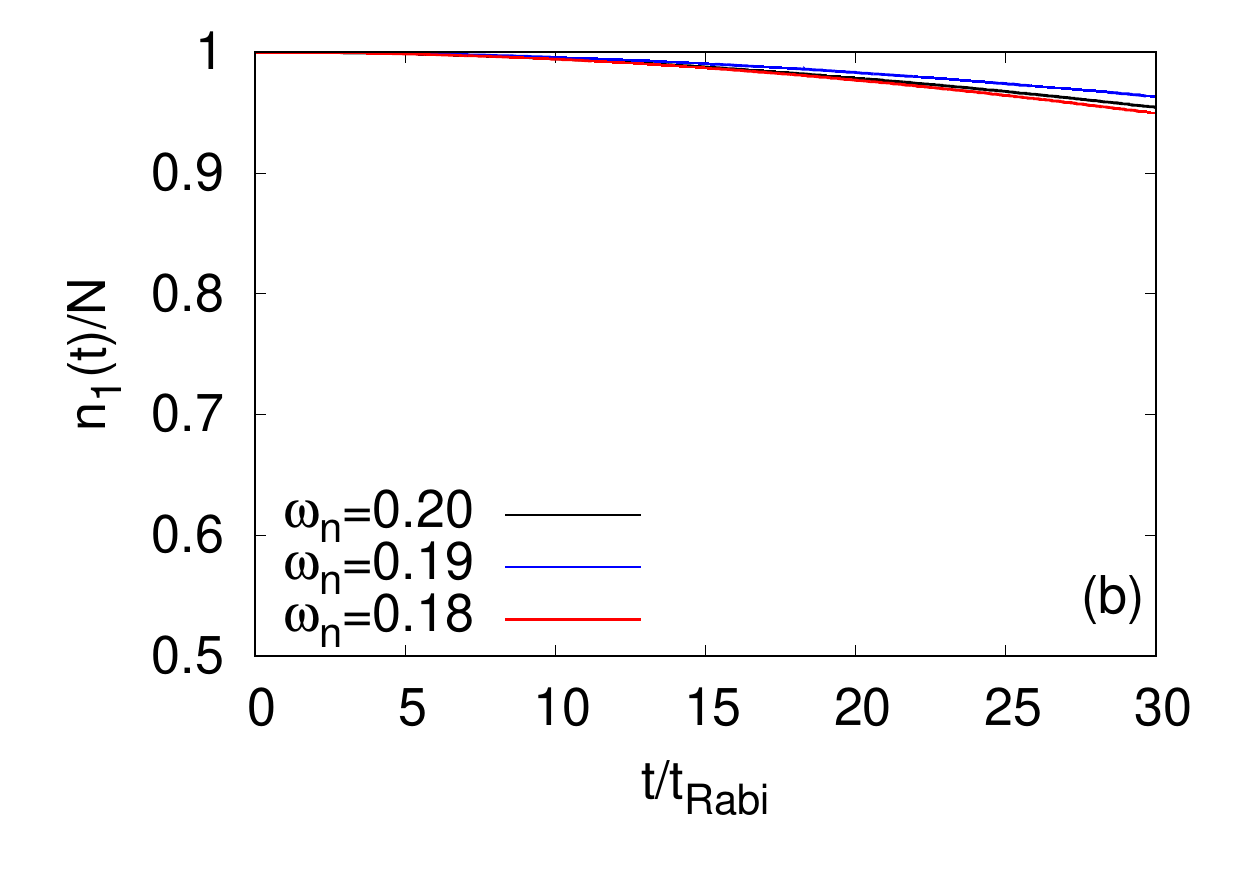}}\\
\caption{Time-dependent occupation per particle of the first natural orbital, $\dfrac{n_1(t)}{N}$,  in a transversely-asymmetric 2D double-well potential. The number of bosons is $N=10$  and  the interaction parameter $\Lambda=0.01\pi$. Panels (a) and (b) are for $\Psi_G$ and $\Psi_X$, respectively.   The data have been obtained  with $M = 6$ time-dependent orbitals.    We show here  dimensionless quantities. Color codes are explained in each panel.}
\label{Fig11}
\end{figure*}

 Now, we demonstrate  the effect of growing degree of correlations on the loss of coherence of the bosonic clouds, $\Psi_G$ and $\Psi_X$, at the transversal resonant tunneling  and also at its vicinity. In Fig.~\ref{Fig11}, we display the time-dependent occupation of the first natural orbital,  namely the condensate fraction, $\dfrac{n_1(t)}{N}$. As observed in the longitudinal resonant tunneling, here also, $\dfrac{n_1(t)}{N}$ decays with time with  a weak oscillatory background. As $\Psi_G$ lies in the lowest band, it feels the barrier more  and yields comparatively strong background oscillations than $\Psi_X$. It is observed that both  states lose their coherence faster at the resonant value of $\omega_n$ and the loss of coherence becomes gradually slower away from the resonant tunneling. The loss of coherence at the resonant tunneling are seen to be almost 35\% and 5\% for   $\Psi_G$ and $\Psi_X$, respectively, at the maximum time ($t=30t_{Rabi}$) considered here.  Side by side, the loss of coherence in the system is accompanied by a  growing degree of fragmentation, leading to increase of the occupancy of higher natural orbitals.   See the supplemental materials for the detailed analysis of the mechanism of fragmentation in transversal resonant tunneling as well as to verification of convergence.

To   further characterize the many-body correlations at  the transversal resonant tunneling condition for $\Psi_G$ and $\Psi_X$, we graphically present the dynamical behavior of the variances of  the position and  momentum operators along the $x$- and $y$-directions, and the angular-momentum operator. Also, the many-body results are compared with the corresponding mean-field results. These physical quantities at other values of $\omega_n$, i.e., 0.18 and 0.20 for $\Psi_G$ and  0.19 and 0.20 for $\Psi_X$,  along with their convergences are demonstrated in the supplemental materials.

Fig.~\ref{Fig12} displays the dynamics of the many-particle position and momentum variances per particle along the $x$-direction, $\dfrac{1}{N}\Delta_{\hat{X}}^2(t)$ and $\dfrac{1}{N}\Delta_{{\hat{P}_X}}^2(t)$, respectively,  at the transversal resonant tunneling  conditions of $\Psi_G$ and $\Psi_X$. The panels show that the  fragmentation developed in the system yields the deviation of the many-body results compared to the corresponding mean-field one. This deviation is  larger for $\Psi_G$ in comparison with $\Psi_X$, which is consistent with the corresponding survival probability. For $\Psi_G$, the  many-body dynamics of  $\dfrac{1}{N}\Delta_{\hat{X}}^2(t)$ and $\dfrac{1}{N}\Delta_{{\hat{P}_X}}^2(t)$ almost reach at its equilibrium when the density oscillations collapse. It is found  that the frequency of oscillations is practically the same for the mean-field and many-body dynamics of $\dfrac{1}{N}\Delta_{\hat{X}}^2(t)$ and $\dfrac{1}{N}\Delta_{{\hat{P}_X}}^2(t)$. As we have seen in the dynamics of the survival probability, the oscillation frequency of both  quantities is higher for $\Psi_X$ compared to $\Psi_G$.  Moreover, both for $\Psi_G$ and $\Psi_X$, we find that the frequency of oscillations of the survival probability is twice the frequency of oscillations of $\dfrac{1}{N}\Delta_{\hat{X}}^2(t)$ while they are  practically identical   for $\dfrac{1}{N}\Delta_{{\hat{P}_X}}^2(t)$.  An interesting many-body feature is observed for $\Psi_G$ in the time evolution of $\dfrac{1}{N}\Delta_{\hat{X}}^2(t)$. Here, we find  that there are two types of oscillations  taking place, namely, one with high amplitude and the second with a small amplitude. This type of oscillations may arise due to the transition of bosons  from the lower to higher band, which requires a detailed analysis, e.g.  by a many-body linear-response theory \cite{Theisen2016},  which goes beyond the scope of the present work. The  time evolution of $\dfrac{1}{N}\Delta_{{\hat{P}_X}}^2(t)$ is accompanied by  two  types of oscillations,  smaller amplitude with  higher frequency and larger amplitude with lower frequency, but is mostly dominated by the former one which is coming from the  breathing mode oscillations.

\begin{figure*}[!h]
{\includegraphics[trim = 0.1cm 0.5cm 0.1cm 0.2cm, scale=.60]{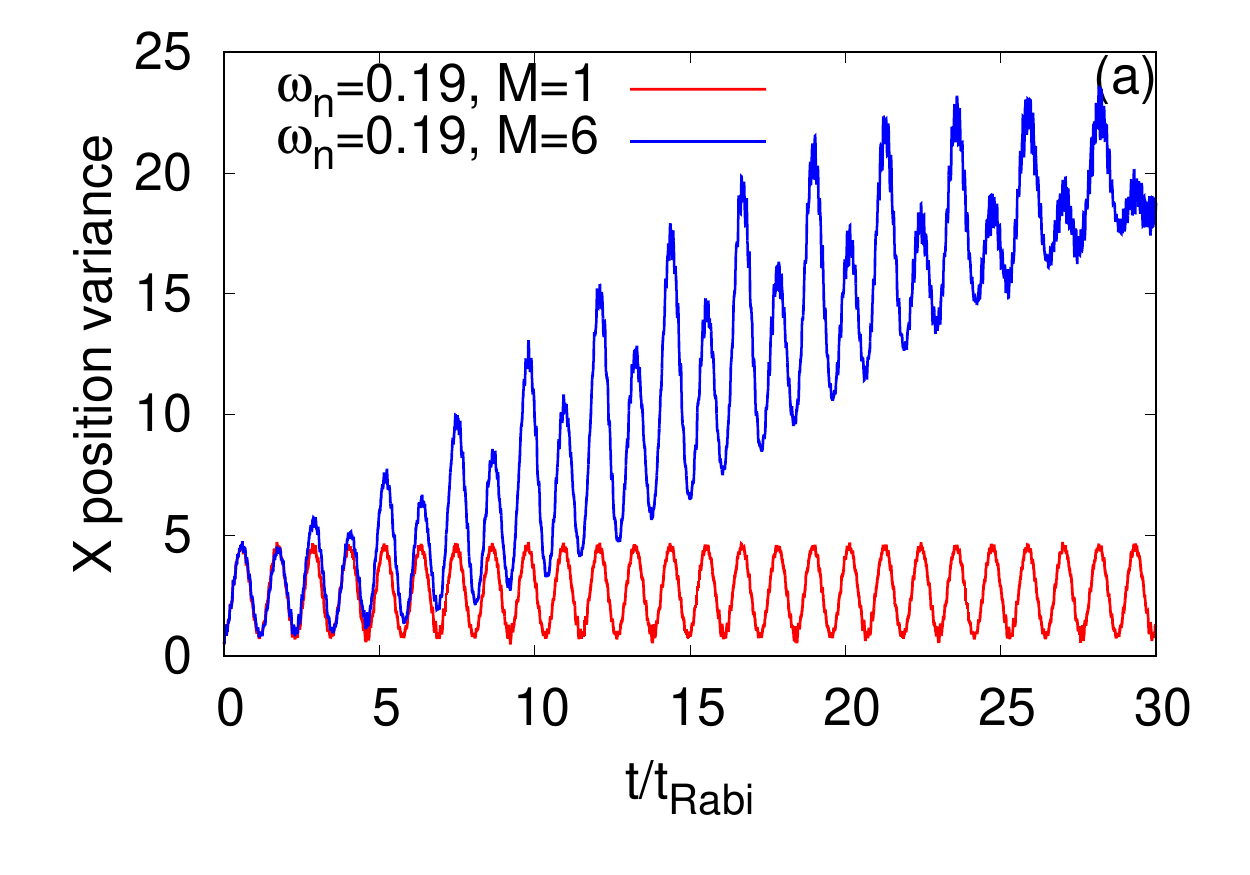}}
{\includegraphics[trim = 0.1cm 0.5cm 0.1cm 0.2cm, scale=.60]{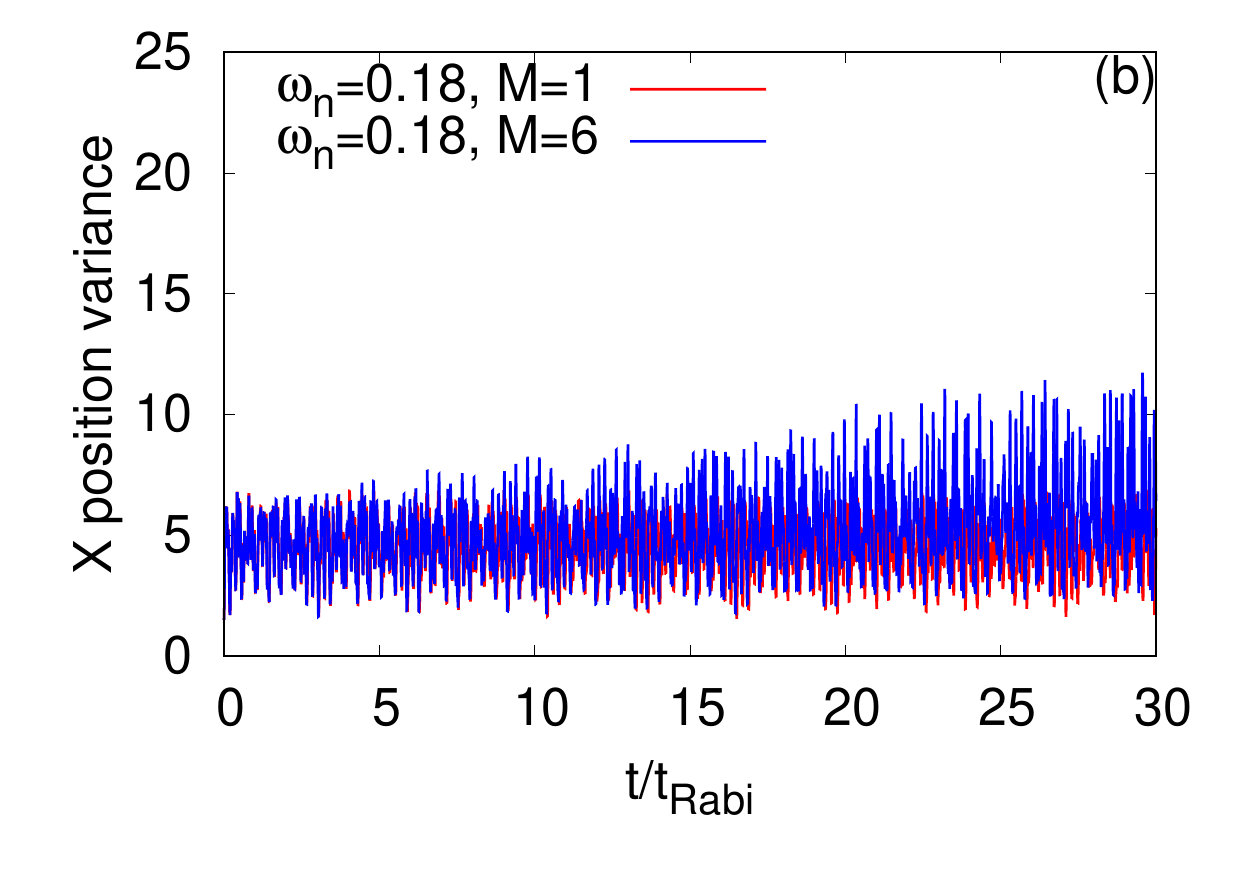}}\\
{\includegraphics[trim = 0.1cm 0.5cm 0.1cm 0.2cm, scale=.60]{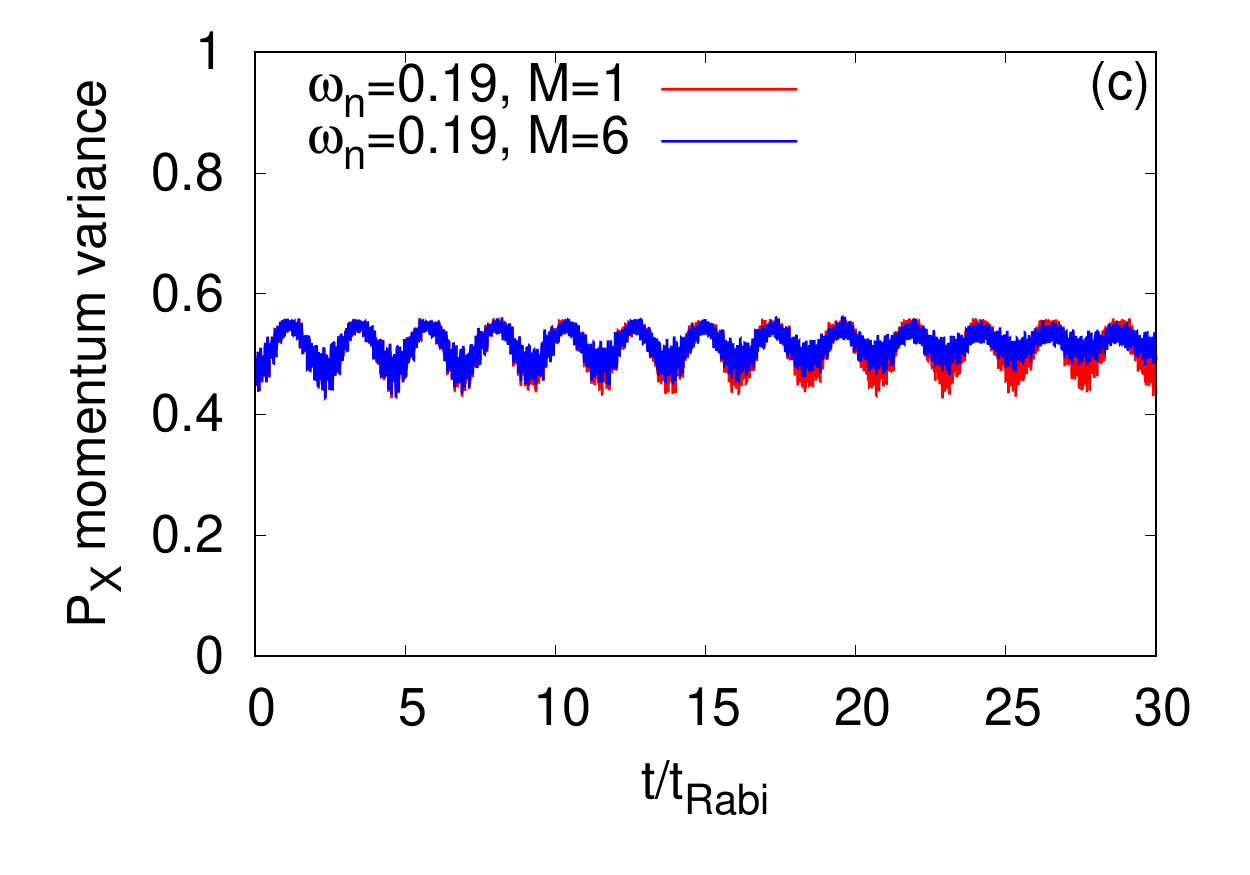}}
{\includegraphics[trim = 0.1cm 0.5cm 0.1cm 0.2cm, scale=.60]{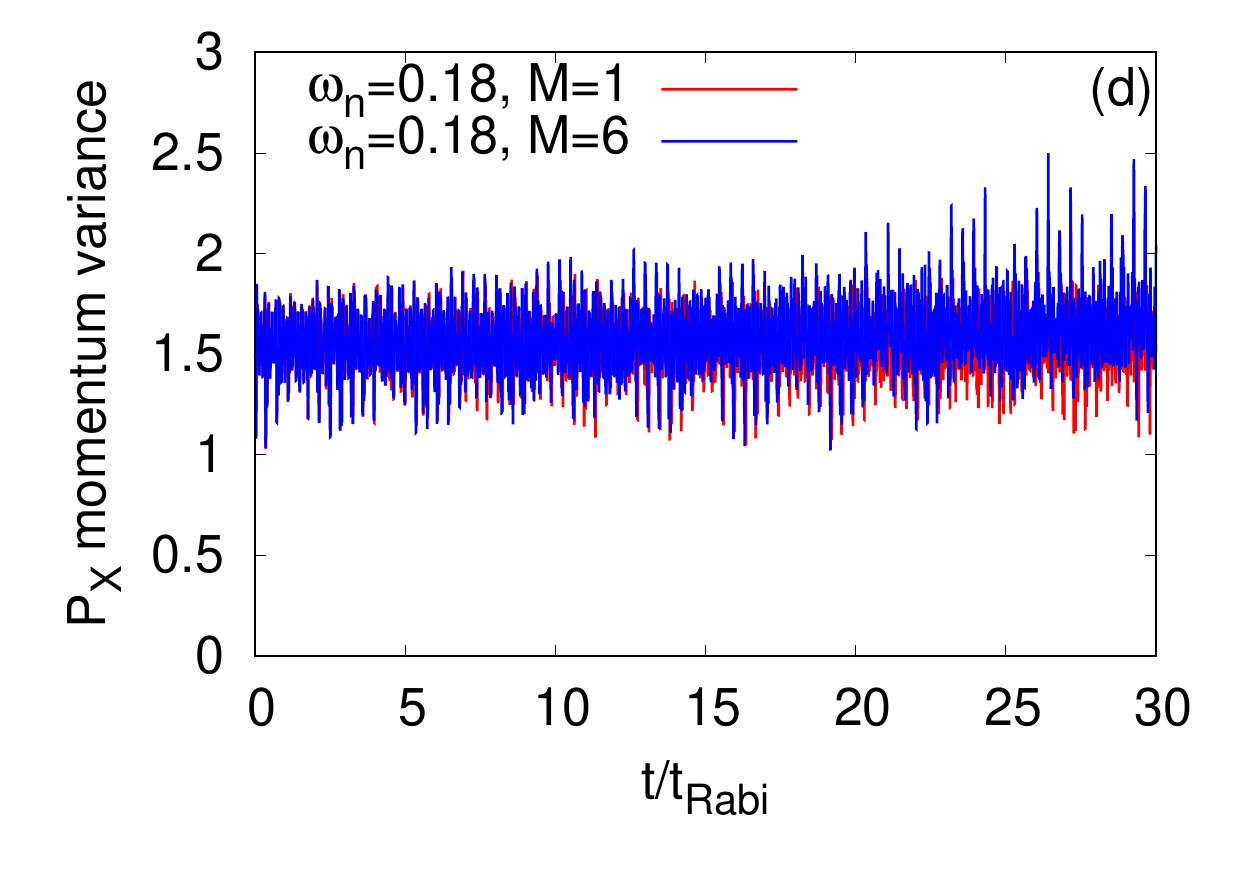}}\\
\caption{Time-dependent position and momentum variances   per particle along the $x$-direction in a transversely-asymmetric 2D double-well potential for $\Psi_G$ (left column, $\omega_n=0.19$) and $\Psi_X$ (right column, $\omega_n=0.18$) under transversal resonant tunneling conditions.  The number of bosons is $N=10$  and the interaction parameter $\Lambda=0.01\pi$.   $M=1$ signifies the mean-field results.   The many-body dynamics are computed  with $M= 6$ time-dependent orbitals.   We show here dimensionless quantities. Color codes are explained in each panel.}
\label{Fig12}
\end{figure*}

\begin{figure*}[!h]
{\includegraphics[trim = 0.1cm 0.5cm 0.1cm 0.2cm, scale=.60]{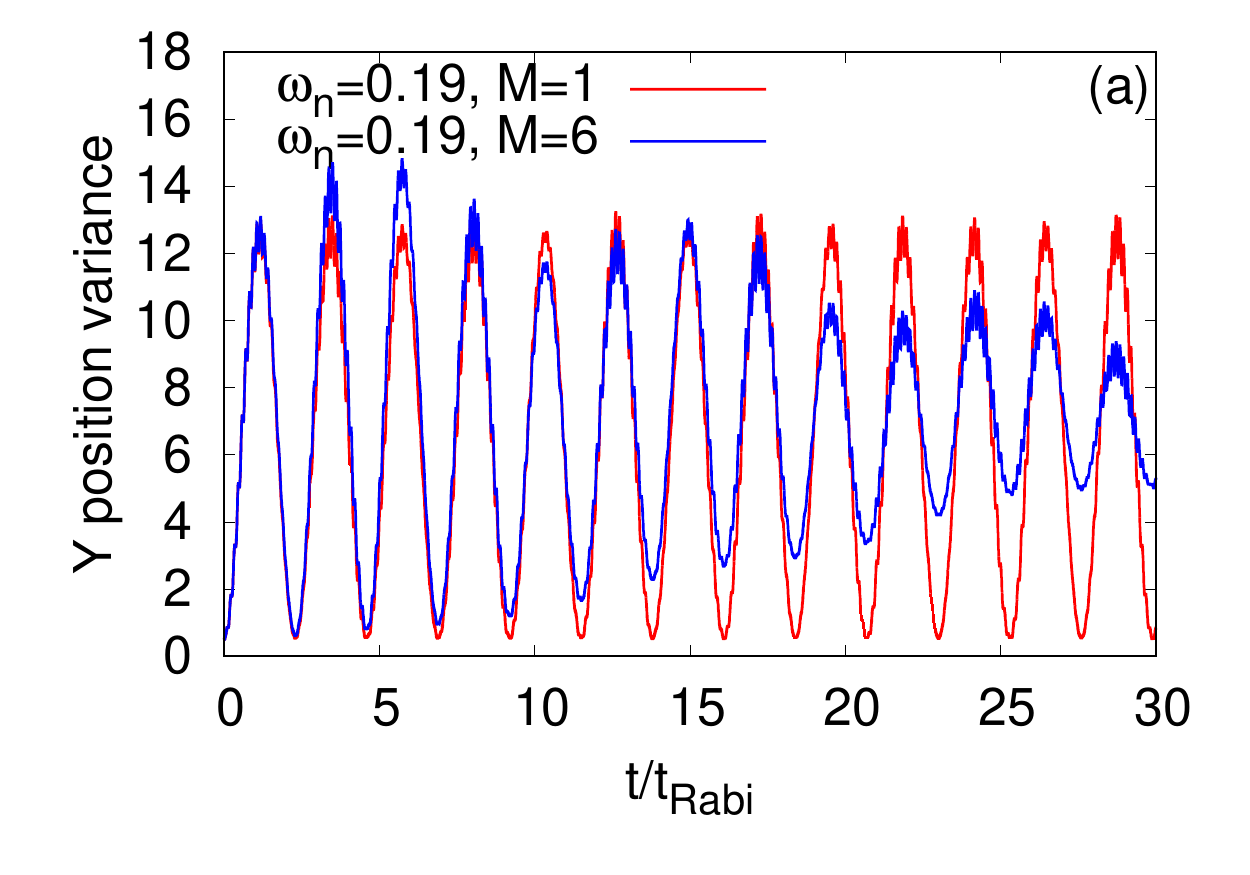}}
{\includegraphics[trim = 0.1cm 0.5cm 0.1cm 0.2cm, scale=.60]{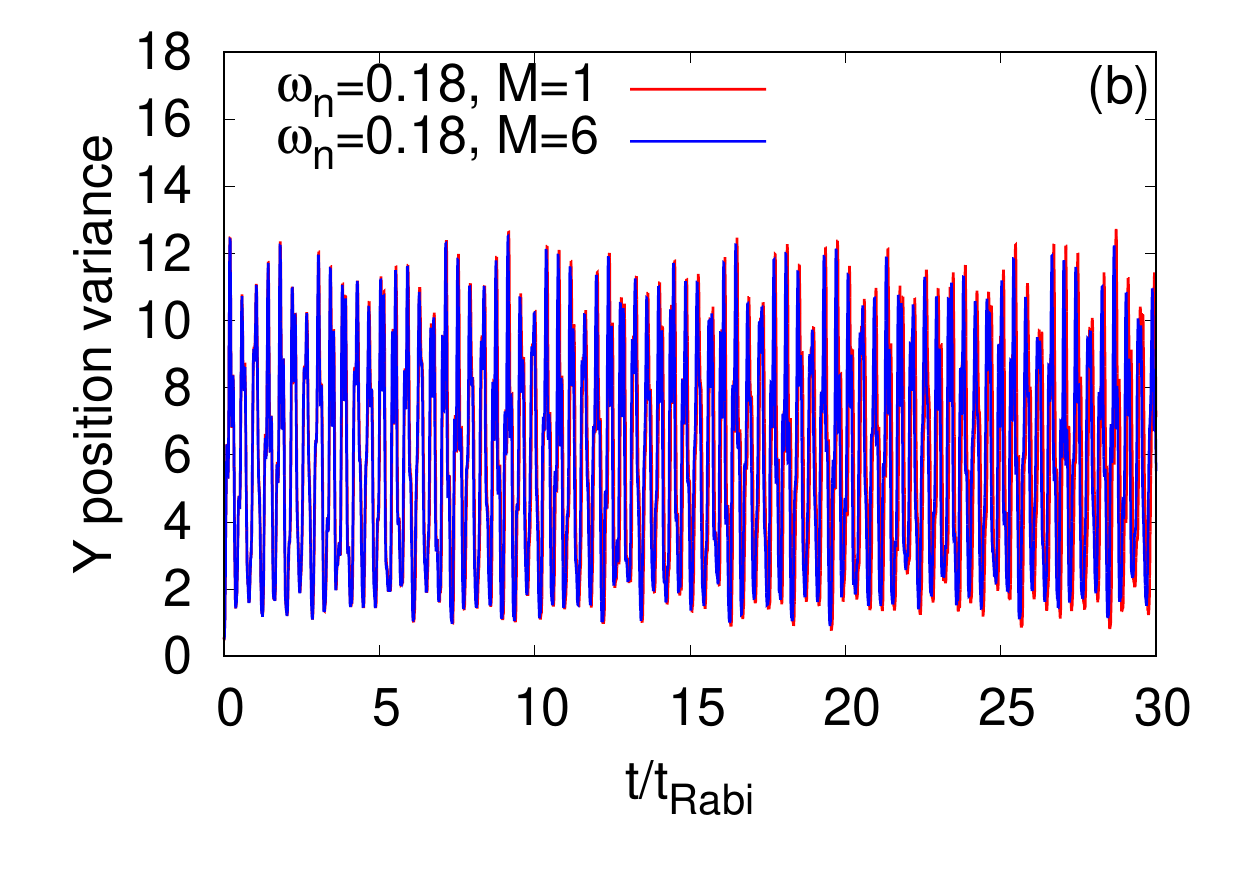}}\\
{\includegraphics[trim = 0.1cm 0.5cm 0.1cm 0.2cm, scale=.60]{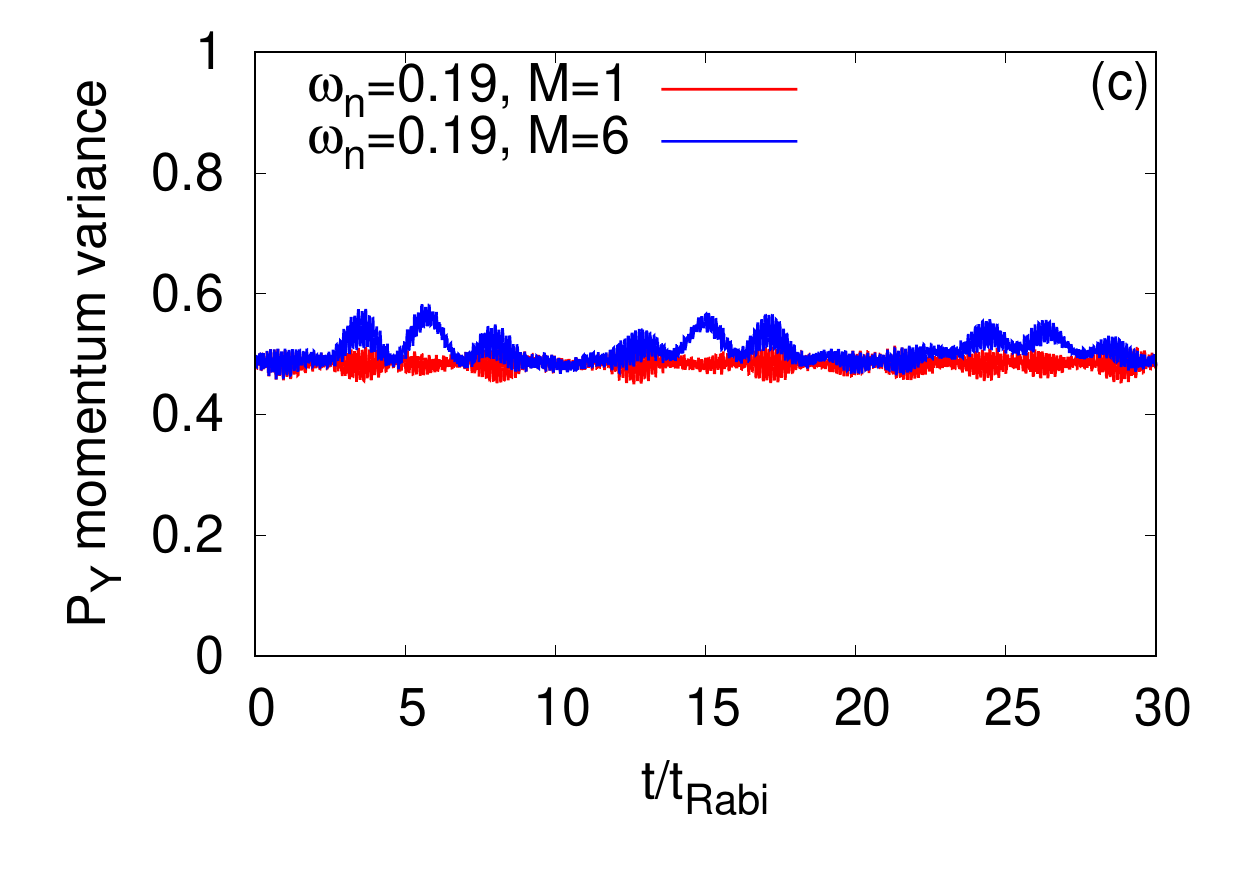}}
{\includegraphics[trim = 0.1cm 0.5cm 0.1cm 0.2cm, scale=.60]{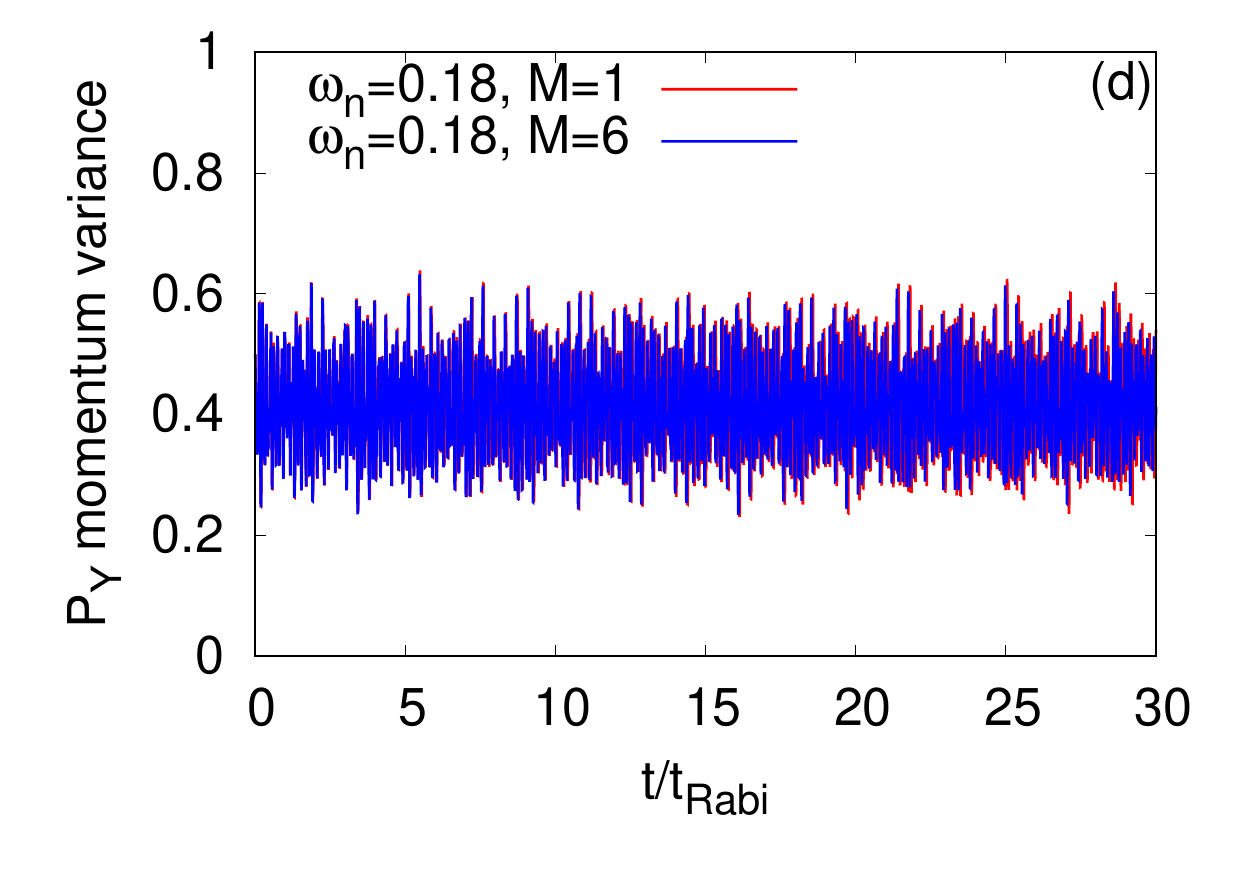}}\\
\caption{Time-dependent position and momentum variance  per particle along the $y$-direction in a transversely-asymmetric 2D double-well potential for $\Psi_G$ (left column, $\omega_n=0.19$) and $\Psi_X$ (right column, $\omega_n=0.18$) under transversal resonant tunneling conditions.  The number of bosons is $N=10$  and the interaction parameter $\Lambda=0.01\pi$.   $M=1$ signifies the mean-field results.   The many-body dynamics are computed  with $M= 6$ time-dependent orbitals.   We show here dimensionless quantities. Color codes are explained in each panel.}
\label{Fig13}
\end{figure*}

Now, let us discuss the role played by  the transversal resonant tunneling of $\Psi_G$ and $\Psi_X$ on the time evolution of the position and momentum variances along the $y$-direction, $\dfrac{1}{N}\Delta_{{\hat{Y}}}^2(t)$ and $\dfrac{1}{N}\Delta_{{\hat{P}_Y}}^2(t)$, respectively (see Fig.~\ref{Fig13}). Unlike in the longitudinal resonant tunneling where the  position and momentum variances along the $y$-direction of $\Psi_G$  have very small fluctuations in amplitude  (of the order of $10^{-3}$, see Fig. S4 in the supplemental materials), here, in the transversal resonant tunneling, we observe comparatively large fluctuations which are clearly more  prominent for $\dfrac{1}{N}\Delta_{{\hat{Y}}}^2(t)$. For both the states, $\Psi_G$ and $\Psi_X$, it is found that the frequency of oscillation of $\dfrac{1}{N}\Delta_{{\hat{Y}}}^2(t)$ overlaps with the corresponding frequency of the survival probability. The amplitude of the many-body time evolution of  $\dfrac{1}{N}\Delta_{{\hat{Y}}}^2(t)$ decays with time,   originating from the growing degree of fragmentation which is more evident for $\Psi_G$. It is clear from  Fig.~\ref{Fig13} that $\dfrac{1}{N}\Delta_{{\hat{Y}}}^2(t)$ is dominated by the density oscillations while  $\dfrac{1}{N}\Delta_{{\hat{P}_Y}}^2(t)$ by  the breathing oscillations.  We notice that, for $\Psi_G$, the transversal resonant tunneling gives rise to a beating pattern in the many-body dynamics of $\dfrac{1}{N}\Delta_{{\hat{Y}}}^2(t)$ and $\dfrac{1}{N}\Delta_{{\hat{P}_Y}}^2(t)$. This beating pattern may be the consequence of the combination of different breathing frequencies. A dedicate study of many-body excitations in the transversely- asymmetric 2D double-well potential could resolve the many-boson states.

\begin{figure*}[!h]
{\includegraphics[trim = 0.1cm 0.5cm 0.1cm 0.2cm, scale=.60]{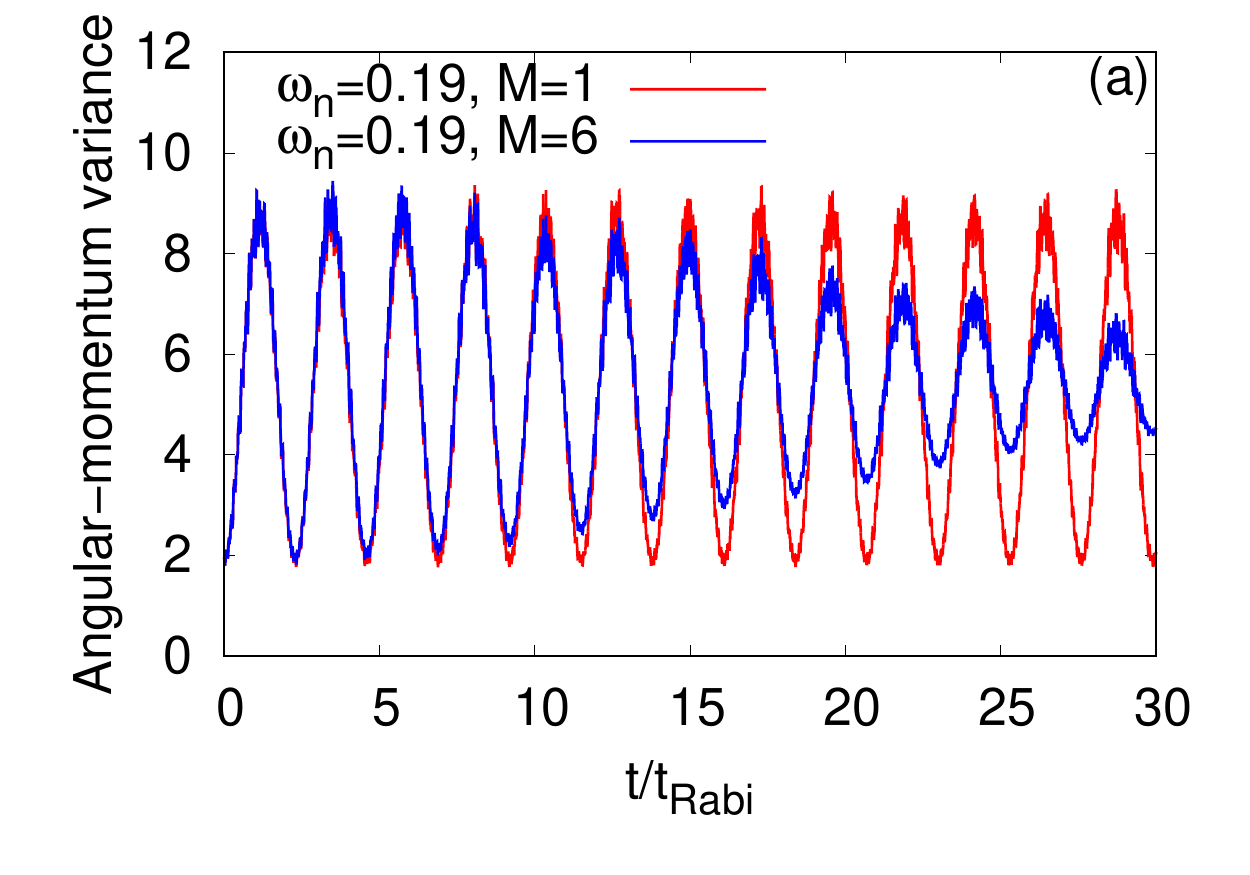}}
{\includegraphics[trim = 0.1cm 0.5cm 0.1cm 0.2cm, scale=.60]{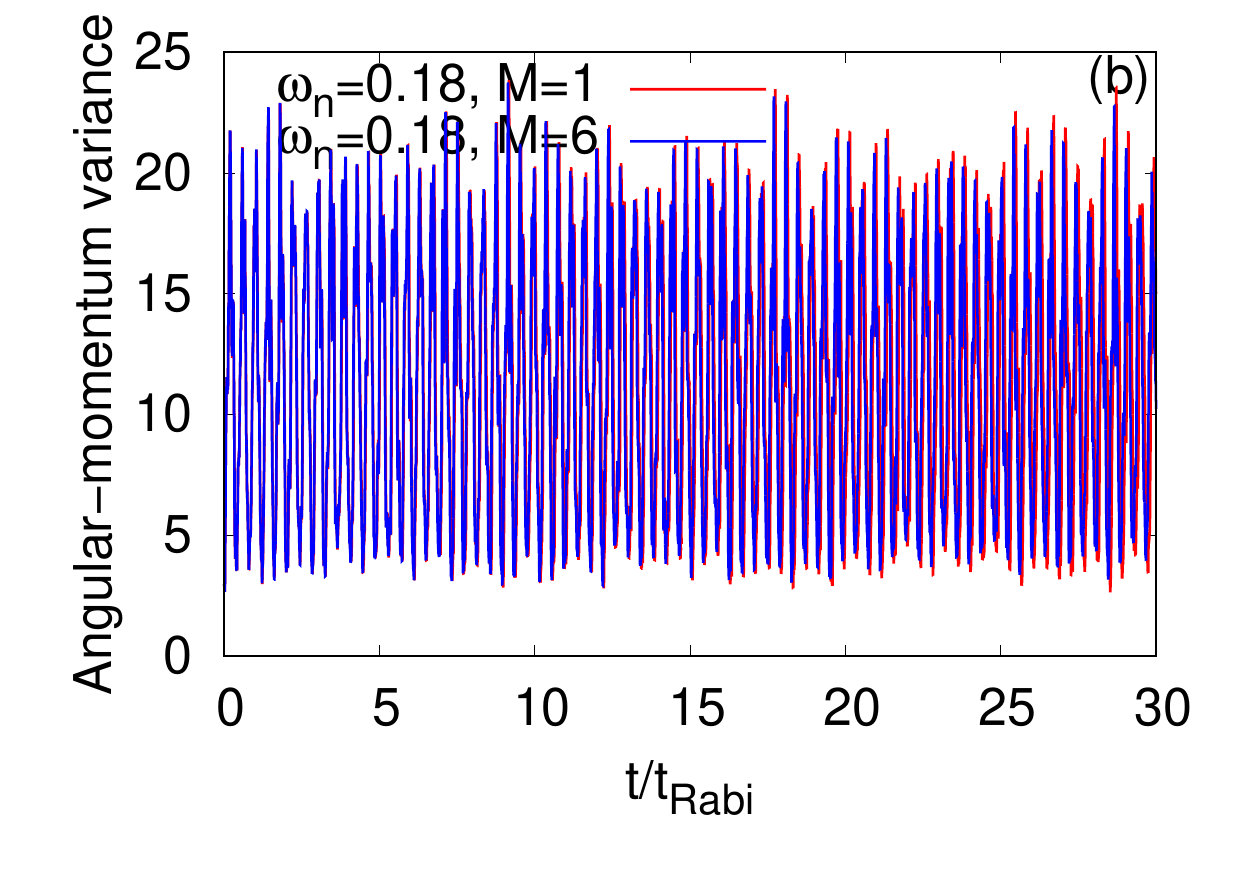}}\\
\caption{Time-dependent variance per particle of the $z$- component of the angular-momentum operator, $\dfrac{1}{N}\Delta_{\hat{L}_Z}^2(t)$, in a transversely-asymmetric 2D double-well potential for $\Psi_G$ (left column, $\omega_n=0.19$) and $\Psi_X$ (right column, $\omega_n=0.18$) under transversal resonant tunneling conditions.  The number of bosons is $N=10$  and the interaction parameter $\Lambda=0.01\pi$.   $M=1$ signifies the mean-field results.   The many-body dynamics are computed  with $M= 6$ time-dependent orbitals.   We show here dimensionless quantities. Color codes are explained in each panel.}
\label{Fig14}
\end{figure*}

Proceeding, we examine the time evolution of the angular-momentum variance per particle, $\dfrac{1}{N}\Delta_{{\hat{L}_Z}}^2(t)$, for $\Psi_G$ and $\Psi_X$ at the transversal resonant tunneling scenario (see Fig.~\ref{Fig14}). The mean-field dynamics of $\dfrac{1}{N}\Delta_{{\hat{L}_Z}}^2(t)$ for $\Psi_G$ showcases smooth undulations and  it is uneven for $\Psi_X$  which is consistent with the survival probabilities of the corresponding states. The amplitude of the oscillations of the mean-field $\dfrac{1}{N}\Delta_{{\hat{L}_Z}}^2(t)$ of $\Psi_X$ is larger  compared to $\Psi_G$, as one can see in panels (a) and (b) of Fig.~\ref{Fig14}. For both states, we notice that the frequency of the oscillations of $\dfrac{1}{N}\Delta_{{\hat{L}_Z}}^2(t)$ overlaps with the corresponding frequency of $\dfrac{1}{N}\Delta_{{\hat{Y}}}^2(t)$. The growing degree of the fragmentation causes a decay in amplitude of the many-body $\dfrac{1}{N}\Delta_{{\hat{L}_Z}}^2(t)$ which is significantly more prominent for $\Psi_G$ and hardly visible for $\Psi_X$.

\section{SUMMARY AND CONCLUSIONS}
The paradigm of a bosonic Josephson junction, in which bosons can tunnel back and forth between two potential wells, is shown to be very rich at resonant tunneling condition in two spatial dimensions. We have investigated the dynamical behavior of a few intricate coherent bosonic clouds at  resonant tunneling conditions in a two-dimensional asymmetric Josephson junction. In particular, we  examine two types of resonant tunnelings, longitudinal and transversal. For the longitudinal resonant tunneling, the initial bosonic clouds are considered as the ground and transversely-excited states, whereas for transversal resonant tunneling, they are taken as the ground and longitudinally-excited states.  The excited states are considered in such a way that we can explore  the impact of  longitudinal resonant tunneling on the transversely-excited state and analogously, transversal resonant tunneling on the longitudinally-excited state.  The dynamical behavior is analyzed   by solving the full many-body Schr\"odinger equation and presenting the time evolution of the survival probability, depletion or fragmentation, and the many-particle position, momentum, and angular momentum variances.   To characterize the influence of the many-body correlations on the  tunneling dynamics of the different shapes of the bosonic clouds, we compare the mean-field and many-body results. 

Focusing on the longitudinal resonant tunneling scenario, we verify that a gradual change of asymmetry parameter $c$ either from the values of resonant tunnelings, or the value for which one gets the symmetric double-well potential,  decreases the tunneling probability of the bosons.   The many-body correlations exhibit the collapse of the density oscillations which is manifested   in the dynamics of  the survival probability.  The collapse rate becomes    slower when one moves away from the symmetric double-well potential to  the first longitudinal resonant tunneling condition ($c=0.25$) and further slower toward the  second  longitudinal resonant tunneling condition ($c=0.5$). Moreover, at  $c=0.5$  we  find a partial revival pattern in the mean-field  as well as the many-body survival probabilities.  Although the rate of collapse of density oscillation for $\Psi_Y$ is slower compared to  $\Psi_G$ at a fixed value of $c$, the presence of transverse excitation  in $\Psi_Y$ makes it more fragmented than  $\Psi_G$ at a given time.  Compared to the symmetric double-well potential, we  observe that both  initial states become less fragmented at $c=0.25$ and further less fragmented at $c=0.5$.

Referring to  the transversal resonant tunneling scenario,   we scan the frequency along the transverse direction in the right well, $\omega_n$, starting from the value $\omega_n=1$ (symmetric 2D double-well) and reaching down to  $\omega_n=0.17$, and find that the first resonant tunneling in the transverse direction  occurs at $\omega_n=0.19$ and 0.18 for $\Psi_G$ and $\Psi_X$, respectively. This resonant tunneling is achieved when the  bosons tunnel from  $\Psi_G$ of the left well to $(y^2-1)\Psi_G$ of the right well. Similarly,  for $\Psi_X$, the bosons tunnel from $\Psi_X$ to $(y^2-1)\Psi_X$.  The time evolution of the many-body survival probability of $\Psi_G$ and $\Psi_X$ signifies that the rate of collapse of density oscillations  is faster for $\Psi_G$ compared  to $\Psi_X$. This collapse of density oscillations occurs due to  the growing degree of many-body correlations which is graphically shown in terms of the loss of coherence in the system.

To obtain further information of the time-dependent many-particle wavefunction, we present the many-particle variances of the observables, $\dfrac{1}{N}\Delta_{{\hat{X}}}^2(t)$, $\dfrac{1}{N}\Delta_{{\hat{Y}}}^2(t)$, $\dfrac{1}{N}\Delta_{{\hat{P}_X}}^2(t)$, $\dfrac{1}{N}\Delta_{{\hat{P}_Y}}^2(t)$, and $\dfrac{1}{N}\Delta_{{\hat{L}_Z}}^2(t)$, both at the longitudinal and transversal resonant tunneling conditions. Also, we demonstrate a possible interconnection of the above-mentioned quantities  with the survival probability and loss of coherence.  It  is  observed  that  the growing degree of fragmentation impacts  the  time-evolution  of  each variance for the longitudinal and transversal resonant tunneling differently. We notice that, in general, at the resonant tunneling, effects of breathing dynamics are amplified   in the dynamics of variances. Focusing on the longitudinal resonant tunneling at $c=0.5$,  a signature of partial revival is observed in the dynamics of $\dfrac{1}{N}\Delta_{{\hat{X}}}^2(t)$, $\dfrac{1}{N}\Delta_{{\hat{P}_X}}^2(t)$, and $\dfrac{1}{N}\Delta_{{\hat{L}_Z}}^2(t)$. Also, in the  longitudinal resonant tunneling, $\dfrac{1}{N}\Delta_{{\hat{Y}}}^2(t)$  shows a very small-amplitude fluctuations of the order of $10^{-3}$, which is contrary to the transversal resonant tunneling where the amplitude of $\dfrac{1}{N}\Delta_{{\hat{Y}}}^2(t)$ is significantly high, of the order of $10^1$. Moreover, the breathing oscillations in the dynamics of $\dfrac{1}{N}\Delta_{{\hat{P}_Y}}^2(t)$ at the  longitudinal resonant tunneling is barely visible, while at the transversal resonant tunneling it is appreciable. Interestingly, at the transversal resonant tunneling,  we notice a beating pattern in the dynamics of $\dfrac{1}{N}\Delta_{{\hat{Y}}}^2(t)$ and $\dfrac{1}{N}\Delta_{{\hat{P}_Y}}^2(t)$ which may arise due to the mixture of different breathing frequencies.

The present work can inspire several promising and interesting future research directions. An immediate extension would be the dynamics of even more intricate  bosonic structures, e.g., vortex state or a mixture of vortex states induced by Raman process using light with orbital angular momentum \cite{Schmiegelow2016, Bhowmik2018a, Bhowmik2016, Bhowmik2018b, Bhowmik2020_vector} or by synthetic magnetic fields \cite{Price2016} in the longitudinal and transversal resonant tunneling conditions. Also, to apply a linear response theory \cite{Theisen2016} to  accurately calculate the different breathing frequencies involved in the time evolution of different quantities at the resonant tunneling scenario would be challenging. Finally, in the long run we could envision that additional geometries would open up in three spatial dimensions. 
\clearpage

\section*{Appendix A: Many-particle variance}

The variance of an observable, $\hat{A}$, is determined  by the combination of the expectation values of $\hat{A}$ and  $\hat{A}^2$. Here the expectation value of $\hat{A}$ $=\sum_{j=1}^N \hat{a}({r_j})$ depends on the one-body operators while the expectation of  $\hat{A}^2$ is a mixture of one- and two-body operators,  ${\hat{A}}^2=\sum_{j=1}^N\hat{a}^2({r_j})+\sum_{j<k}2\hat{a}({r_j})\hat{a}({r_k})$.   The variance  can be written as \cite{Alon2019b}

\setcounter{equation}{0}
\renewcommand{\theequation}{A.\arabic{equation}}

\begin{eqnarray}\label{A1}
\dfrac{1}{N}\Delta_{\hat{A}}^2&(t)&=\dfrac{1}{N}[\langle\Psi(t)|\hat{A}^2|\Psi(t)\rangle-\langle\Psi(t)|\hat{A}|\Psi(t)\rangle^2] \nonumber \\
&=& \dfrac{1}{N}\Bigg\{\sum_j n_j(t)\int d\textbf{r}\phi_j^*(\textbf{r}; t)\hat{a}^2({\textbf{r}})\phi_j(\textbf{r}; t)-\left[\sum_j n_j(t)\int d\textbf{r}\phi_j^*(\textbf{r}; t)\hat{a}({\textbf{r}})\phi_j(\textbf{r}; t)\right]^2 \nonumber \\ &+& \sum_{jpkq}\rho_{jpkq}(t)\left[\int d\textbf{r}\phi_j^*(\textbf{r}; t)\hat{a}({\textbf{r}})\phi_k(\textbf{r}; t)\right] \left[\int d\textbf{r}\phi_p^*(\textbf{r}; t)\hat{a}({\textbf{r}})\phi_q(\textbf{r}; t)\right]\Bigg\},
\end{eqnarray}
where $\{\phi_j(\textbf{r}; t)\}$ are the natural orbitals, $\{n_j(t)\}$  the natural occupations, and $\rho_{jpkq}(t)$ are the elements of the   reduced two-particle density matrix, $\rho(\textbf{r}_1, \textbf{r}_2, \textbf{r}_1^\prime, \textbf{r}_2^\prime; t)= \sum \limits_{jpkq}\rho_{jpkq}(t)\phi_j^*(\textbf{r}_1^\prime; t) \phi_p^*(\textbf{r}_2^\prime; t)$ 
$\phi_k(\textbf{r}_1; t) \phi_q(\textbf{r}_2; t).$  For one-body operators which are local in  position space, the variance described in Eq~\ref{A1} boils down to \cite{Lode2020}
 
\begin{equation}\label{A2}
\dfrac{1}{N}\Delta_{\hat{A}}^2(t)= \int d\textbf{r}\dfrac{\rho(\textbf{r};t)}{N}\hat{a}^2({\textbf{r}})-N\left[\int \dfrac{\rho(\textbf{r};t)}{N}\hat{a}({\textbf{r}}) \right]^2 + \int d\textbf{r}_1 d\textbf{r}_2 \dfrac{\rho^{(2)}(\textbf{r}_1, \textbf{r}_2, \textbf{r}_1, \textbf{r}_2; t)}{N}a(\textbf{r}_1)a(\textbf{r}_2),
\end{equation} 
where $\rho(\textbf{r};t)$ is the time-dependent density.  Eq~\ref{A1} describes the variances when  the center-of-mass of the bosonic clouds are at the origin and the wavefunction is $\Psi(0,0)$ (or denoted in short by $\Psi$).   As the initial states considered in this work are prepared at the position $(a,b)=(-2,0)$, here we present a general form  of variances by incorporating   the translated wavefunction, i.e., $\Psi(a,b)$ (or  denoted in short by $\Psi_{ab}$).  For   the position and momentum operators, the variances do not change due to the translated wavefunction  \cite{Alon2019b}, therefore,  $\dfrac{1}{N}\Delta_{{\hat{X}}}^2\Big|_{\Psi_{ab}}=\dfrac{1}{N}\Delta_{{\hat{X}}}^2\Big|_{\Psi}$,  $\dfrac{1}{N}\Delta_{{\hat{Y}}}^2\Big|_{\Psi_{ab}}=\dfrac{1}{N}\Delta_{{\hat{Y}}}^2\Big|_{\Psi}$, $\dfrac{1}{N}\Delta_{{\hat{P}_X}}^2\Big|_{\Psi_{ab}}=\dfrac{1}{N}\Delta_{{\hat{P}_X}}^2\Big|_{\Psi}$, and  $\dfrac{1}{N}\Delta_{{\hat{P}_Y}}^2\Big|_{\Psi_{ab}}=\dfrac{1}{N}\Delta_{{\hat{P}_Y}}^2\Big|_{\Psi}$. But, for the angular-momentum variance, the situation becomes intricate  and expressed as  \cite{Alon2019b}
 
\begin{eqnarray}\label{A3}
\dfrac{1}{N}\Delta_{{\hat{L}_Z}}^2\Big|_{\Psi_{ab}}&=&\dfrac{1}{N}\Delta_{{\hat{L}_Z}}^2\Big|_{\Psi}+\dfrac{1}{N}a^2\Delta_{{\hat{P}_Y}}^2\Big|_{\Psi}+\dfrac{1}{N}b^2\Delta_{{\hat{P}_X}}^2\Big|_{\Psi} \nonumber \\
&+& \dfrac{1}{N}\Big\{a [\langle \Psi|{\hat{L}_Z}{\hat{P}_Y}+{\hat{P}_Y}{\hat{L}_Z}|\Psi\rangle - 2\langle\Psi|{\hat{L}_Z}|\Psi\rangle\langle|\Psi|{\hat{P}_Y}|\Psi\rangle]\nonumber \\
&-& b [\langle \Psi|{\hat{L}_Z}{\hat{P}_X}+{\hat{P}_X}{\hat{L}_Z}|\Psi\rangle - 2\langle\Psi|{\hat{L}_Z}|\Psi\rangle\langle|\Psi|{\hat{P}_X}|\Psi\rangle]\nonumber \\
&-& 2ab [\langle \Psi|{\hat{P}_Y}{\hat{P}_X}|\Psi\rangle - \langle\Psi|{\hat{P}_Y}|\Psi\rangle\langle|\Psi|{\hat{P}_X}|\Psi\rangle]\Big\}.
\end{eqnarray}
Eq. ~(\ref{A3}) is used in the main text to evaluate the  angular-momentum variance of the different initial conditions.  For a wavefunction $\Psi$ with particular spatial symmetry the evaluation of ~(\ref{A3})  can simplify.

\section*{Appendix B: Role of the width of  interparticle interaction potential on the dynamics}
We have presented an extensive investigation leading to be  a wealth of results on the physics of longitudinal and transversal resonant  tunneling of a many-particle system in two spatial dimensions, both at the mean-field and many-body levels of theory. We have used a model potential of finite width and find it instructive and important to demonstrate the robustness of our findings to this parameter. In order to verify whether the width of the interparticle interaction potential, $\sigma$, affects the dynamical behavior at the longitudinal and transversal resonant conditions, we recomputed all the properties discussed in this work for two additional smaller widths, $\sigma=0.25/\sqrt{\pi}$ and $\sigma=0.25$, both at the mean-field and many-body levels. It is noted that the main text explores the dynamics with $\sigma=0.25\sqrt{\pi}$ at both the resonant tunneling scenarios.   In the mean-field dynamics, we find that the dynamical behavior of  all the properties obtained for $\sigma=0.25/\sqrt{\pi}$ and $\sigma=0.25$ fall on top of the corresponding results computed for $\sigma=0.25\sqrt{\pi}$. As for an example, we plot the dynamics of a sensitive quantity,  $\dfrac{1}{N}\Delta_{{\hat{X}}}^2(t)$, see Fig.~\ref{fig_A0}.  The results for both resonant scenarios manifest that all the  mean-field quantities discussed in this work  are independent of $\sigma$. 

\renewcommand{\thefigure}{B\arabic{figure}}

\setcounter{figure}{0}
\begin{figure*}[!h]
\centering
{\includegraphics[trim = 0.1cm 0.5cm 0.1cm 0.2cm, scale=.60]{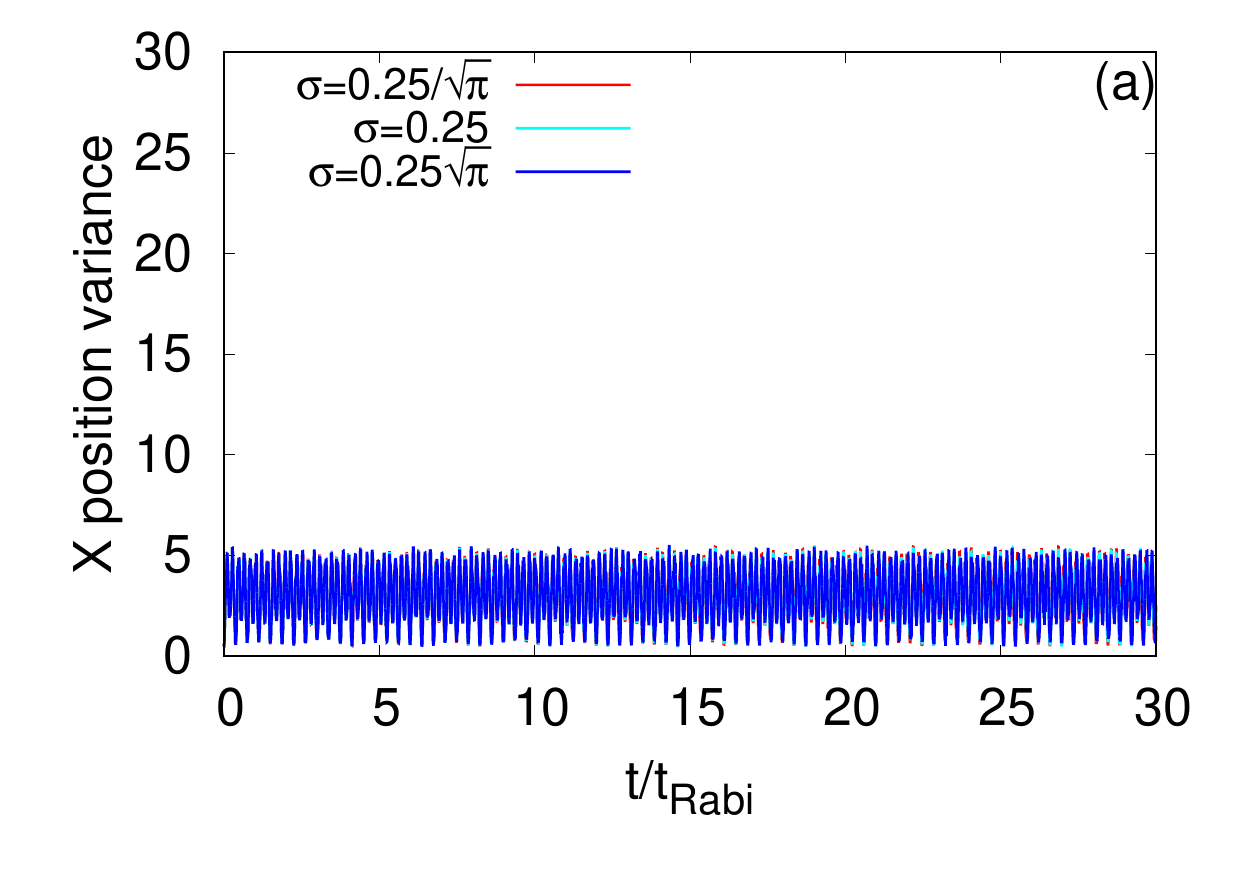}}
{\includegraphics[trim = 0.1cm 0.5cm 0.1cm 0.2cm, scale=.60]{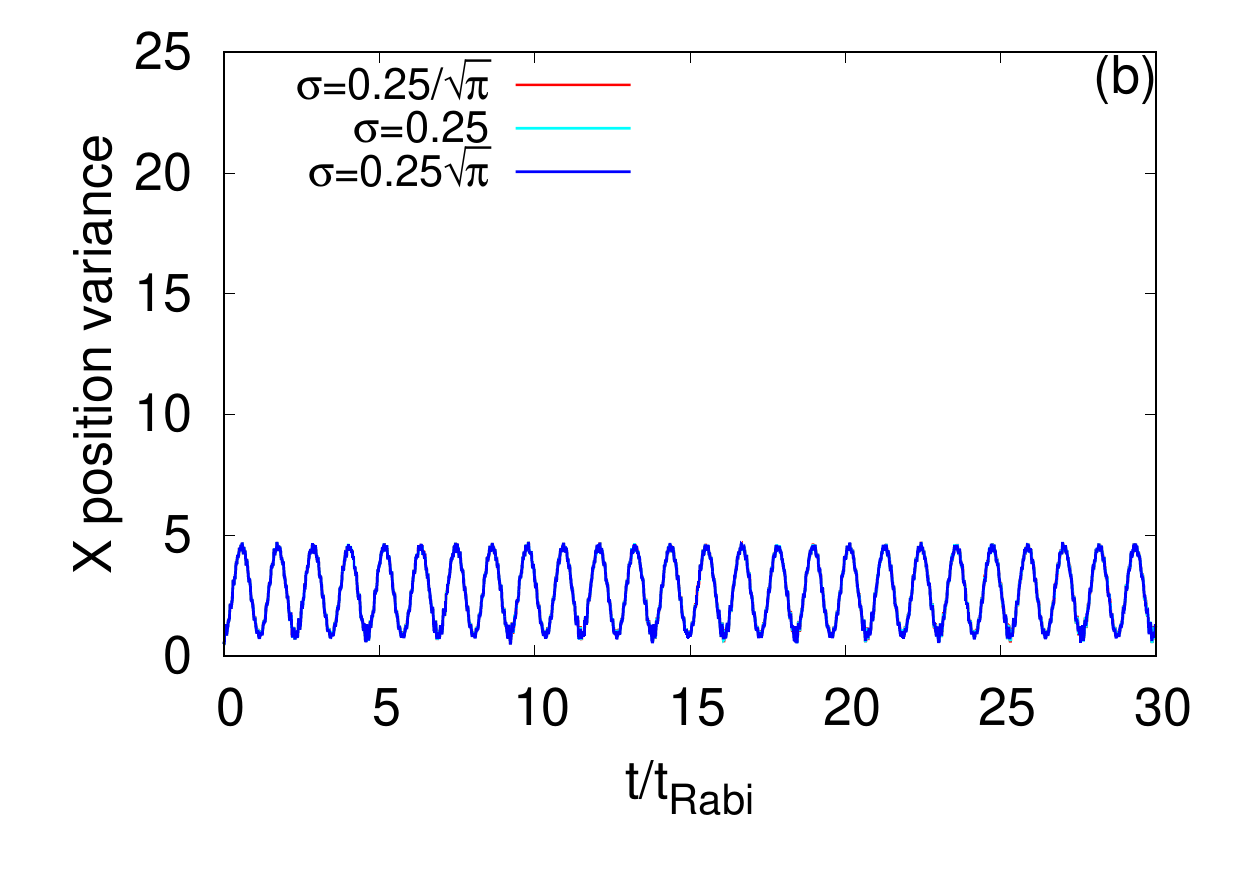}}\\
\caption{\label{fig_A0}Time-dependent mean-field position variance per particle along the $x$-direction, $\dfrac{1}{N}\Delta_{{\hat{X}}}^2(t)$, for the ground state at the (a) longitudinal ($c=0.25$) and (b) transversal ($\omega_n=0.19$) resonant scenarios. The three different  widths of interparticle  interaction potential are $\sigma=0.25/\sqrt{\pi}$, $\sigma=0.25$, and $\sigma=0.25\sqrt{\pi}$.  The interaction parameter is $\Lambda=0.01\pi$.  The  mean-field results are found to be independent on $\sigma$.  We show here   dimensionless quantities. Color codes are explained in each panel.}
\end{figure*}

To present the role of width of $\sigma$ at the many-body dynamics, we select the dynamical behavior of the  most sensitive quantities of the ground state at the first longitudinal resonant condition ($c=0.25$) and transversal resonant condition ($\omega_n=0.19$).

\begin{figure*}[!h]
\centering
{\includegraphics[trim = 0.1cm 0.5cm 0.1cm 0.2cm, scale=.60]{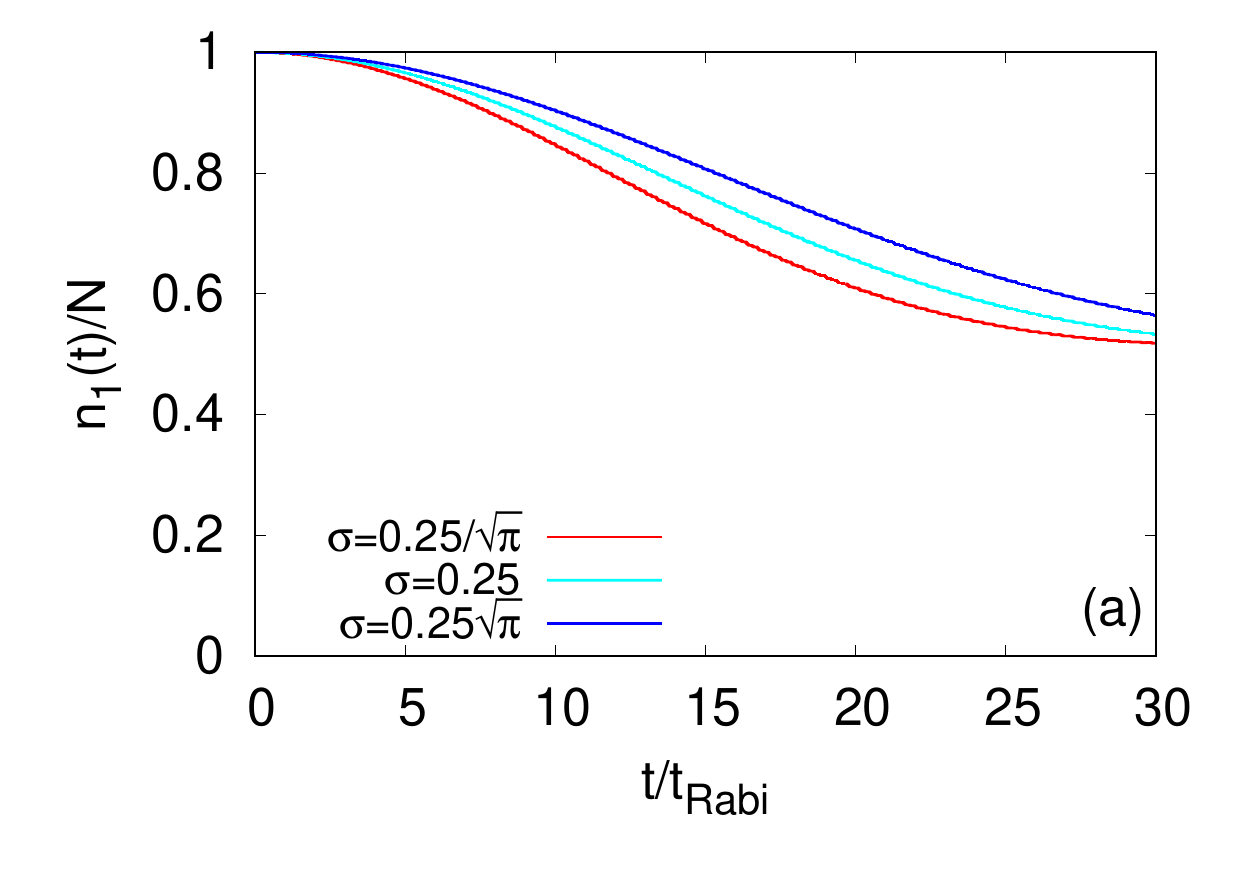}}
{\includegraphics[trim = 0.1cm 0.5cm 0.1cm 0.2cm, scale=.60]{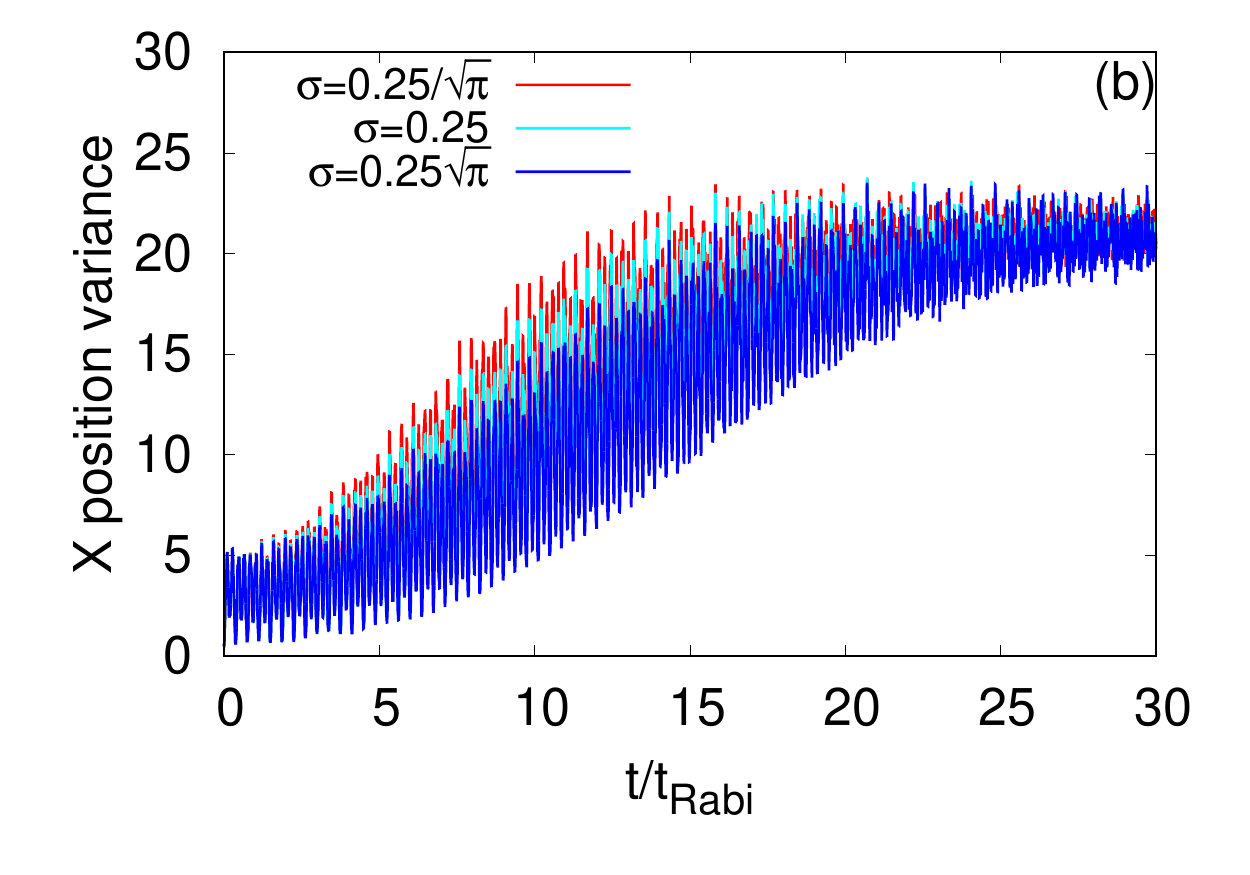}}\\
{\includegraphics[trim = 0.1cm 0.5cm 0.1cm 0.2cm, scale=.60]{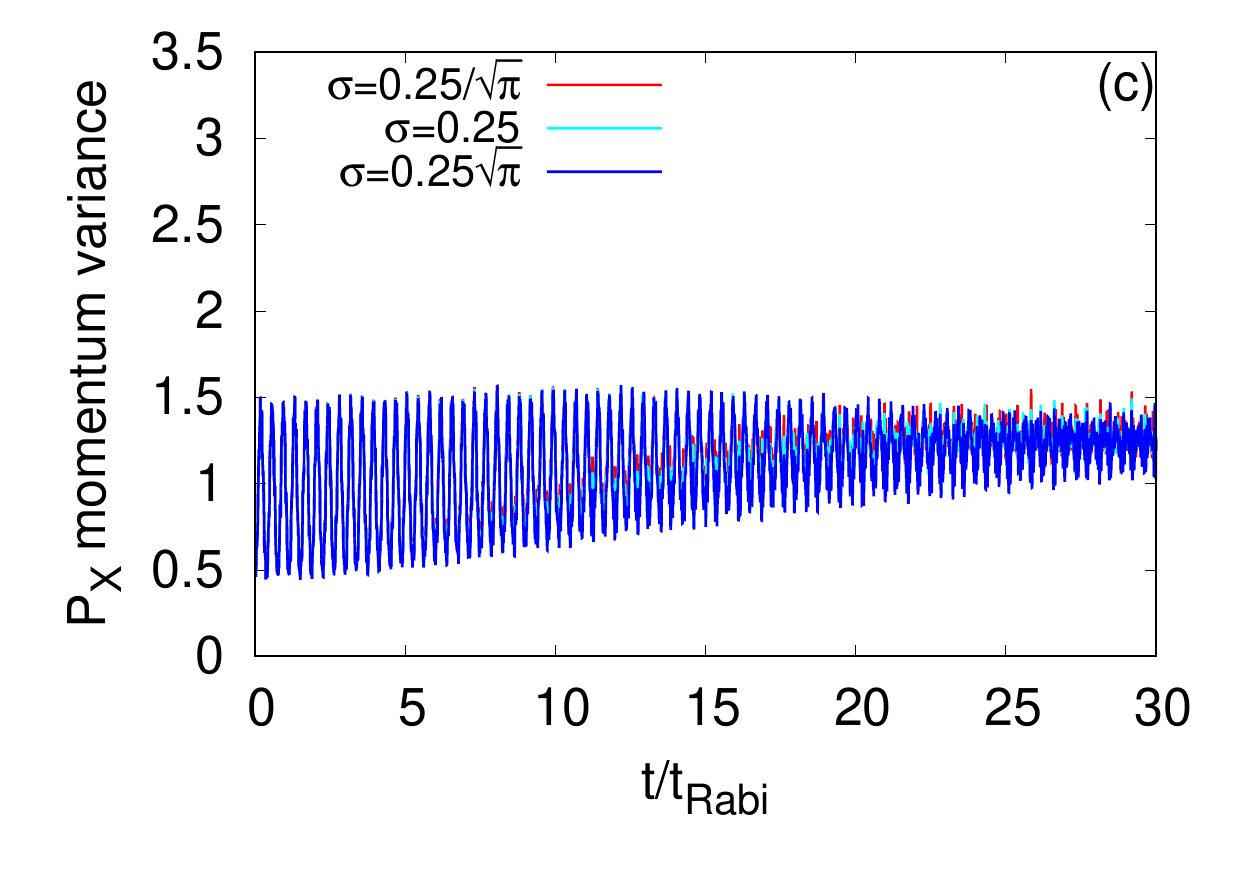}}
{\includegraphics[trim = 0.1cm 0.5cm 0.1cm 0.2cm, scale=.60]{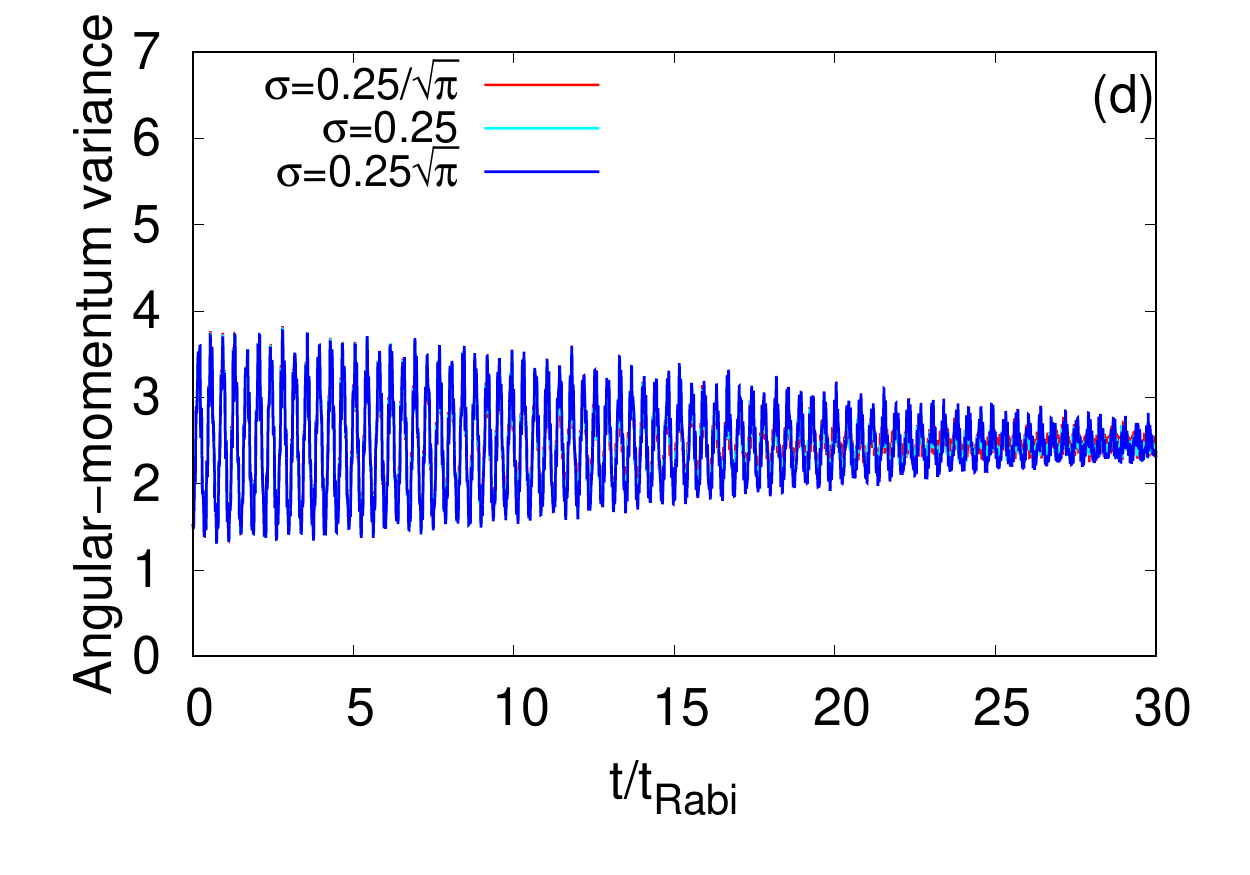}}\\
\caption{\label{fig_A1} Dependency of the many-body dynamics of  ground state at the longitudinal resonant condition, $c=0.25$, for the three different  widths of interparticle  interaction potential, i.e., $\sigma=0.25/\sqrt{\pi}$, $\sigma=0.25$, and $\sigma=0.25\sqrt{\pi}$.  The dynamics is shown by (a) $\dfrac{n_1(t)}{N}$, (b) $\dfrac{1}{N}\Delta_{{\hat{X}}}^2(t)$, (c) $\dfrac{1}{N}\Delta_{{\hat{P}_X}}^2(t)$, and (d) $\dfrac{1}{N}\Delta_{{\hat{L}_Z}}^2(t)$. The dynamics are computed with $N=10$  bosons and the interaction parameter $\Lambda=0.01\pi$. The many-body dynamics are computed with $M= 6$ time-dependent orbitals. The quantitative many-body results are found to be weakly dependent on $\sigma$. with We show here   dimensionless quantities. Color codes are explained in each panel.}
\end{figure*}

In Fig.~\ref{fig_A1}, we present the many-body dynamics of $\dfrac{n_1(t)}{N}$,  $\dfrac{1}{N}\Delta_{{\hat{X}}}^2(t)$,  $\dfrac{1}{N}\Delta_{{\hat{P}_X}}^2(t)$, and  $\dfrac{1}{N}\Delta_{{\hat{L}_Z}}^2(t)$ at $c=0.25$. From Fig.~\ref{fig_A1} (a), one can observe that the rate of loss of coherence is slightly faster for $\sigma=0.25/\sqrt{\pi}$ compared to other values of $\sigma$.  It is found that decreasing the width of $\sigma$ of the interparticle interaction potential gradually diminishes the quantitative difference between the dynamical behaviors of a particular quantity.   Interestingly, the variances which are known to be sensitive many-body quantities due to depletion are hardly influenced by the width of $\sigma$. Although there is  a small quantitative difference in the long-time fragmentation dynamics for the choices of $\sigma$, the many-body physics described for the longitudinal resonant scenario is robust which can be seen  in the dynamical behavior of the different quantum mechanical properties, see Fig.~\ref{fig_A1} (b), (c), and (d). 

\begin{figure*}[!h]
\centering
{\includegraphics[trim = 0.1cm 0.5cm 0.1cm 0.2cm, scale=.60]{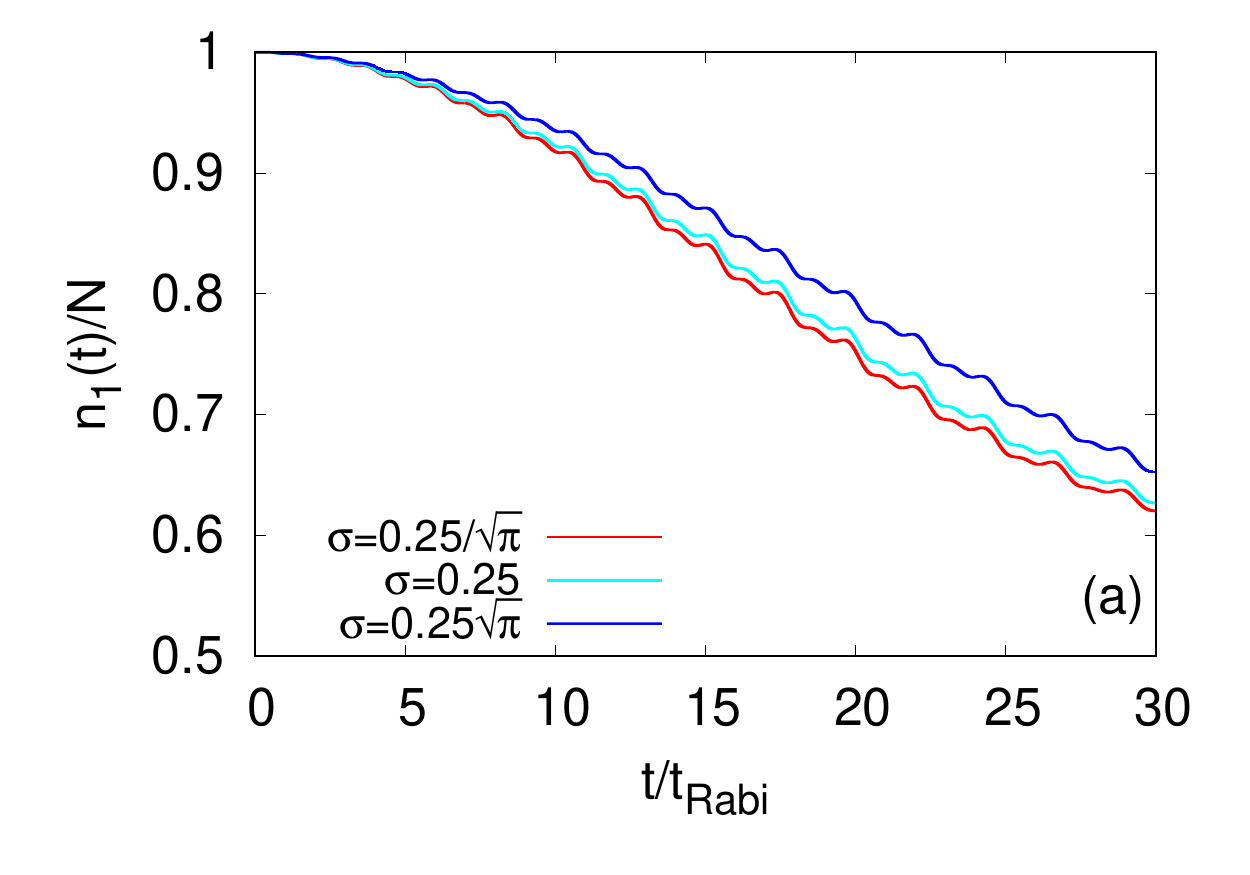}}
{\includegraphics[trim = 0.1cm 0.5cm 0.1cm 0.2cm, scale=.60]{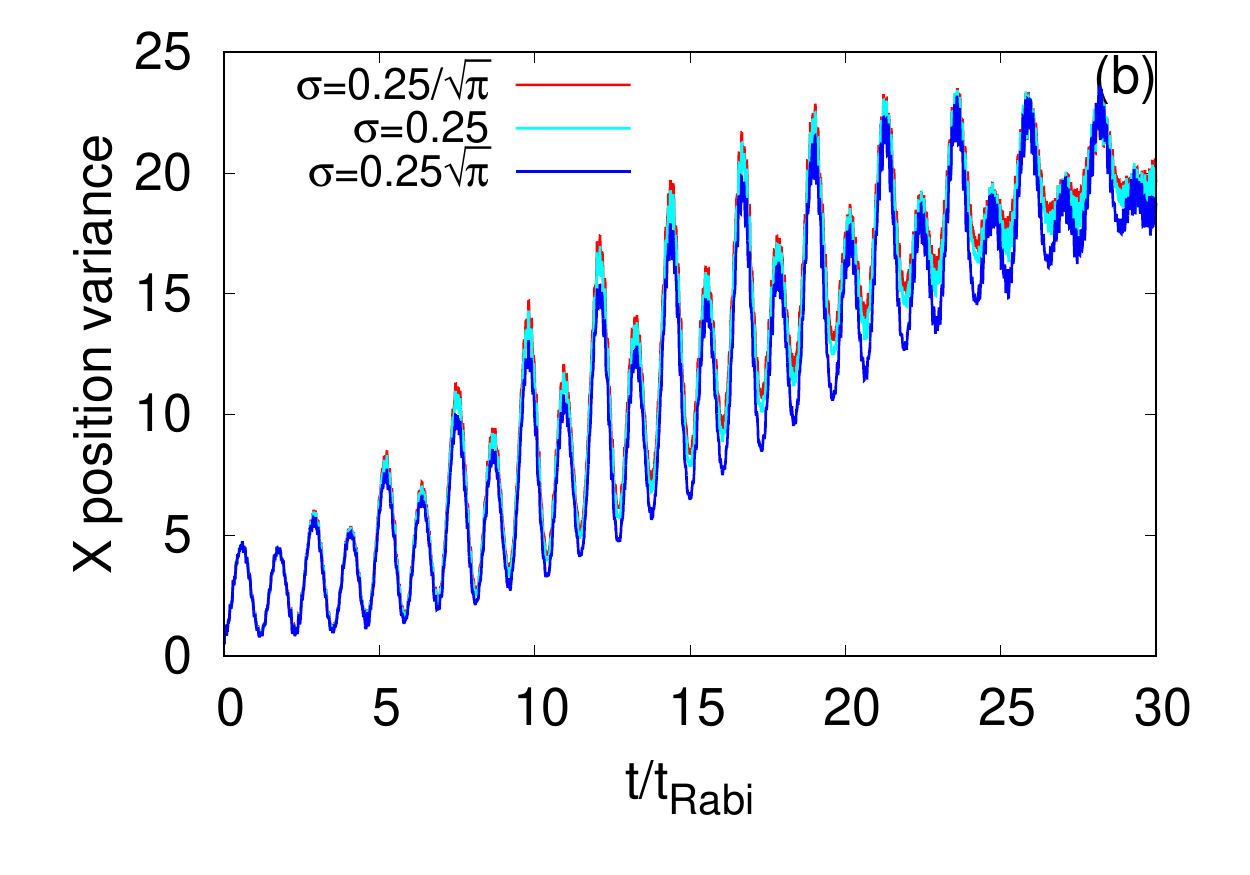}}\\
{\includegraphics[trim = 0.1cm 0.5cm 0.1cm 0.2cm, scale=.60]{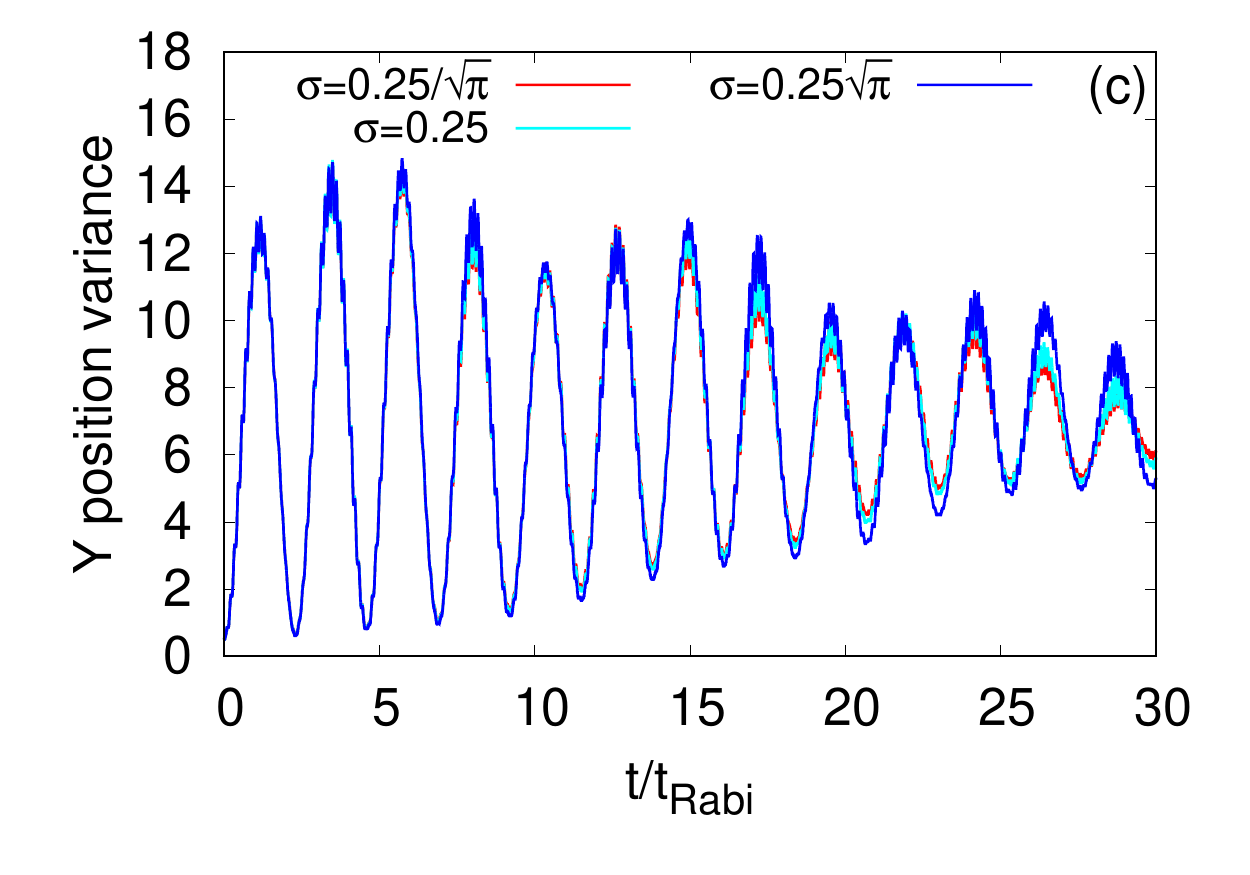}}
{\includegraphics[trim = 0.1cm 0.5cm 0.1cm 0.2cm, scale=.60]{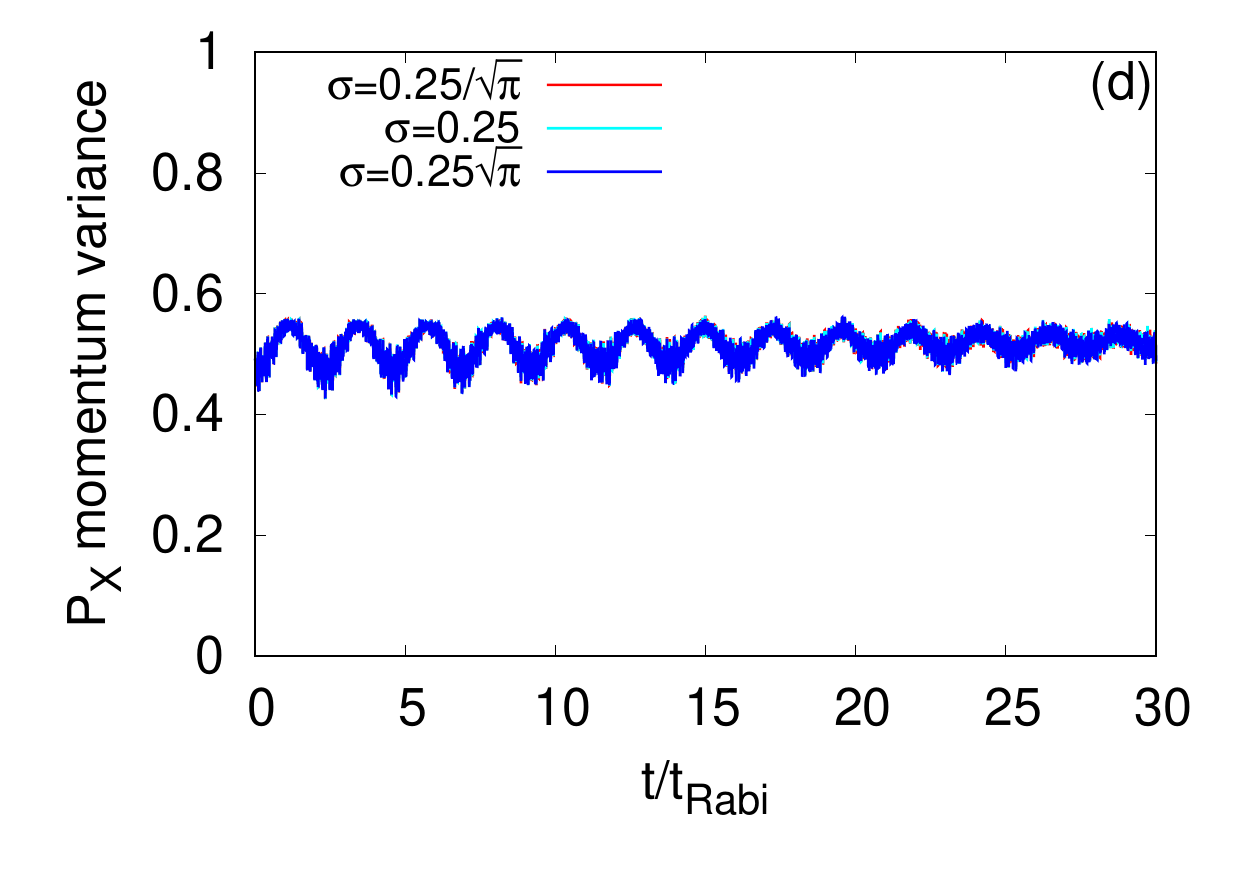}}\\
{\includegraphics[trim = 0.1cm 0.5cm 0.1cm 0.2cm, scale=.60]{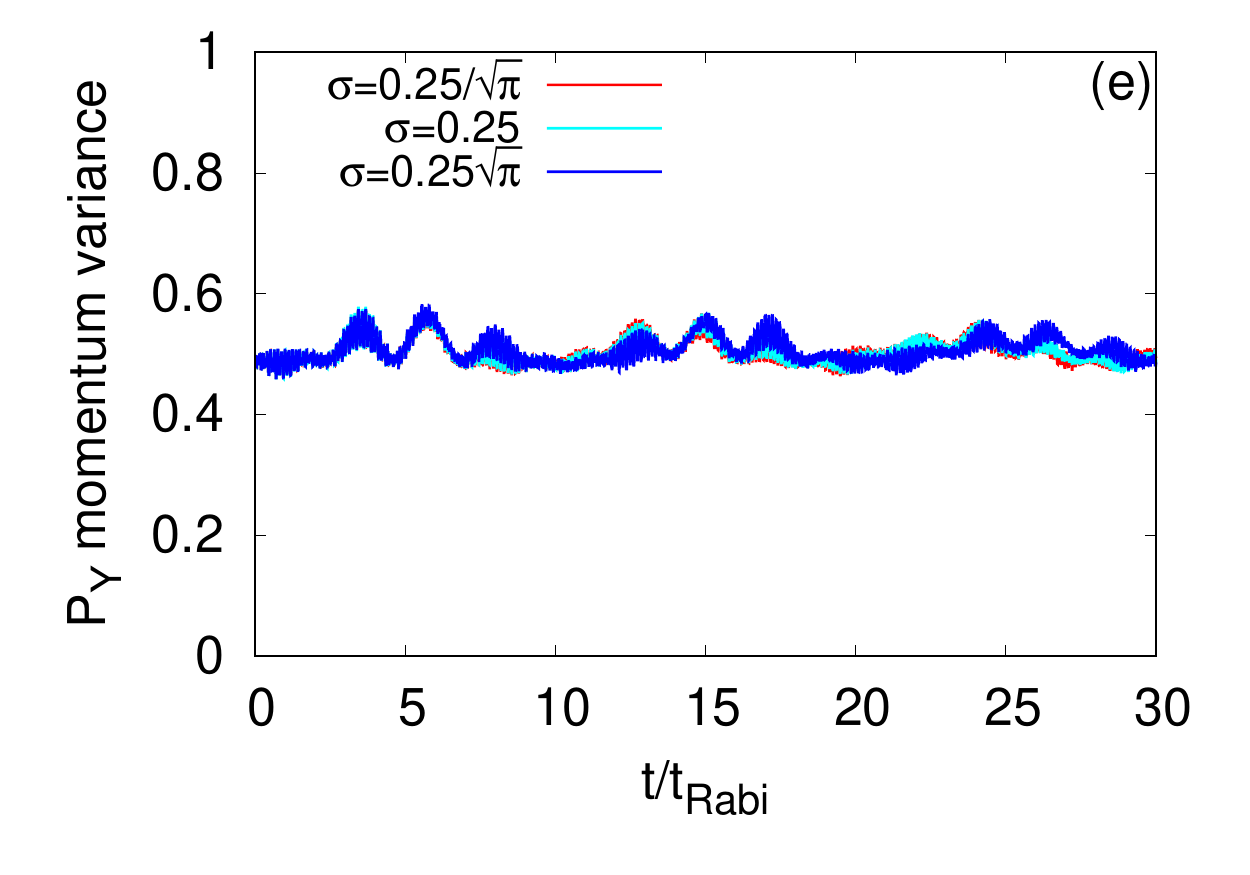}}
{\includegraphics[trim = 0.1cm 0.5cm 0.1cm 0.2cm, scale=.60]{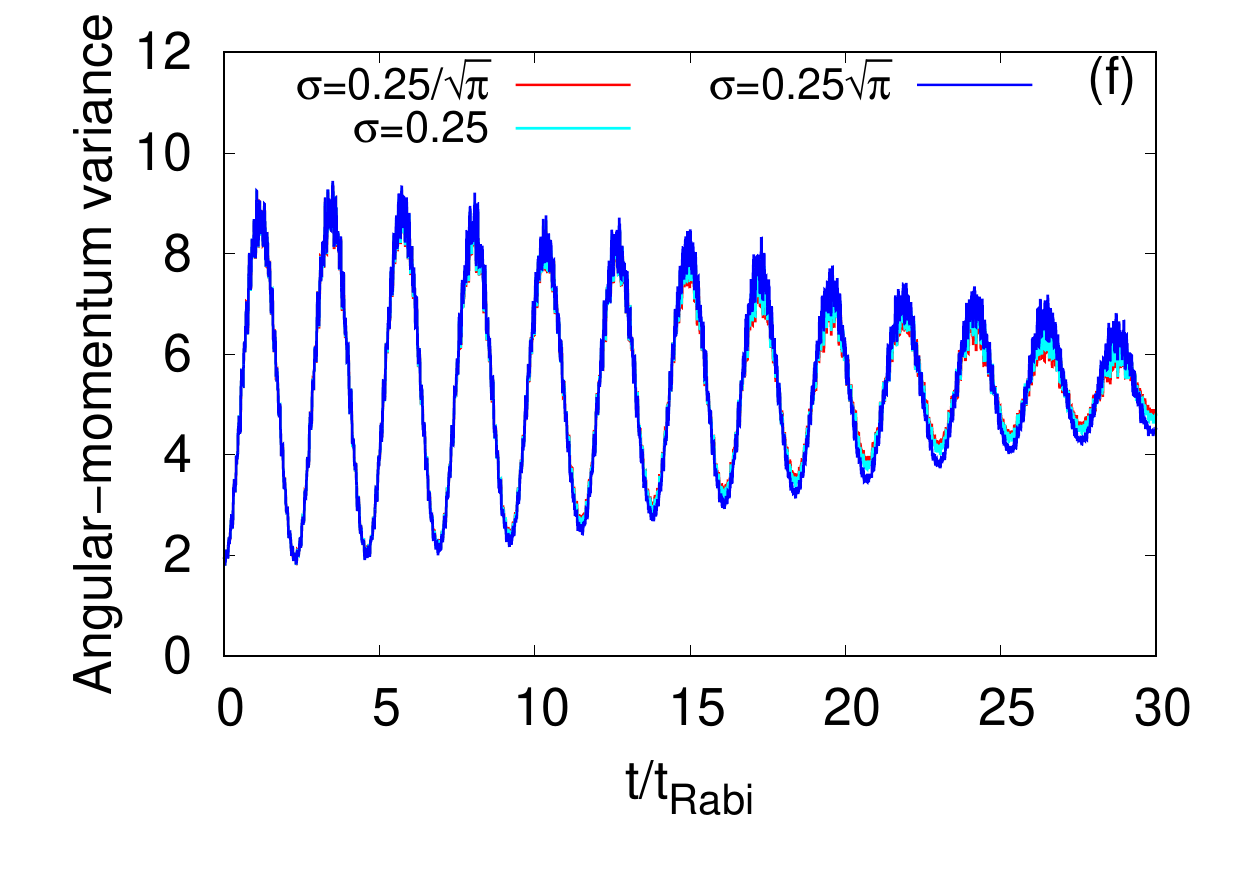}}\\
\caption{\label{fig_A2}Dependency of the many-body dynamics of  ground state at the transversal resonant condition, $\omega_n=0.19$, for the three different  widths of interparticle  interaction potential, i.e., $\sigma=0.25/\sqrt{\pi}$, $\sigma=0.25$, and $\sigma=0.25\sqrt{\pi}$.  The dynamics is shown by (a) $\dfrac{n_1(t)}{N}$,  (b) $\dfrac{1}{N}\Delta_{{\hat{X}}}^2(t)$, (c) $\dfrac{1}{N}\Delta_{{\hat{Y}}}^2(t)$ (d) $\dfrac{1}{N}\Delta_{{\hat{P}_X}}^2(t)$, (e) $\dfrac{1}{N}\Delta_{{\hat{P}_Y}}^2(t)$, and (f) $\dfrac{1}{N}\Delta_{{\hat{L}_Z}}^2(t)$. The dynamics are computed with $N=10$  bosons and the interaction parameter $\Lambda=0.01\pi$. The many-body dynamics are computed with $M= 6$ time-dependent orbitals. The quantitative many-body results are found to be weakly dependent on $\sigma$. We show here   dimensionless quantities. Color codes are explained in each panel.}
\end{figure*}

Fig.~\ref{fig_A2} displays the role of $\sigma$ at the transversal resonant scenario of the ground state. Here we present the dynamics of $\dfrac{n_1(t)}{N}$,   $\dfrac{1}{N}\Delta_{{\hat{X}}}^2(t)$,  $\dfrac{1}{N}\Delta_{{\hat{Y}}}^2(t)$  $\dfrac{1}{N}\Delta_{{\hat{P}_X}}^2(t)$,  $\dfrac{1}{N}\Delta_{{\hat{P}_Y}}^2(t)$, and  $\dfrac{1}{N}\Delta_{{\hat{L}_Z}}^2(t)$. Similar to the longitudinal resonant scenario, here also, we notice that the development of the fragmentation is slightly quicker for $\sigma=0.25/\sqrt{\pi}$ and slightly slower for $\sigma=0.25\sqrt{\pi}$. Moreover, there is no qualitative difference in the tunneling dynamics at the transversal resonant scenario for the different choices of $\sigma$. The beating pattern in the dynamics of $\dfrac{1}{N}\Delta_{{\hat{P}_Y}}^2(t)$ is also observed at $\sigma=0.25/\sqrt{\pi}$ and $\sigma=0.25$. All in all, we demonstrate and observe that the width of the interparticle interaction potential does not qualitatively affect the mean-field and many-body  physics of the tunneling dynamics at the resonant tunneling scenario.

%) ============================================================================
% === REFERENCES =============================================================
% ============================================================================

%\bibliographystyle{abbrv}

%\bibliography{bib}\

\section*{ACKNOWLEDGMENTS}
This research was supported by the Israel Science Foundation (Grants No. 600/15 and
1516/19). AB acknowledges Sudip Kumar Haldar for some helpful discussions. Computation
time on the High Performance Computing system Hive of the Faculty of
Natural Sciences at University of Haifa, and the Hawk at the High Performance Computing
Center Stuttgart (HLRS) is gratefully acknowledged.

\clearpage
\section*{Supplemental material for Longitudinal and transversal resonant tunneling  of interacting bosons in a  two-dimensional Josephson junction: Mean-field and many-body dynamics}
In this supplemental material, we provide the convergences of the quantities discussed in the main text along with further complementary results and their analysis.  The convergences of the results   are shown with respect to the  number of time-dependent orbitals and discrete-variable-representation grid points. The results are calculated by applying the multiconfigurational time-dependent Hartree for bosons (MCTDHB) method \cite{Streltsov_Alon_2007, Alon_Streltsov_2008, Lode_Madsen_2020} and for our numerical computations, we use the numerical implementation in 
\cite{Streltsov_Streltsova_2015, Streltsov_Cederbaum_2019}. We demonstrate the details of the mechanism of fragmentation  in terms of the occupancy of the higher natural orbitals for both  resonant tunneling conditions which were not discussed in the main text. Moreover, we graphically display the   position and momentum variances \cite{Klaiman_Alon_2015, Klaiman_Streltsov_2016} along the $y$-direction for the longitudinal resonant tunneling at the mean-field and many-body levels (not shown in the main text). We have found in the main text that for $\Psi_G$ and $\Psi_X$, the transversal resonant conditions are satisfied at $\omega_n=0.19$ and $\omega_n=0.18$, respectively, and demonstrated the variances only at the resonant values of $\omega_n$. Here, we present all the many-body variances \cite{Alon_OE_2019, Sakmann_Schmiedmayer_2018}, discussed in the main text, along with their convergences at $\omega_n=0.18$, $0.19$, and $0.20$ for both the initial states. Some conclusions are drawn from the differences between the variances at and slightly off resonance.   $M=1$ represents  the mean-field level. Here, the results are displayed for $M=6$ and $M=10$ time-dependent orbitals for $\Psi_G$ and $\Psi_X$, and $M=10$ and $M=12$ time-dependent orbitals for $\Psi_Y$. The fully converged results imply that $M=6$ time-dependent orbitals are required to accurately present the dynamics of  $\Psi_G$ and $\Psi_X$ in the asymmetric 2D double-well potential, while for $\Psi_Y$, one requires $M=10$ time-dependent orbitals for the time considered in this work \cite{Bhowmik_Haldar_2020}. It has been checked that the size of the 2D box considered in this work is adequate  for both longitudinal and transversal scenarios.  Section-V and section-VI describe   the longitudinal and transversal resonant tunneling scenarios, respectively.  

\section{Convergences of quantities in longitudinal resonant tunneling}
Here,  the convergences of the many-body quantities  in the longitudinal resonant tunneling condition are illustrated for the ground $(\Psi_G)$ and transversely-excited $(\Psi_Y)$ states. The computations performed in the main text are with $M=6$ and $10$ time-dependent orbitals for  $\Psi_G$ and $\Psi_Y$, respectively. Here, to testify the convergence with the orbital numbers, we repeat our computations with $M=10$ and $12$ time-dependent orbitals for  $\Psi_G$ and $\Psi_Y$, respectively. The  bosonic clouds consist of $N=10$ bosons  with the interaction parameter $\Lambda=0.01\pi$.  Also, the many-body
Hamiltonian is represented by $64\times64$ exponential discrete-variable-representation grid points in a
box size $[-10, 10) \times [-10, 10)$. Here, we demonstrate the numerical convergences for the loss of coherence, fragmentation, and the variances of the position, momentum, and angular-momentum operators.

\renewcommand{\thefigure}{S\arabic{figure}}

\setcounter{figure}{0}
\begin{figure*}[!h]
\centering
{\includegraphics[trim = 0.1cm 0.5cm 0.1cm 0.2cm, scale=.60]{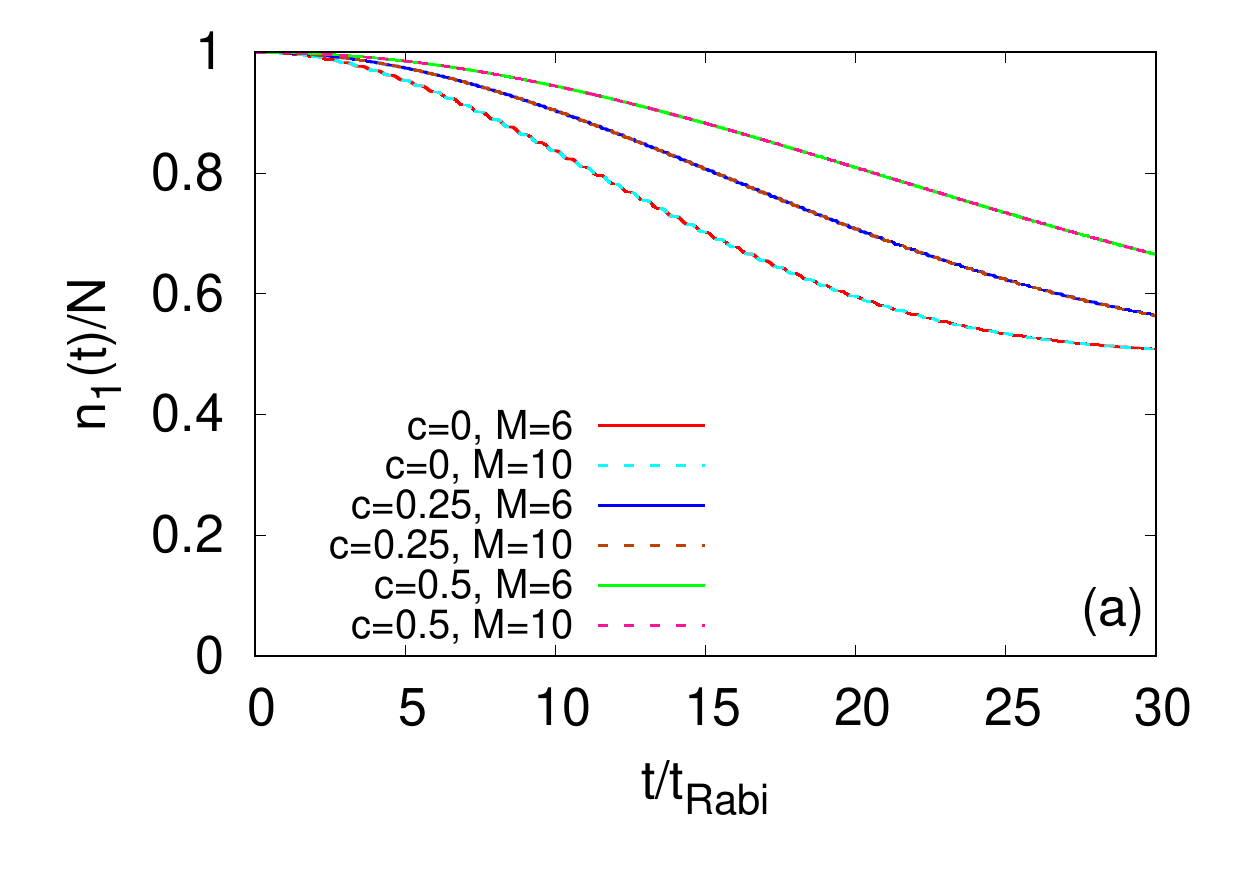}}
{\includegraphics[trim = 0.1cm 0.5cm 0.1cm 0.2cm, scale=.60]{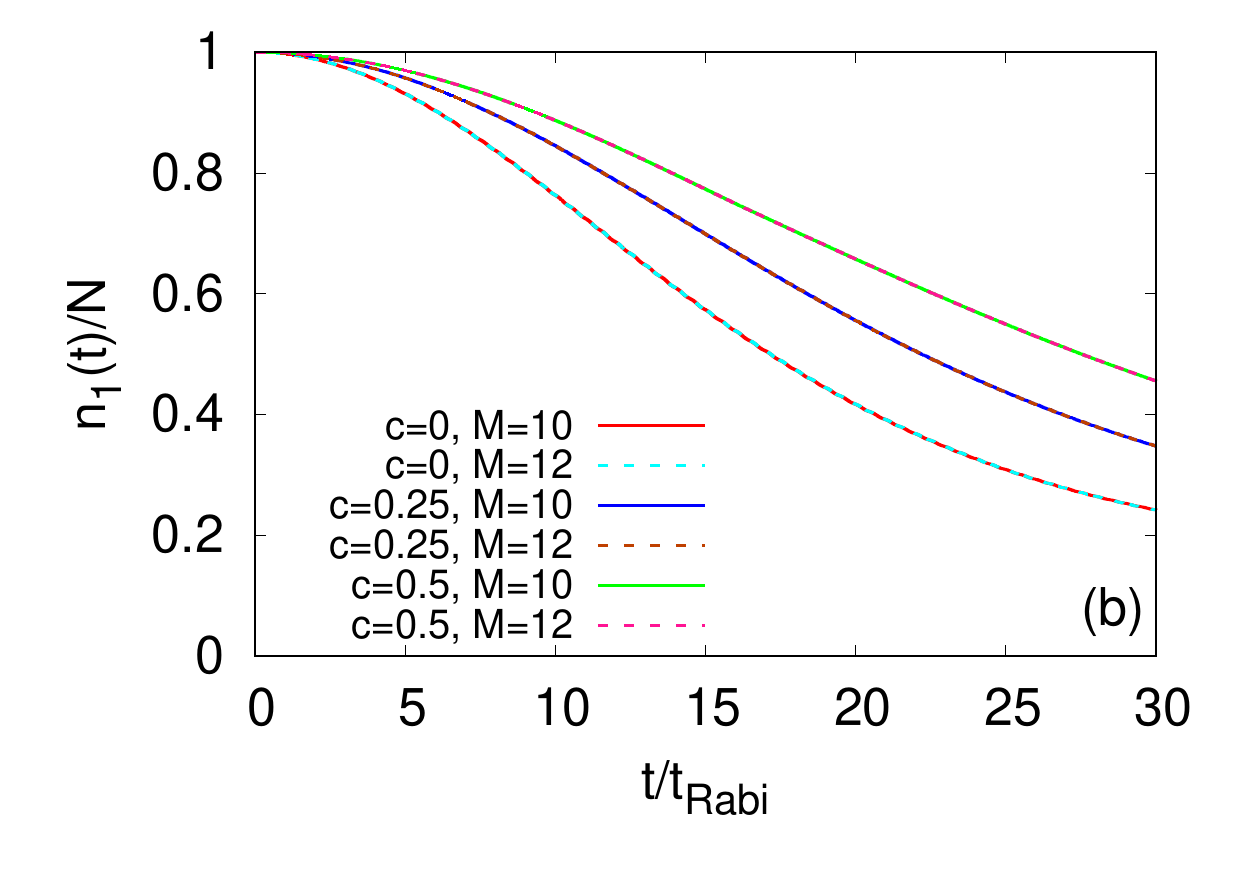}}\\
\caption{\label{figS1}Convergence of  the  occupation number per particle of the first natural orbital, $\dfrac{n_1(t)}{N}$, with the number of time-dependent orbitals for the initial states (a) $\Psi_G$ and (b) $\Psi_Y$ in the longitudinally-asymmetric 2D double-well potential. The  bosonic clouds consist of $N=10$ bosons  with the interaction parameter $\Lambda=0.01\pi$. The asymmetry parameters are $c=0$, 0.25, and 0.5.  The convergences is verified with $M=6$,  $10$ time-dependent orbitals for  the state $\Psi_{G}$. While we demonstrate the convergence of the results for $\Psi_{Y}$  using $M=10$,  $12$ time-dependent orbitals.   We show here   dimensionless quantities. Color codes are explained in each panel.}
\end{figure*}

\begin{figure*}[!h]
\centering
{\includegraphics[trim = 0.1cm 0.5cm 0.1cm 0.2cm, scale=.60]{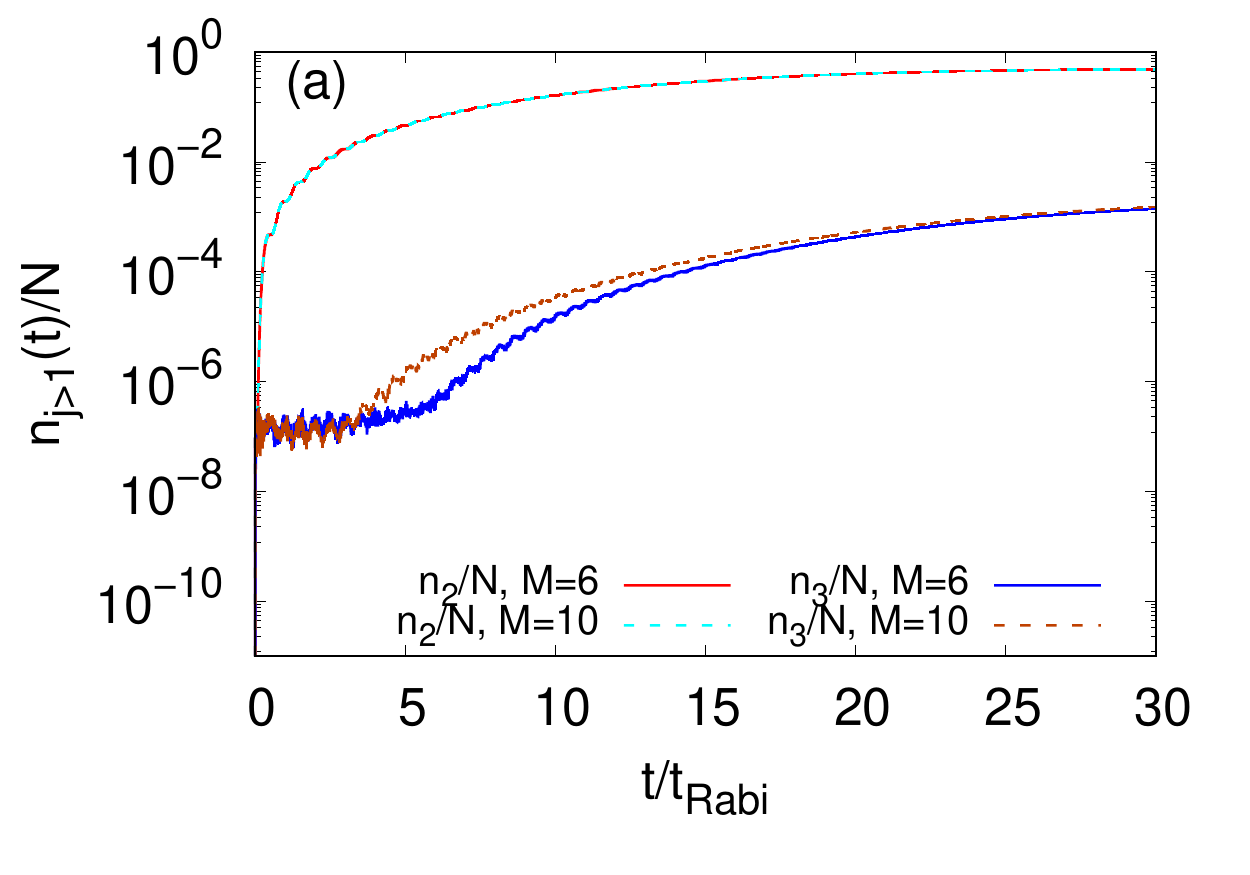}}
{\includegraphics[trim = 0.1cm 0.5cm 0.1cm 0.2cm, scale=.60]{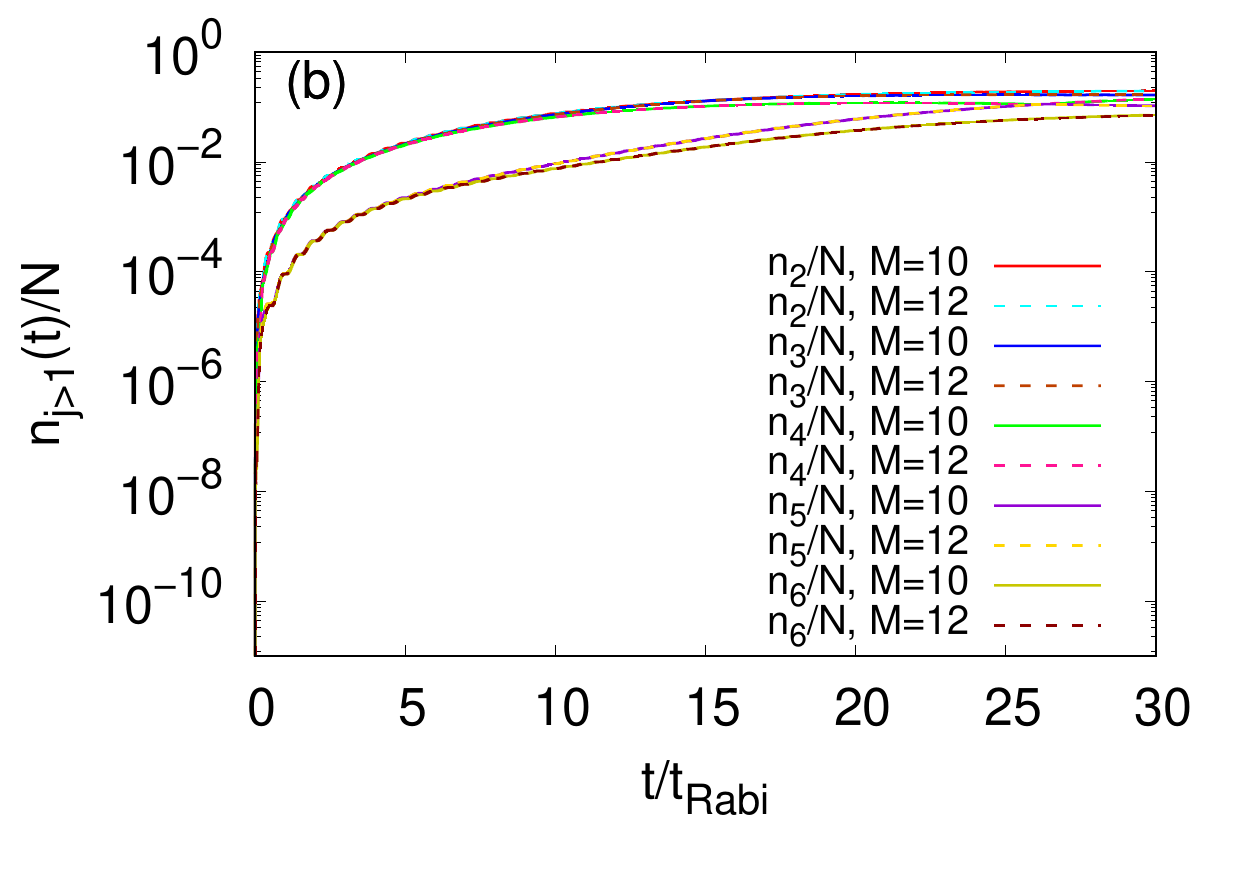}}\\
{\includegraphics[trim = 0.1cm 0.5cm 0.1cm 0.2cm, scale=.60]{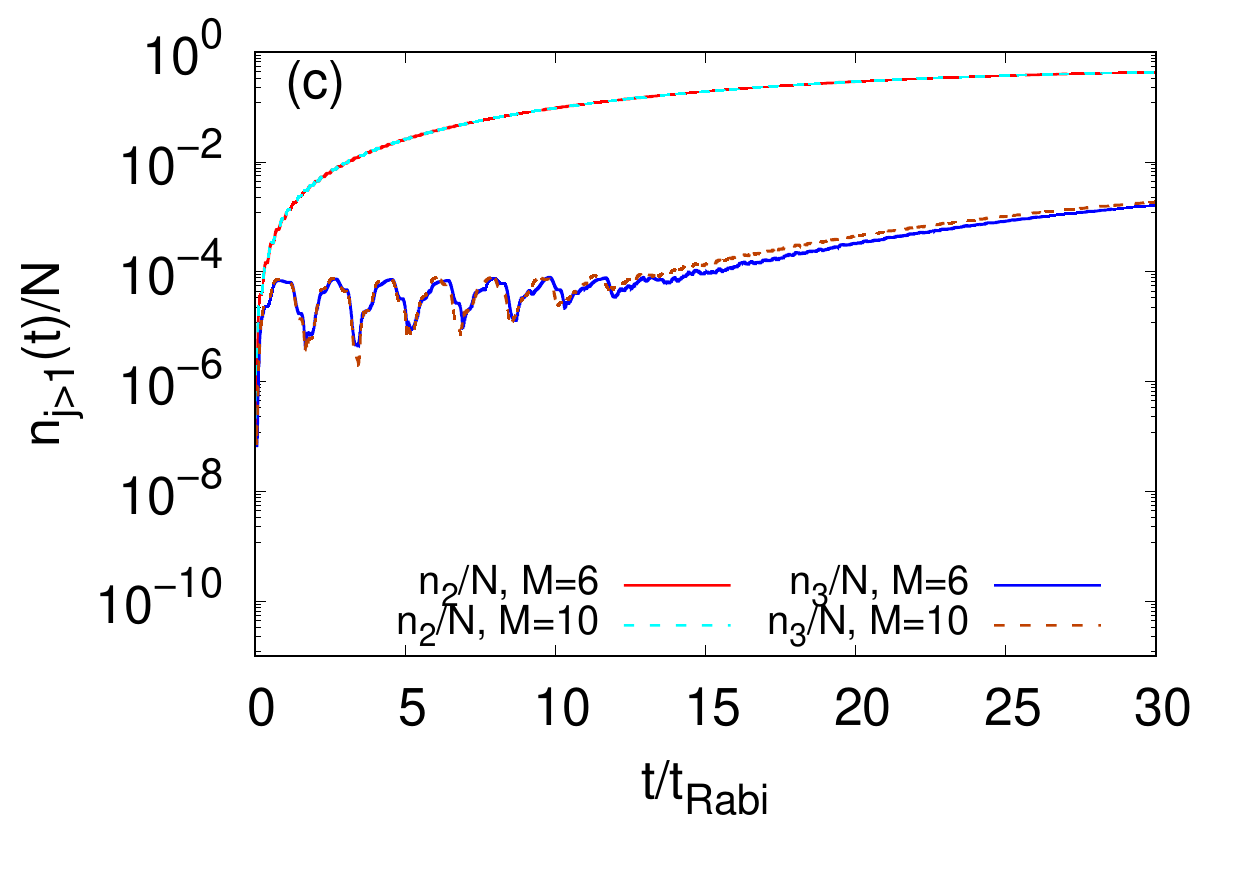}}
{\includegraphics[trim = 0.1cm 0.5cm 0.1cm 0.2cm, scale=.60]{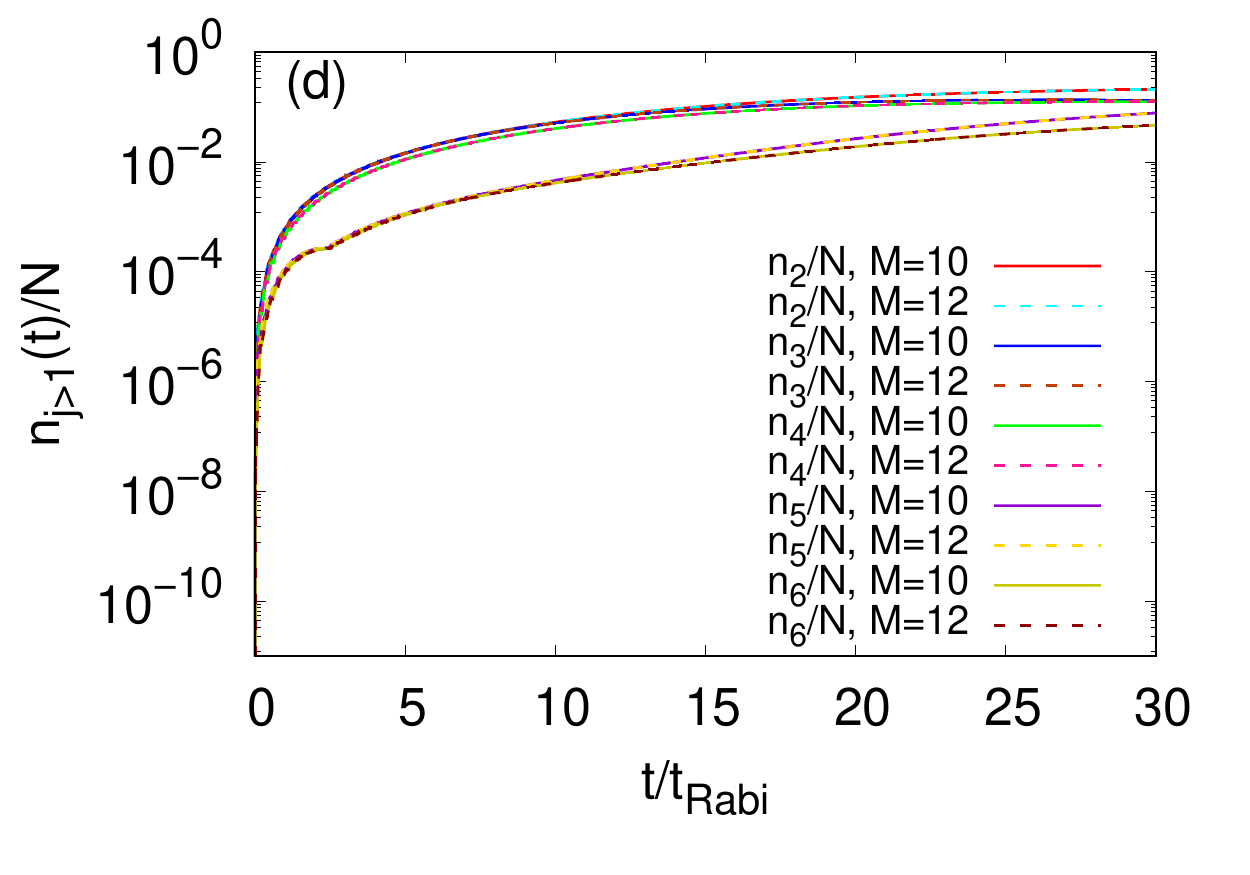}}\\
{\includegraphics[trim = 0.1cm 0.5cm 0.1cm 0.2cm, scale=.60]{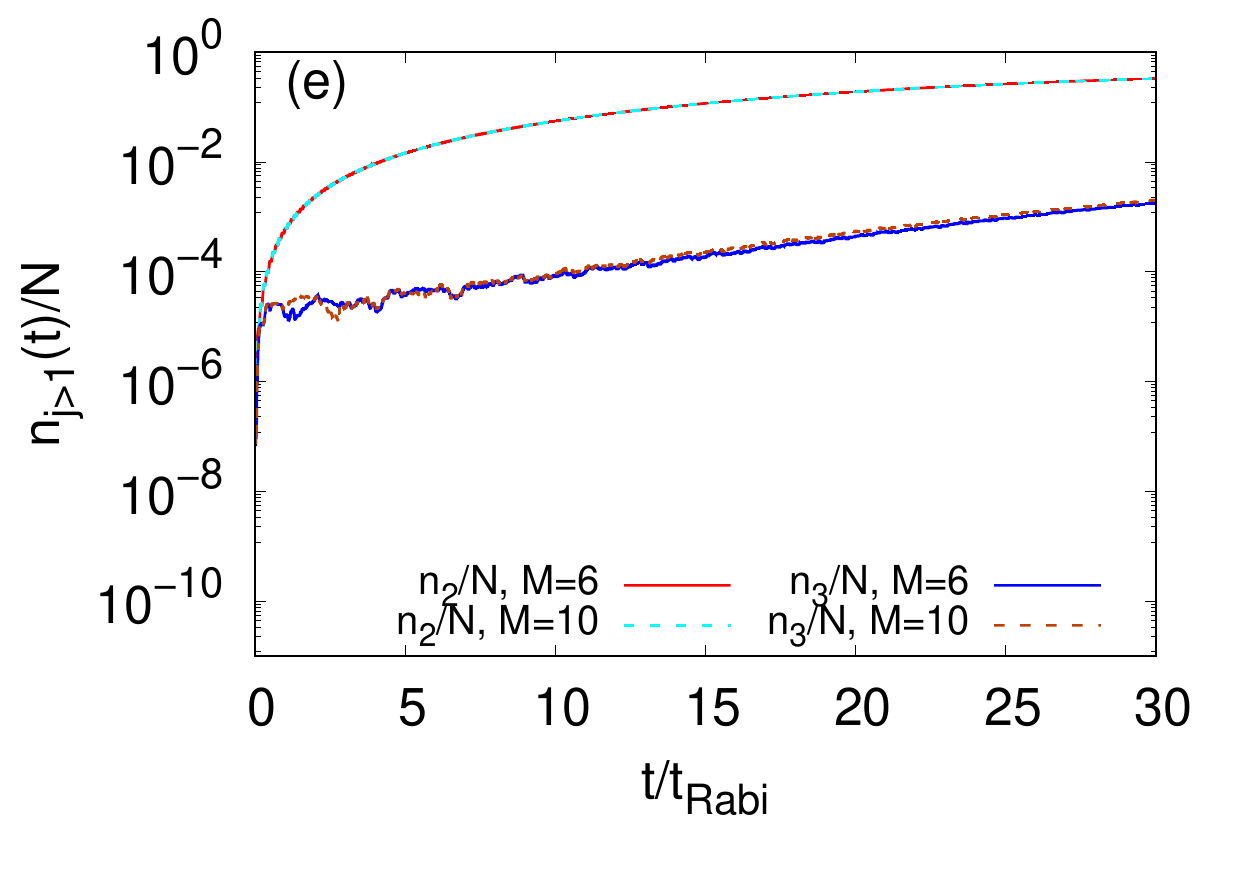}}
{\includegraphics[trim = 0.1cm 0.5cm 0.1cm 0.2cm, scale=.60]{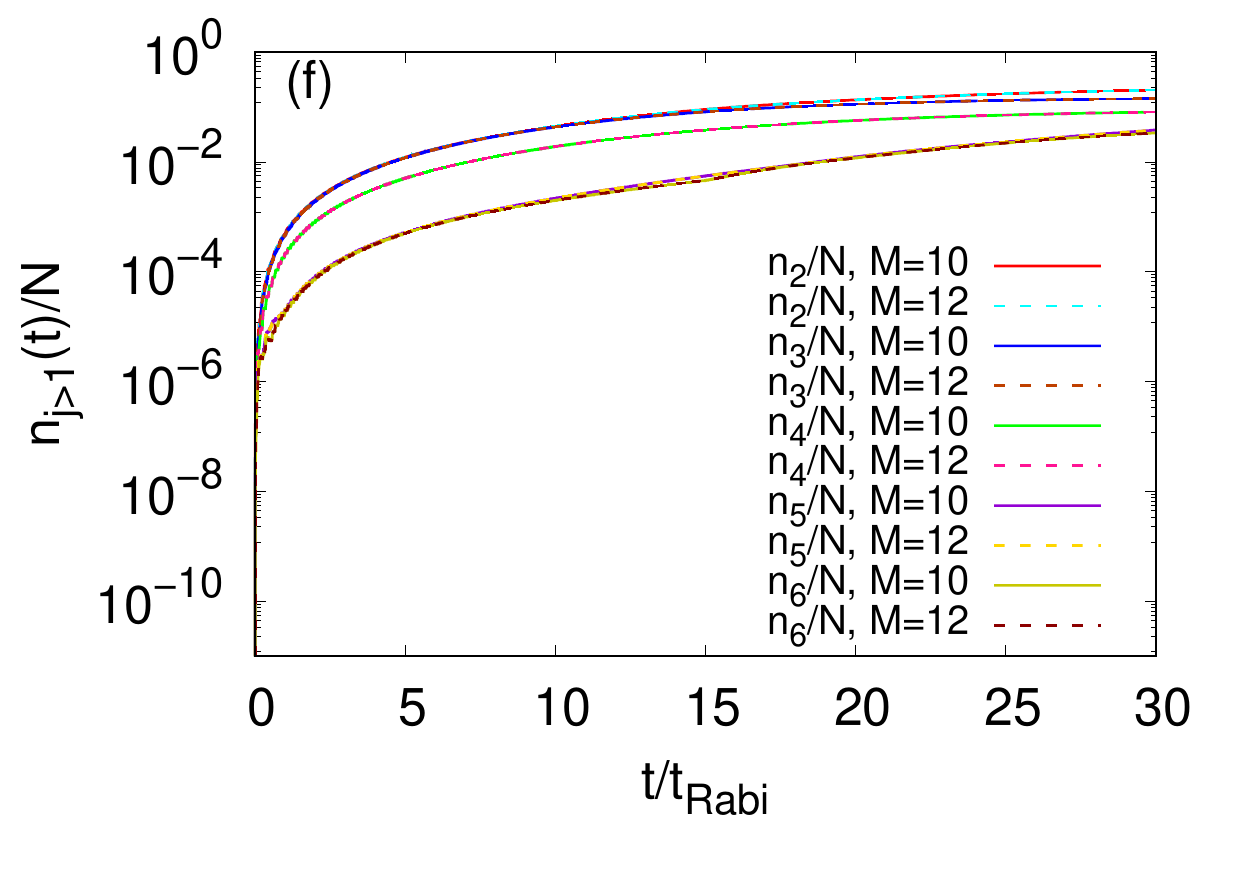}}\\
\caption{\label{figS2}Convergence of the natural occupation number per particle, $\dfrac{n_{j>1}(t)}{N}$, with the number of time-dependent orbitals for the initial states $\Psi_G$ (left column) and  $\Psi_Y$ (right column) in the longitudinally-asymmetric 2D double-well potential. The  bosonic clouds consist of $N=10$ bosons  with the interaction parameter $\Lambda=0.01\pi$. The results for the asymmetry parameters $c=0$, 0.25, and 0.5 are presented row-wise.  The convergence is verified with $M=6$,  $10$ time-dependent orbitals for  the state $\Psi_{G}$. While we demonstrate the convergence of the results for $\Psi_{Y}$  using $M=10$,  $12$ time-dependent orbitals.  We show here   dimensionless quantities. Color codes are explained in each panel.}
\end{figure*}

\begin{figure*}[!h]
\centering
{\includegraphics[trim = 0.1cm 0.5cm 0.1cm 0.2cm, scale=.60]{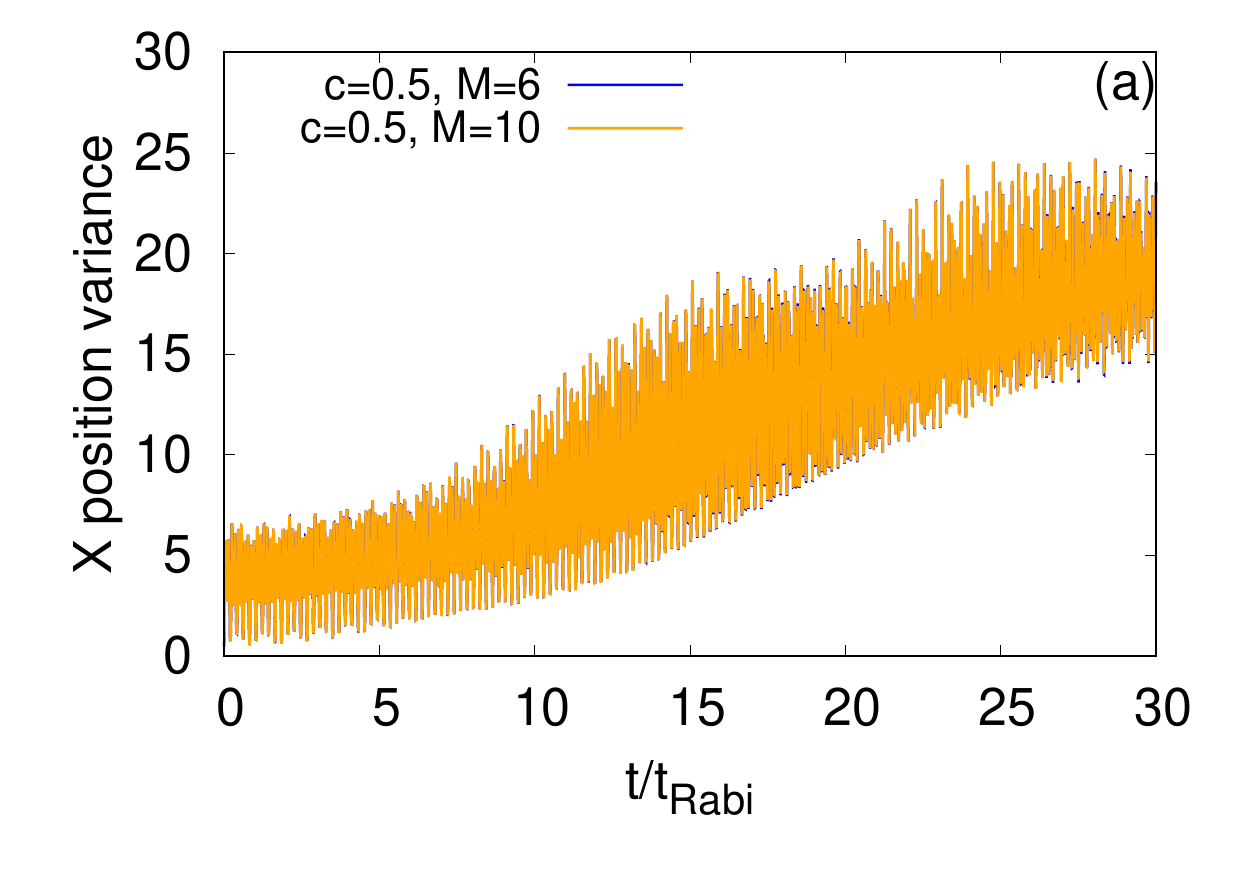}}
{\includegraphics[trim = 0.1cm 0.5cm 0.1cm 0.2cm, scale=.60]{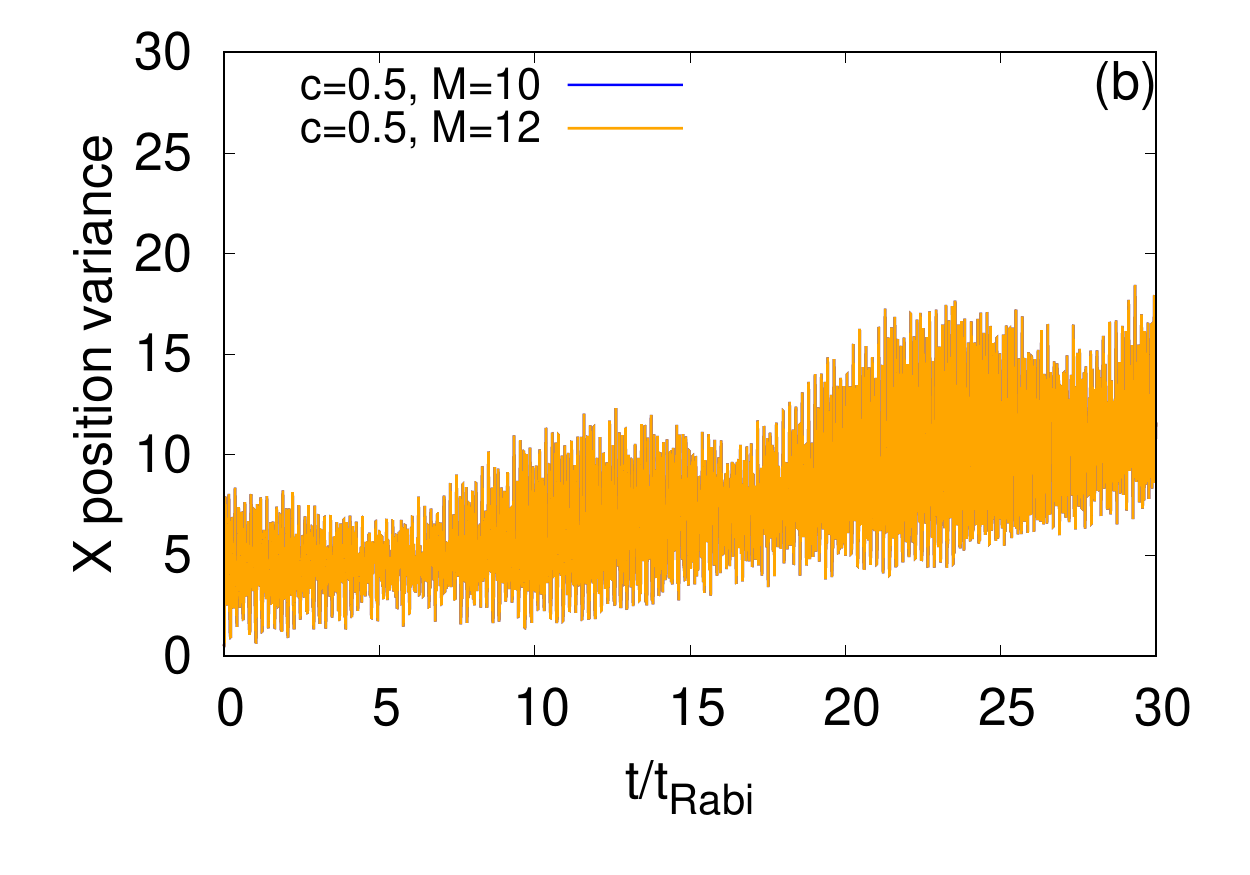}}\\
\vglue 0.25 truecm
{\includegraphics[trim = 0.1cm 0.5cm 0.1cm 0.2cm, scale=.60]{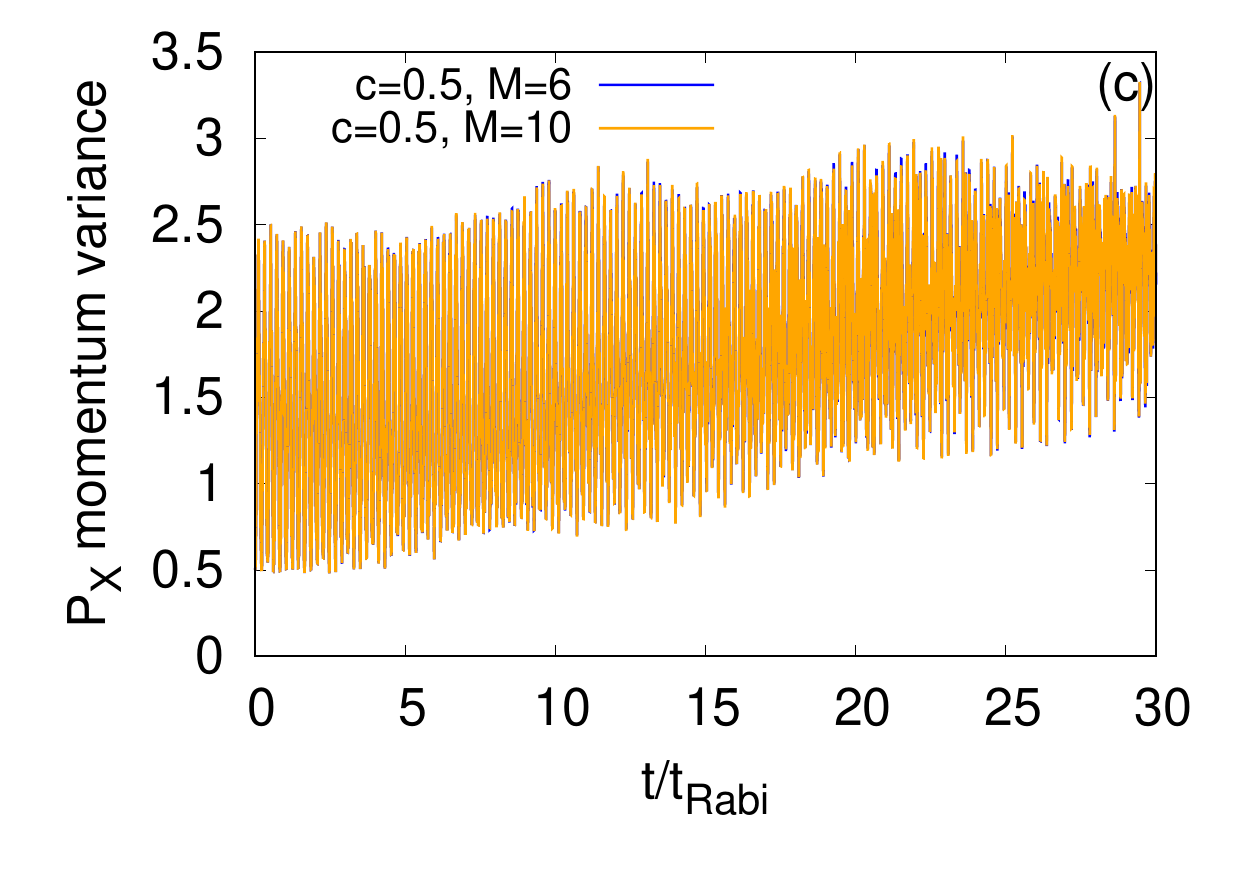}}
{\includegraphics[trim = 0.1cm 0.5cm 0.1cm 0.2cm, scale=.60]{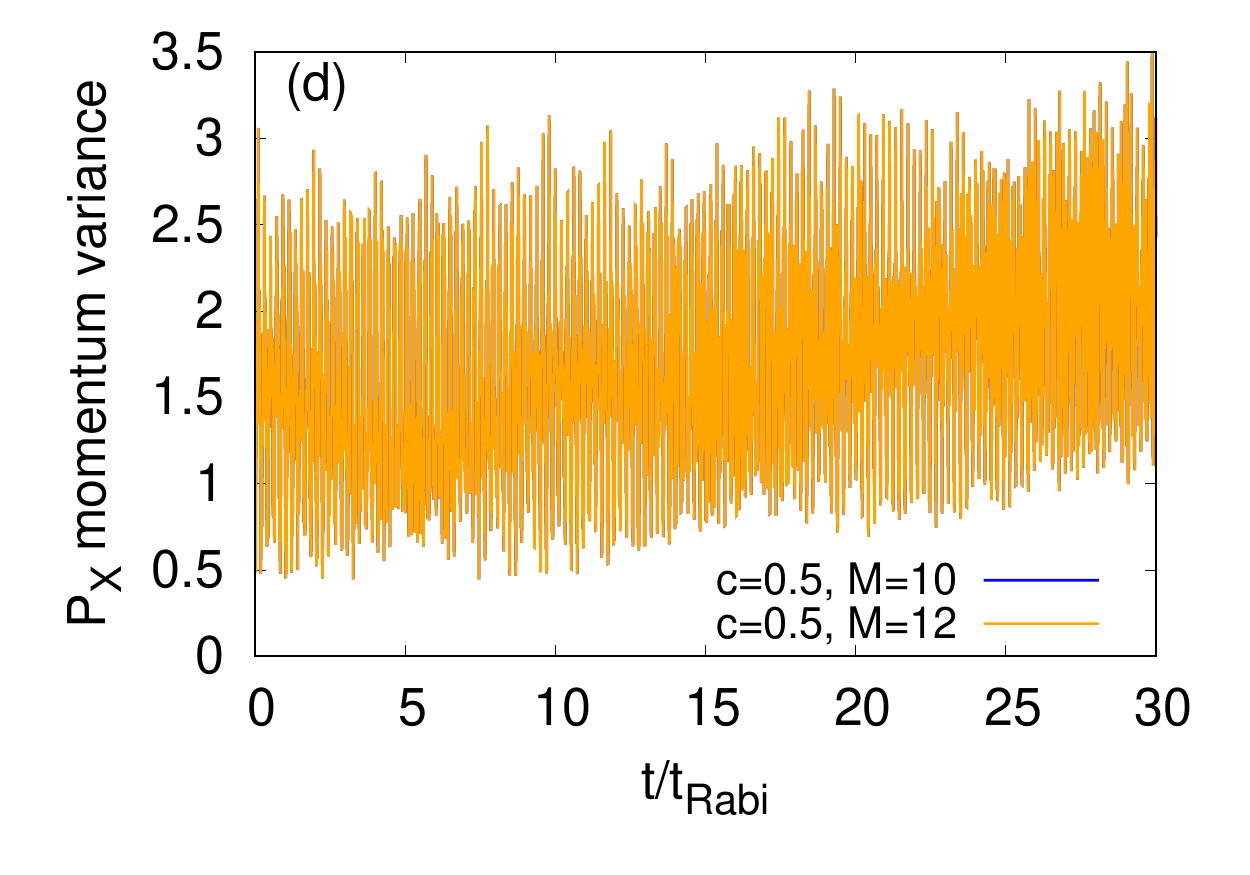}}\\
\caption{\label{figS3}Convergences of the time-dependent many-body position (1st row) and momentum (2nd row) variances per particle along the $x$-direction, $\dfrac{1}{N}\Delta_{\hat{X}}^2(t)$ and $\dfrac{1}{N}\Delta_{\hat{P}_X}^2(t)$, respectively, with the number of time-dependent orbitals for the initial states $\Psi_G$ (left column) and  $\Psi_Y$ (right column) in the  longitudinally-asymmetric 2D double-well potential.  The  bosonic clouds consist of $N=10$ bosons  with the interaction parameter $\Lambda=0.01\pi$.  The results are for the asymmetry parameter $c=0.5$.  The convergence is verified with $M=6$,  $10$ time-dependent orbitals for  the state $\Psi_{G}$. While we demonstrate the convergence of the results for $\Psi_{Y}$  using $M=10$,  $12$ time-dependent orbitals.    We show here   dimensionless quantities. Color codes are explained in each panel.}
\end{figure*}

Fig.~\ref{figS1} represents the convergence of a basic quantity -- loss of coherence -- of the bosonic clouds, $\Psi_{G}$ and $\Psi_{Y}$, in terms of the occupation numbers per particle of the first natural orbitals, $\dfrac{n_1(t)}{N}$. The figure showcases the results for all the asymmetry parameters, i.e., $c=0$, $0.25$, and $0.5$, discussed in the main text. For both  states, $\dfrac{n_1(t)}{N}$ decays with time with small background oscillations due to gradually increasing quantum correlations. The plots show that $\dfrac{n_1(t)}{N}$ computed from a larger
number of time-dependent orbitals falls on top of the corresponding results calculated using a smaller number of time-dependent orbitals. The overlapping of the curves  exhibits the convergence of $\dfrac{n_1(t)}{N}$ with the  number of orbitals.

Having analyzed the convergence of $\dfrac{n_1(t)}{N}$, we investigate the further details of how the  coherence is lost in the system by examining how fragmentation is developed.  Fig.~\ref{figS2} displays the time evolution  of   the occupancies of the higher natural orbitals  per particle,  $\dfrac{n_{j>1}(t)}{N}$, for $\Psi_{G}$ and $\Psi_{Y}$ with their convergences with the number of time-dependent orbitals. Hence, convergence of the occupation numbers in the dynamics considered in this work is demonstrated from large to small (top to bottom).

 As presented in the discussion of loss of coherence, here also,  we plot $\dfrac{n_{j>1}(t)}{N}$ for the  asymmetry parameters, $c=0$, $0.25$, and $0.5$. The higher natural orbitals which have  significant amount of occupancies for $\Psi_{G}$ (second and third natural orbitals)  and $\Psi_{Y}$ (second to sixth natural orbitals) are shown in the figure for all asymmetry parameters.  To demonstrate the convergence with the time-dependent orbital numbers, we present  the fragmentation dynamics with $M=6$, $10$ time-dependent orbitals  for $\Psi_{G}$ and $M=10$, $12$ time-dependent orbitals for $\Psi_{Y}$. Here it is observed that as time passes by  the occupations of all the higher natural orbitals gradually increase (Note the logarithmic scale).  Moreover, $\dfrac{n_3(t)}{N}$ of $\Psi_{G}$  are oscillatory first and then increasing, clearly visible at $c=0.25$. We notice that  only two higher natural orbitals are considerably occupied for $\Psi_{G}$ but for $\Psi_{Y}$, the amount of occupancies of the five higher natural orbitals are rather significant. The comparison of the fragmentation dynamics of $\Psi_{G}$ and $\Psi_{Y}$ exhibits that whenever  the transverse excitations exist in the
system, more natural orbitals are required to accurately capture the dynamical behavior.  However, the occupancies of the higher natural orbitals computed from a larger number of time-dependent orbitals overlap with the  respective results found from a smaller number of time-dependent orbitals, indicating the convergence of the fragmentation dynamics for $\Psi_{G}$ and $\Psi_{Y}$. The convergences of the occupancies of all the orbitals demonstrated till now automatically imply the convergence of the most basic quantity, i.e., survival probability, see Fig. 3 in the main text.

\begin{figure*}[!h]
\centering
{\includegraphics[trim = 0.1cm 0.5cm 0.1cm 0.2cm, scale=.60]{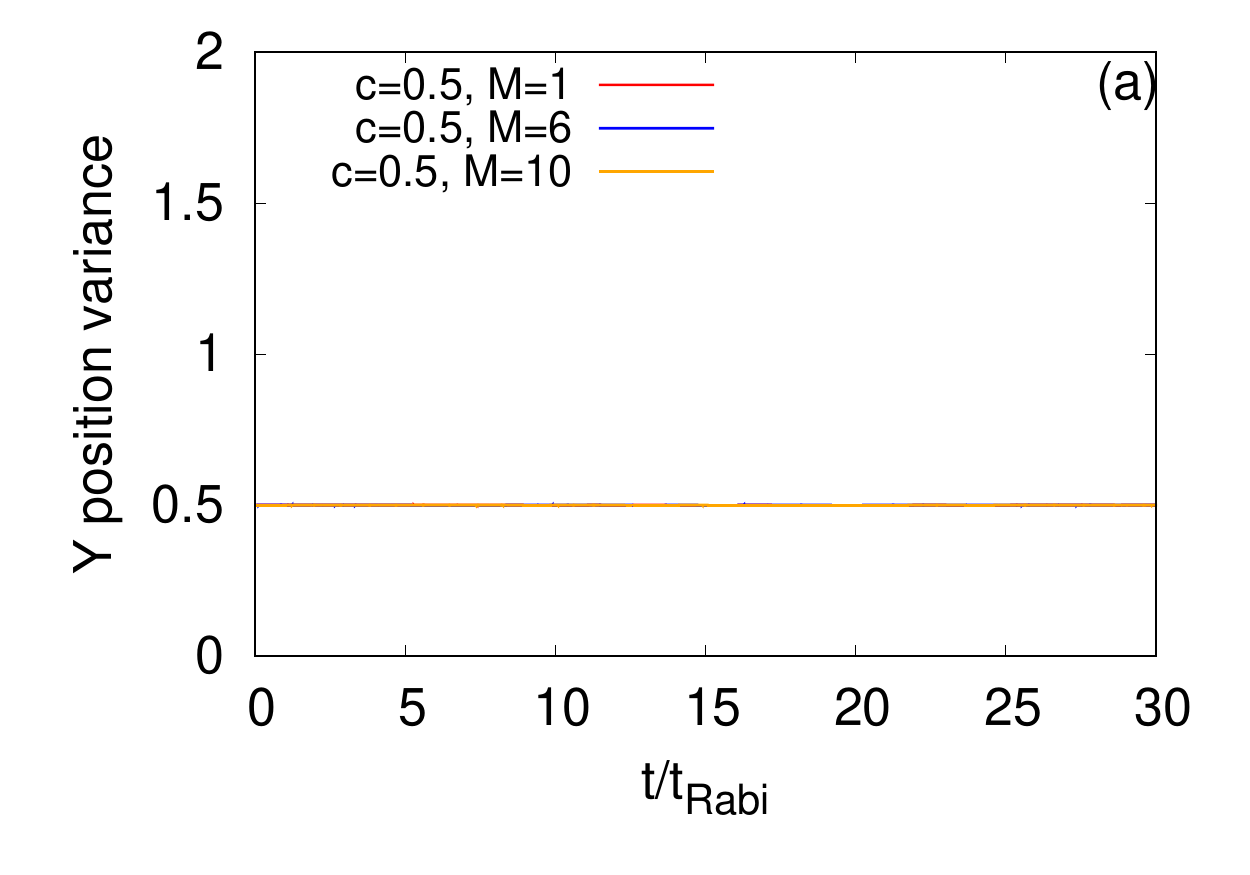}}
{\includegraphics[trim = 0.1cm 0.5cm 0.1cm 0.2cm, scale=.60]{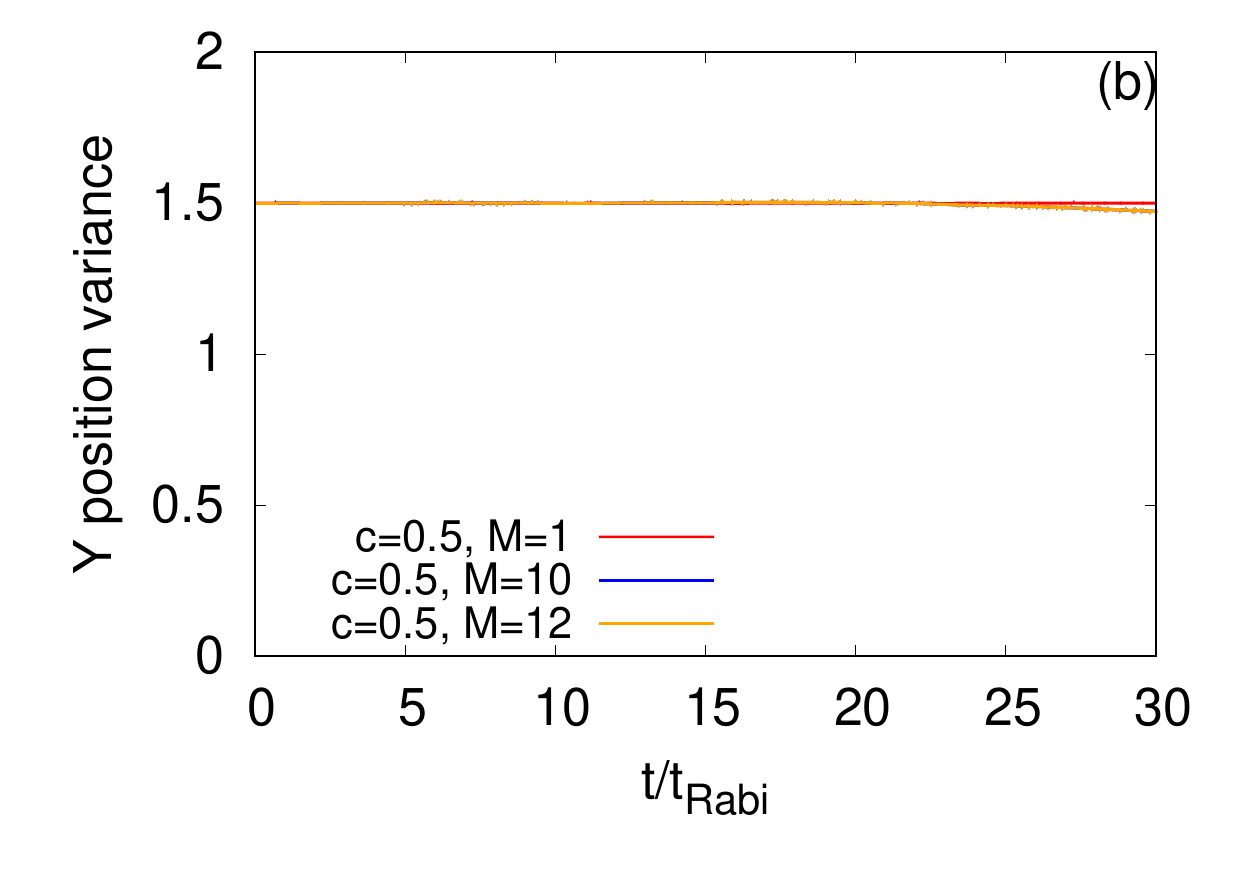}}\\
\vglue 0.25 truecm
{\includegraphics[trim = 0.1cm 0.5cm 0.1cm 0.2cm, scale=.60]{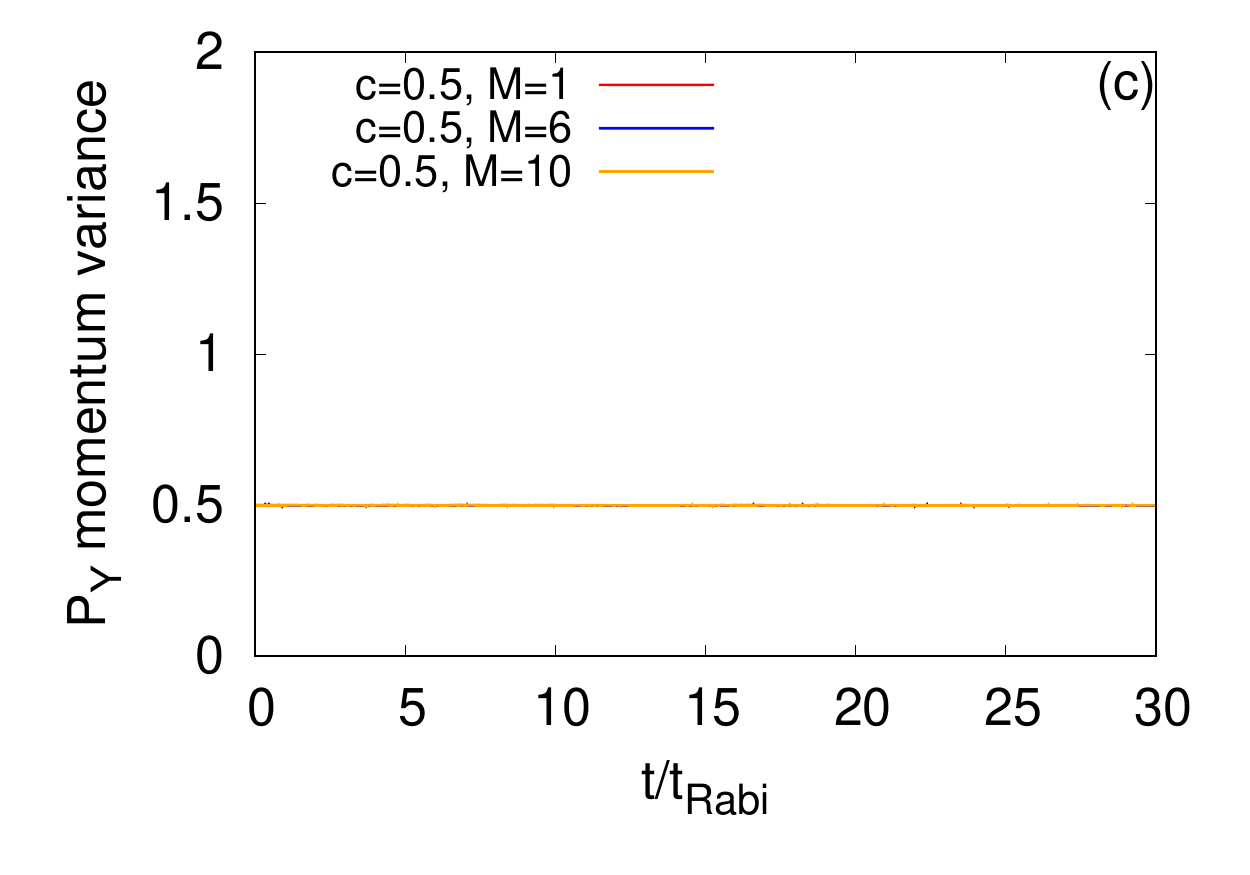}}
{\includegraphics[trim = 0.1cm 0.5cm 0.1cm 0.2cm, scale=.60]{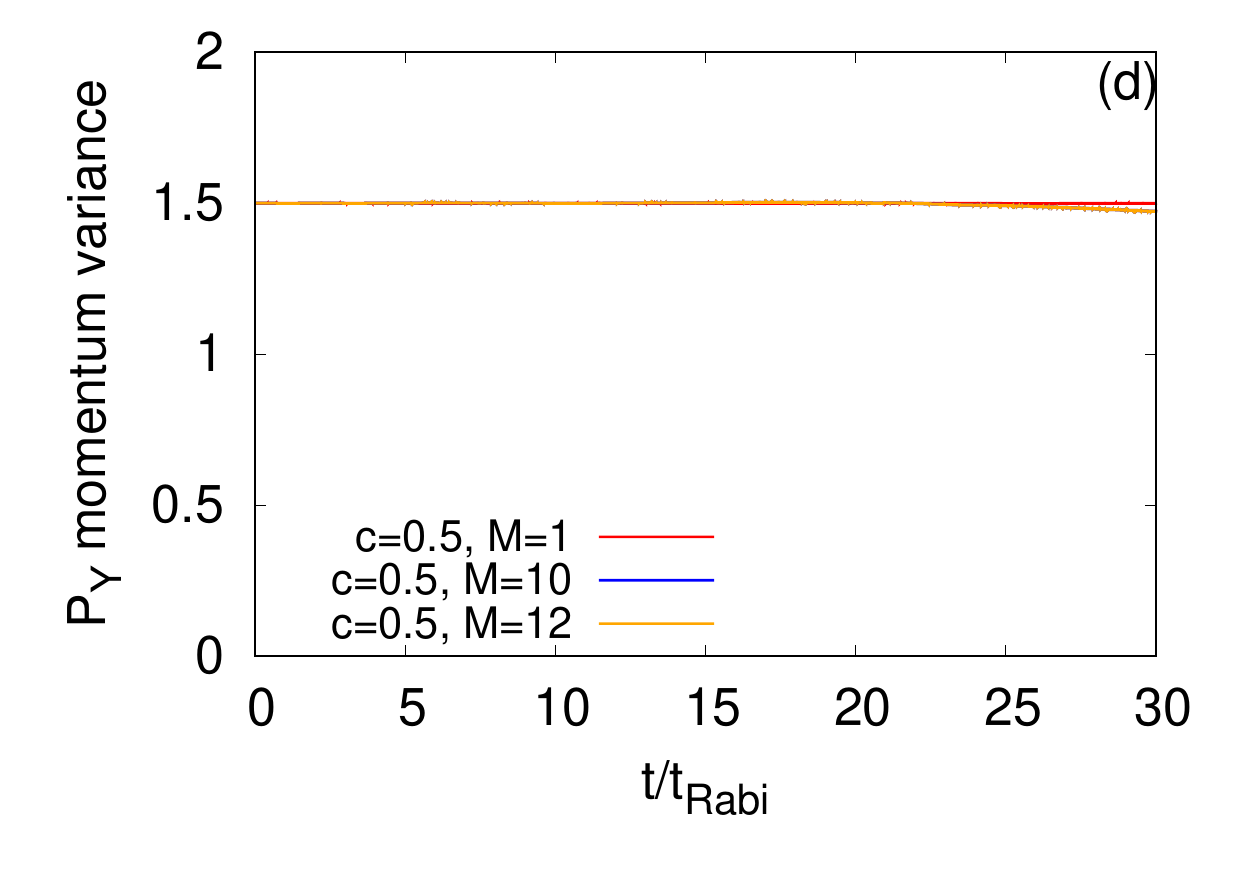}}\\
\caption{\label{figS4}Convergences of the time-dependent many-body position (1st row) and momentum (2nd row) variances per particle along the $y$-direction, $\dfrac{1}{N}\Delta_{\hat{Y}}^2(t)$ and $\dfrac{1}{N}\Delta_{\hat{P}_Y}^2(t)$, respectively, with the number of time-dependent orbitals for the initial states $\Psi_G$ (left column) and  $\Psi_Y$ (right column)  in the  longitudinally-asymmetric 2D double-well potential.  The  bosonic clouds consist of $N=10$ bosons  with the interaction parameter $\Lambda=0.01\pi$.  The results are for the asymmetry parameter $c=0.5$.  $M=1$ shows the mean-field results. The convergence is verified with $M=6$,  $10$ time-dependent orbitals for  the state $\Psi_{G}$. While we demonstrate the convergence of the results for $\Psi_{Y}$  using $M=10$,  $12$ time-dependent orbitals.    We show here   dimensionless quantities. Color codes are explained in each panel.}
\end{figure*}

The time evolution of the many-body position and momentum variances per particle along the $x$-direction, $\dfrac{1}{N}\Delta_{\hat{X}}^2(t)$ and $\dfrac{1}{N}\Delta_{\hat{P}_X}^2(t)$, respectively, along with their numerical convergences with respect to the orbital numbers are illustrated in Fig.~\ref{figS3}. Here, we checked and verified the convergences of  $\dfrac{1}{N}\Delta_{\hat{X}}^2(t)$ and $\dfrac{1}{N}\Delta_{\hat{P}_X}^2(t)$ for all the asymmetry parameters discussed in the main text. But, to demonstrate the convergence, we choose only the second resonant tunneling condition, i.e.,  $c=0.5$. In the main text, we discussed that, at $c=0.5$, $\dfrac{1}{N}\Delta_{\hat{X}}^2(t)$ and $\dfrac{1}{N}\Delta_{\hat{P}_X}^2(t)$ have two kinds of oscillations, namely, small frequency with large amplitude (due to the density oscillation) and high frequency with small amplitude (due to the breathing mode oscillation). The panels in Fig.~\ref{figS3} show that both  variances are  well  converged with the orbital numbers. Even the oscillation which has high frequency with small amplitude, found with lower orbital numbers, overlaps with the corresponding results calculated from the higher orbital numbers.

As discussed in the main text, the many-particle position and momentum variances per particle along the $y$-direction, $\dfrac{1}{N}\Delta_{\hat{Y}}^2(t)$ and $\dfrac{1}{N}\Delta_{\hat{P}_Y}^2(t)$, respectively, have  tiny   fluctuations, of the order of $10^{-3}$ for all the  asymmetry parameters discussed in this work. Here we graphically show   the time evolution of $\dfrac{1}{N}\Delta_{\hat{Y}}^2(t)$ and $\dfrac{1}{N}\Delta_{\hat{P}_Y}^2(t)$ both at the mean-field and many-body levels along with their many-body convergences at $c=0.5$, see Fig.~\ref{figS4}. We observe that the almost frozen $\dfrac{1}{N}\Delta_{\hat{Y}}^2(t)$ and $\dfrac{1}{N}\Delta_{\hat{P}_Y}^2(t)$ for both  states, $\Psi_G$ and $\Psi_Y$, are fully converged with the  number of orbitals.

\begin{figure*}[!h]
\centering
{\includegraphics[trim = 0.1cm 0.5cm 0.1cm 0.2cm, scale=.60]{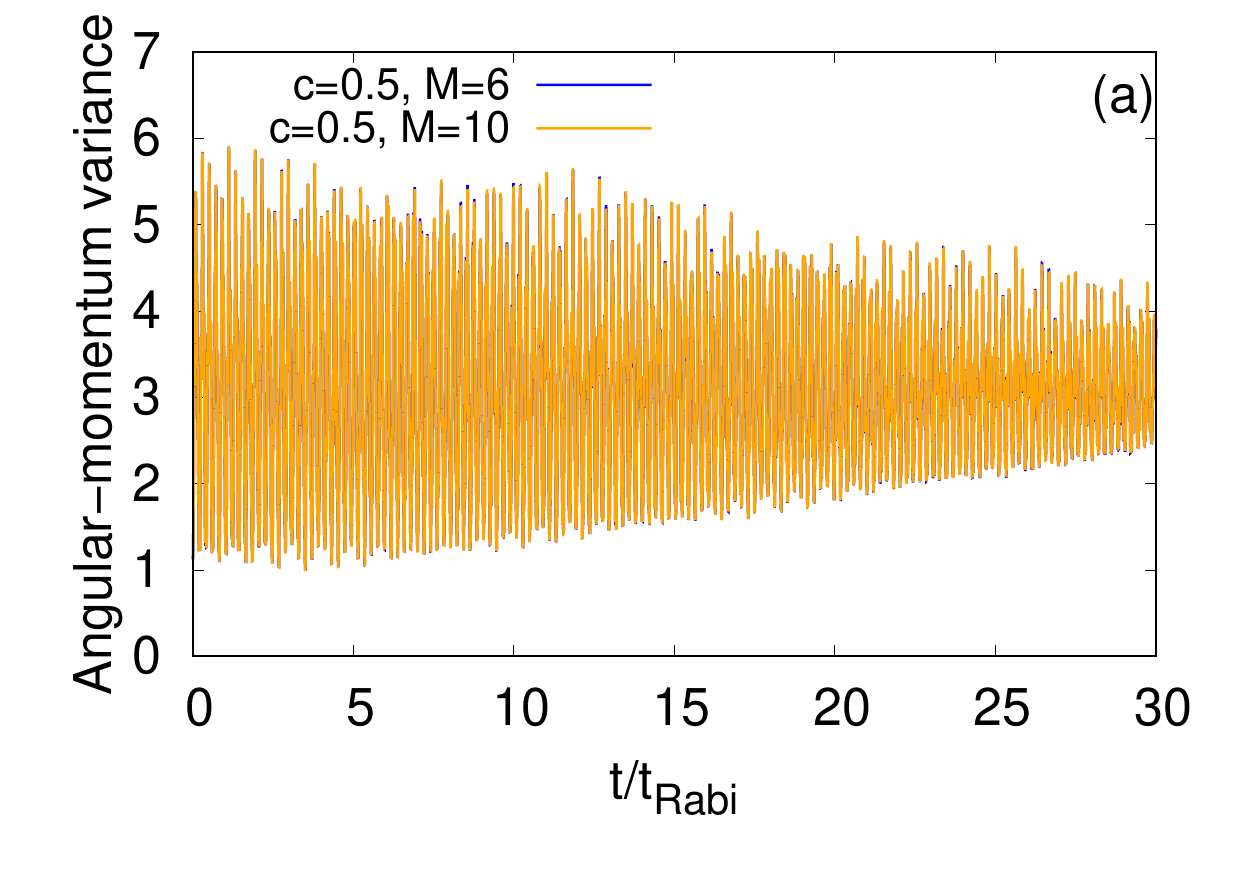}}
{\includegraphics[trim = 0.1cm 0.5cm 0.1cm 0.2cm, scale=.60]{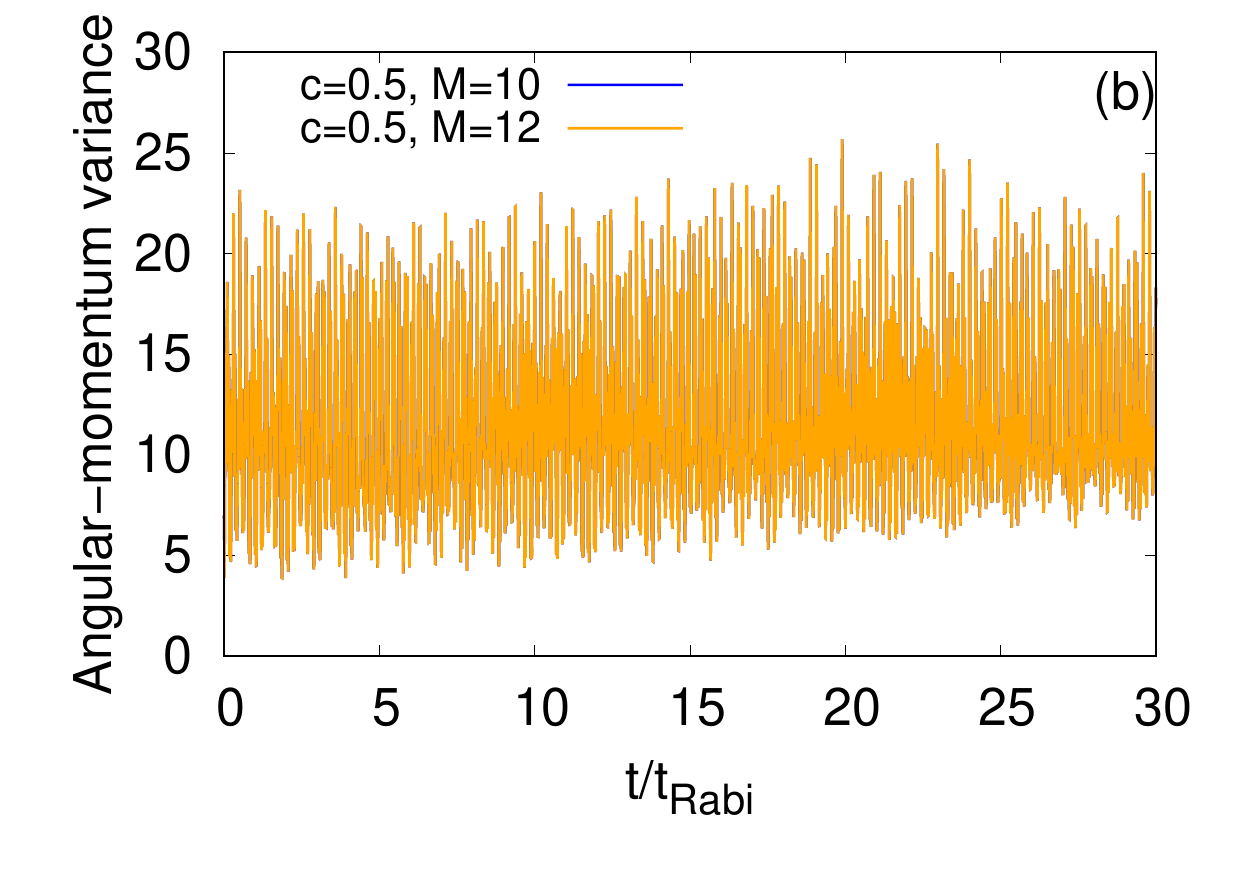}}\\
\caption{\label{figS5}Convergence of the time-dependent variance per particle of the $z$- component of the angular-momentum operator, $\dfrac{1}{N}\Delta_{\hat{L}_Z}^2(t)$,  with the number of time-dependent orbitals for the initial states (a) $\Psi_G$  and  (b) $\Psi_Y$ in the longitudinally-asymmetric 2D double-well potential.  The  bosonic clouds consist of $N=10$ bosons  with the interaction parameter $\Lambda=0.01\pi$.  The results are for the asymmetry parameter $c=0.5$.  The convergence is verified with $M=6$,  $10$ time-dependent orbitals for  the state $\Psi_{G}$. While we demonstrate the convergence of the results for $\Psi_{Y}$  using $M=10$,  $12$ time-dependent orbitals.    We show here   dimensionless quantities. Color codes are explained in each panel.}
\end{figure*}

Fig.~\ref{figS5} depicts the convergence of the many-body angular-momentum variance per particle, $\dfrac{1}{N}\Delta_{\hat{L}_Z}^2(t)$, for both  initial states, $\Psi_G$ and $\Psi_Y$.  As mentioned in the dynamics of $\dfrac{1}{N}\Delta_{\hat{X}}^2(t)$ and $\dfrac{1}{N}\Delta_{\hat{P}_X}^2(t)$ at $c=0.5$, here also, we observe the combined effect of the density and breathing-mode oscillations.  There is no visible difference of $\dfrac{1}{N}\Delta_{\hat{L}_Z}^2(t)$ found when computed from  the smaller and larger numbers of orbitals, which corresponds to the convergence of $\dfrac{1}{N}\Delta_{\hat{L}_Z}^2(t)$ with the number of orbitals.

\begin{figure*}[!h]
\centering
{\includegraphics[trim = 0.1cm 0.5cm 0.1cm 0.2cm, scale=.60]{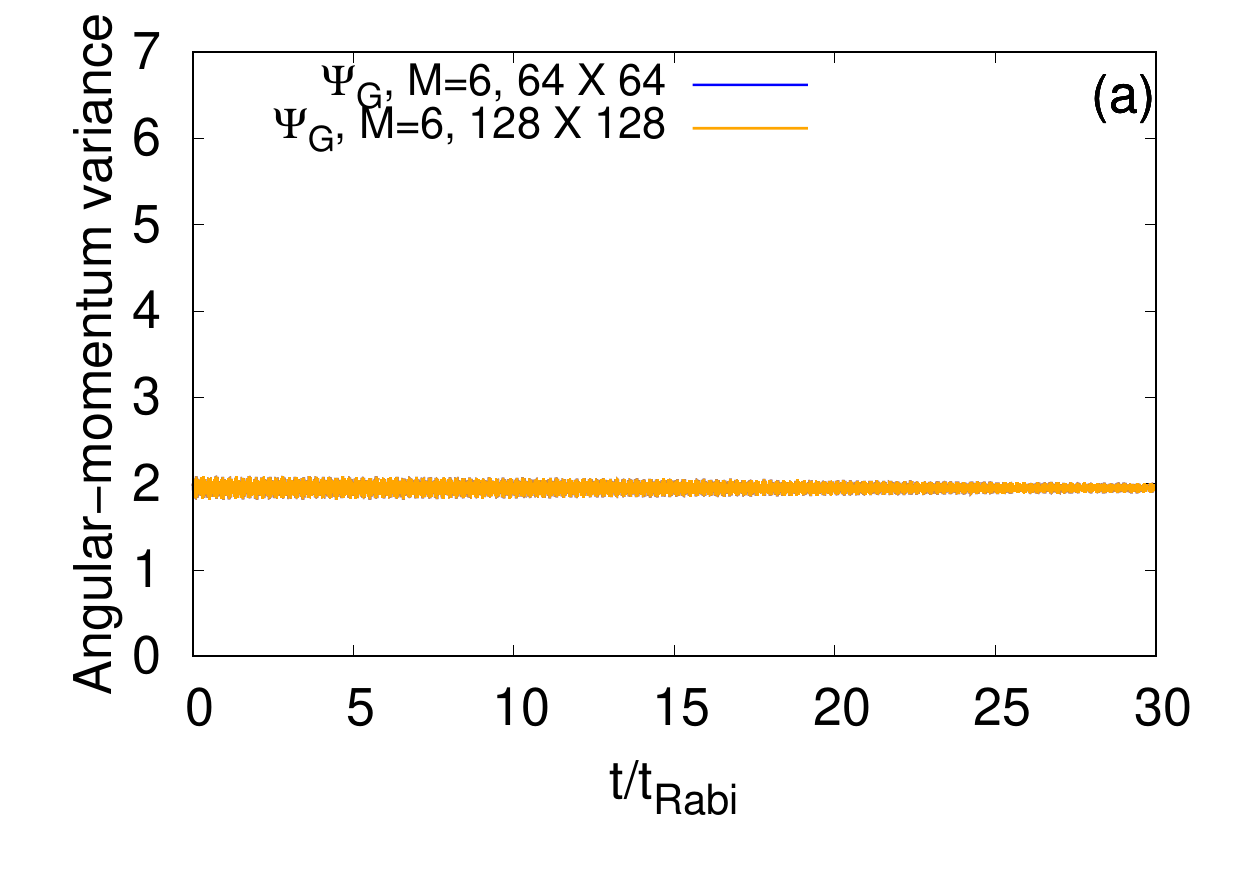}}
{\includegraphics[trim = 0.1cm 0.5cm 0.1cm 0.2cm, scale=.60]{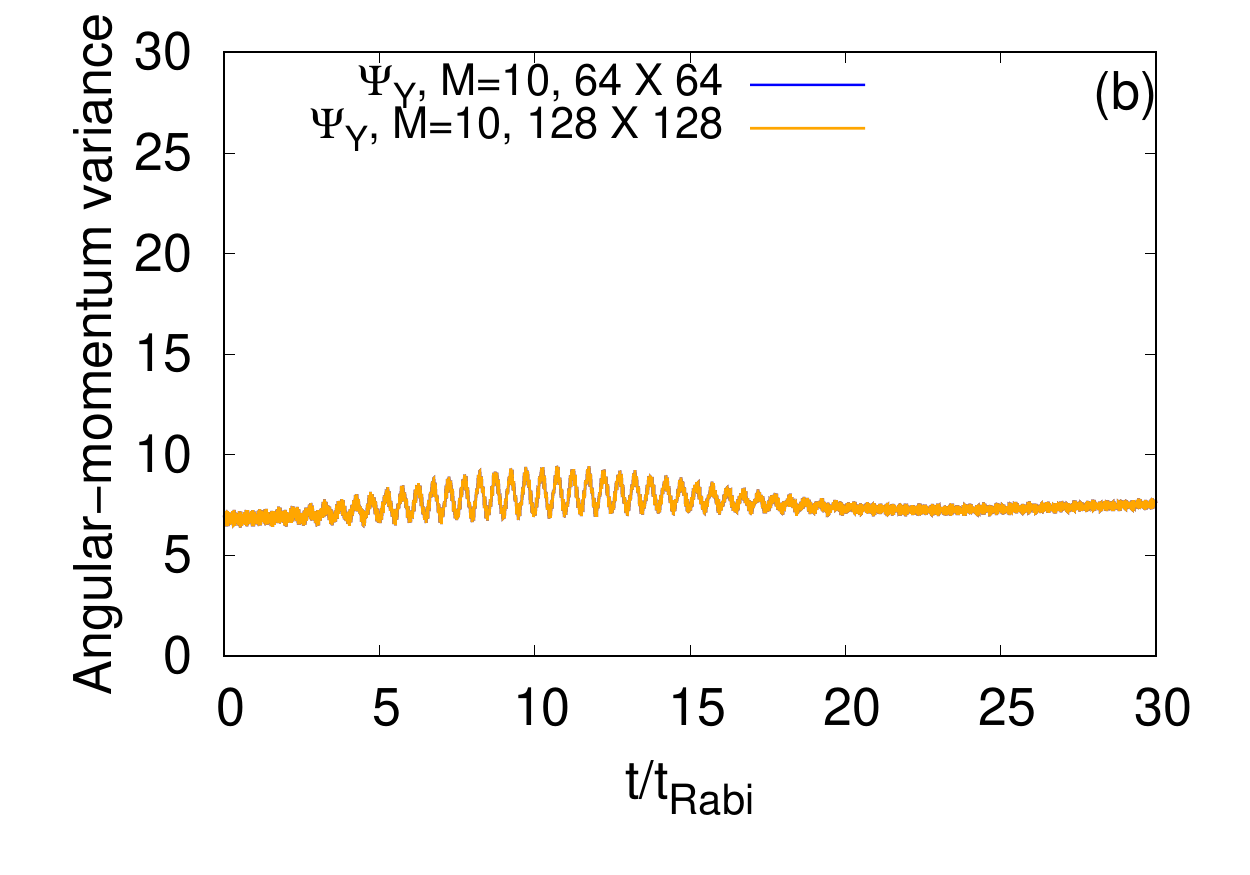}}\\
{\includegraphics[trim = 0.1cm 0.5cm 0.1cm 0.2cm, scale=.60]{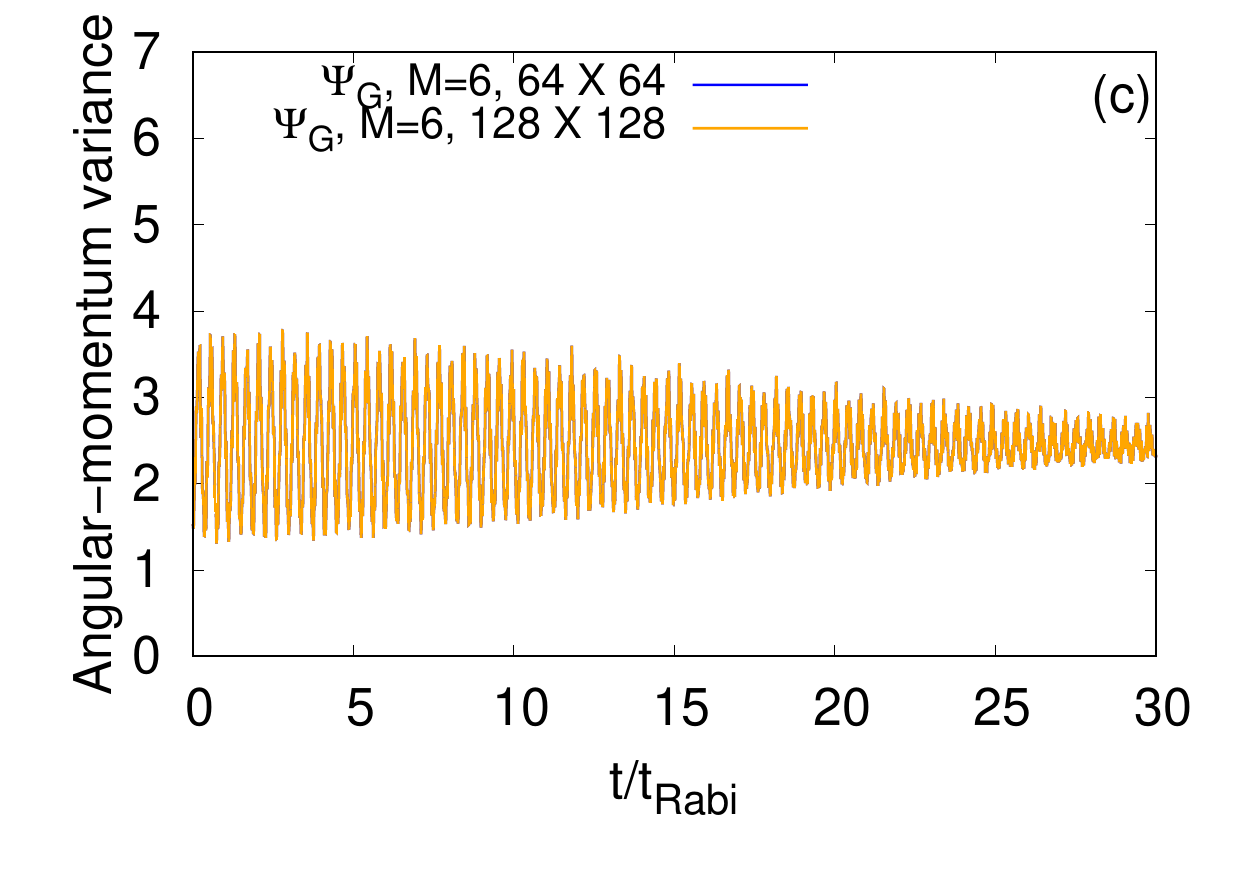}}
{\includegraphics[trim = 0.1cm 0.5cm 0.1cm 0.2cm, scale=.60]{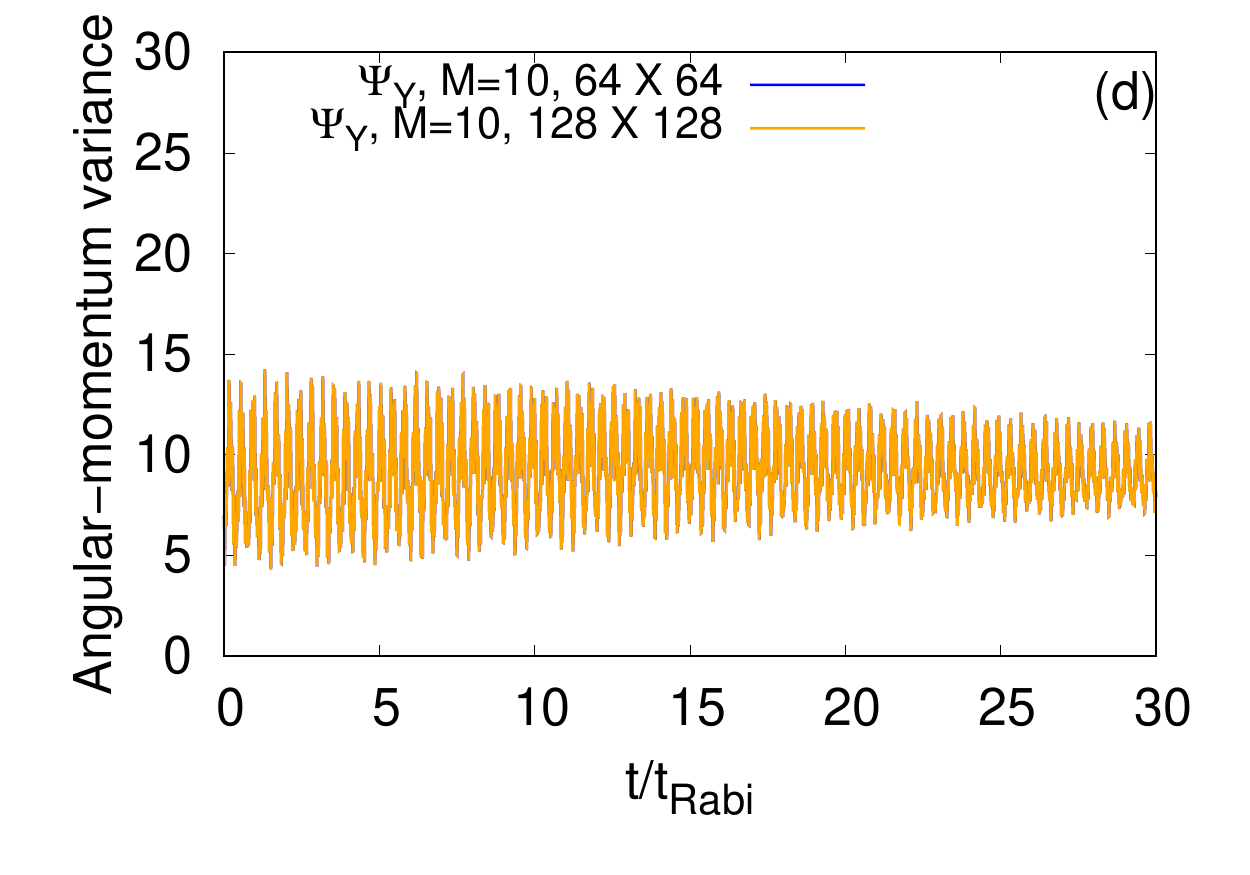}}\\
{\includegraphics[trim = 0.1cm 0.5cm 0.1cm 0.2cm, scale=.60]{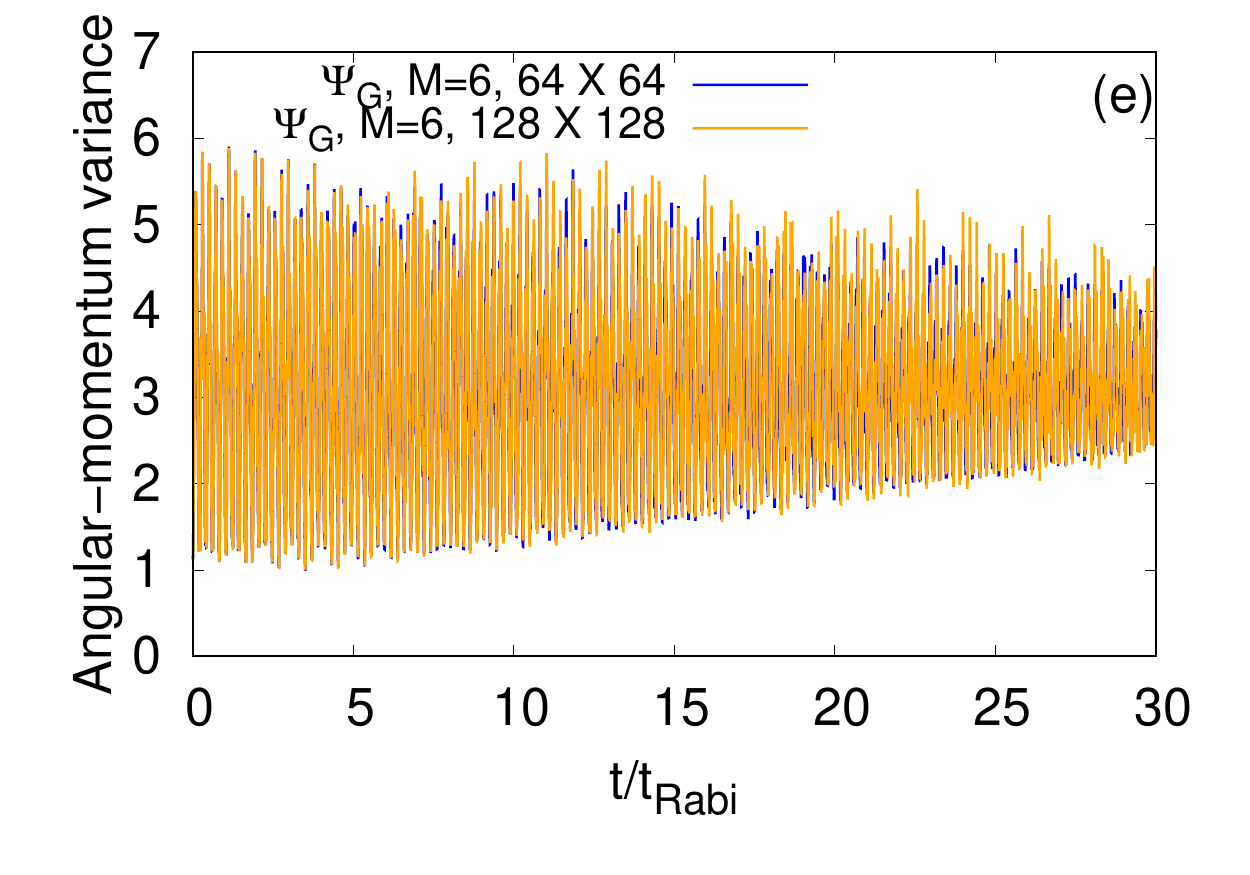}}
{\includegraphics[trim = 0.1cm 0.5cm 0.1cm 0.2cm, scale=.60]{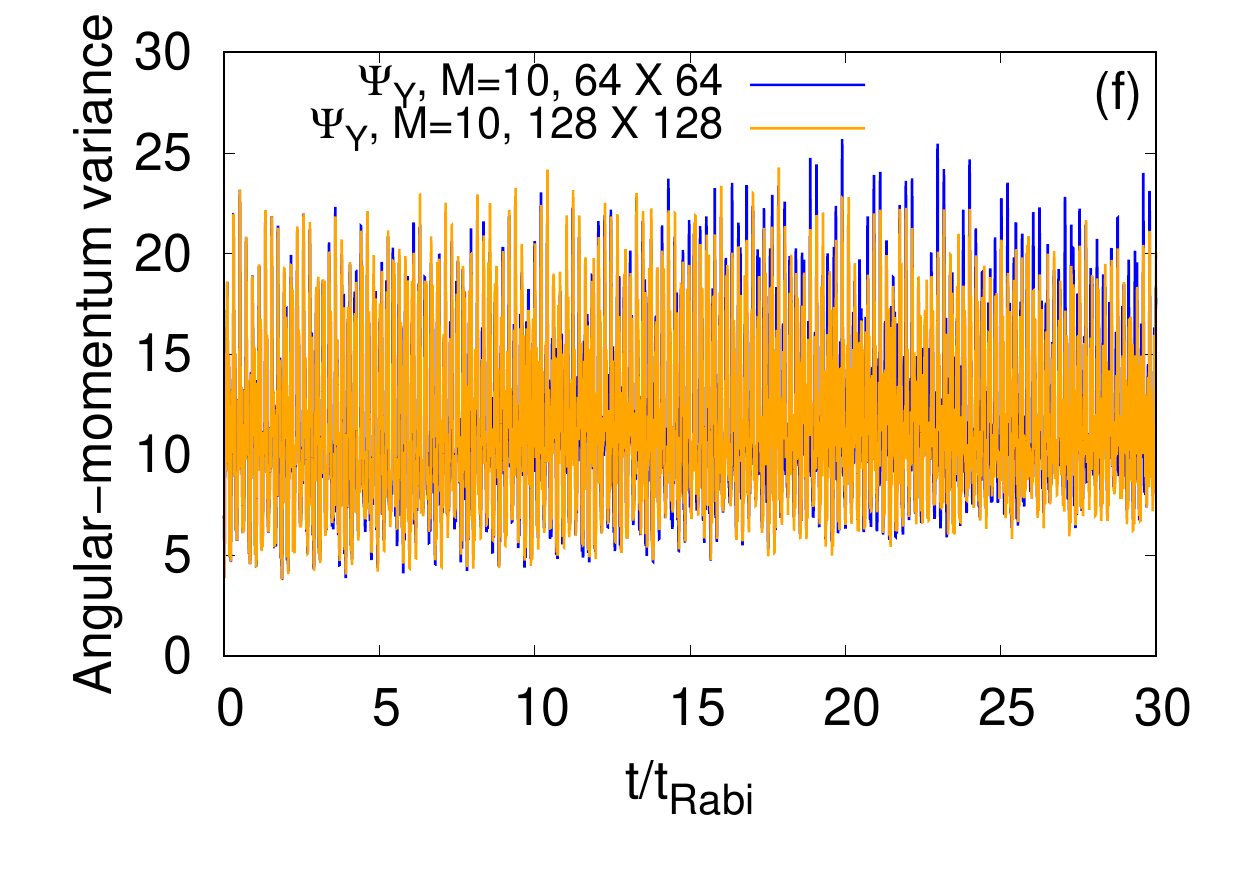}}\\
\caption{\label{figSA} Convergence of the time-dependent variance per particle of the $z$- component of the angular-momentum operator, $\dfrac{1}{N}\Delta_{\hat{L}_Z}^2(t)$,  with the number of grid points for the initial states  $\Psi_G$ (left column) and   $\Psi_Y$ (right column) in the longitudinally-asymmetric 2D double-well potential. The  bosonic clouds consist of $N=10$ bosons  with the interaction parameter $\Lambda=0.01\pi$.  The results for the asymmetry parameters $c=0$, 0.25, and 0.5 are presented row-wise.   The convergence is verified with $64\times 64$ and $128\times 128$ grid points.   We show here   dimensionless quantities. Color codes are explained in each panel.}
\end{figure*}

Now, we demonstrate the convergence of the results described in the main text with the number of grid points. So far all the results for the longitudinal resonant tunneling scenario were computed with $64 \times 64$ grid points. To  check the convergence with the grid points, we recomputed  all the quantities  with $128\times 128$ grid points for all asymmetry parameters. In order to display the convergence of our results, we choose the most sensitive quantity, $\dfrac{1}{N}\Delta_{\hat{L}_Z}^2(t)$ presented in this work. We have verified that all quantities, the survival probability, occupation numbers, and the variances, $\dfrac{1}{N}\Delta_{{\hat{X}}}^2(t)$,  $\dfrac{1}{N}\Delta_{{\hat{Y}}}^2(t)$  $\dfrac{1}{N}\Delta_{{\hat{P}_X}}^2(t)$, and  $\dfrac{1}{N}\Delta_{{\hat{P}_Y}}^2(t)$, are fully converged with the grid density.  The overlapping curves of $\dfrac{1}{N}\Delta_{\hat{L}_Z}^2(t)$ with increasing the grid density signifies that the results are fully converged for  $64 \times 64$ grid points for all values of $c$ for $\Psi_G$, and at $c=0$ and 0.25 for $\Psi_Y$. The small difference at loner times in the results for the excited state in Fig.~\ref{figSA} (f) represents that $\dfrac{1}{N}\Delta_{\hat{L}_Z}^2(t)$ is well converged with $64 \times 64$ grid points at $c=0.5$.

\clearpage

\section{Convergences of quantities in transversal resonant tunneling}
Similar to the longitudinal resonant tunneling, now we demonstrate the convergence of the loss of coherence, fragmentation, and the variances of position, momentum, and angular-momentum operators  with respect to the number of time-dependent orbitals and the density of the discrete-variable-representation grid points. Here,  the initial states are $\Psi_G$ and the longitudinally-excited state  $\Psi_X$. In the main text, all the physical quantities in the transversal resonant tunneling  for $\Psi_G$ and $\Psi_X$ are computed using $M=6$ time-dependent orbitals. To verify the convergence with respect to the number of  orbitals, we recomputed all the quantities with $M=10$ time-dependent orbitals. For the transversal resonant tunneling, the many-body Hamiltonian is represented by $128\times 128$ exponential discrete-variable-representation grid points in a box size of $[-10,10)\times [-10,10)$.

\begin{figure*}[!h]
\centering
{\includegraphics[trim = 0.1cm 0.5cm 0.1cm 0.2cm, scale=.60]{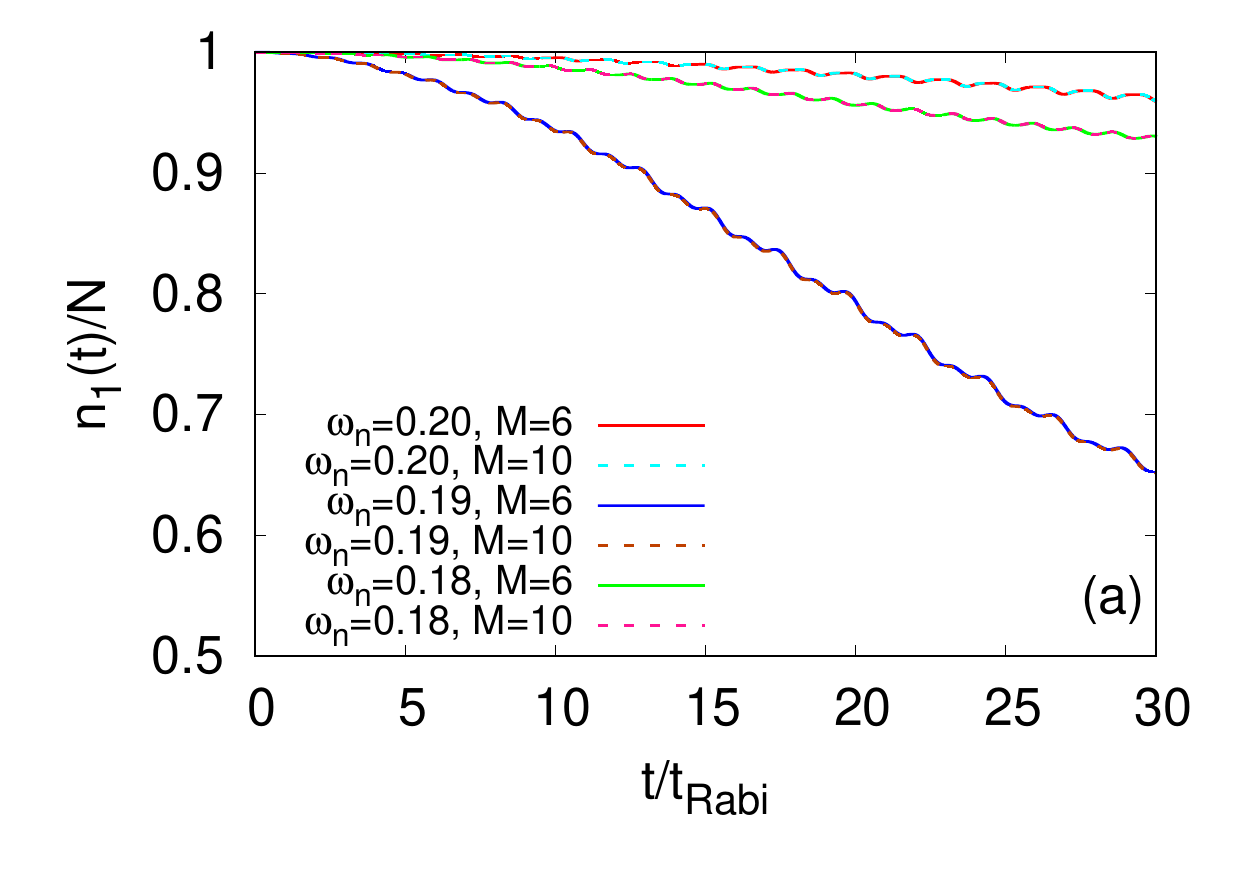}}
{\includegraphics[trim = 0.1cm 0.5cm 0.1cm 0.2cm, scale=.60]{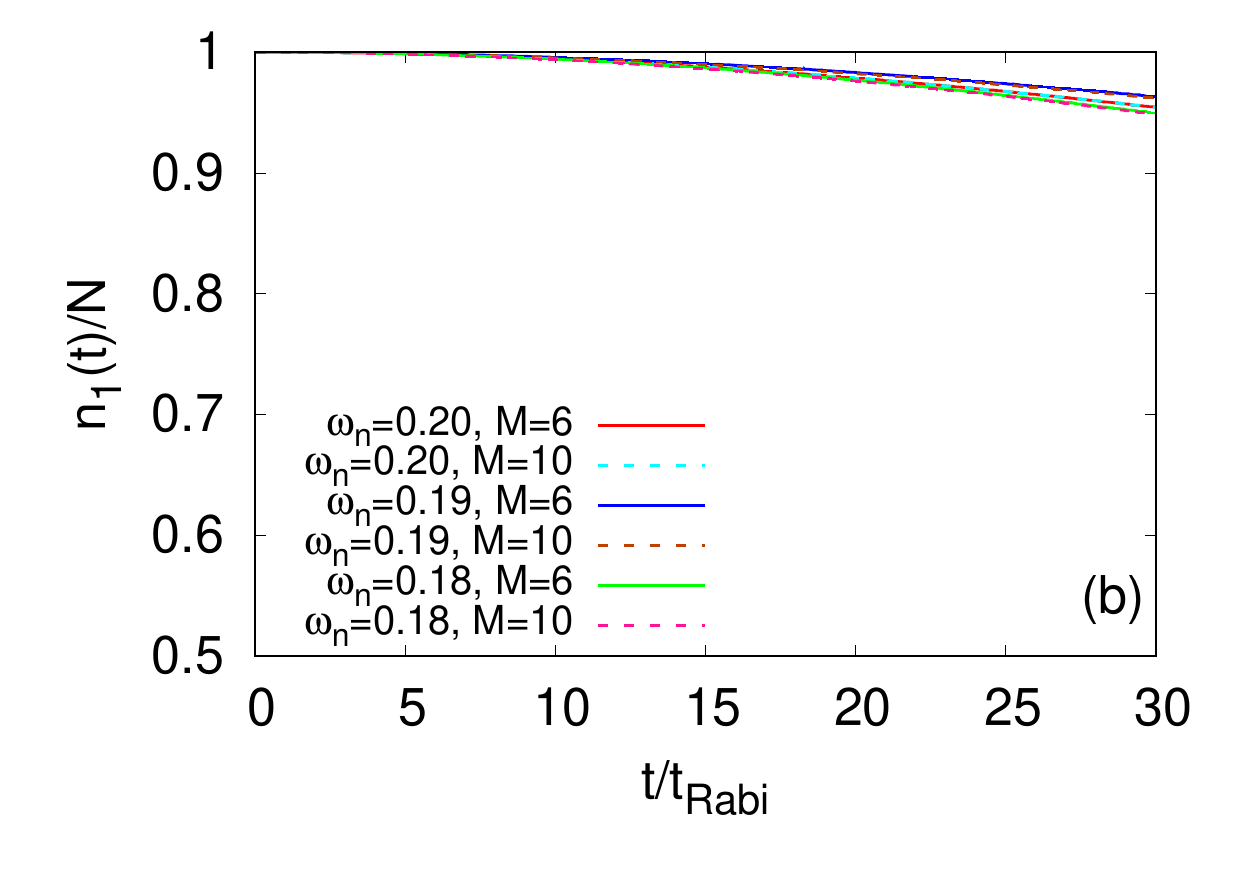}}\\
\caption{\label{figS6} Convergence of  the  occupation number per particle of the first natural orbital, $\dfrac{n_1(t)}{N}$, with the number of time-dependent orbitals for the initial states (a) $\Psi_G$ and (b) $\Psi_X$ in the transversely-asymmetric 2D double-well potential. The  bosonic clouds consist of $N=10$ bosons  with the interaction parameter $\Lambda=0.01\pi$.  The frequencies  are $\omega_n=0.20$, 0.19, and 0.18.  The convergence is verified with $M=6$ and  $M=10$ time-dependent orbitals for both  states.     We show here   dimensionless quantities. Color codes are explained in each panel.}
\end{figure*}

Fig.~\ref{figS6} depicts the numerical convergence  of the loss of coherence, represented as the occupation numbers per particle of the first natural orbital, $\dfrac{n_1(t)}{N}$, for $\Psi_G$ and  $\Psi_X$ with the number of orbitals $M$.  The frequency of the wider right well $\omega_n$, see Eq.(2.5) and Fig. 1(b) of the main text, is considered as $\omega_n=0.20$, $0.19$, and $0.18$.  We remind that at $\omega_n=0.19$ and 0.18, we find the transeversal resonant tunneling scenario for $\Psi_G$ and  $\Psi_X$, respectively.  As discussed in the main text for $M=6$ orbitals,  here  also for $M=10$ orbitals,  $\dfrac{n_1(t)}{N}$ decays with an oscillatory background for both  initial states and falls on top of the corresponding results found from $M=6$ orbitals, indicating the convergence of  condensate fraction  with the number of orbitals.

\begin{figure*}[!h]
\centering
{\includegraphics[trim = 0.1cm 0.5cm 0.1cm 0.2cm, scale=.60]{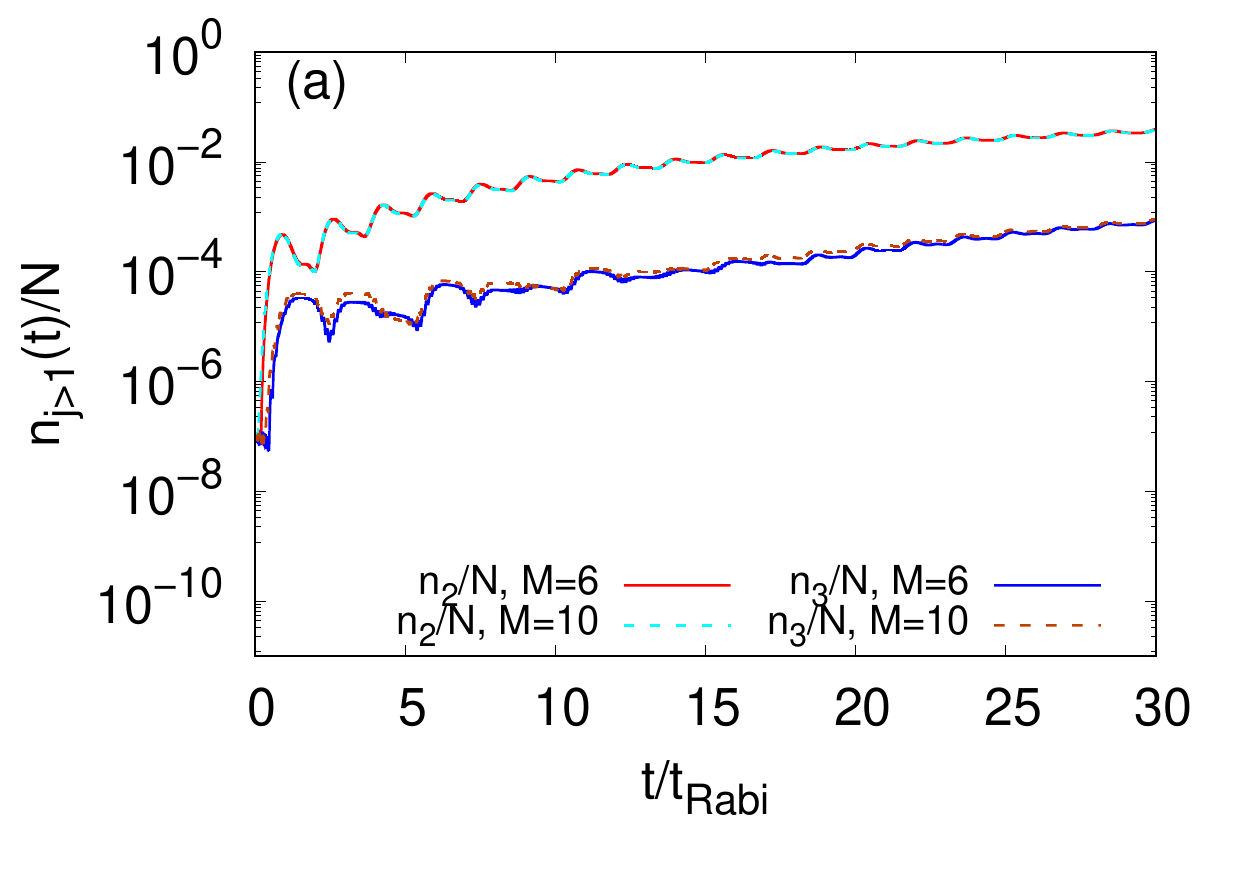}}
{\includegraphics[trim = 0.1cm 0.5cm 0.1cm 0.2cm, scale=.60]{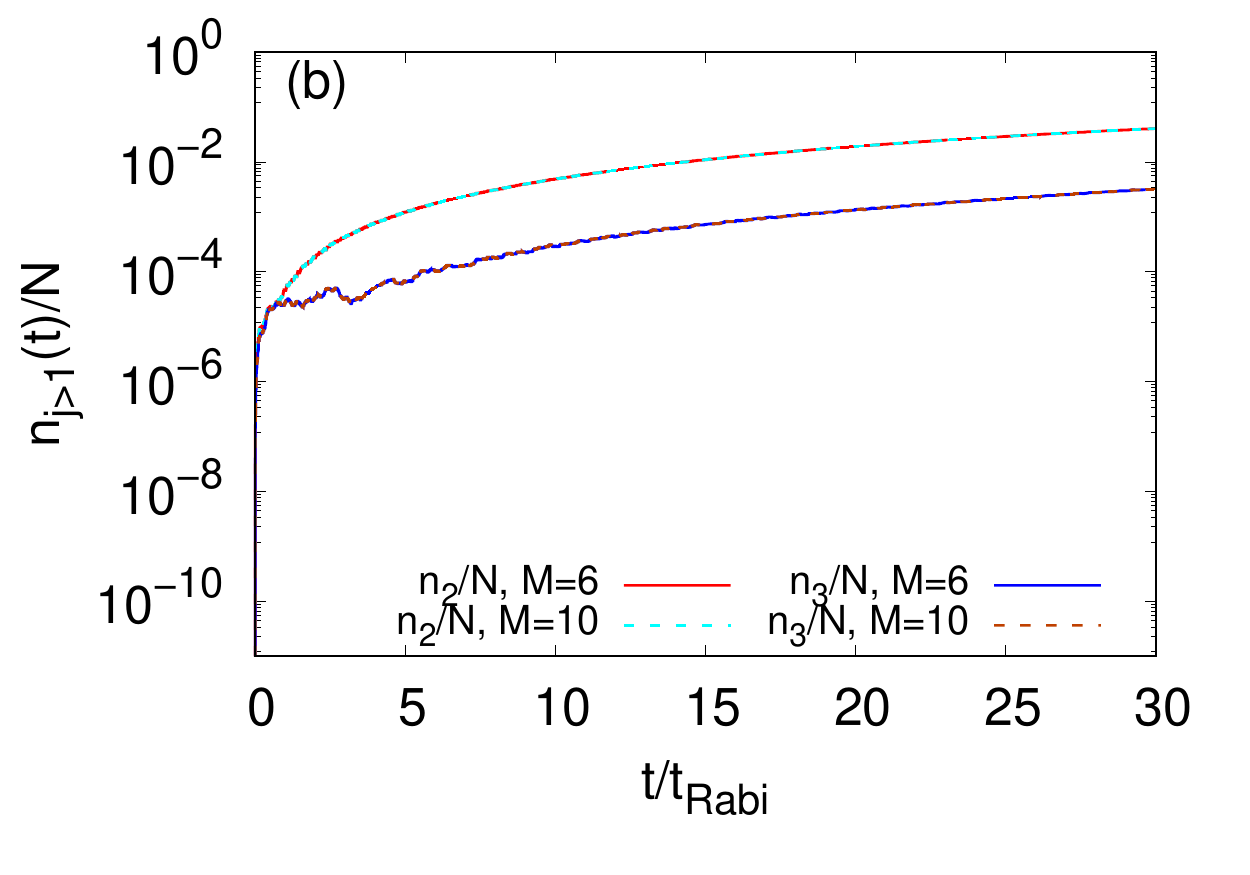}}\\
{\includegraphics[trim = 0.1cm 0.5cm 0.1cm 0.2cm, scale=.60]{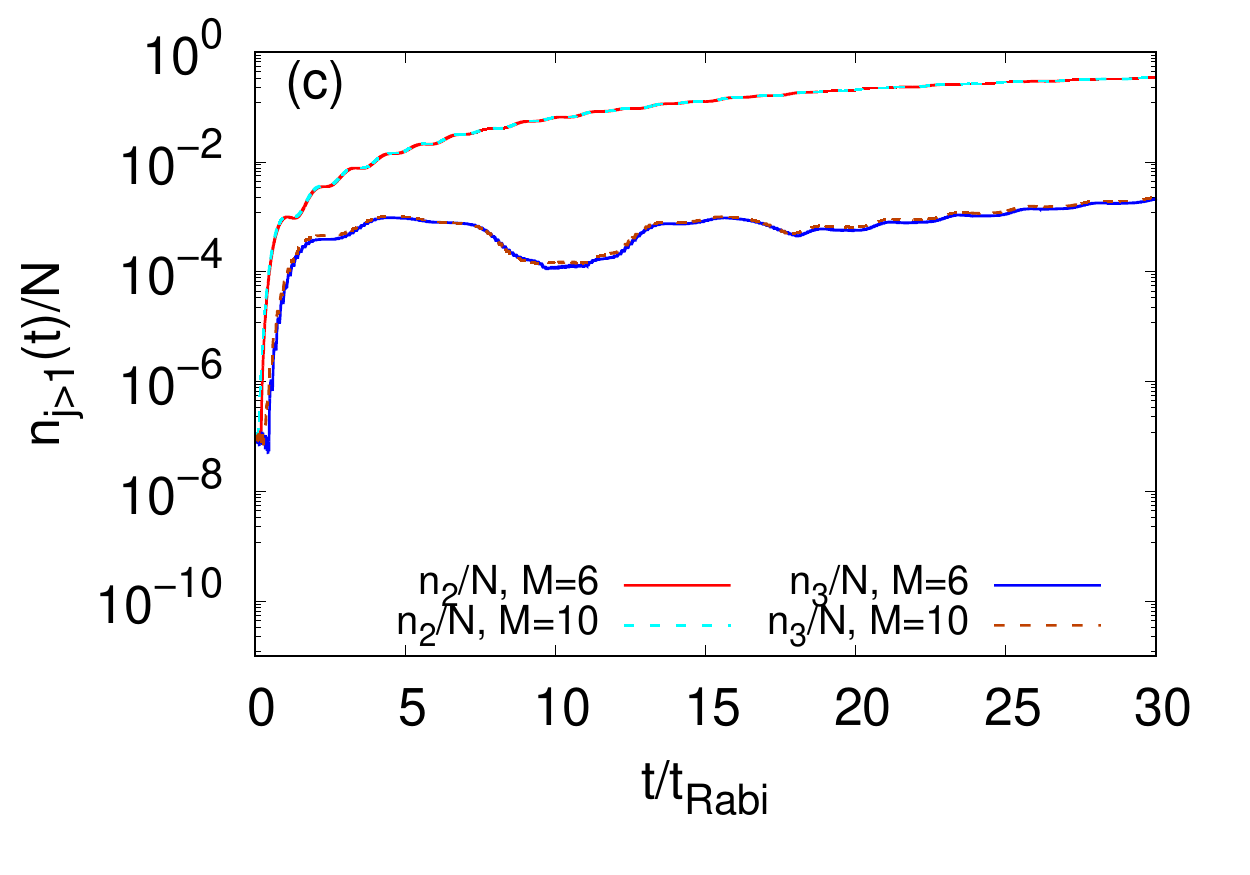}}
{\includegraphics[trim = 0.1cm 0.5cm 0.1cm 0.2cm, scale=.60]{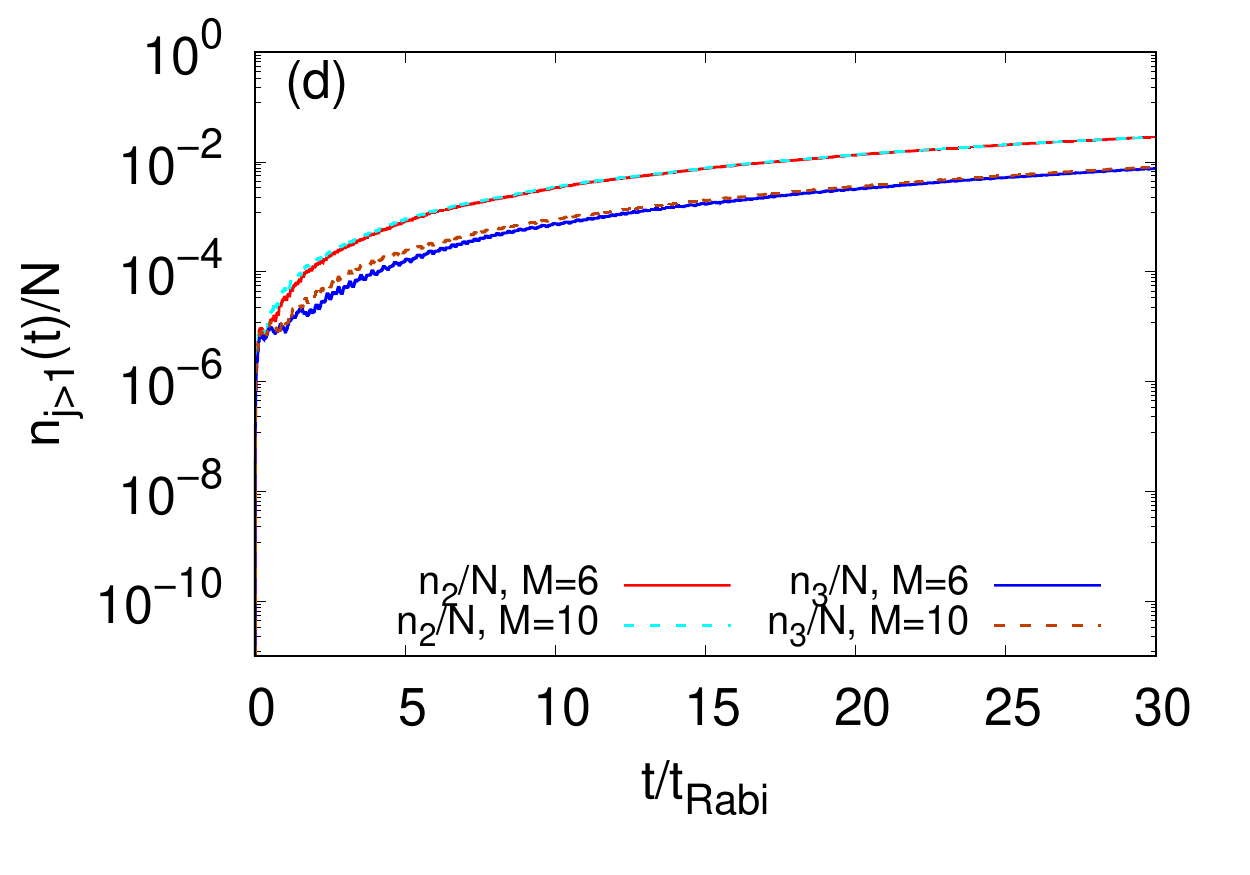}}\\
{\includegraphics[trim = 0.1cm 0.5cm 0.1cm 0.2cm, scale=.60]{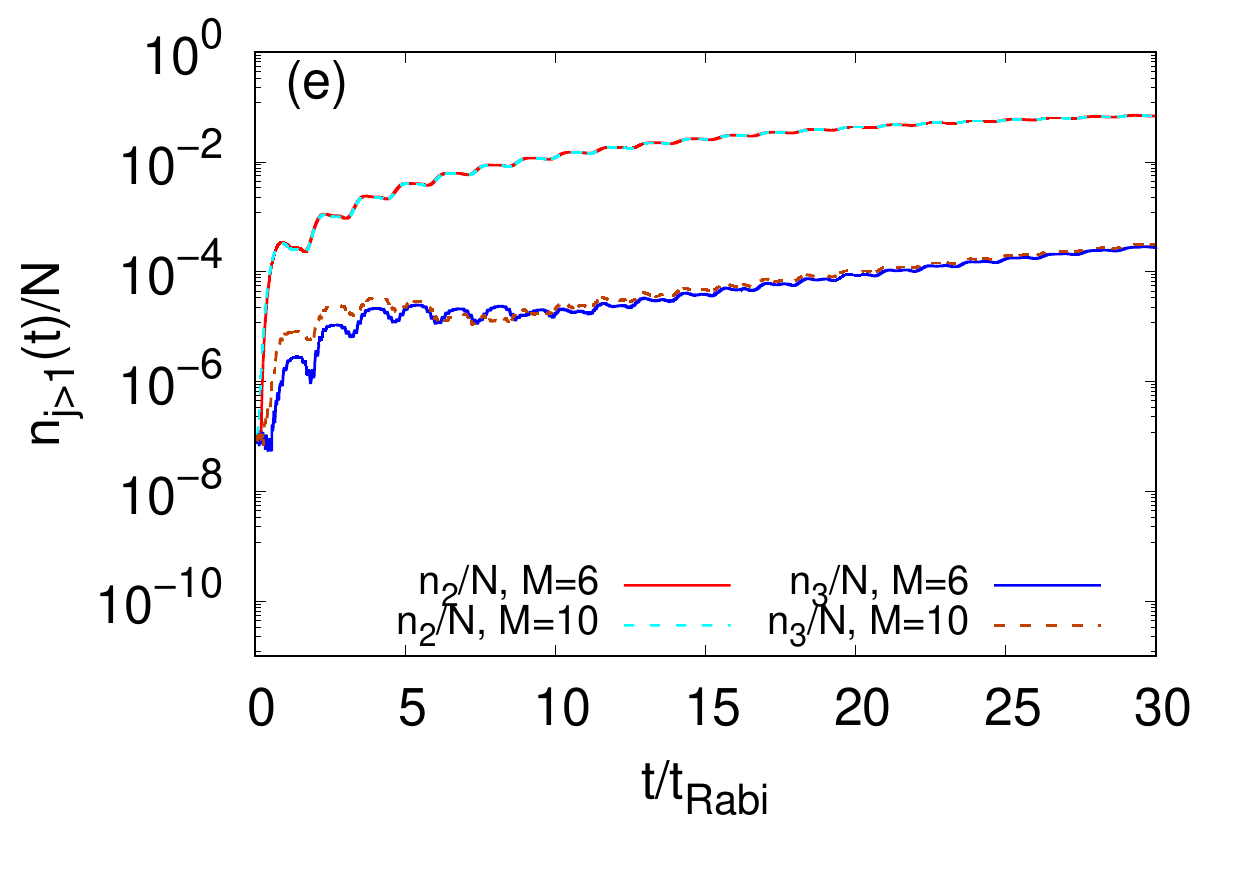}}
{\includegraphics[trim = 0.1cm 0.5cm 0.1cm 0.2cm, scale=.60]{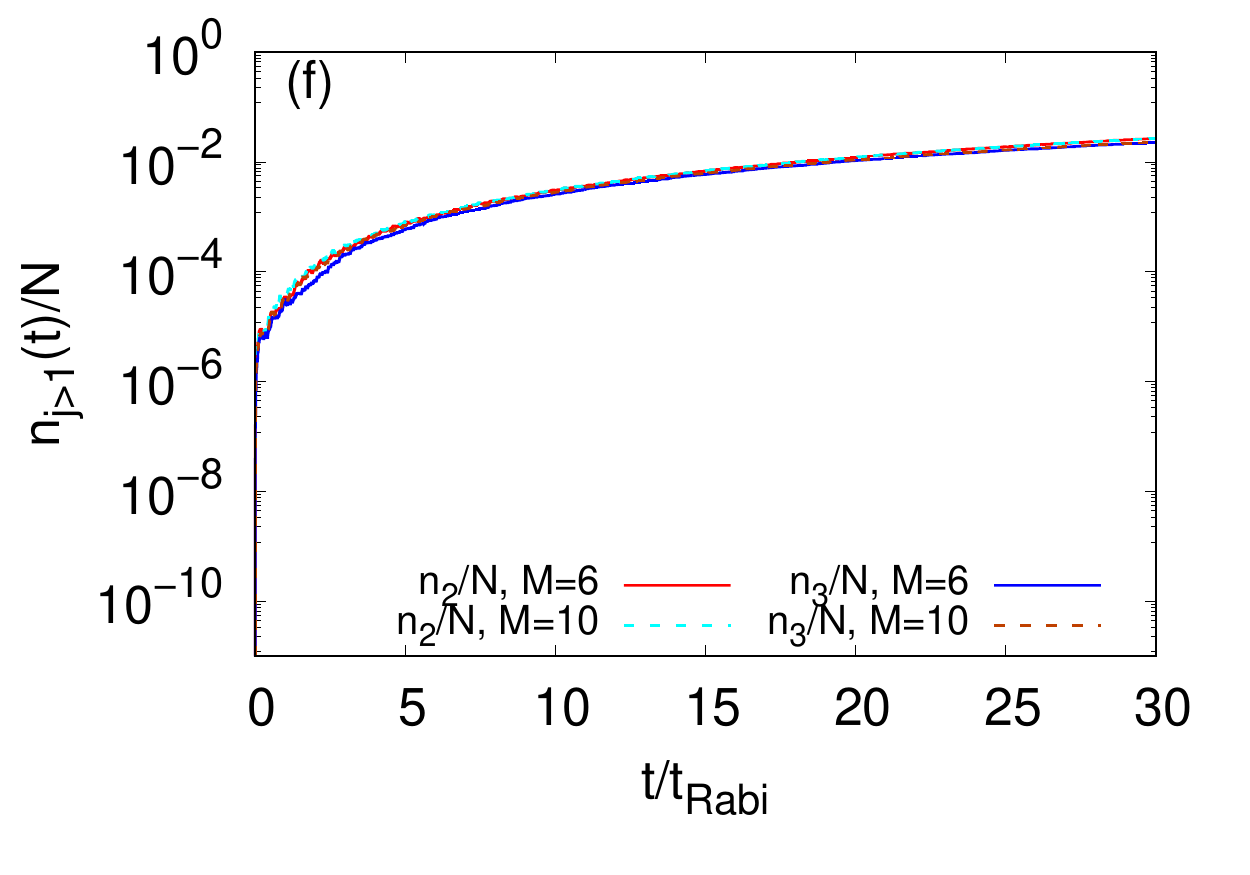}}\\
\caption{\label{figS7}Convergence of the natural occupation number per particle, $\dfrac{n_{j>1}(t)}{N}$, with the number of time-dependent orbitals for the initial states $\Psi_G$ (left column) and  $\Psi_X$ (right column) in the transversely-asymmetric 2D double-well potential. The  bosonic clouds consist of $N=10$ bosons  with the interaction parameter $\Lambda=0.01\pi$.  The results for the frequencies $\omega_n=0.20$, 0.19, and 0.18 are presented row-wise.   The convergence is verified with $M=6$ and  $M=10$ time-dependent orbitals for  both states.    We show here   dimensionless quantities. Color codes are explained in each panel.}
\end{figure*}

In order to learn how the fragmentation develops in the transversal resonant tunneling scenario and at its vicinity, we explore the  time-dependent occupancies of higher natural orbitals per particle, $\dfrac{n_{j>1}(t)}{N}$, along with their convergences with the number of orbitals for $\Psi_G$  and  $\Psi_X$. We graphically present only $\dfrac{n_{2}(t)}{N}$ and $\dfrac{n_{3}(t)}{N}$ in Fig.~\ref{figS7} as they have occupations in appreciable amount. Among the three different values of $\omega_n$, $\dfrac{n_{2}(t)}{N}$ is maximally occupied at $\omega_n=0.19$ for  $\Psi_G$ and at $\omega_n=0.18$ for  $\Psi_X$, as these values of $\omega_n$ produce the transversal resonant tunneling for the respective states. We observe the oscillatory nature in the the dynamics of $\dfrac{n_{2}(t)}{N}$ and $\dfrac{n_{3}(t)}{N}$ for  $\Psi_G$, whereas it is hardly visible for $\Psi_X$.

\begin{figure*}[!h]
\centering
{\includegraphics[trim = 0.1cm 0.5cm 0.1cm 0.2cm, scale=.60]{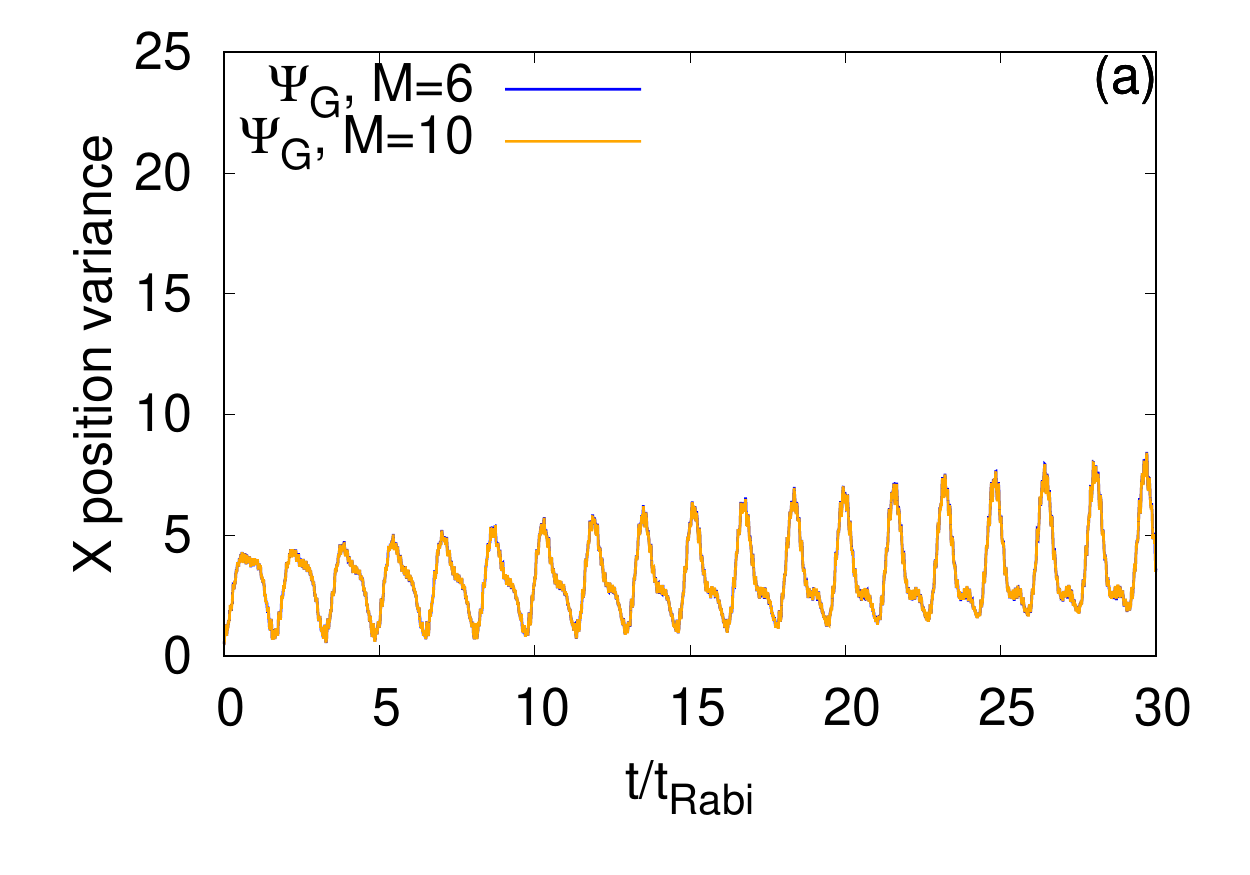}}
{\includegraphics[trim = 0.1cm 0.5cm 0.1cm 0.2cm, scale=.60]{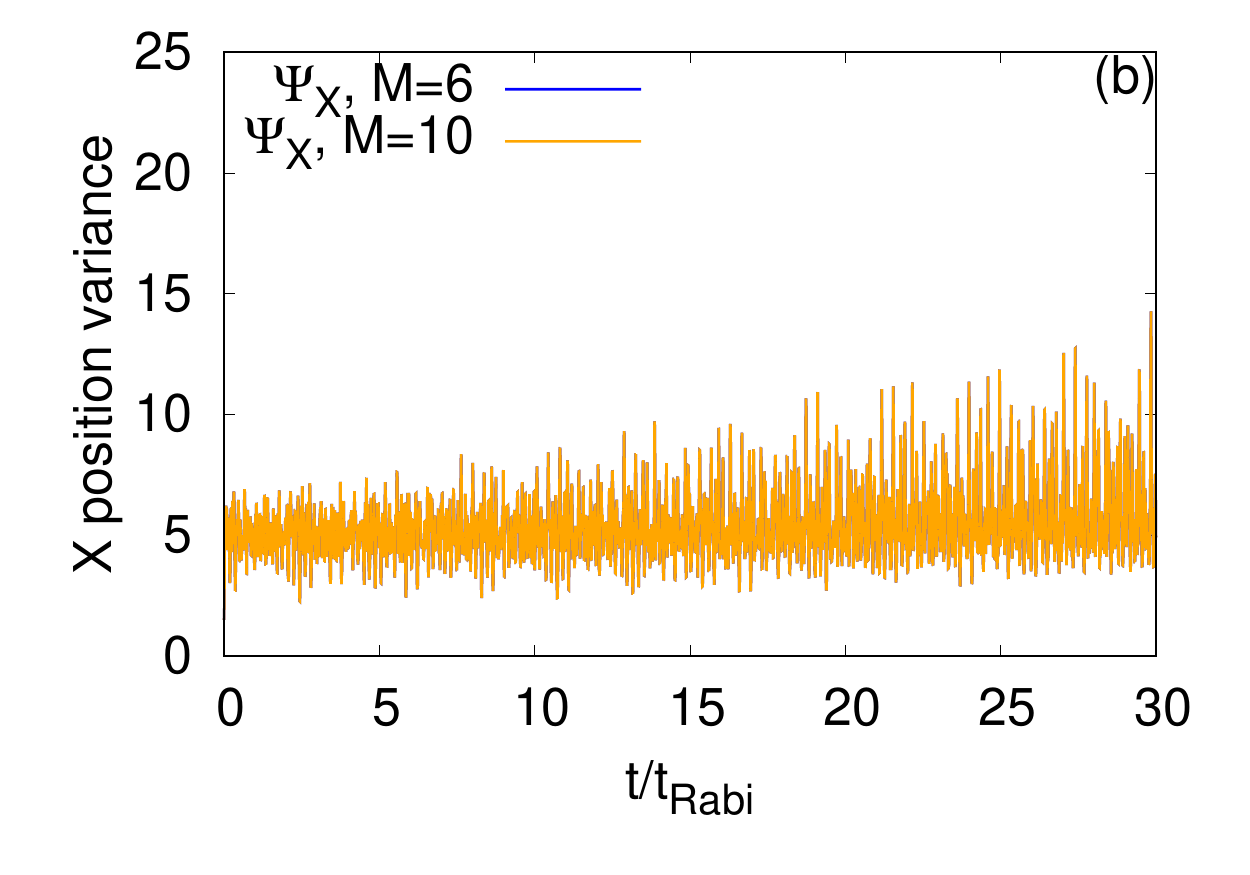}}\\
\vglue 0.25 truecm
{\includegraphics[trim = 0.1cm 0.5cm 0.1cm 0.2cm, scale=.60]{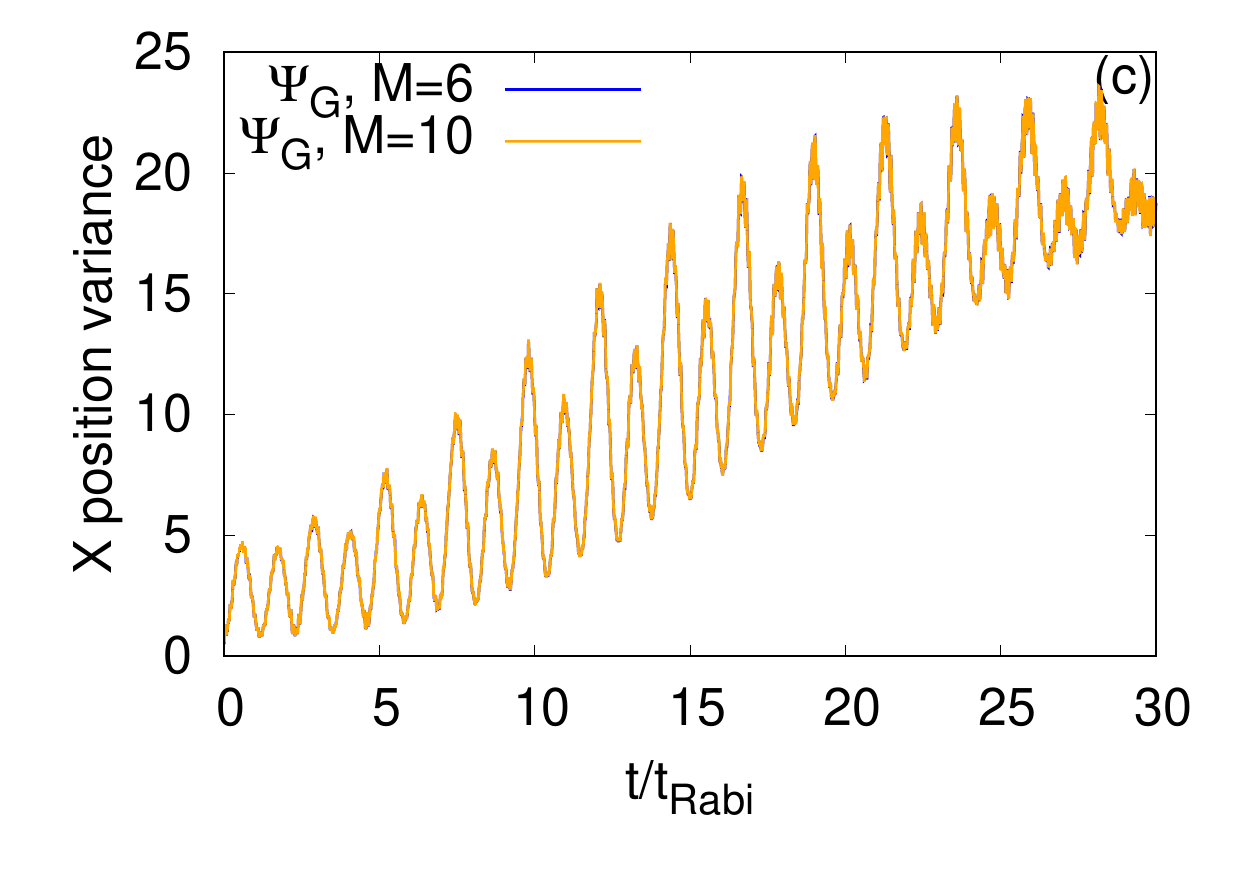}}
{\includegraphics[trim = 0.1cm 0.5cm 0.1cm 0.2cm, scale=.60]{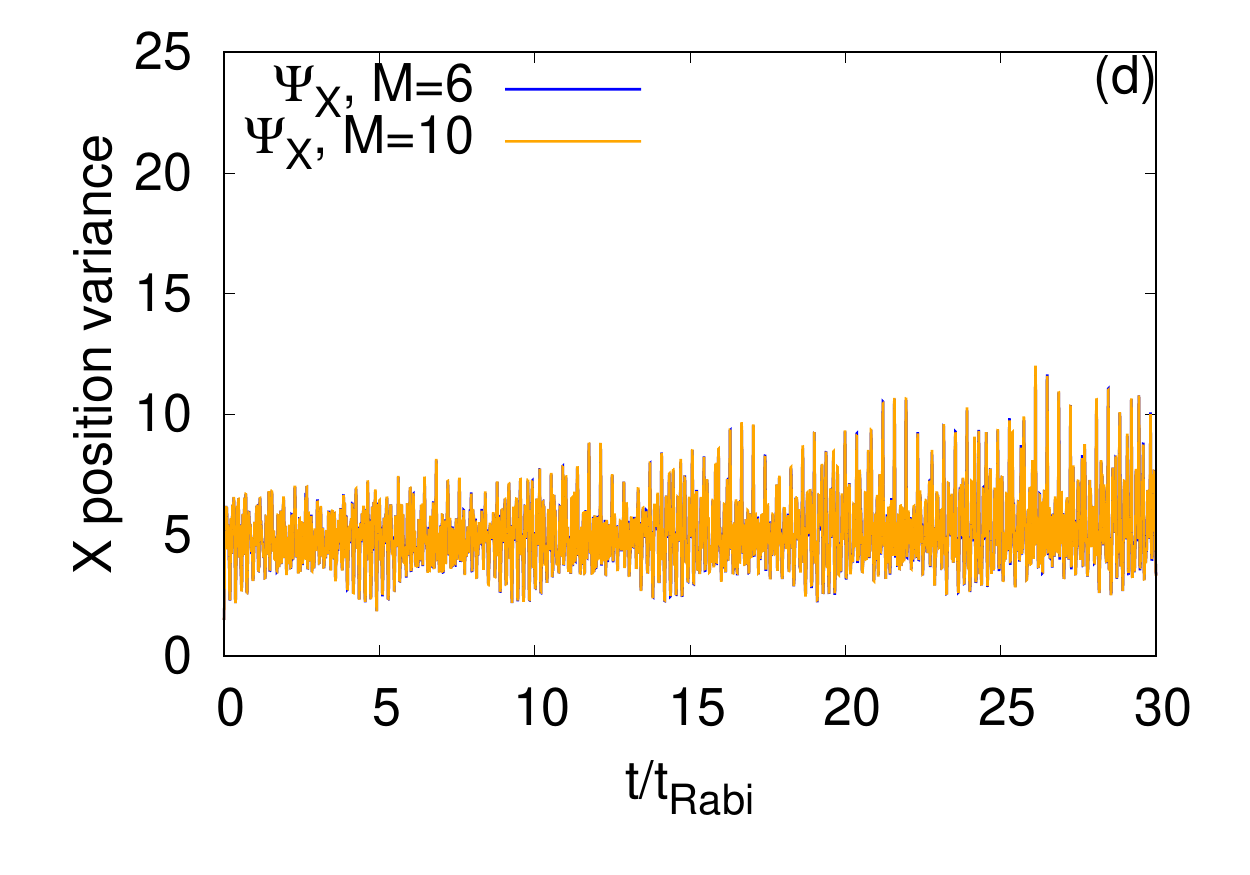}}\\
\vglue 0.25 truecm
{\includegraphics[trim = 0.1cm 0.5cm 0.1cm 0.2cm, scale=.60]{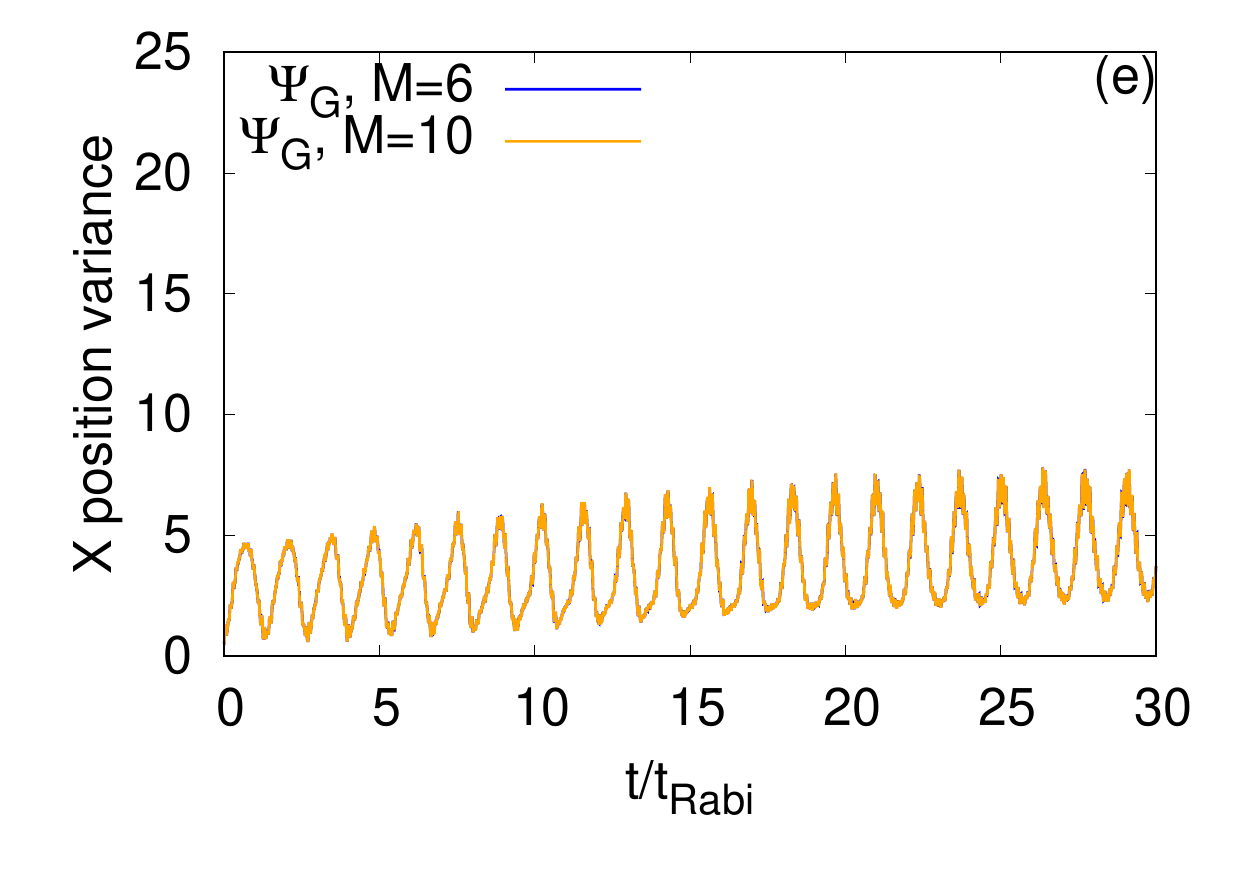}}
{\includegraphics[trim = 0.1cm 0.5cm 0.1cm 0.2cm, scale=.60]{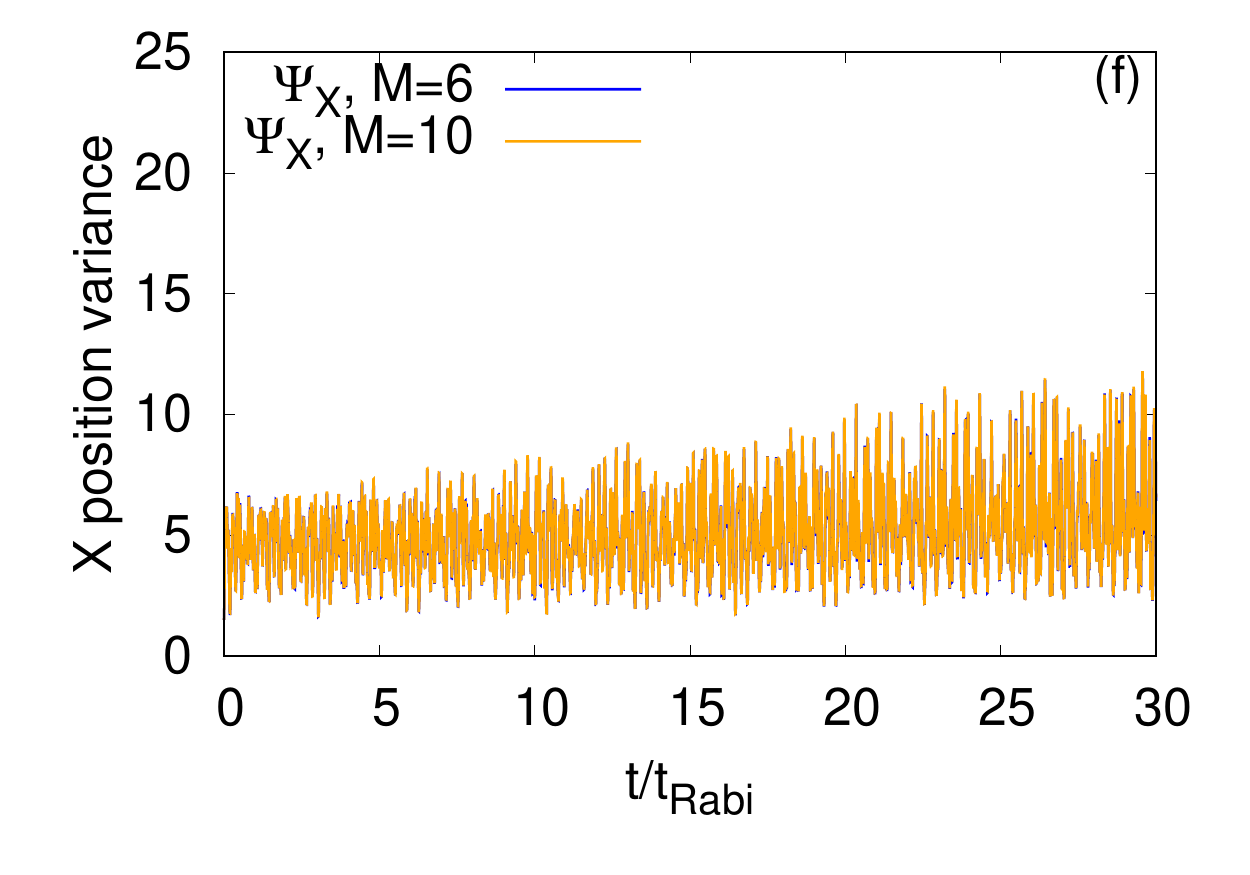}}\\
\caption{\label{figS8}Convergence of the time-dependent many-body position variance per particle along the $x$-direction, $\dfrac{1}{N}\Delta_{\hat{X}}^2(t)$, with the number of time-dependent orbitals for the initial states $\Psi_G$ (left column) and  $\Psi_X$ (right column) in the transversely-asymmetric 2D double-well potential.  The  bosonic clouds consist of $N=10$ bosons  with the interaction parameter $\Lambda=0.01\pi$. The results for the frequencies $\omega_n=0.20$, 0.19, and 0.18 are presented row-wise. The many-body results are computed  with $M=6$ time-dependent orbitals. The convergence is verified with $M=10$ time-dependent orbitals.    We show here   dimensionless quantities. Color codes are explained in each panel.}
\end{figure*}

\begin{figure*}[!h]
\centering
{\includegraphics[trim = 0.1cm 0.5cm 0.1cm 0.2cm, scale=.60]{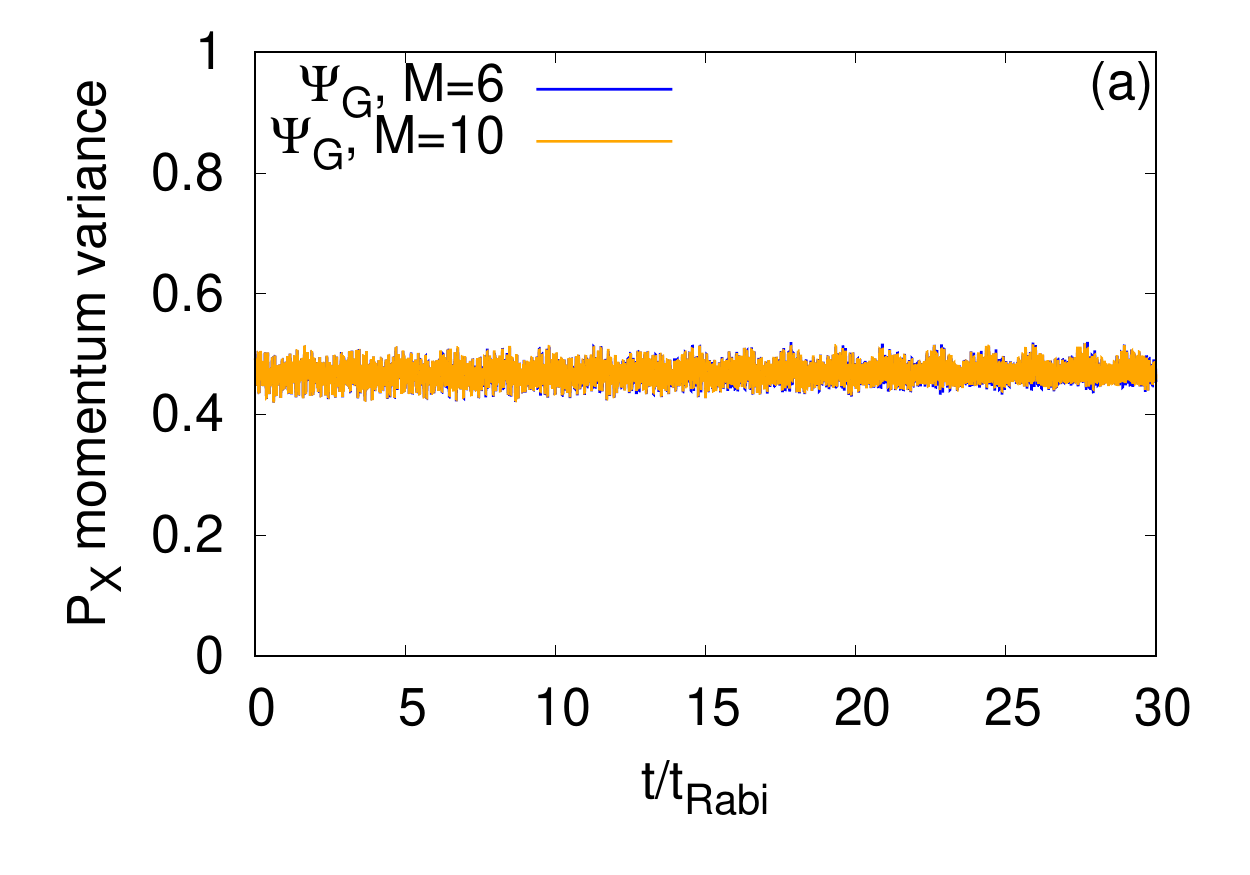}}
{\includegraphics[trim = 0.1cm 0.5cm 0.1cm 0.2cm, scale=.60]{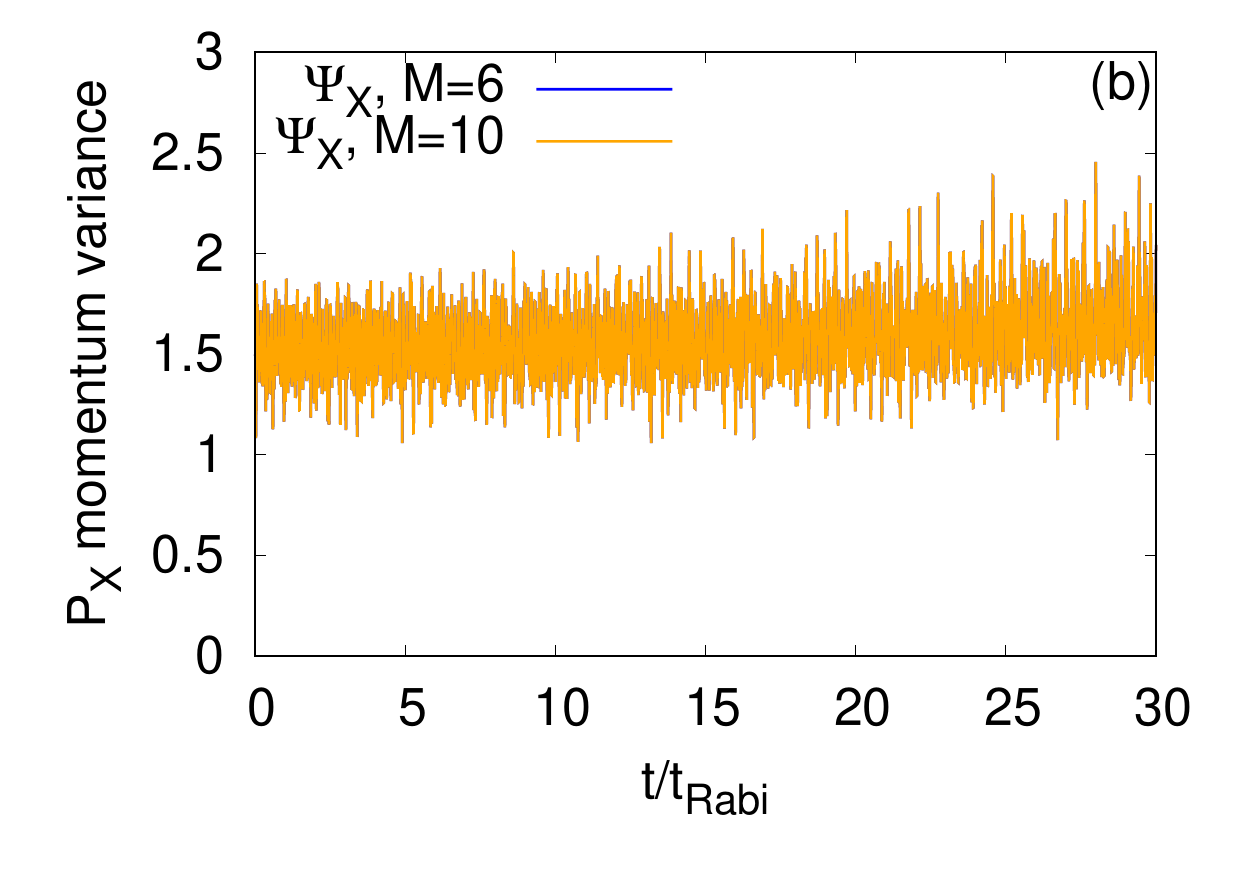}}\\
\vglue 0.25 truecm
{\includegraphics[trim = 0.1cm 0.5cm 0.1cm 0.2cm, scale=.60]{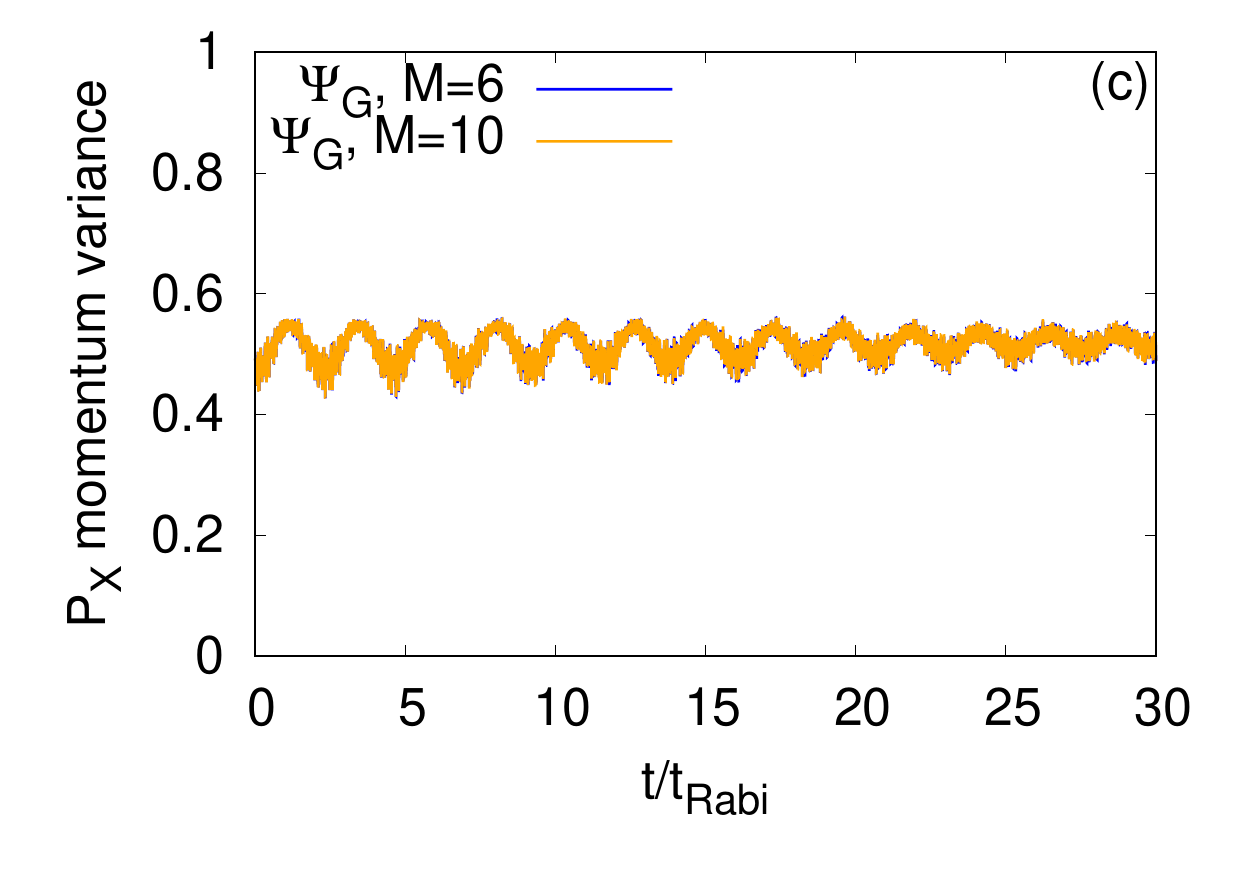}}
{\includegraphics[trim = 0.1cm 0.5cm 0.1cm 0.2cm, scale=.60]{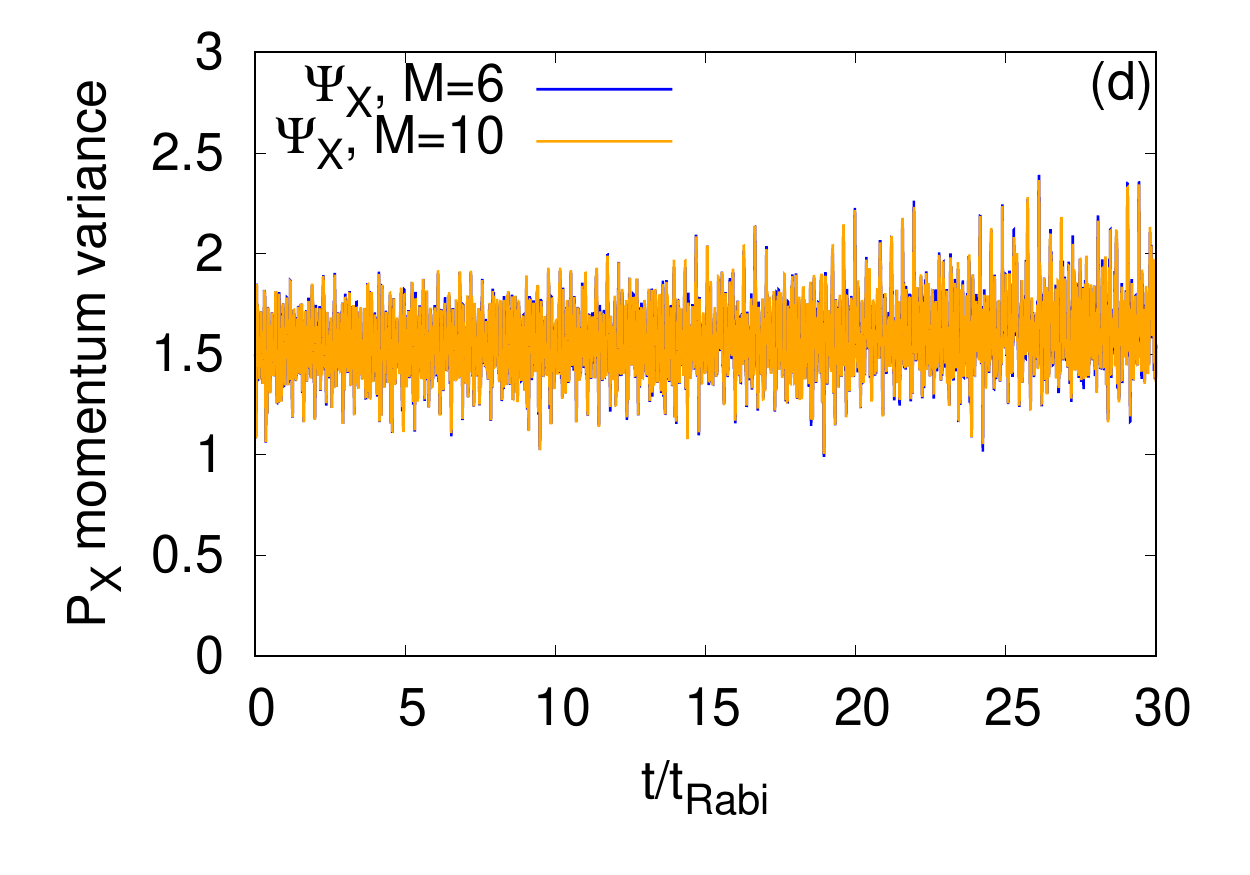}}\\
\vglue 0.25 truecm
{\includegraphics[trim = 0.1cm 0.5cm 0.1cm 0.2cm, scale=.60]{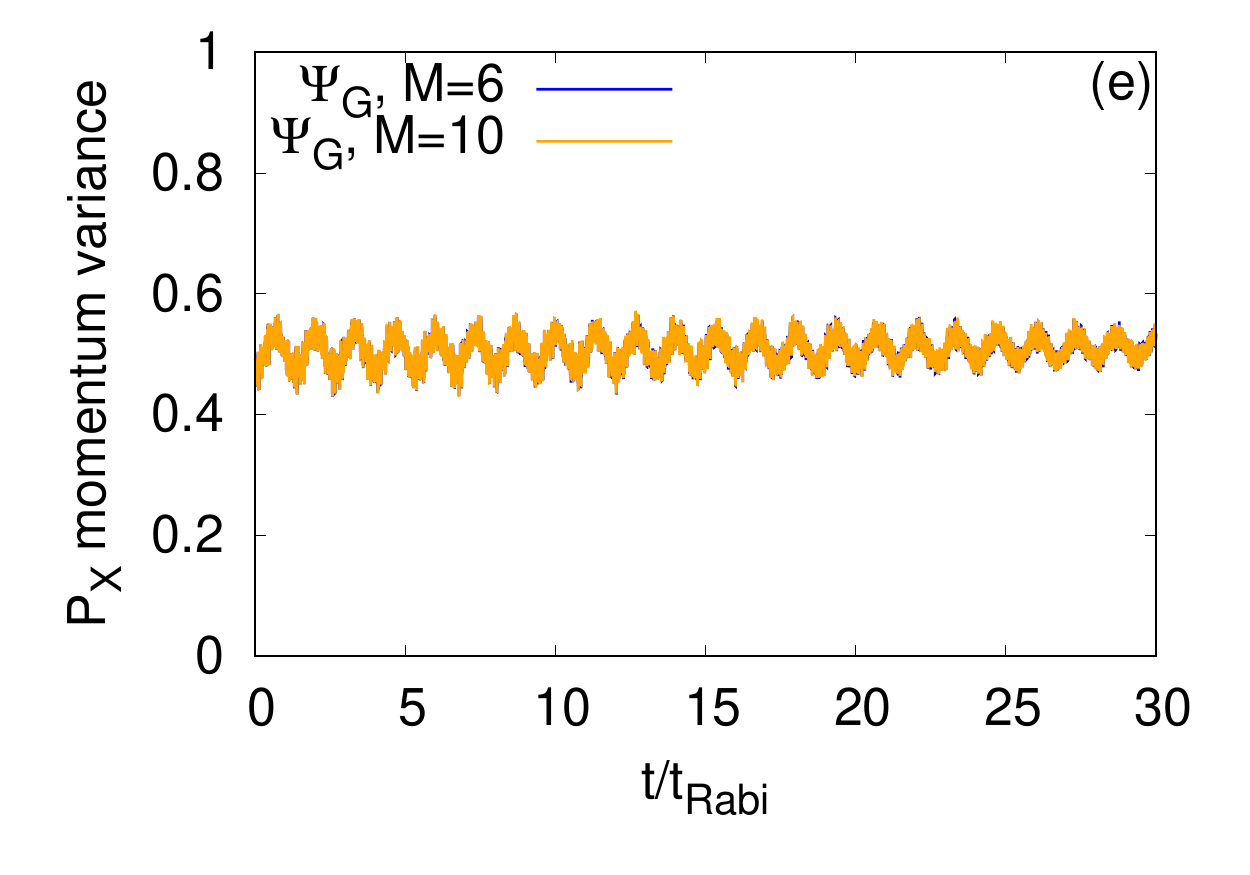}}
{\includegraphics[trim = 0.1cm 0.5cm 0.1cm 0.2cm, scale=.60]{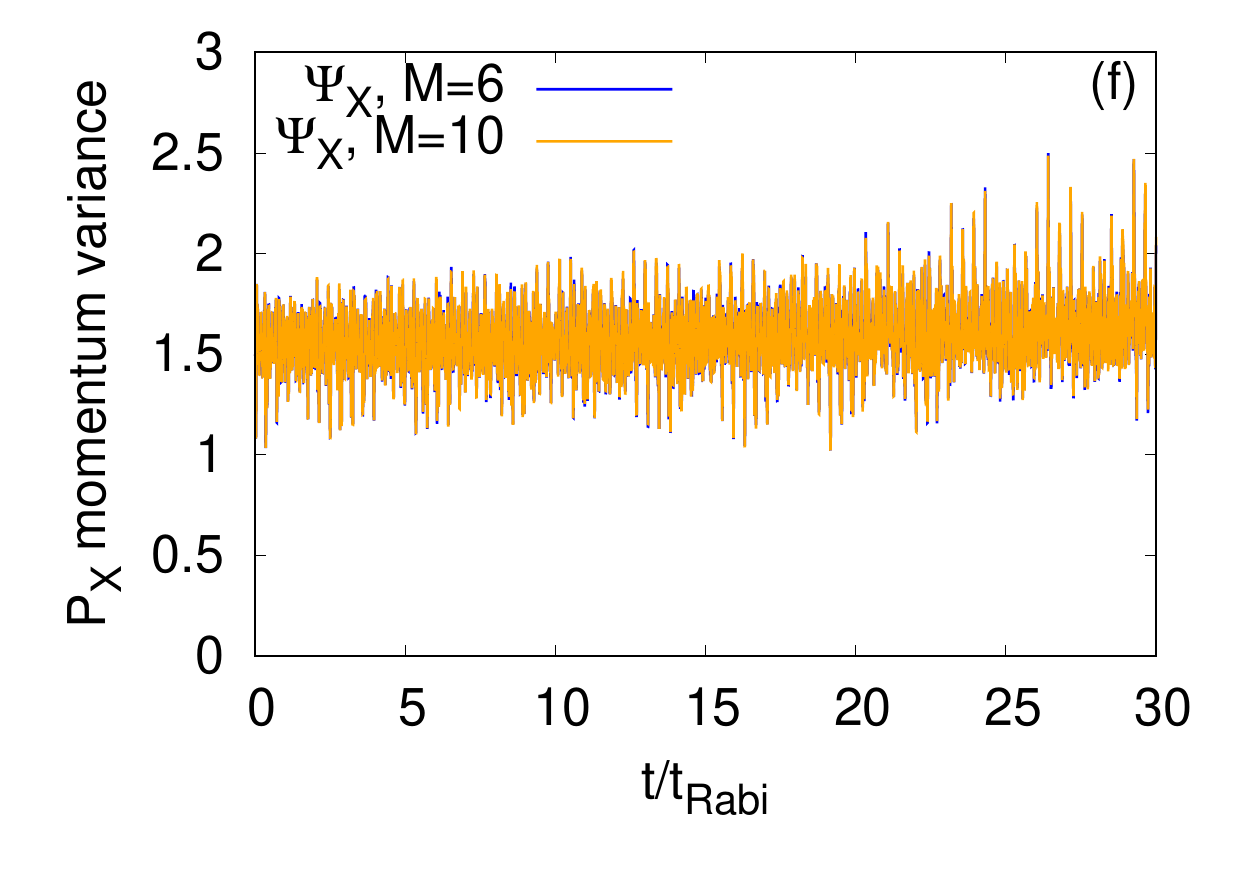}}\\
\caption{\label{figS9}Convergence of the time-dependent many-body momentum variance per particle along the $x$-direction, $\dfrac{1}{N}\Delta_{\hat{P}_X}^2(t)$, with the number of time-dependent orbitals for the initial states $\Psi_G$ (left column) and  $\Psi_X$ (right column)  in the transversely-asymmetric 2D double-well potential. The  bosonic clouds consist of $N=10$ bosons  with the interaction parameter $\Lambda=0.01\pi$.  The results for the frequencies $\omega_n=0.20$, 0.19, and 0.18 are presented row-wise. The many-body results are computed  with $M=6$ time-dependent orbitals. The convergence is verified with $M=10$ time-dependent orbitals.   We show here   dimensionless quantities. Color codes are explained in each panel.}
\end{figure*}

Figs.~\ref{figS8} and \ref{figS9} show the many-body position and momentum variances per particle along the $x$-direction, $\dfrac{1}{N}\Delta_{\hat{X}}^2(t)$ and $\dfrac{1}{N}\Delta_{\hat{P}_X}^2(t)$, respectively, as well as  their numerical convergences with number of orbitals. Here we analyze the many-body $\dfrac{1}{N}\Delta_{\hat{X}}^2(t)$ and $\dfrac{1}{N}\Delta_{\hat{P}_X}^2(t)$ at $\omega_n=0.20$, 0.19, and 0.18 for both $\Psi_G$  and  $\Psi_X$.   $\dfrac{1}{N}\Delta_{\hat{X}}^2(t)$ and $\dfrac{1}{N}\Delta_{\hat{P}_X}^2(t)$ computed with $M=10$ orbitals fall on top of the corresponding quantities found from $M=6$ orbitals,  signifying the convergences of these quantities for both   states. For $\Psi_X$, we observe marginal differences in terms of  frequency and amplitude for both the quantities,  $\dfrac{1}{N}\Delta_{\hat{X}}^2(t)$ and $\dfrac{1}{N}\Delta_{\hat{P}_X}^2(t)$, when computed at $\omega_n=0.20$, 0.19, and 0.18. These marginal differences  are consistent with the corresponding survival probabilities discussed in the main text.   The  differences are negligible for both the quantities at the  three different frequencies for $\Psi_X$ as it lies in the first excited band and feels a  smaller barrier when it tunnels. The frequency of oscillations of $\dfrac{1}{N}\Delta_{\hat{X}}^2(t)$ for $\Psi_X$ is practically half   the corresponding oscillation frequency of the survival probabilities.

 Unlike $\Psi_X$, we find significantly different many-body dynamics of  $\dfrac{1}{N}\Delta_{\hat{X}}^2(t)$ and  $\dfrac{1}{N}\Delta_{\hat{P}_X}^2(t)$ for $\Psi_G$  at $\omega_n=0.20$, 0.19, and 0.18. In order to characterize the nature of  $\dfrac{1}{N}\Delta_{\hat{X}}^2(t)$ for $\Psi_G$, we notice  that the average value of $\dfrac{1}{N}\Delta_{\hat{X}}^2(t)$ increases  as time passes by, which is found to be maximal at the resonant value $\omega_n=0.19$.  As demonstrated for $\Psi_G$  in the main text, the oscillation frequency of $\dfrac{1}{N}\Delta_{\hat{X}}^2(t)$ is twice  the oscillation frequency of the survival probability at $\omega_n=0.19$. Dissimilar  to the resonant value of $\omega_n$, here at $\omega_n=0.18$ and 0.20, we  find that the oscillation frequencies of the survival probability and $\dfrac{1}{N}\Delta_{\hat{X}}^2(t)$ practically overlap with each other.   At $\omega_n=0.19$, $\dfrac{1}{N}\Delta_{\hat{X}}^2(t)$ displays two different amplitudes of oscillations.  This signature can also be seen  at $\omega_n=0.20$ (hardly visible). The time evolution of $\dfrac{1}{N}\Delta_{\hat{P}_X}^2(t)$ for $\Psi_G$ is mostly dominated by the breathing mode oscillations. As time progresses, the amplitude of the oscillations of  $\dfrac{1}{N}\Delta_{\hat{P}_X}^2(t)$ decreases for $\Psi_G$ at all values of $\omega_n$ considered in this work.

\begin{figure*}[!h]
\centering
{\includegraphics[trim = 0.1cm 0.5cm 0.1cm 0.2cm, scale=.60]{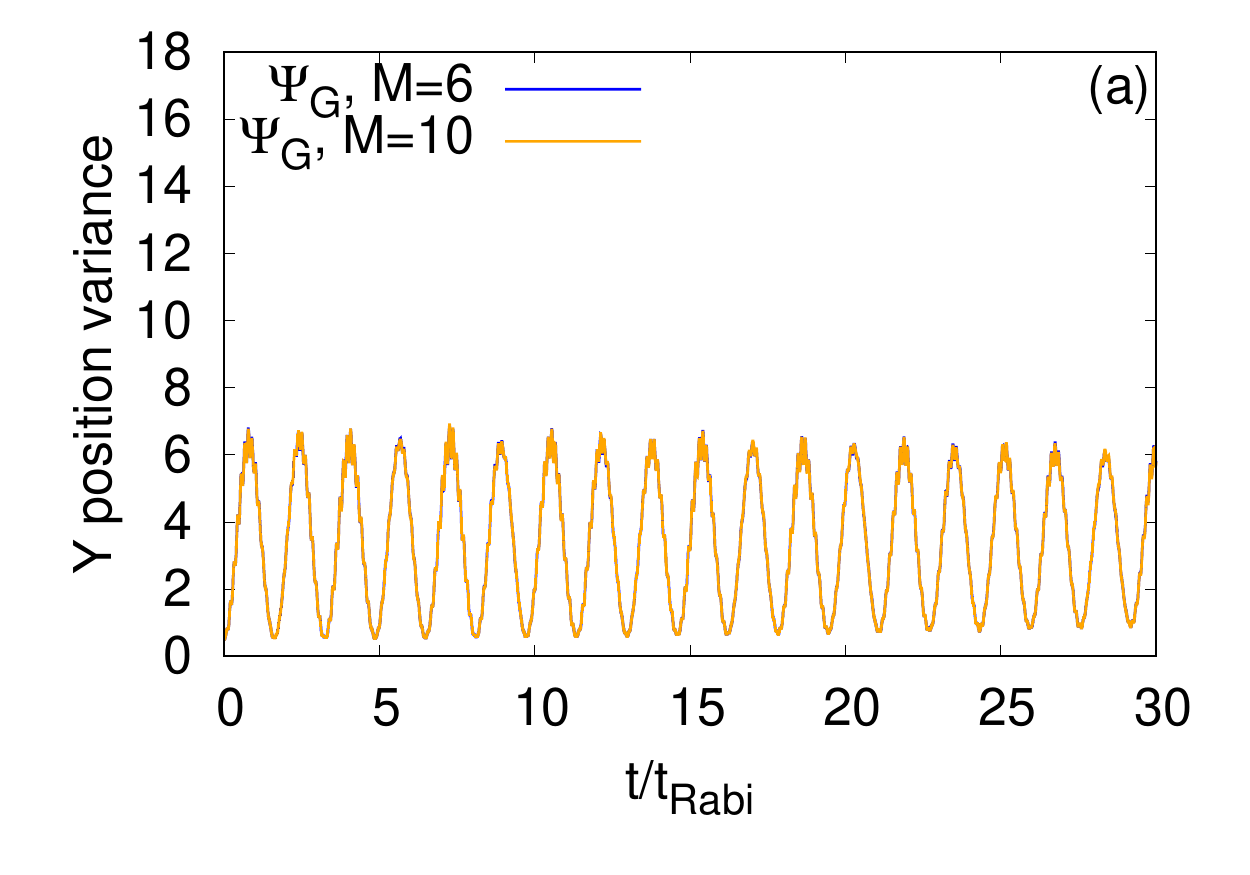}}
{\includegraphics[trim = 0.1cm 0.5cm 0.1cm 0.2cm, scale=.60]{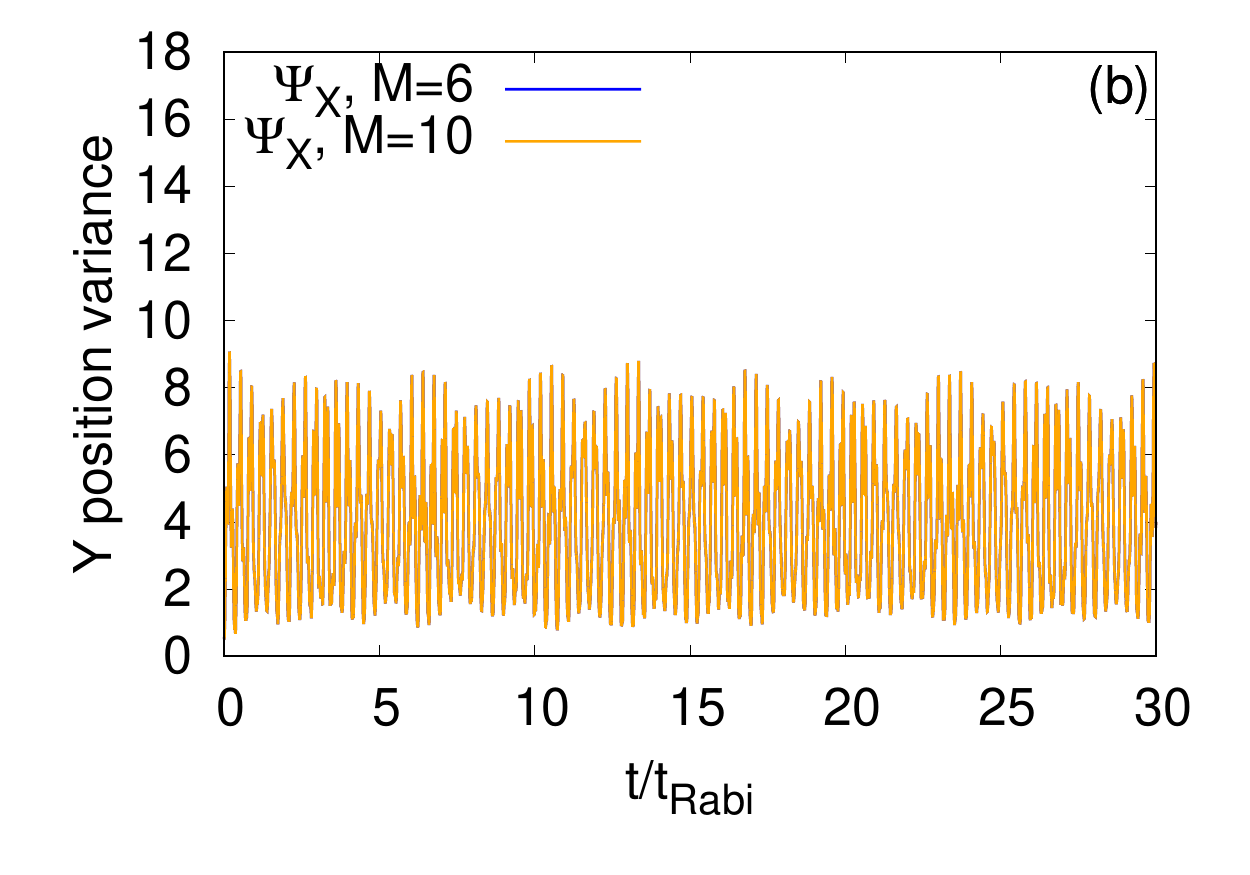}}\\
\vglue 0.25 truecm
{\includegraphics[trim = 0.1cm 0.5cm 0.1cm 0.2cm, scale=.60]{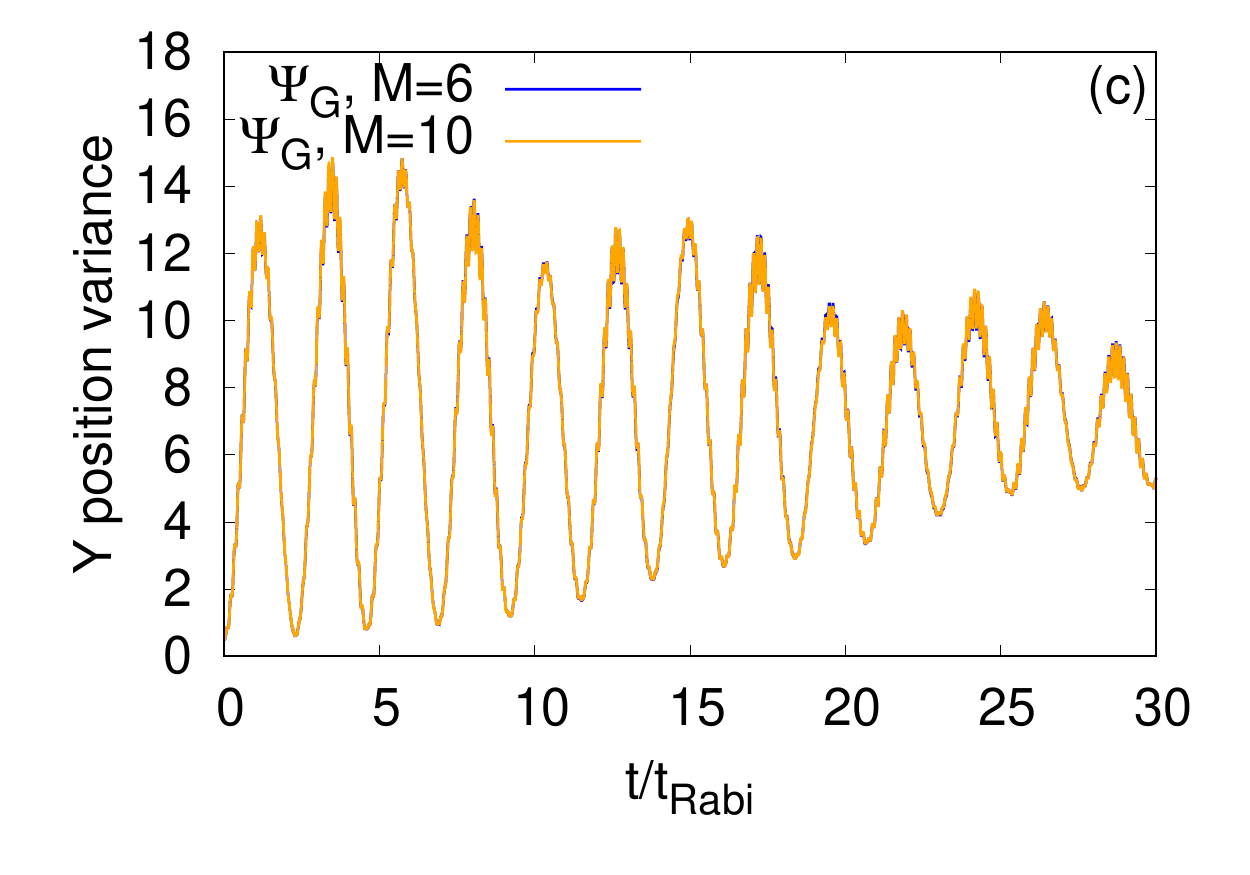}}
{\includegraphics[trim = 0.1cm 0.5cm 0.1cm 0.2cm, scale=.60]{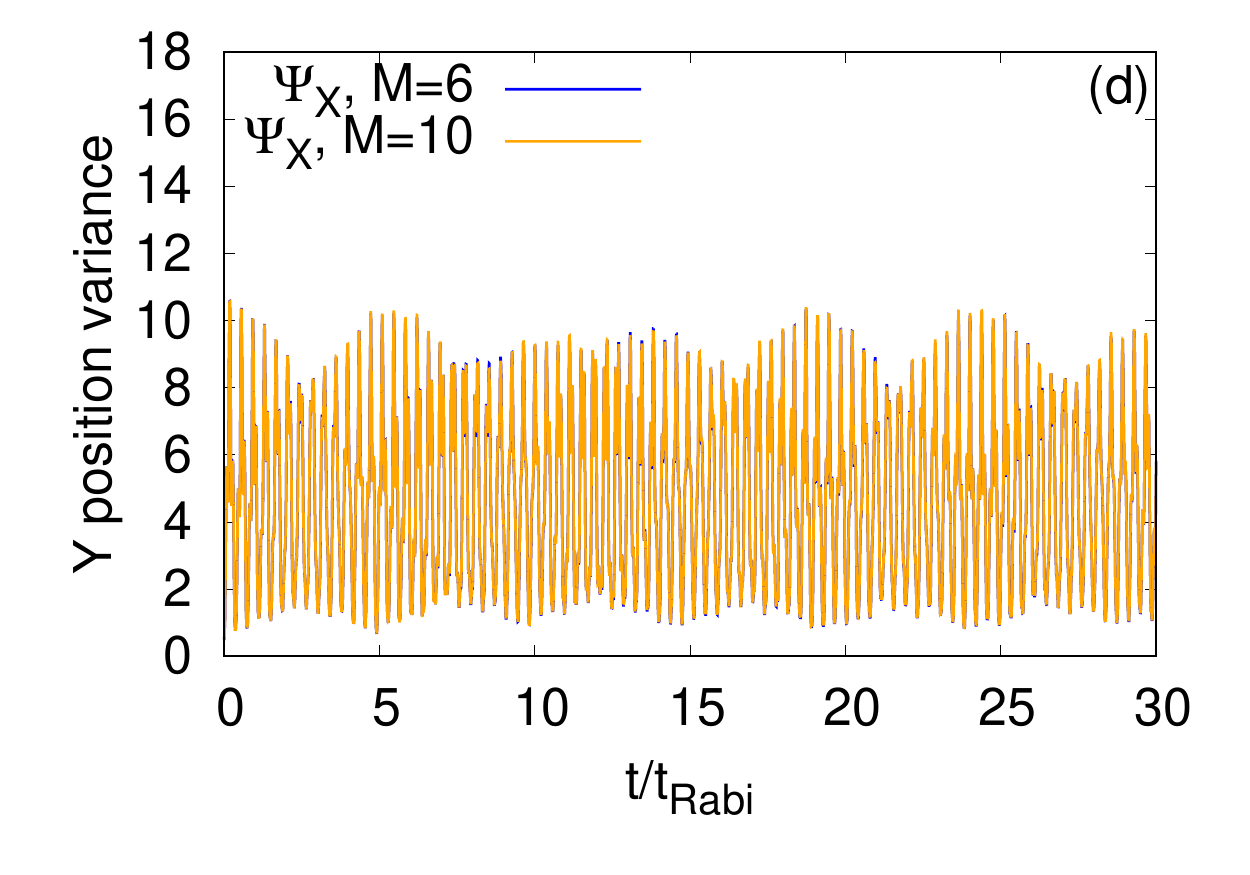}}\\
\vglue 0.25 truecm
{\includegraphics[trim = 0.1cm 0.5cm 0.1cm 0.2cm, scale=.60]{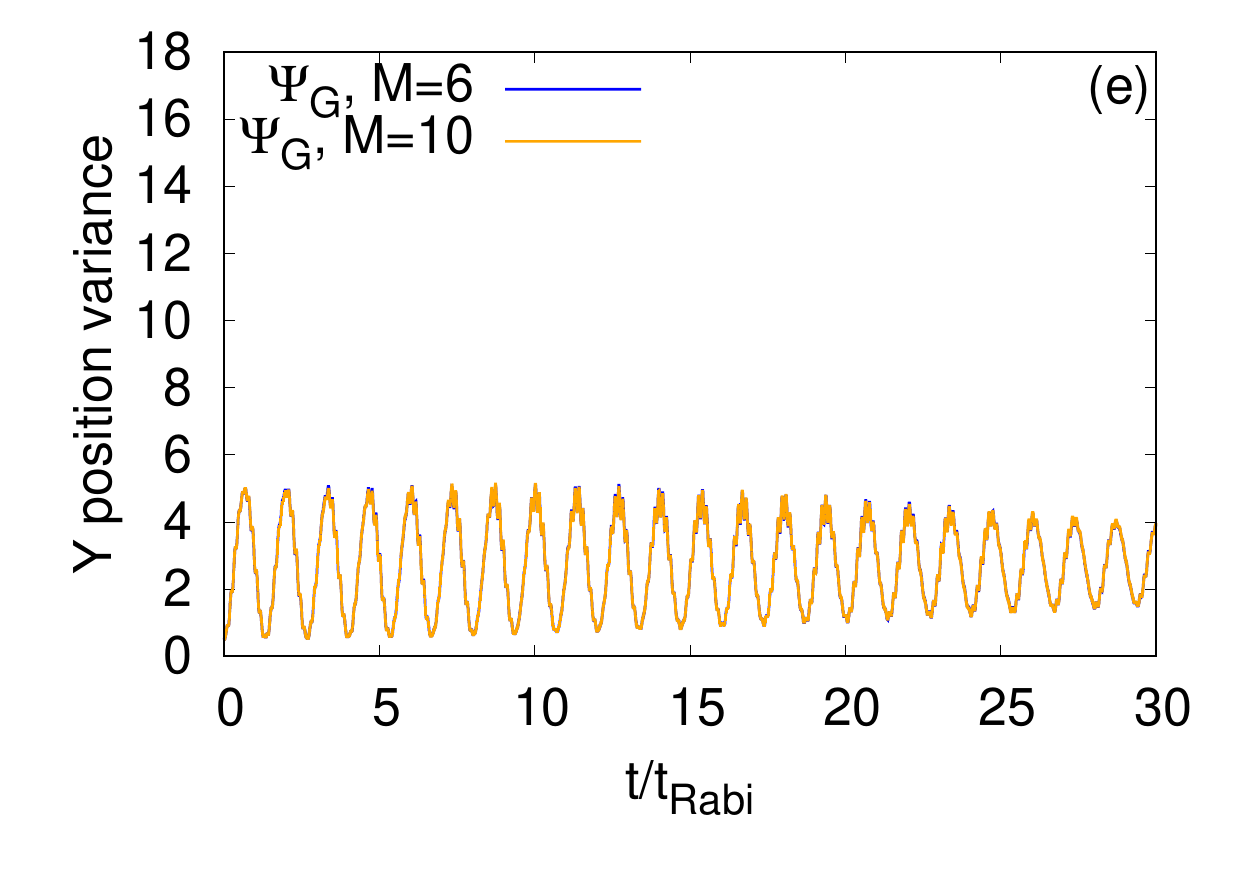}}
{\includegraphics[trim = 0.1cm 0.5cm 0.1cm 0.2cm, scale=.60]{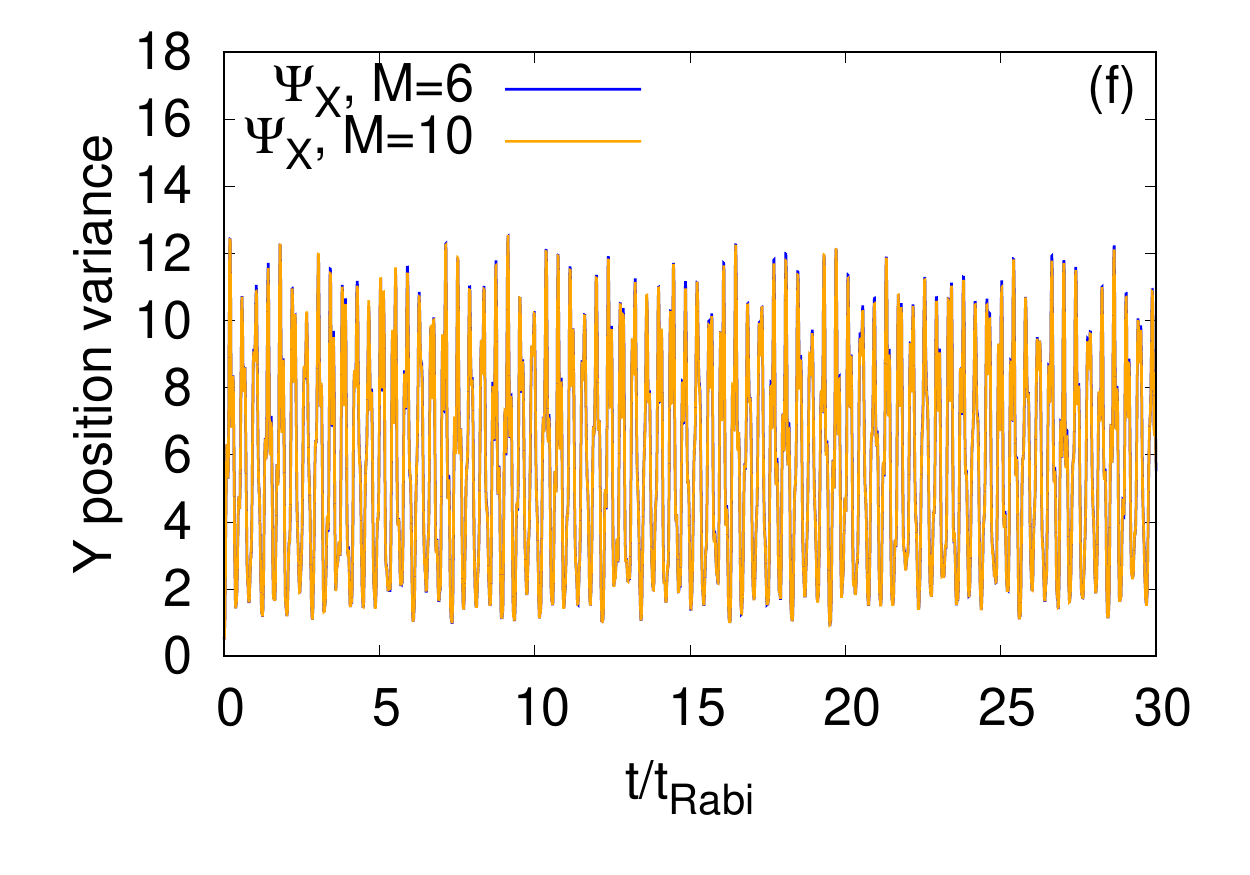}}\\
\caption{\label{figS10}Convergence of the time-dependent many-body position variance per particle along the $y$-direction, $\dfrac{1}{N}\Delta_{\hat{Y}}^2(t)$, with the number of time-dependent orbitals for the initial states $\Psi_G$ (left column) and  $\Psi_X$ (right column) in the transversely-asymmetric 2D double-well potential.  The  bosonic clouds consist of $N=10$ bosons  with the interaction parameter $\Lambda=0.01\pi$.  The results for the frequencies $\omega_n=0.20$, 0.19, and 0.18 are presented row-wise. The many-body results are computed  with $M=6$ time-dependent orbitals. The convergence is verified with $M=10$ time-dependent orbitals.   We show here   dimensionless quantities. Color codes are explained in each panel.}
\end{figure*}

\begin{figure*}[!h]
\centering
{\includegraphics[trim = 0.1cm 0.5cm 0.1cm 0.2cm, scale=.60]{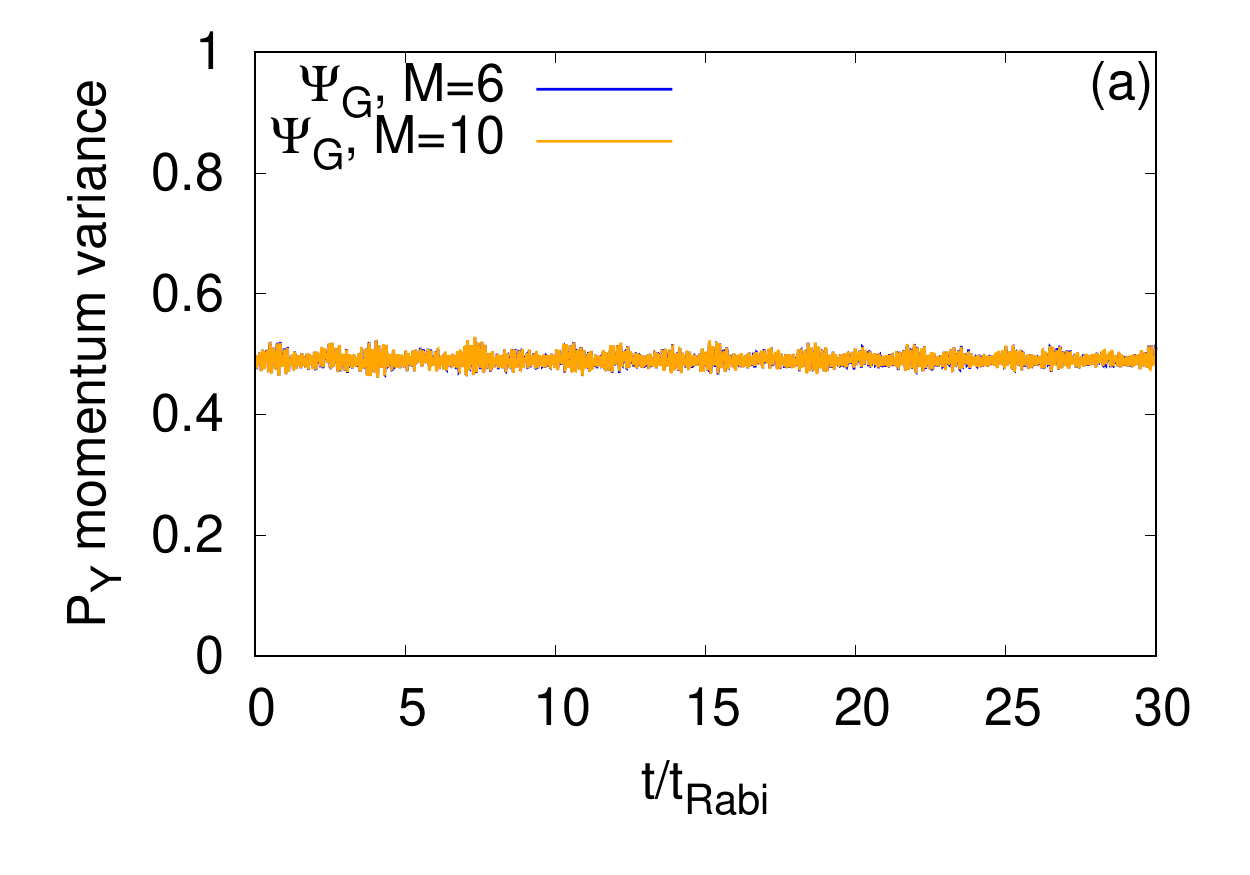}}
{\includegraphics[trim = 0.1cm 0.5cm 0.1cm 0.2cm, scale=.60]{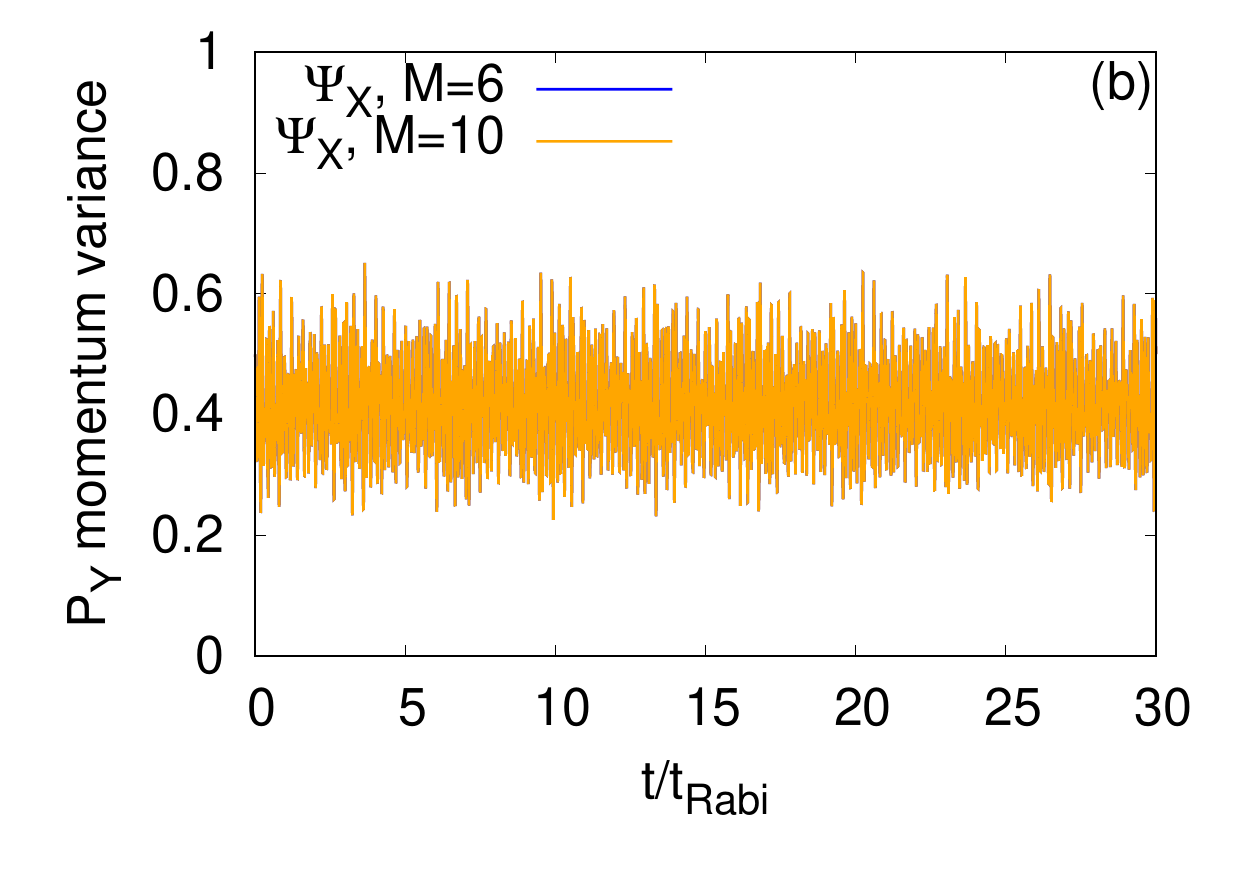}}\\
\vglue 0.25 truecm
{\includegraphics[trim = 0.1cm 0.5cm 0.1cm 0.2cm, scale=.60]{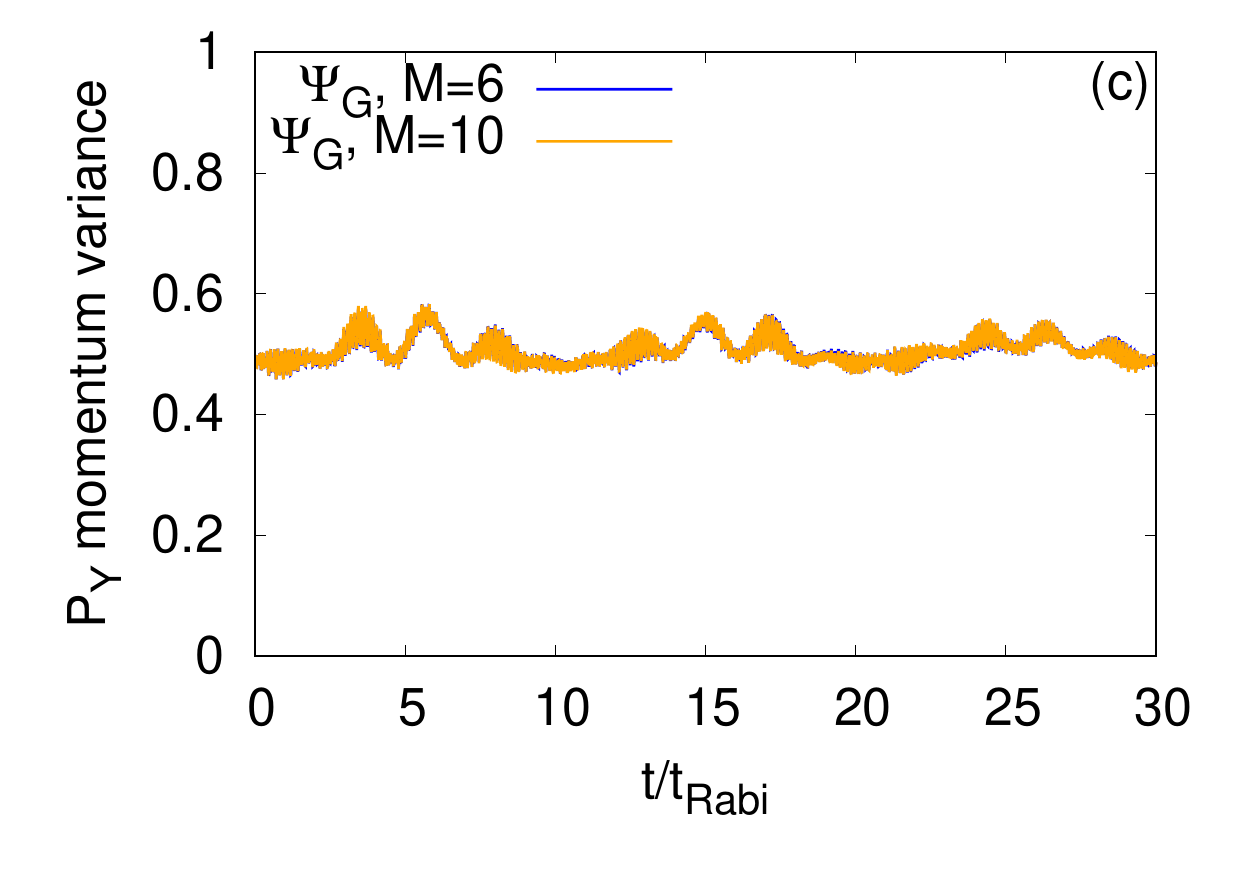}}
{\includegraphics[trim = 0.1cm 0.5cm 0.1cm 0.2cm, scale=.60]{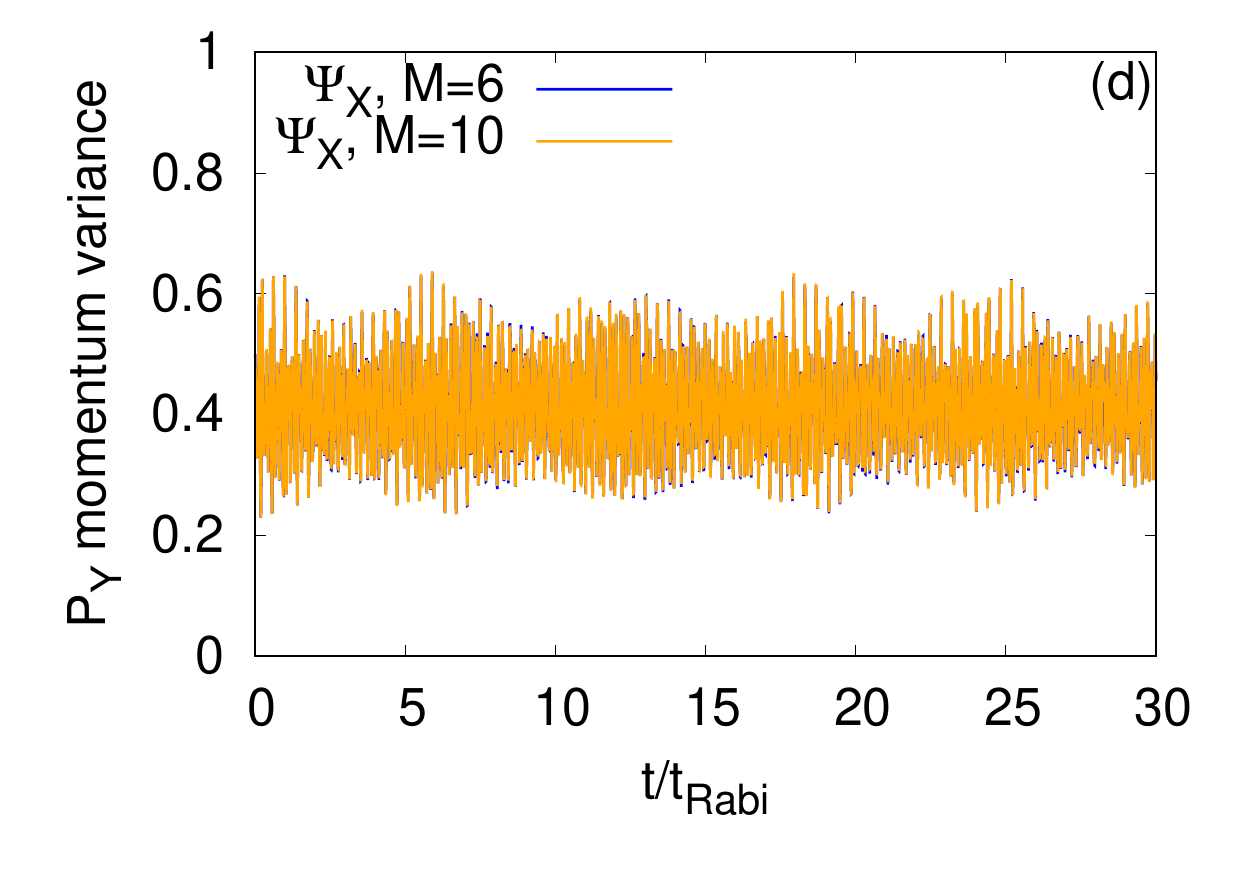}}\\
\vglue 0.25 truecm
{\includegraphics[trim = 0.1cm 0.5cm 0.1cm 0.2cm, scale=.60]{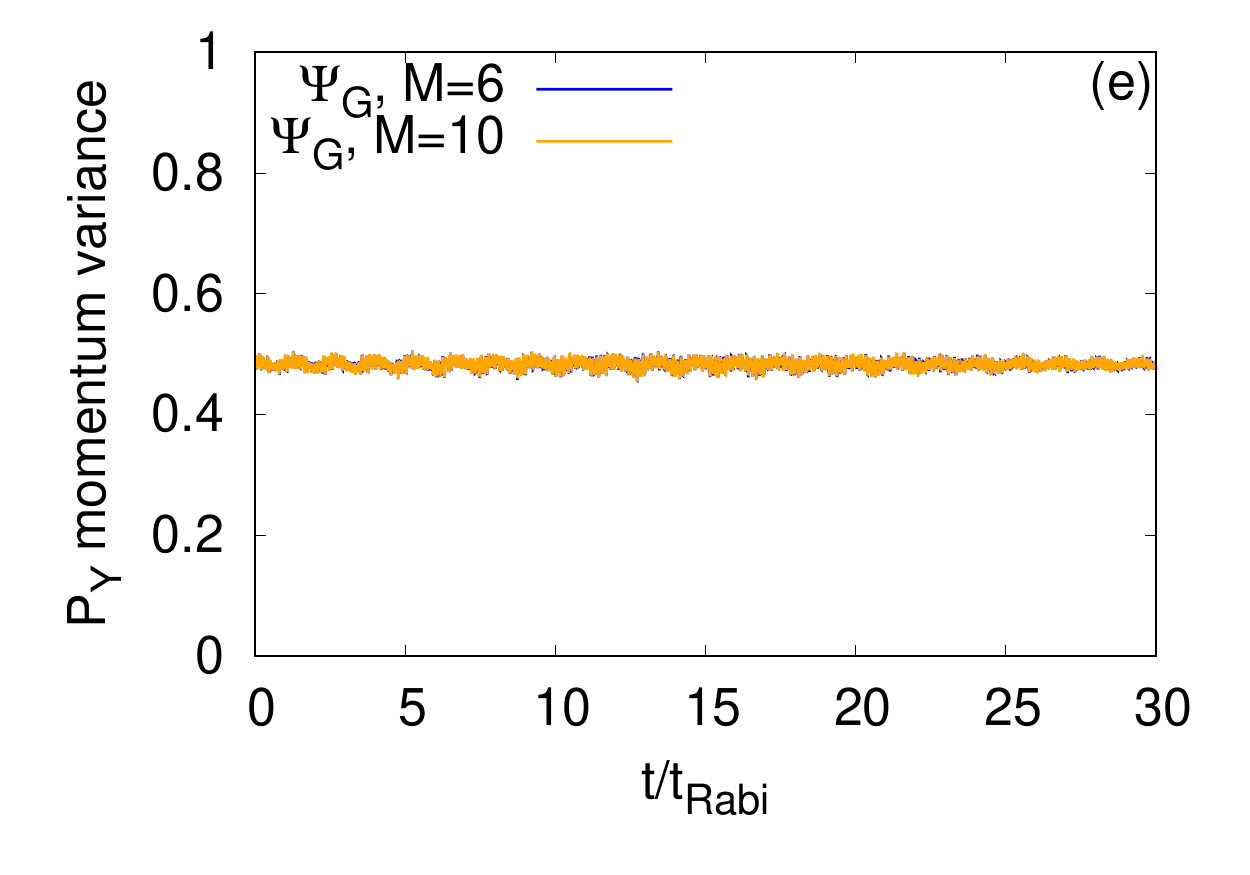}}
{\includegraphics[trim = 0.1cm 0.5cm 0.1cm 0.2cm, scale=.60]{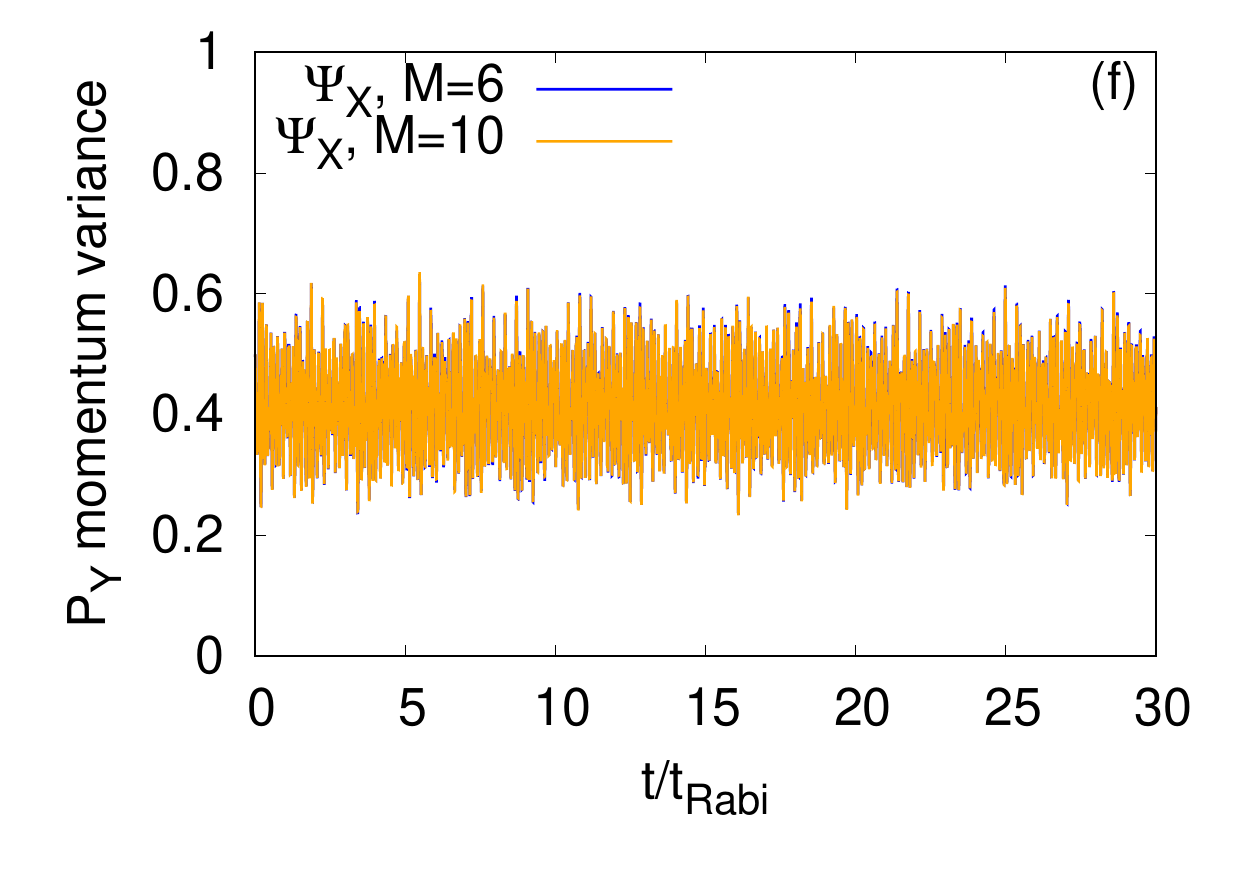}}\\
\caption{\label{figS11}Convergence of the time-dependent many-body momentum variance per particle along the $y$-direction, $\dfrac{1}{N}\Delta_{\hat{P}_Y}^2(t)$, with the number of time-dependent orbitals for the initial states $\Psi_G$ (left column) and  $\Psi_X$ (right column) in the transversely-asymmetric 2D double-well potential.  The  bosonic clouds consist of $N=10$ bosons  with the interaction parameter $\Lambda=0.01\pi$.  The results for the frequencies $\omega_n=0.20$, 0.19, and 0.18 are presented row-wise. The many-body results are computed  with $M=6$ time-dependent orbitals. The convergence is verified with $M=10$ time-dependent orbitals.   We show here   dimensionless quantities. Color codes are explained in each panel.}
\end{figure*}

In Figs.~\ref{figS10} and \ref{figS11}, we display the many-body position and momentum variances per particle along the $y$-direction, $\dfrac{1}{N}\Delta_{\hat{Y}}^2(t)$ and $\dfrac{1}{N}\Delta_{\hat{P}_Y}^2(t)$, respectively, for $\Psi_G$ and $\Psi_X$ at $\omega_n=0.20$, 0.19, and 0.18, along with their numerical convergences with the number of orbitals. For both  states,  $\dfrac{1}{N}\Delta_{\hat{Y}}^2(t)$ and $\dfrac{1}{N}\Delta_{\hat{P}_Y}^2(t)$  computed using $M=6$ and $M=10$ time-dependent orbitals fall on top of each other,  signifying the convergence of the quantities with the number of orbitals. For $\Psi_G$, the amplitude of the many-body time evolution of $\dfrac{1}{N}\Delta_{\hat{Y}}^2(t)$ decays with the growing degree of correlations, see Fig.~\ref{figS10}, and the decay rate is maximal at $\omega_n=0.19$ and minimal at $\omega_n=0.20$. At the resonant value for $\Psi_G$, $\omega_n=0.19$, the amplitude of $\dfrac{1}{N}\Delta_{\hat{Y}}^2(t)$ reaches upto around 15, while at the other frequencies they are comparatively small, i.e., around 4.5 at $\omega_n=0.18$ and around 6.5 at $\omega_n=0.20$. For the slower rate of growth of the fragmentation for $\Psi_X$ in comparison to $\Psi_G$, the decay rate of the many-body $\dfrac{1}{N}\Delta_{\hat{Y}}^2(t)$ is hardly visible even at the resonant value of $\omega_n$, see Fig.~\ref{figS10}. By comparing   the many-body $\dfrac{1}{N}\Delta_{\hat{Y}}^2(t)$ at different frequencies for $\Psi_X$, we observe that at the resonant condition for $\Psi_X$, i.e., $\omega_n=0.18$,  the amplitude of oscillation reaches upto 12 whereas at  $\omega_n=0.19$ and $\omega_n=0.20$, the respective amplitudes reach  upto around 10.5 and 9, respectively. 

Focusing on the momentum variance along the $y$-direction for $\Psi_G$, we notice that at $\omega_n=0.18$ and $\omega_n=0.20$, $\dfrac{1}{N}\Delta_{\hat{P}_Y}^2(t)$ have the amplitudes of  fluctuations  of the order of $10^{-2}$ and at the resonant value the respective fluctuations of amplitude become $10^{-1}$. In case of $\Psi_X$, the amplitude of  fluctuations of $\dfrac{1}{N}\Delta_{\hat{P}_Y}^2(t)$ for $\Psi_X$ is  large compared to $\Psi_G$ at each value of $\omega_n$. We notice that the amplitude of   $\dfrac{1}{N}\Delta_{\hat{P}_Y}^2(t)$ for $\Psi_X$ fluctuates between approximately 0.23 to 0.60  at all three frequencies.

\begin{figure*}[!h]
\centering
{\includegraphics[trim = 0.1cm 0.5cm 0.1cm 0.2cm, scale=.60]{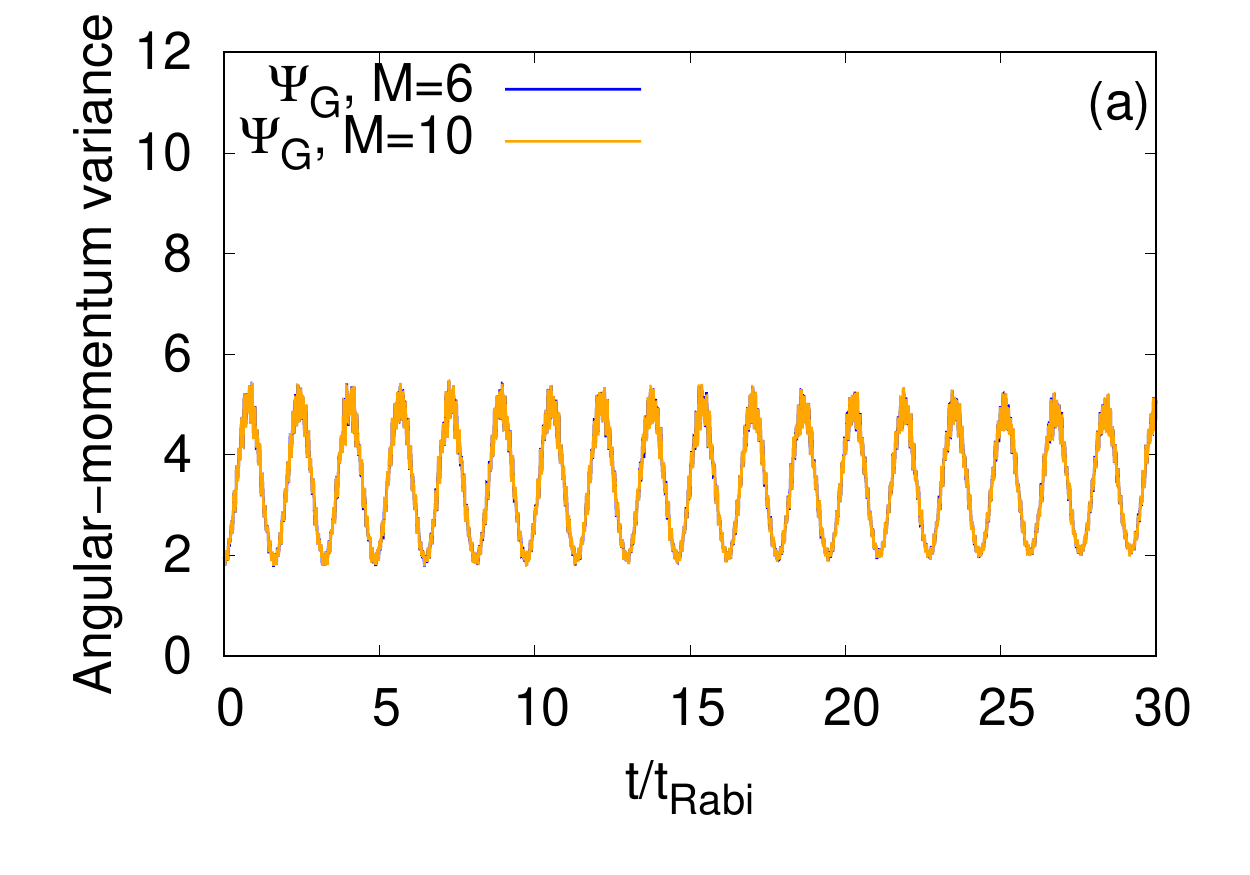}}
{\includegraphics[trim = 0.1cm 0.5cm 0.1cm 0.2cm, scale=.60]{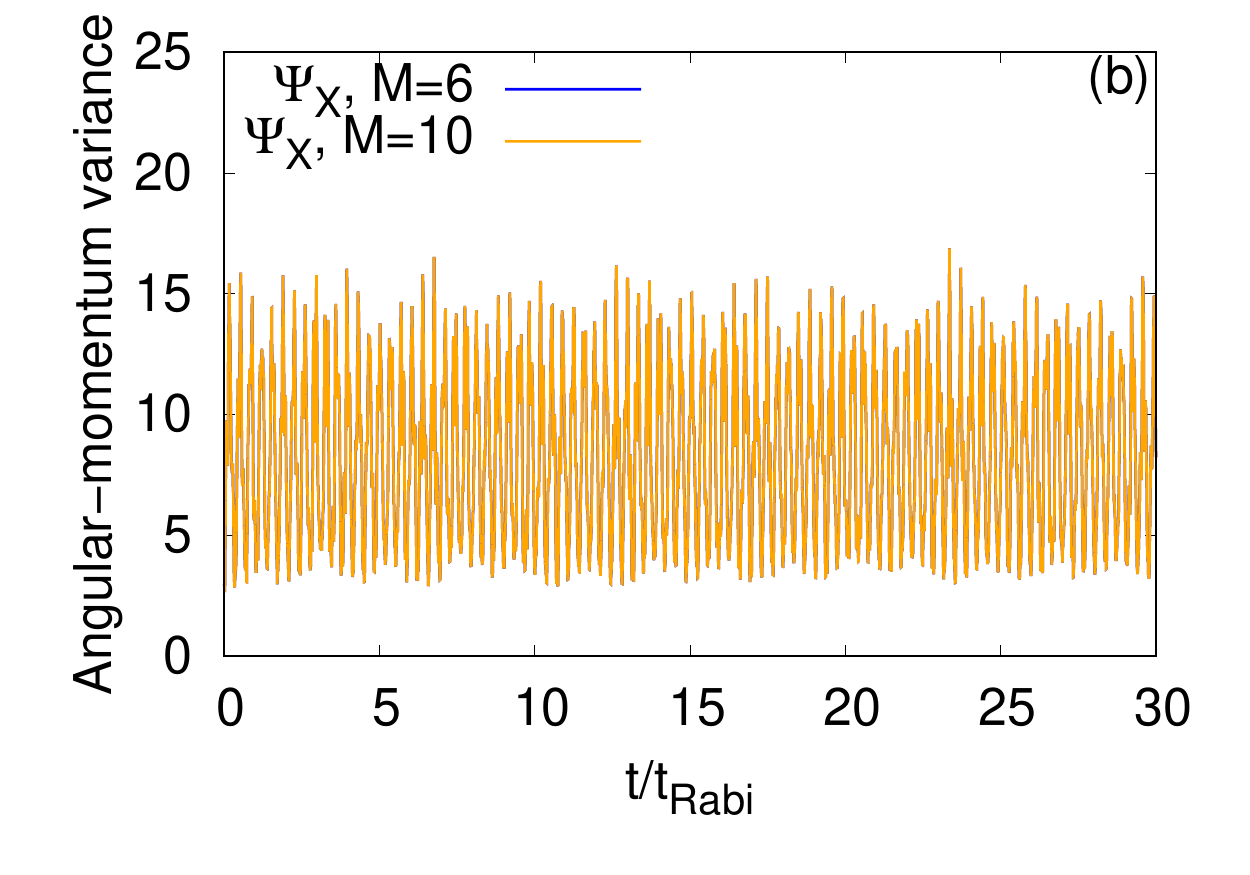}}\\
\vglue 0.25 truecm
{\includegraphics[trim = 0.1cm 0.5cm 0.1cm 0.2cm, scale=.60]{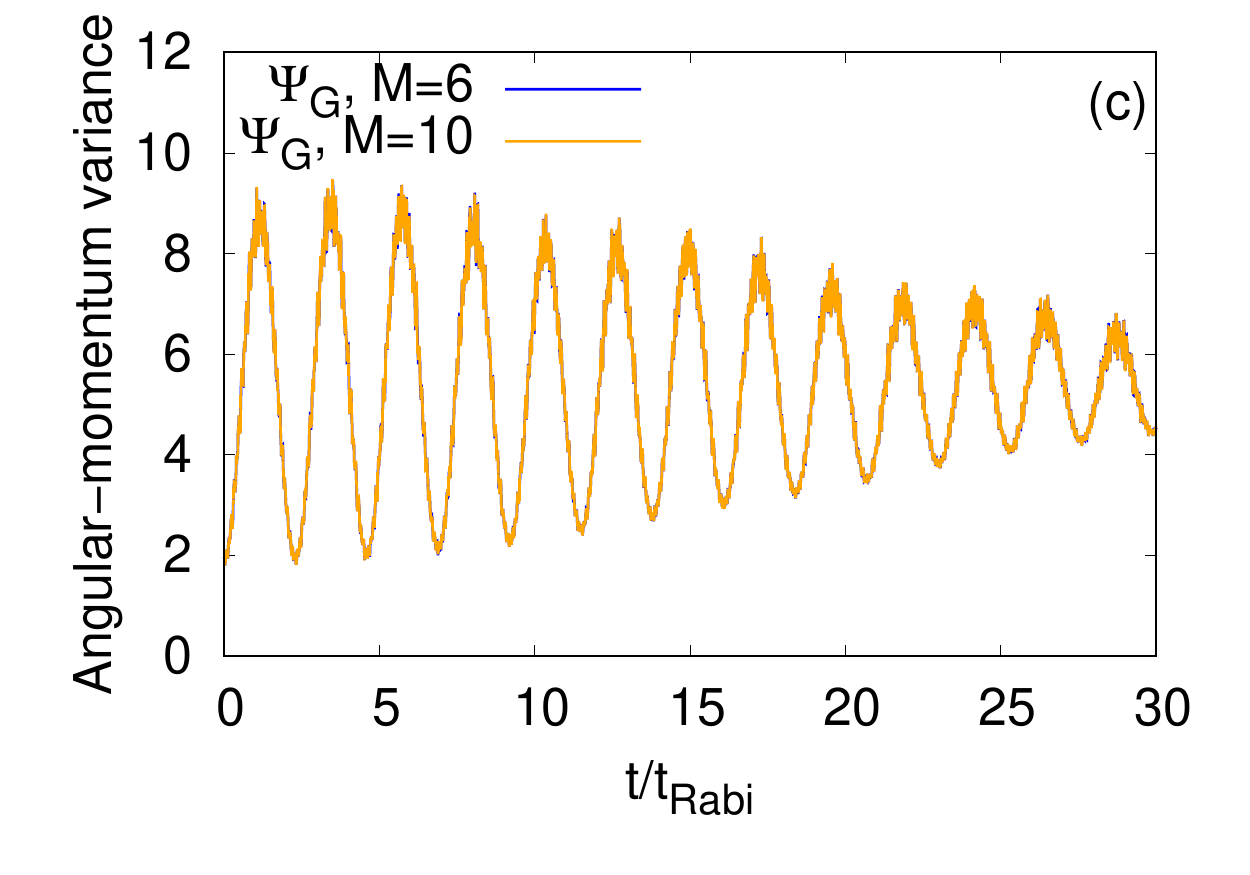}}
{\includegraphics[trim = 0.1cm 0.5cm 0.1cm 0.2cm, scale=.60]{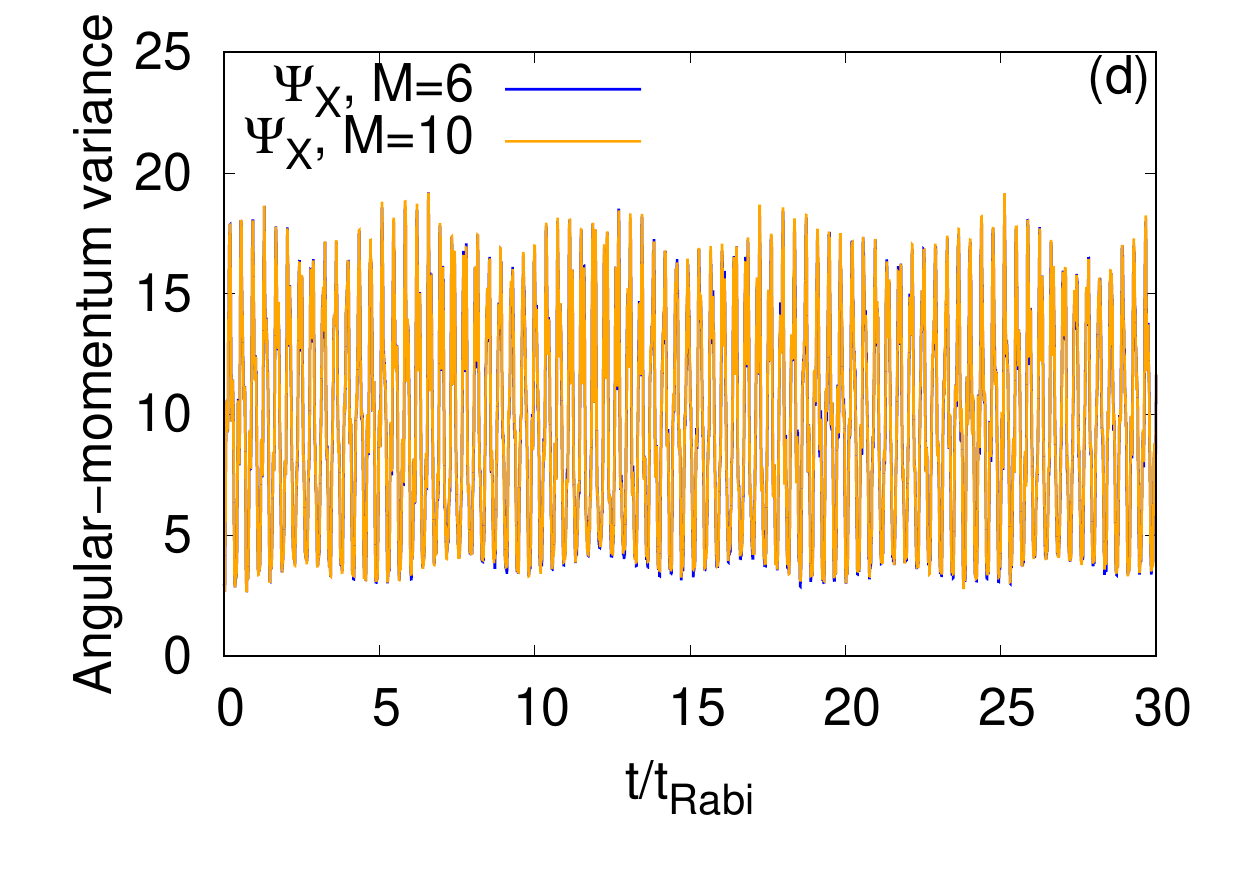}}\\
\vglue 0.25 truecm
{\includegraphics[trim = 0.1cm 0.5cm 0.1cm 0.2cm, scale=.60]{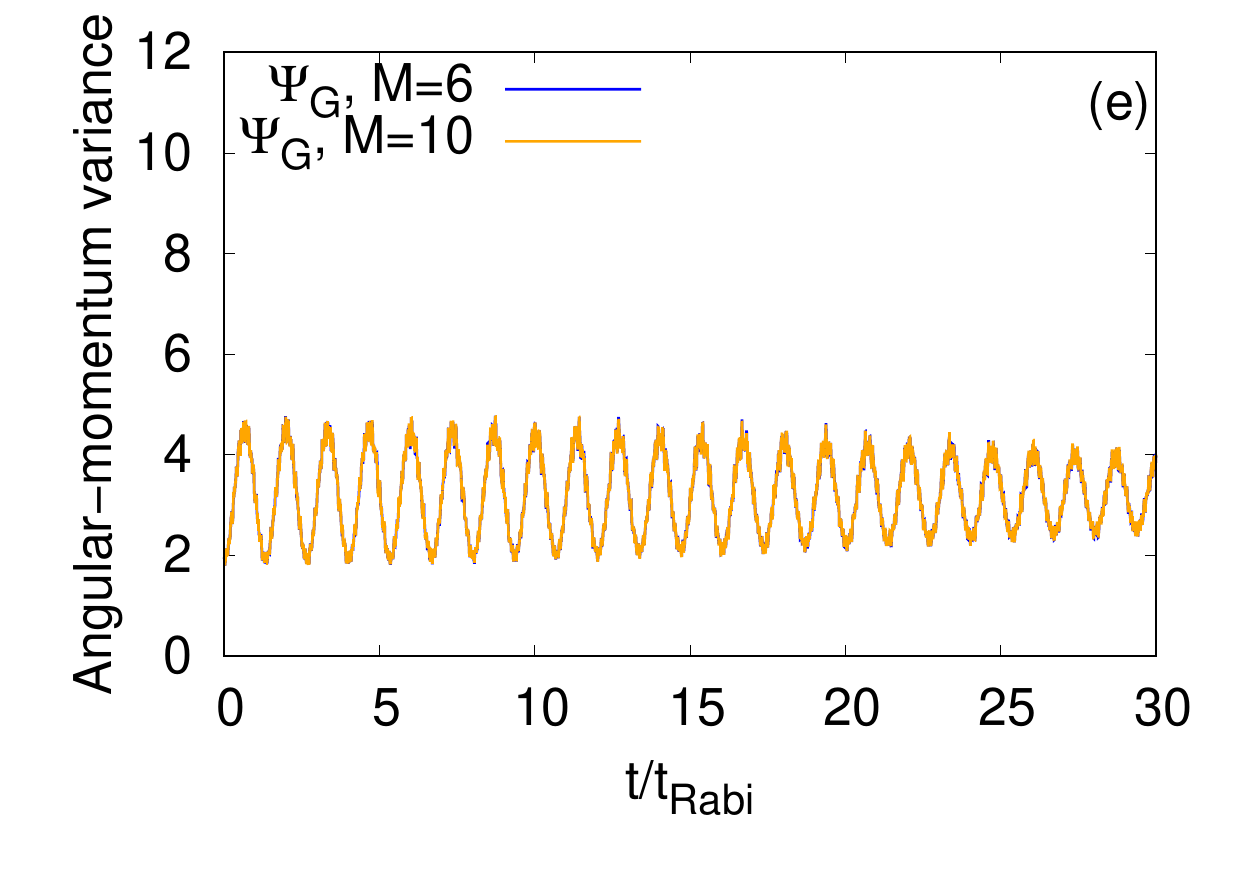}}
{\includegraphics[trim = 0.1cm 0.5cm 0.1cm 0.2cm, scale=.60]{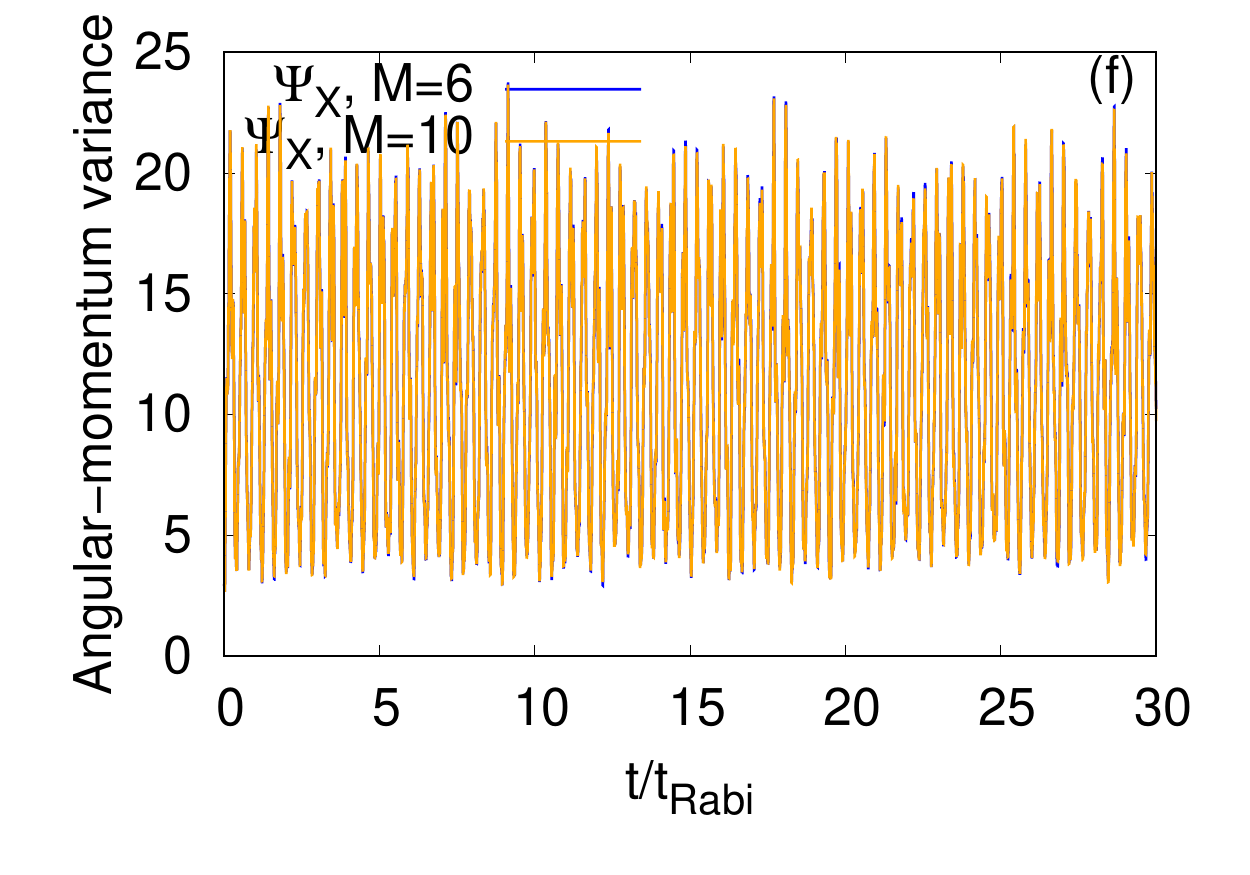}}\\
\caption{\label{figS12}Convergence of the time-dependent variance per particle of the $z$- component of the angular-momentum operator, $\dfrac{1}{N}\Delta_{\hat{L}_Z}^2(t)$, with the number of time-dependent orbitals for the initial states $\Psi_G$ (left column) and  $\Psi_X$ (right column) in the transversely-asymmetric 2D double-well potential.  The  bosonic clouds consist of $N=10$ bosons  with the interaction parameter $\Lambda=0.01\pi$.  The results for the frequencies $\omega_n=0.20$, 0.19, and 0.18 are presented row-wise. The many-body results are computed  with $M=6$ time-dependent orbitals. The convergence is verified with $M=10$ time-dependent orbitals.   We show here   dimensionless quantities. Color codes are explained in each panel.}
\end{figure*}

Fig.~\ref{figS12} collects the numerical convergence of the angular-momentum variance per particle, $\dfrac{1}{N}\Delta_{\hat{L}_Z}^2(t)$, for $\Psi_G$ and $\Psi_X$ at $\omega_n=0.20$, 0.19, and 0.18. As discussed for $\dfrac{1}{N}\Delta_{\hat{Y}}^2(t)$,  here also the amplitude of  $\dfrac{1}{N}\Delta_{\hat{L}_Z}^2(t)$ decays with time for $\Psi_G$.  The decay rate is the slowest at $\omega_n=0.20$ and fastest at $\omega_n=0.19$. But for $\Psi_X$,  $\dfrac{1}{N}\Delta_{\hat{L}_Z}^2(t)$ does not follow the trend of  a  decaying amplitude as for $\Psi_G$. In general, we observe that at the resonant value of $\omega_n$, $\dfrac{1}{N}\Delta_{\hat{L}_Z}^2(t)$ takes maximum value compared to other $\omega_n$ frequencies considered here. Moreover, comparing the many-body time evolution of $\dfrac{1}{N}\Delta_{\hat{L}_Z}^2(t)$, computed using $M=6$ and 10 orbitals, signifies that the results are fully converged with $M=6$ orbitals.

\begin{figure*}[!h]
\centering
{\includegraphics[trim = 0.1cm 0.5cm 0.1cm 0.2cm, scale=.60]{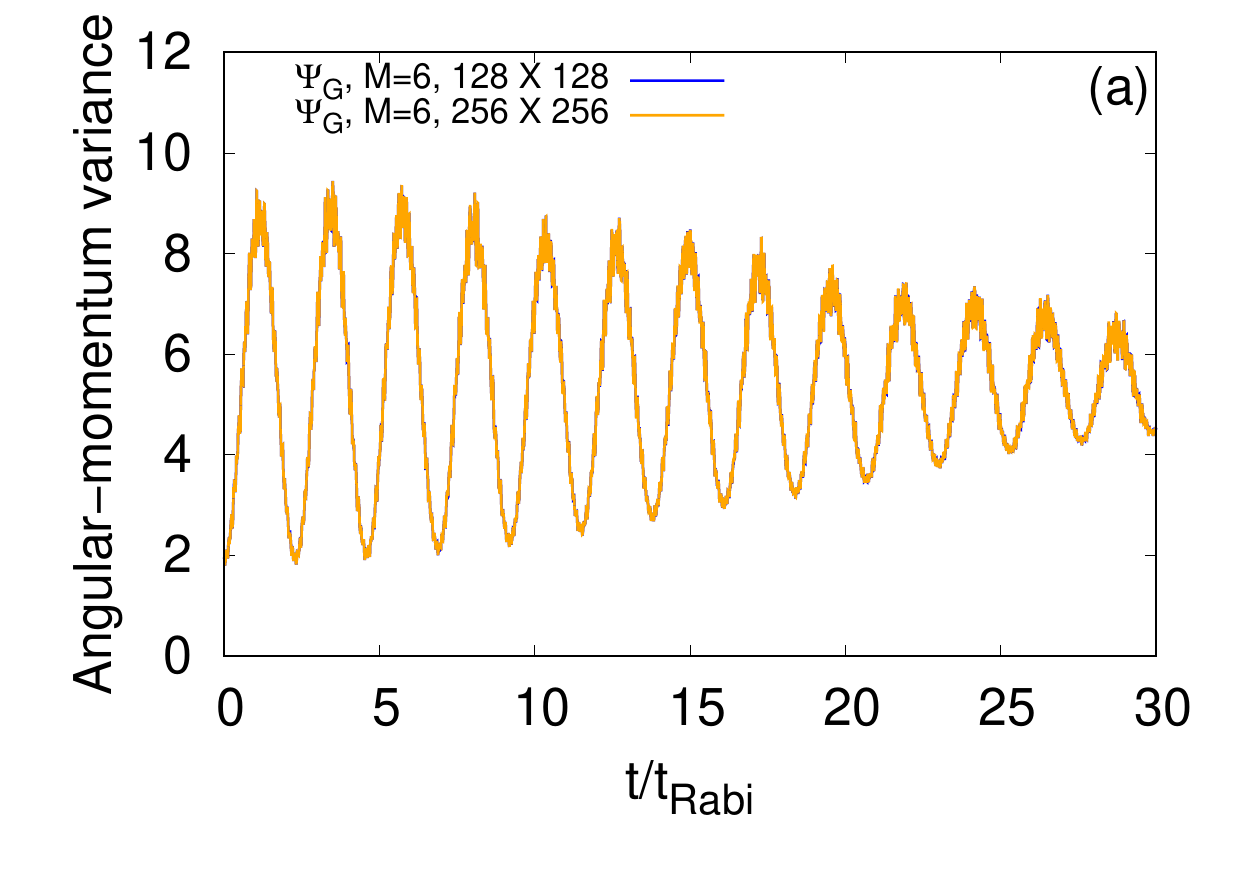}}
{\includegraphics[trim = 0.1cm 0.5cm 0.1cm 0.2cm, scale=.60]{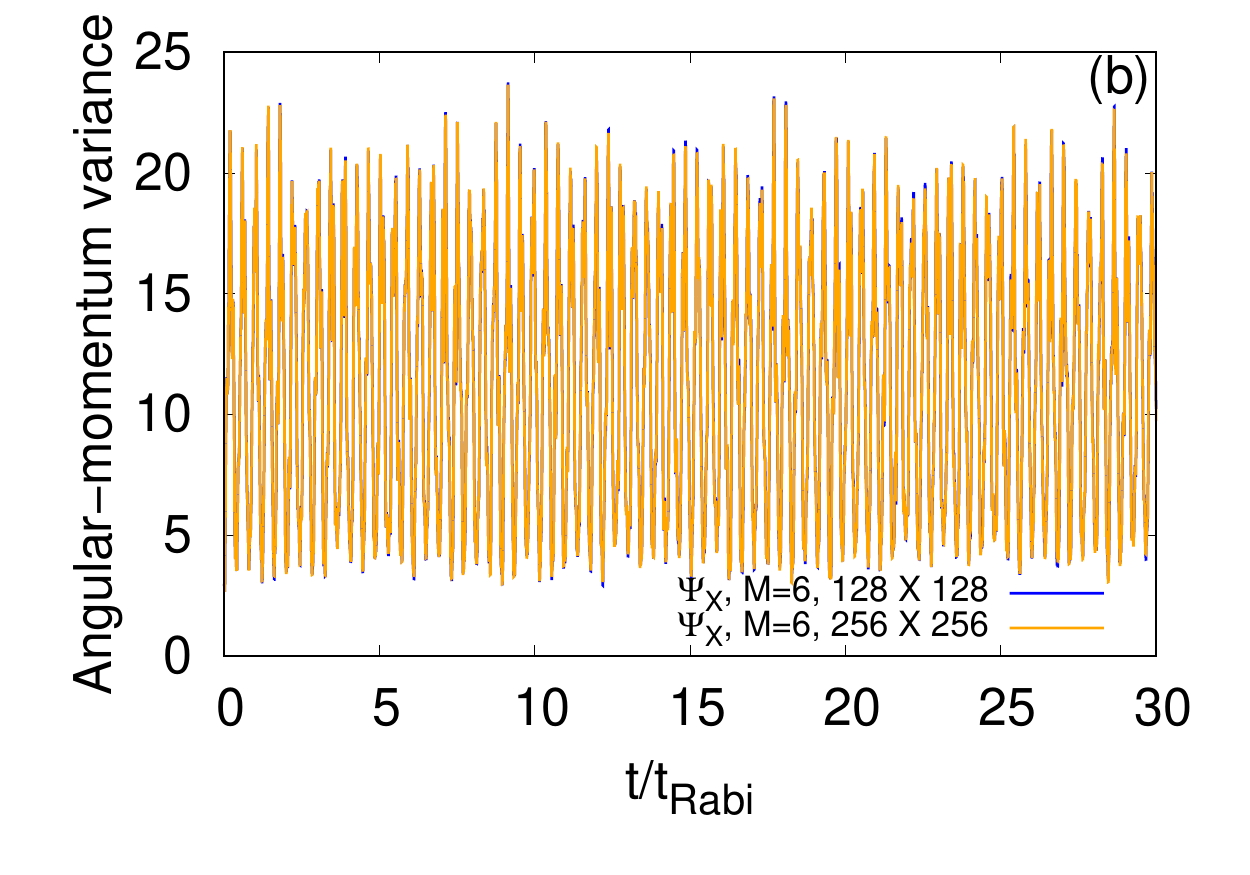}}\\
\caption{\label{figS13}Convergence of the time-dependent variance per particle of the $z$- component of the angular-momentum operator, $\dfrac{1}{N}\Delta_{\hat{L}_Z}^2(t)$, with the he number of grid points.   Initial states are (a) $\Psi_G$ ($\omega_n=0.19$) and (b) $\Psi_X$ ($\omega_n=0.18$) in transversely-asymmetric 2D double-well potential.  The  bosonic clouds consist of $N=10$ bosons  with the interaction parameter $\Lambda=0.01\pi$.    The convergence is demonstrated with $128\times 128$ and $256\times 256$ grid points.    We show here   dimensionless quantities. Color codes are explained in each panel.}
\end{figure*}

The main text and  the supplemental material so far describe  all quantities, i.e., the survival probability, loss of coherence, fragmentation,  and the variances for the transversal resonant scenario  with $128\times 128$ grid points.  In order to verify the convergence with the grid points, we repeat our computation with $256\times 256$ grid points for both objects, $\Psi_G$ and $\Psi_X$ and of course the computations become demanding due to the increased grid density.   To demonstrate the convergence with the grid points, we display $\dfrac{1}{N}\Delta_{\hat{L}_Z}^2(t)$  which is the most sensitive quantity discussed in this work.  Fig.~\ref{figS13} displays the convergence of the many-body $\dfrac{1}{N}\Delta_{\hat{L}_Z}^2(t)$ at $\omega_n=0.19$ for $\Psi_G$ and $\omega_n=0.18$ for $\Psi_X$, as there is no visible effect with increasing the  number of grid points. Of course all other quantities are similarly converged.

\clearpage

%) ============================================================================
% === REFERENCES =============================================================
% ============================================================================

%\bibliographystyle{abbrv}

%\bibliography{bib}\

\end{document}